%% file: vortices-in-dense-QCD-v10-arxiv.tex
\documentclass[preprint]{ptephy}

\preprintnumber{XXXX-XXXX}

\usepackage{bm,amsmath,amssymb,enumerate}

\include{definitions-v7}

\begin{document}

\title{Vortices and Other Topological Solitons\\ 
in Dense Quark Matter}

\author{
\name{Minoru Eto}{1}, 
\name{Yuji Hirono}{2,3,4}, 
\name{Muneto Nitta}{5}\footnote{corresponding author}, and 
\name{Shigehiro Yasui}{6}
\thanks{These authors contributed equally to this work}}

\address{
\affil{1}{Department of Physics, Yamagata University, Kojirakawa 1-4-12, Yamagata, Yamagata 990-8560, Japan
}
\affil{2}{Department of Physics, University of Tokyo, 
Hongo~7-3-1, Bunkyo-ku, Tokyo 113-0033, Japan}
\affil{3}{Theoretical Research Division, Nishina Center, RIKEN, Wako 351-0198, Japan
}
\affil{4}{Department of Physics, Sophia University, Tokyo 102-8554, Japan
}
\affil{5}{Department of Physics, and Research and Education Center for Natural Sciences, Keio University, Hiyoshi 4-1-1, Yokohama, Kanagawa 223-8521, Japan}
\affil{6}{KEK Theory Center, Institute of Particle 
and Nuclear Studies,
High Energy Accelerator Research Organization (KEK),
1-1 Oho, Tsukuba, Ibaraki 305-0801, Japan}
\email{nitta@phys-h.keio.ac.jp}
}

\begin{abstract}

Dense quantum chromodynamic matter accommodates various kind of topological solitons such
 as vortices, domain walls, monopoles, kinks, boojums and so on.
In this review, we discuss various properties of topological solitons in
 dense quantum chromodynamics (QCD) and their phenomenological implications.
A particular emphasis is placed on the topological solitons in the color-flavor-locked (CFL) phase,
which exhibits both superfluidity and superconductivity.
The properties of topological solitons are discussed in terms of  
effective field theories such as 
the Ginzburg-Landau theory, the chiral Lagrangian,  
or the Bogoliubov--de Gennes equation.
The most fundamental 
string-like topological excitations in the CFL phase 
are the non-Abelian vortices,  
which are 1/3 quantized superfluid vortices 
and color magnetic flux tubes. 
These vortices are created 
at a phase transition by the Kibble-Zurek mechanism or
when 
the CFL phase is realized in 
compact stars, which rotate rapidly.
The interaction between vortices 
is found to be 
repulsive and consequently 
a vortex lattice is formed in rotating CFL matter. 
Bosonic and fermionic zero-energy modes 
are trapped in the core of a non-Abelian vortex 
and propagate along it as gapless excitations. 
The former 
consists of translational zero modes (a Kelvin mode) 
with a quadratic dispersion and
${\mathbb C}P^2$ Nambu-Goldstone gapless modes 
with a linear dispersion,  
associated with the CFL symmetry 
spontaneously broken in the core of a vortex, 
while the latter is Majorana fermion zero modes
belonging to the triplet of the symmetry 
remaining in the core of a vortex. 
The low-energy effective theory of 
the bosonic zero modes is constructed as 
a non-relativistic free complex scalar field
and a relativistic ${\mathbb C}P^2$ model 
in 1+1 dimensions. 
The effects of strange quark mass, electromagnetic interactions 
and non-perturbative quantum corrections 
are taken into account in the ${\mathbb C}P^2$ effective theory.
Various topological objects associated with 
non-Abelian vortices are studied; 
colorful boojums at the CFL interface, 
the quantum color magnetic monopole confined by vortices, 
which 
supports the notion of 
quark-hadron duality,  
and 
Yang-Mills instantons inside a non-Abelian vortex 
as lumps are discussed. 
The interactions between a non-Abelian vortex and quasi-particles 
such as phonons, gluons, mesons, and photons are studied. 
As a consequence of the interaction with photons, 
a vortex lattice behaves as a cosmic polarizer. 
As a remarkable consequence of Majorana fermion zero modes, 
non-Abelian vortices 
are shown to behave as a novel kind of non-Abelian anyon.
In the order parameters of 
chiral symmetry breaking, we discuss fractional and integer 
axial domain walls, Abelian and non-Abelian axial vortices, 
axial wall-vortex composites, and Skyrmions. 
\end{abstract}

\vspace{-2cm}

\subjectindex{xxxx, xxx}

\maketitle

\newpage
\tableofcontents
\newpage

\include{intro-v9}
\include{gl-v9}

\include{vortices-v9}

\include{dynamics-v9}

\include{eff-th-v9}

\include{interaction-v9}

\include{boojum-v9}
\include{fermion-v9}

\include{na-statistics-v9}
\include{global-v9}
\include{other-v9}

\include{conclusion-v9}

\begin{appendix}
\include{susy-v9}
\include{toric-v9}
\include{eff_act_appendix-v9}
\include{dual-v9}

\include{appendix-fermion-v9}
\end{appendix}

\newpage

\newpage
\bibliographystyle{ptephy}
\bibliography{dense-QCD-v2}

\vfill\pagebreak

\end{document}

%% file: definitions-v7.tex
\makeatletter
\@addtoreset{equation}{section}

\makeatother

\def\ee{\end{eqnarray}}
\def\Lag{\mathcal{L}}
\def\p{\partial}

\def\D{\mathcal{D}}

\def\=:{=\hspace{-.7em}\raisebox{1.1ex}{.}\hspace{.1em}\raisebox{-0.2ex}{.} }

\def\ee{\end{eqnarray}}
\def\Lag{\mathcal{L}}

\def\p{\partial}

\def\D{\mathcal{D}}

\def\=:{=\hspace{-.7em}\raisebox{1.1ex}{.}\hspace{.1em}\raisebox{-0.2ex}{.} }

\newcommand {\beq}{\begin{eqnarray}}
\newcommand {\eeq}{\end{eqnarray}}
\newcommand {\non}{\nonumber\\}

\newcommand {\1}[1]{\frac{1}{#1}}

\newcommand {\ph}{\varphi}

\newcommand {\Ph}{\Phi}

\newcommand {\Sig}{\Sigma}

\newcommand {\del}{\partial}

\newcommand {\tr}{{\rm Tr}\,}

\newcommand{\hs}[1]{\hspace{#1 mm}}

\newcommand{\SU}{SU}

\newcommand{\U}{U}
\def\Tr{{\rm Tr}}
\def\Lag{\mathcal{L}}

\newcommand\B{{\rm B}}

\newcommand{\diff}{\mathrm{d}}

\newcommand{\Slash}[1]{{\ooalign{\hfil/\hfil\crcr$#1$}}}
\newcommand\+{\dagger}  
\newcommand\diag{\operatorname{diag}}
\def\Tr{\mathop{\mathrm{Tr}}}

\newcommand\Lcal{\mathcal{L}}

\newcommand\Dcal{\mathcal{D}}
\newcommand\e{\mathrm{e}}

%% file: intro-v9.tex
\section{Introduction}\label{sec:intro}

Topological solitons are a subject of considerable interest in
condensed matter physics \cite{Mermin:1979zz}. 
Their properties 
have been studied extensively and it has been found that they play quite important roles phenomenologically.
One such example is in superfluidity. 
Superfluidity 
emerges in a wide variety of physical systems such 
as helium superfluids \cite{volovik2009universe}
or ultracold atomic gases
\cite{pethick2002bose,pitaevskii2003bose,ueda2010fundamentals}.
Superfluids are known to accommodate {\it quantized vortices} as
topological solitons, which are 
important degrees of freedom to investigate
the dynamics of superfluids 
\cite{donnelly1991quantized,volovik2009universe,
Tsubota:2010,Tsubota2013191}.
The observation of quantized vortices has worked as the evidence of superfluidity for 
ultracold atomic gases 
such as Bose-Einstein condensates (BECs) 
\cite{Abo-Shaeer:2001} and Fermi gases in the 
BEC/Bardeen-Cooper-Schrieffer(BCS) crossover 
regime \cite{Zwierlein:2005}.
A rotating superfluid is threaded with numerous vortices and they form a vortex lattice. 
Vortices are also created at phase transitions
by the Kibble-Zurek mechanism
\cite{Kibble:1976sj,Hindmarsh:1994re,Zurek:1985qw,Zurek:1996sj}. 
Superfluid vortices also play pivotal roles in quantum turbulence 
in superfluid helium and atomic BECs 
\cite{Vinen:2007,0953-8984-21-16-164207,Tsubota2013191,
doi:10.1146/annurev-conmatphys-062910-140533}. 
In lower dimensions, 
vortices are essential in Berezinskii-Kosterlitz-Thouless (BKT) transitions \cite{Berezinskii1971,Berezinskii1972,Kosterlitz:1973xp}.

\begin{figure}[ht]
\begin{tabular}{cc}
\centering
\includegraphics[width=6cm]{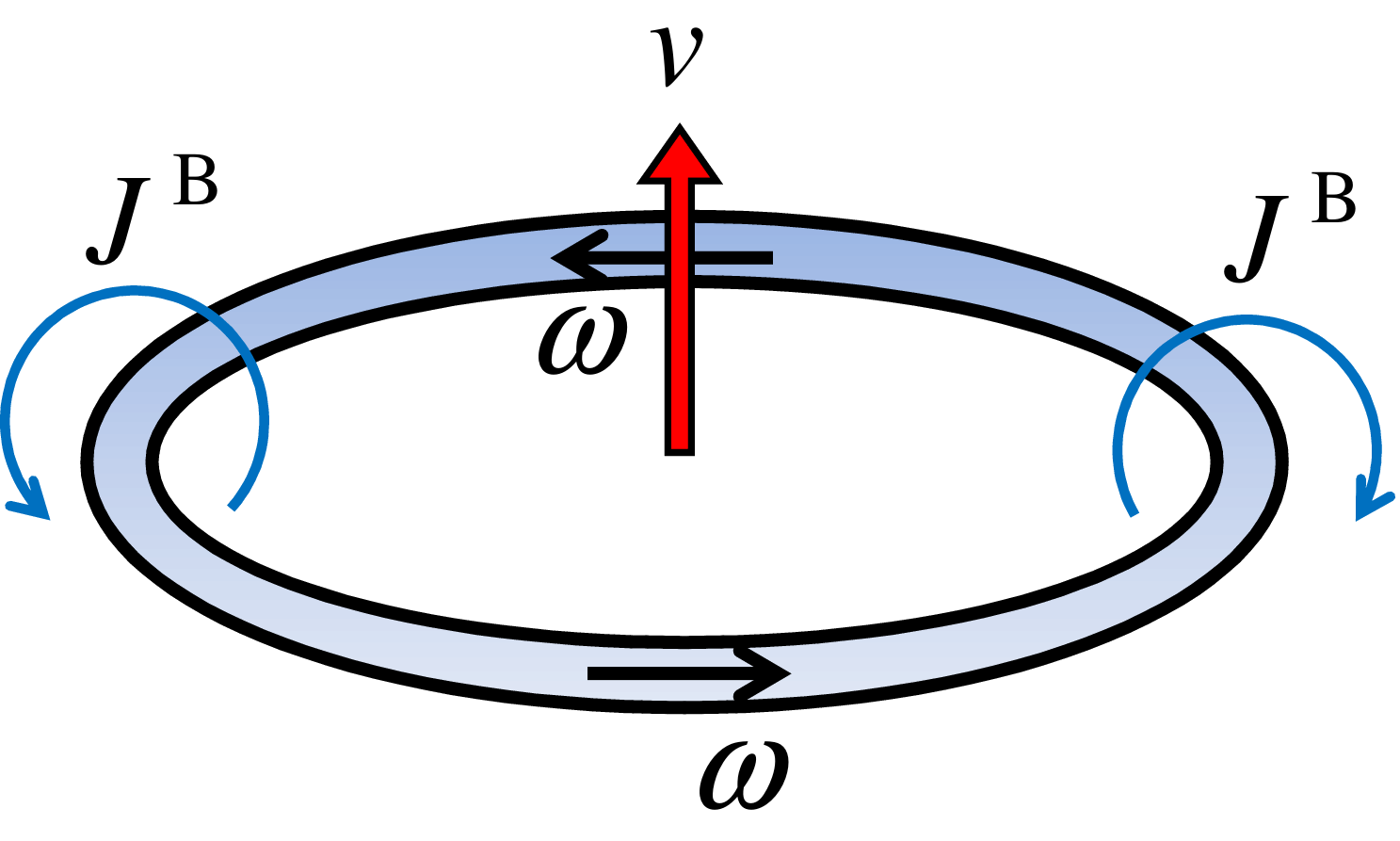}
&
\includegraphics[width=7cm]{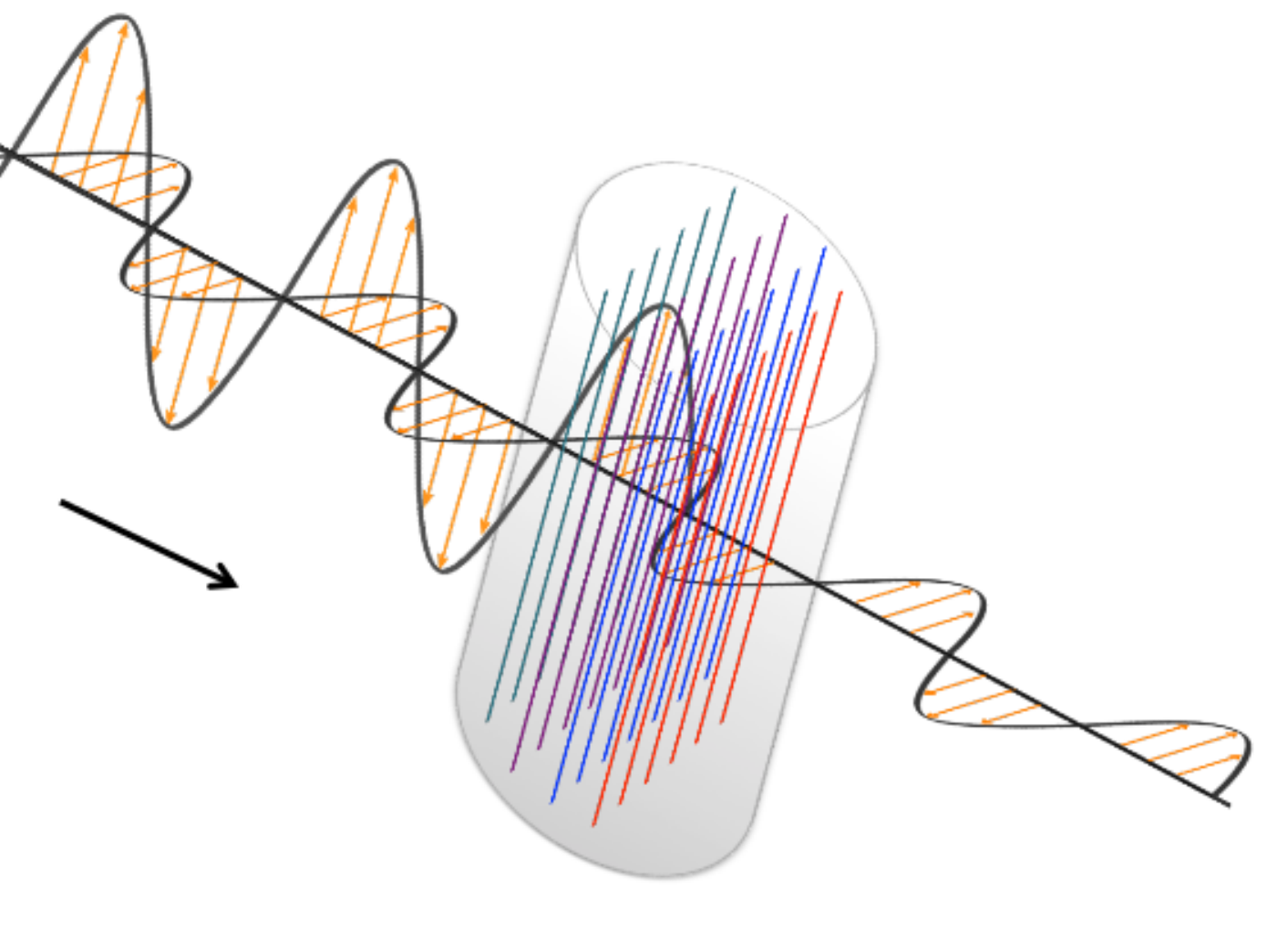}
\\
(a) Vortex ring (Sec.~\ref{sec:intervortex-force})
 & (b) Vortex lattice and photons (Sec.~\ref{sec:int-em})
\\
\includegraphics[width=8cm]{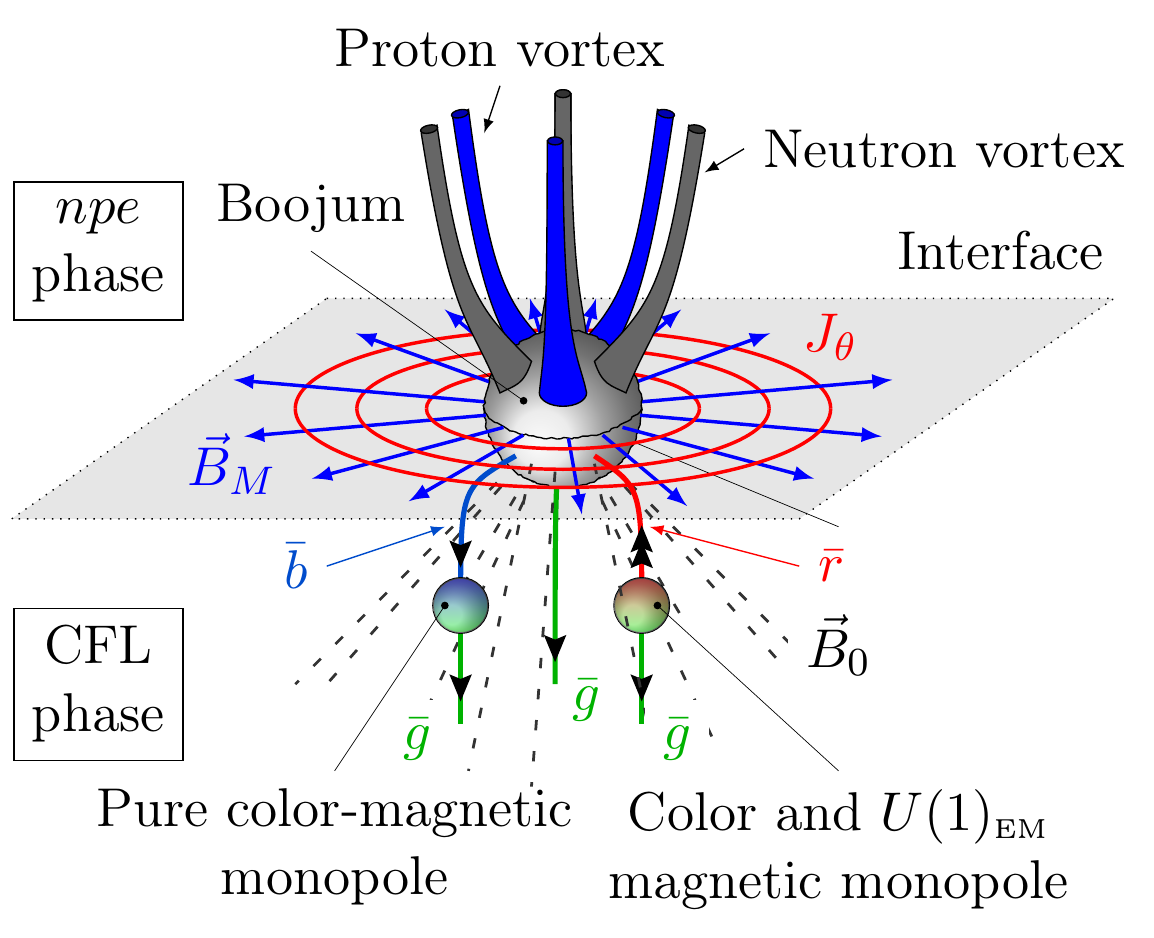} 
&
\includegraphics[width=7cm]{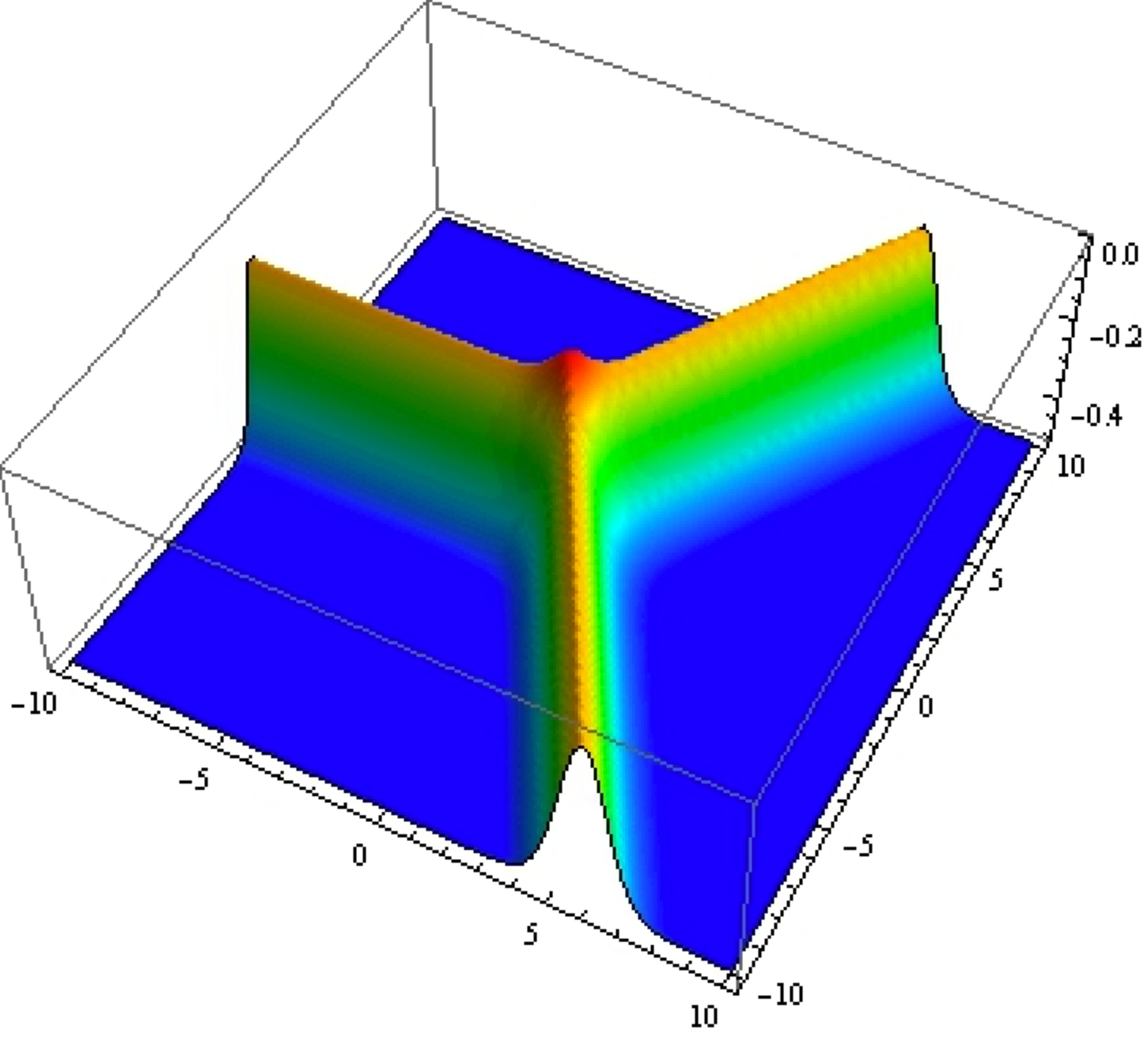}
\\
(c) Colorful boojum (Sec.~\ref{sec:boojum}) & (d) Axial domain wall junction (Sec.~\ref{sec:composite-wall-vortex})
\\
\end{tabular}
\caption{
Examples of topological solitons in dense QCD. For details, see the
 relevant sections.
}
\label{fig:soliton-examples}
\end{figure}

Topological solitons also manifest themselves in the condensed matter
physics of quantum chromodynamics (QCD), which is the theory of the
strong interaction.
The stability of topological solitons is closely related to the
structure of the vacuum. 
QCD matter exhibits a rich variety of phases at finite temperatures
and/or baryon densities \cite{Fukushima:2010bq}. 
Depending on the symmetry breaking patterns, QCD matter accommodates
various kind of topological solitons (see Fig.~\ref{fig:soliton-examples} for some examples), some of which are listed in Table
\ref{tab:solitons}.
Since topological solitons affect the bulk properties of the matter, 
it is important to investigate their basic properties 
and phenomenological implications.
They could affect the properties of the matter created in heavy-ion
collisions or the matter realized inside compact stars. 
Theoretical studies suggest that quark matter is expected to exhibit color superconductivity, triggered by quark-quark pairings,
at high baryon densities and low temperatures 
\cite{Barrois:1977xd,Bailin:1983bm,Iwasaki:1994ij,Rapp:1997zu,Alford:1997zt}. 
It has been predicted in Refs.~\cite{Alford:1997zt,Alford:1998mk} that 
the ground state is the color-flavor-locked (CFL) phase 
at asymptotically high densities, in which the three light flavors 
(up, down, and strange) of quarks 
contribute symmetrically to the pairing. 
There are also various other phases, such as 
the two-flavor superconducting (2SC) phase \cite{Bailin:1983bm},  
the kaon condensation (CFL+K) phase \cite{Bedaque:2001je}, 
the crystalline superconducting phase \cite{Casalbuoni:2003wh,Anglani:2013gfu}, 
and the magnetic CFL (MCFL) phase \cite{Ferrer:2005vd,Ferrer:2011ig,Ferrer:2012wa}.
For reviews of the phase structure of QCD matter, see
Refs.~\cite{Rajagopal:2000wf,Alford:2001dt,Ren:2004nn,Alford:2007xm,
Fukushima:2010bq}.

\begin{center}
\begin{table}[ht]
\begin{tabular}{|c|c|c|c|c|c|}
\hline
{\large Topological Solitons}
 & {{\large Phase}} %
 & {{\large OPS}} 
 & Type
 & {{\large Homotopy}}  
 & {\large Sec.} 
\\
 & 
 & {{\large $\mathcal M$}} 
 &
 & {{\large $\pi_n (\mathcal M)$}} 
 &
 \\ \hline \hline
$U(1)_{\rm B}$ superfluid vortex  & CFL  & $U(1)_{\rm B}$ &
	   D & $\pi_1(\mathcal M) \simeq \mathbb Z$ &  \ref{sec:U(1)B}
\\ \hline
NA semi-superfluid vortex &CFL & 
$U(3)_{\rm C-F+B}$ & D 
& $\pi_1( \mathcal M) \simeq
	 \mathbb Z$ & \ref{sec:NA-vortices}
 \\  \hline
confined monopole &  
CFL & $\mathbb Z_{3}$ (in $\mathbb C P^2$) & D & $\pi_0 (\mathcal M) \simeq \mathbb Z_3 $ &  \ref{sec:monopoles}
\\ \hline
trapped YM instanton & 
CFL & $\mathbb C P^2$ & T & $\pi_2(\mathcal M) \simeq \mathbb Z$ &  \ref{sec:instanton-in-vortex}
\\ \hline
$U(1)_{\rm A}$ axial vortex  & CFL  & $U(1)_{\rm A}$ & D & $\pi_1(\mathcal M) \simeq \mathbb Z$  &   \ref{sec:U(1)A}
\\ \hline
NA  axial vortex &CFL & $U(3)_{\rm L-R+A}$ & D & $\pi_1( \mathcal M) \simeq 
	 \mathbb Z$ & \ref{sec:gv_NA} \\  \hline
$U(1)_{\rm A}$ integer axial wall & CFL  &  $U(1)_{\rm A}$ &  T & $\pi_1(\mathcal M) \simeq \mathbb Z$&  \ref{sec:domain-wall-massive}
\\ \hline
$U(1)_{\rm A}$ fractional axial wall & CFL  &   $\mathbb Z_3$ in $U(1)_{\rm A}$& D &
$\pi_0 (\mathcal M) \simeq \mathbb Z_3 $ & \ref{sec:fractional-SG}
\\ \hline
Skyrmion (qualiton) & CFL  & $U(3)_{\rm L-R+A}$ & T & $\pi_3(\mathcal M) \simeq \mathbb
	     Z$  &  \ref{sec:skyrmion} 
\\ \hline
boojum & CFL edge & $U(3)_{\rm C-F+B}$ & D & relative &  \ref{sec:boojum}
\\ \hline
2SC domain wall & 2SC & $U(1)_{\rm A}$ & T &  $\pi_1(\mathcal M) \simeq \mathbb Z$ & \ref{sec:2SC}
\\ \hline
$U(1)_{\rm Y}$ supercond. vortex & CFL+K  & $U(1)_{\rm Y}$  &  
D & $\pi_1(\mathcal M) \simeq
	     \mathbb Z$ &  \ref{sec:CFLK}
\\ \hline
(drum) vorton & CFL+K  &$U(1)_{\rm Y}$, $U(1)_{\rm EM}$& D & -- &   \ref{sec:CFLK}
\\ \hline
kaon domain wall  & CFL+K &  $U(1)_{\rm Y}$ & T & $\pi_1(\mathcal M) \simeq \mathbb Z$ &  \ref{sec:CFLK}
\\ \hline
\end{tabular} 
\caption{
Topological solitons in dense QCD. 
For each topological soliton, we have summarized the phase in which it appears,
 the relevant order parameter space, 
and relevant homotopy group. 
The types of topological solitons are classified into 
``defects'' or ``textures'',  
denoted by D and T, respectively. 
The former are characterized by a map from the boundary 
$\partial {\bf R}^n \simeq S^{n-1}$
of space ${\bf R}^n$ to the order parameter space (OPS) 
and consequently by the homotopy group $\pi_{n-1} ({\cal M})$, 
while the latter are 
characterized by a map from the whole  
space ${\bf R}^n$ to the OPS 
and consequently by the homotopy group 
$\pi_n ({\cal M})$, 
where $n$ counts the codimensions of the solitons. 
Confined monopole and trapped instanton imply 
a monopole and instanton inside a non-Abelian vortex, 
where the order parameter space 
is ${\mathbb C}P^2$.
For vortons in the CFL +K phase,
 $U(1)_{\rm EM}$ symmetry is broken only inside 
a $U(1)_{\rm Y}$ vortex. 
The section in which each topological soliton is explained in this
 review is also indicated.
Note that this table is not a complete list of topological solitons in
 dense QCD.
``NA'' is the abbreviation for ``non-Abelian'' 
and ``YM'' is that for ``Yang-Mills''. 
``relative'' denotes 
a relative homotopy group 
\cite{volovik2009universe}.
}\label{tab:solitons}
\end{table}
\end{center}

Among the various ground states of QCD matter, 
the color-flavor-locked(CFL) phase is an important phase, which is realized at asymptotically high densities where weak coupling theory is
applicable and theoretically controlled calculations are possible.
The CFL phase has an interesting property that it
exhibits both superfluidity and superconductivity, 
because of the spontaneous breaking of the global 
$U(1)_{\rm B}$ baryon number
symmetry and the local $SU(3)_{\rm C}$ color symmetry, 
respectively.
As in a superfluid helium, we can expect the 
existence of topological vortices from consideration of the
ground-state symmetry \cite{Forbes:2001gj,Iida:2002ev}.
Vortices in the CFL phase will be created if the matter is rotated, 
as is observed in rotating superfluids in condensed matter systems.
Thus, if the CFL phase is realized 
in the cores of dense stars,
vortices are inevitably created  
since the stars rotate rapidly.
The superfluid vortices discussed in Refs.~\cite{Forbes:2001gj,Iida:2002ev}  
have integer winding numbers with respect to 
$U(1)_{\rm B}$ symmetry, and they are dynamically unstable, 
since they can decay into a set 
of vortices with lower energies.
It was first pointed out by Balachandran, Digal, and Matsuura 
\cite{Balachandran:2005ev} 
that the stable vortices are so-called non-Abelian
vortices,
which are superfluid vortices as well as color magnetic flux tubes.
Since they carry 1/3 quantized $U(1)_{\rm B}$ circulations,
an integer $U(1)_{\rm B}$ vortex  
decays into three non-Abelian vortices 
with different color fluxes canceled in total 
\cite{Nakano:2007dr}. 
The color magnetic flux tubes studied before \cite{Iida:2002ev} 
are non-topological and 
unstable, and have a color flux triple of a non-Abelian vortex. 
The properties of non-Abelian vortices 
have been studied 
using the Ginzburg-Landau (GL) theory 
\cite{Eto:2009kg,
Nakano:2007dr, 
Sedrakian:2008ay, 
Shahabasyan:2009zz, 
Eto:2009bh, 
Eto:2009tr, 
Hirono:2010gq} 
or in the Bogoliubov--de Gennes (BdG) equation 
\cite{Yasui:2010yw,Fujiwara:2011za}. 
A remarkable property of non-Abelian vortices is 
that both bosonic and fermionic 
zero energy modes are localized in the core of 
a non-Abelian vortex 
and propagate along it as gapless excitations.
Bosonic zero modes are found in the GL theory;  
one is the Kelvin mode, which also appears in vortices in 
superfluids,  
and the other is 
the orientational zero modes, characterizing an orientation 
of a vortex in the internal space \cite{Nakano:2007dr,Eto:2009bh}.
Both of them are Nambu-Goldstone modes;
the Kelvin mode is 
associated with 
two translational symmetries transverse to 
the vortex, and the orientational modes
are associated with the spontaneous symmetry 
breaking of the CFL symmetry $SU(3)_{\rm C+F}$ 
into  its subgroup $[SU(2)\times U(1)]_{\rm C+F}$ 
inside the vortex core.
There is only one Kelvin mode for each vortex   
although two translational symmetries are 
spontaneously broken, 
because it has a quadratic dispersion 
and is a so-called type-II Nambu-Goldstone mode.
The orientational zero modes have 
a linear dispersion and are  
so-called type-I Nambu-Goldstone modes.
The low-energy effective field theory of 
Kelvin modes is a non-relativistic free complex field 
with the first order time derivative in 
the $1+1$ dimensional vortex world-sheet
\cite{Kobayashi:2013gba}. 
The low-energy effective field theory of 
the orientational zero modes
is written as the relativistic ${\mathbb C}P^2$ model 
inside the $1+1$ dimensional vortex world-sheet, 
where ${\mathbb C}P^2 \simeq SU(3)/[SU(2)\times U(1)]$ 
is the target space spanned by the Nambu-Goldstone modes 
\cite{Eto:2009bh}. 
On the other hand, 
Majorana fermion zero modes 
belonging to a triplet of the core symmetry $SU(2)_{\rm C+F}$ 
have been found in the BdG equation 
and the low-energy effective theory in 
the $1+1$ dimensional vortex world-sheet  
has been derived \cite{Yasui:2010yw}.  
The existence of these fermion zero modes 
is ensured by topology, which can be seen as the index theorem
\cite{Fujiwara:2011za}.  
One remarkable consequence of  
the Majorana fermion zero modes is that 
non-Abelian vortices obey 
a novel kind of non-Abelian statistics 
\cite{Yasui:2010yh,Hirono:2012ad} 
if they are parallel or are restricted to $2+1$ dimensions.

While ${\mathbb C}P^2$ zero modes appear as 
Nambu-Goldstone gapless (zero) modes in the $1+1$ dimensional vortex world-sheet, exactly speaking 
they have to be gapped (massive) 
because of the Coleman-Mermin-Wagner theorem 
\cite{Coleman:1973ci,PhysRevLett.17.1133}, 
which prohibits Nambu-Goldstone modes in $1+1$ dimensions.
This problem is solved once non-perturbative effects 
are taken into account in the ${\mathbb C}P^2$ 
vortex world-sheet theory.
There appears quantum mechanically induced potential 
in the $1+1$ dimensional ${\mathbb C}P^2$ model.
As a consequence, 
there appear quantum magnetic monopoles 
confined by non-Abelian vortices \cite{Eto:2011mk,Gorsky:2011hd}. 
They provide a partial proof 
of the quark-hadron duality. 
In the confining phase, 
quarks are confined and monopoles are conjectured to be 
condensed. As a dual of this, in the CFL phase, 
monopoles are confined and quarks are condensed.  
As another example of topological solitons inside a 
non-Abelian vortex world-sheet, 
we introduce
Yang-Mills instantons, which
stably exist inside a non-Abelian vortex 
as lumps or sigma model instantons
in the $d=1+1$ dimensional ${\mathbb C}P^2$ model 
in the vortex world-sheet. 

The interactions between a vortex and quasi-particles 
can be treated as couplings between the $1+1$ dimensional 
vortex effective theory and fields propagating in the bulk. 
The $U(1)_{\rm B}$ phonon field  
can be mapped to a two-form field by a duality 
or the Hubbard-Stratonovich transformation .
The interaction between a vortex and the  $U(1)_{\rm B}$ phonons 
can be described as point particle interaction 
between a vortex and the two-form field \cite{Hirono:2010gq}, 
as is usually done for superfluid vortices 
or global cosmic strings. 
On the other hand, gluons can be dualized to 
non-Abelian two-form fields \cite{Seo:1979id}.
The interaction between a vortex and gluons 
can be described by the interaction between 
the orientational zero modes and the non-Abelian two-form fields.
Since the $U(1)_{\rm EM}$ electromagnetic symmetry 
is embedded in the flavor $SU(3)_{\rm F}$ symmetry 
that acts on the  ${\mathbb C}P^2$ model as an isometry, 
the interaction with the electromagnetic fields (photons) can be 
incorporated by the $U(1)_{\rm EM}$ gauged ${\mathbb C}P^2$ 
model, where the $U(1)_{\rm EM}$ field propagates in the bulk.
This model is similar to Witten's superconducting strings 
\cite{Witten:1984eb}, in which the $U(1)_{\rm EM}$ bulk gauge 
field couples to a $U(1)$ zero mode on a string 
and consequently 
a persistent superconducting current flows along the string.
In our case, non-Abelian vortices are not superconducting 
because of the difference between the zero modes. 
Instead, an interesting property is found. 
When an electromagnetic wave enters a vortex lattice, 
the magnitude of the electric field parallel to the lattice 
is reduced; consequently, 
the vortex lattice acts as a polarizer \cite{Hirono:2012ki}. 
This may give an observational signal 
if a rotating CFL matter exists.

Apart from the CFL phase or other color superconductors,
it is likely that the $npe$ phase is present 
in the interiors of neutron stars. 
The $npe$ phase is composed of 
a neutron superfluid and a proton superconductor 
\cite{baym1969superfluidity}, 
where the neutron pairing is likely a spin triplet 
at high density.
There, superfluid neutron vortices and superconducting proton 
vortices should exist under a rapid rotation and a strong magnetic field 
of a neutron star, respectively.  
Such vortices are expected to explain the pulsar glitch phenomenon \cite{Anderson:1975zze}. 
If the CFL phase is realized in a neutron star core, 
it may be surrounded by the $npe$ phase. 
Then, one can ask how these vortices in the $npe$ phase 
are connected to the CFL phase.
The endpoints of vortices at the interfaces of various superfluids 
are known as boojums.  
There appear colorful boojums 
at the interface between the CFL phase and 
the $npe$ phase, 
between which 
there may be other phases such 
as the CFL+K, 2SC phases and so on. 
Three neutron vortices and three proton vortices 
meet at the colorful boojum at which 
three non-Abelian vortices in the CFL phase join 
with the total color magnetic flux canceled out. 
This boojum is accompanied by 
colored monopoles appearing  
on non-Abelian vortices due to the presence of 
strange quark mass, 
a Dirac monopole of the massless gauge field,  
and  the surface current of the massive gauge field.

\bigskip
This paper is intended to be a catalog of topological solitons
in dense QCD.
We discuss various properties of topological solitons and their
phenomenological implications.
Particular emphasis is placed on the CFL phase,  
which accommodates a wide variety of topological solitons like vortices,
domain walls, kinks, monopoles and so on.
The paper is organized as follows.

In Section \ref{sec:GL}, 
we review the effective theories
for high density QCD, 
in terms of the Ginzburg-Landau theory (for around the critical
temperature)
and the chiral Lagrangian (for zero temperature).
In Subsection \ref{sec:gl_subsec}, 
the CFL ground state, 
the pattern of the symmetry breaking, 
the order parameter manifold of the CFL phase, 
and the mass spectrum are discussed. 
The effect of finite strange quark mass,
the electromagnetic interaction, 
and 
the mixing of gluons and photons in the CFL 
ground state are also discussed in the GL theory.
We then give the time-dependent Ginzburg-Landau 
(TDGL) Lagrangian.
In Subsection \ref{sec:eff_NG}, 
chiral symmetry breaking, the $U(1)_{\rm B}$ phonons, 
and the CFL mesons are studied 
in the framework of the chiral Lagrangian 
valid at zero temperature.

In Section \ref{sec:vortices}, we turn to vortices 
in the CFL phase. 
In Subsection \ref{sec:Abelian-vortices},
we first study 
Abelian vortices, i.~e.,
 $U(1)_{\rm B}$ superfluid vortices  
and $U(1)_{\rm A}$ axial vortices. 
In Subsection \ref{sec:nt-cmft}, we introduce
non-topological color magnetic flux tubes. 
In Subsection \ref{sec:NA-vortices},
we describe non-Abelian vortices. 
The numerical solutions, 
asymptotic behaviors of the profile functions at large distances, and the tension 
of the minimal stable non-Abelian vortices 
are obtained \cite{Eto:2009kg}. 
We also discuss non-minimal (unstable) non-Abelian vortices.
In Subsection \ref{sec:NA-vortices},
we discuss orientational zero modes 
(collective coordinates)
of a non-Abelian vortex. 
These modes are gapless modes propagating 
along the non-Abelian vortex and 
play important roles in the dynamics 
of non-Abelian vortices.

In Section \ref{sec:dynamics}, 
we study the dynamics of non-Abelian vortices.
In Subsec.~\ref{sec:translational}, 
we construct the effective 
field theory of translational zero modes, known as Kelvin modes, 
and study the dynamics of 
a single vortex string in terms of the low-energy 
effective theory \cite{Kobayashi:2013gba}. 
We also discuss Magnus and inertial forces 
exerted on vortices. 
In Subsection \ref{sec:intervortex-force}, 
the interaction between two non-Abelian vortices 
at large distances is shown to be repulsive, 
independent of orientational modes, 
and 1/3 of that between two Abelian superfluid vortices 
\cite{Nakano:2007dr,Nakano:2008dc}.
We also review the dynamics of two vortices 
and a vortex ring. 
In Subsection \ref{sec:Abelian-vortex-decay}, we discuss decays of a $U(1)_{\rm B}$ vortex and an M$_2$ vortex into three and two non-Abelian vortices, respectively.
In Subsection \ref{sec:lattice}, 
colorful vortex lattices under rotation are discussed. 
In Subsection \ref{sec:relativistic-superfluid}, 
we discuss a relation between relativistic strings 
in relativistic scalar theories and superfluid vortices 
by following Refs.~\cite{Lund:1976ze,Davis:1989gn,Forbes:2001gj},
in which the dynamics of superfluid vortices is 
reproduced in relativistic theory with 
a constant three-form field.

In Section \ref{sec:LEEA}, 
we discuss the dynamics of the orientational zero modes of non-Abelian vortices, 
which is one of the main parts of this review paper.
In Subsection \ref{sec:orientational_eff}, 
we first construct the low-energy effective theory 
of the orientational zero modes of a non-Abelian vortex 
in the limit of zero quark masses and neglecting electromagnetic interactions, 
which is the ${\mathbb C}P^2$ model \cite{Eto:2009bh}.
This model describes gapless propagations of the 
${\mathbb C}P^2$ orientational zero modes along 
the vortex string. 
In Subsection \ref{sec:strange-quark}, we take into account 
the strange quark mass, which induces an effective 
potential in the ${\mathbb C}P^2$ vortex effective theory 
\cite{Eto:2009tr}.
It turns out that all vortices decay into one kind 
immediately by this potential term. 
In Subsection \ref{sec:elemag}, we take into account 
the electromagnetic interactions \cite{Vinci:2012mc}.
First, the tension of the non-Abelian vortex has 
a finite correction, which appears as the effective potential 
in the ${\mathbb C}P^2$ vortex effective theory. 
Second, the low-energy effective vortex theory becomes 
a $U(1)$ gauged ${\mathbb C}P^2$ model. 
As stationary solutions 
of the effective potential, there exist 
the Balachandran-Digal-Matsuura (BDM) vortex 
which is the ground state in the absence of quark masses, 
metastable ${\mathbb C}P^1$ vortices, 
and unstable pure color vortex. 
We present the decay probability of meta stable ${\mathbb C}P^1$ vortices into the BDM vortex through quantum tunneling. 
In Subsection \ref{sec:monopoles}, 
we take into account the quantum corrections 
in the low-energy effective theory in the vortex 
world-sheet   
in the high density limit,  
in which strange quark mass can be neglected. 
There appears quantum mechanically induced potential 
in the low-energy  ${\mathbb C}P^2$ vortex effective theory through non-perturbative effects, 
which is consistent with
the Coleman-Mermin-Wagner theorem 
\cite{Coleman:1973ci,PhysRevLett.17.1133}.
One of the important consequences 
is the appearance of 
quantum monopoles that are confined by non-Abelian vortices \cite{Eto:2011mk,Gorsky:2011hd}. 
They provide a proof of some aspects 
of the quark-hadron duality \cite{Eto:2011mk}, 
i.~e., in the confining phase, 
quarks are confined and monopoles are condensed 
while in the CFL phase, 
monopoles are confined and quarks are condensed.  
In Subsection \ref{sec:instanton-in-vortex}, 
Yang-Mills instantons trapped inside a non-Abelian vortex 
are discussed. They become  lumps or sigma model instantons
in the $d=1+1$ dimensional ${\mathbb C}P^2$ model 
in the vortex world-sheet.

In Section \ref{sec:int}, we study interactions of 
non-Abelian vortices with quasi-particles. 
In Subsection \ref{sec:int-phonon-gluon}, 
the interactions between a non-Abelian vortex 
and phonons and gluons are discussed.
The interaction with phonons is obtained 
by a dual transformation in which 
the phonon field is dualized to a Abelian two-form field,  
while the interaction with gluons 
is achieved \cite{Hirono:2010gq} 
by non-Abelian dual transformation 
in which the gluon field is dualized to a non-Abelian massive 
two-form field \cite{Seo:1979id}. 
The latter provides interaction between 
two non-Abelian vortices at short distances 
intermediated by gluons.
We also give the orientation moduli dependence 
of the intervortex force mediated by exchange 
of massive gluons, which is a new result in this paper.
In Subsection \ref{sec:int-mesons}, 
we also provide the chiral Lagrangian of 
the mesons in the presence of 
a non-Abelian vortex, 
which is also a new result in 
this paper.
In Subsection \ref{sec:int-em}, 
we study the interaction between a non-Abelian vortex 
and photons, 
which is described by a $U(1)$ gauged 
${\mathbb C}P^2$ model. 
As an interesting consequence, we show that 
a lattice of non-Abelian vortices behaves 
as a polarizer \cite{Hirono:2012ki}. 

In Section \ref{sec:boojum}, 
we study the interface of the CFL phase and the confining phase
and find colorful boojums \cite{Cipriani:2012hr}. 
Boojums appear interface of various superfluids 
as the endpoints of vortices. 
In the interface between the CFL phase and 
the confining phase, 
there appears a colorful boojum, 
at which three non-Abelian vortices  
join with canceling the total magnetic flux 
in the CFL phase;  
three neutron vortices and three proton vortices 
join in the $npe$ phase. 
In the CFL phase side, 
non-Abelian vortices decay into one kind by 
the strange quark mass producing color magnetic monopoles, 
and $U(1)_{\rm EM}$ magnetic fluxes of proton vortices 
are decomposed into massless and massive fluxes. 
The former forms a Dirac monopole 
and the latter forms the surface current.
The shape of a colorful boojum is calculated in a simplified model.

In Section \ref{sec:fermion}, 
we study microscopic theory, 
i.~e., the Bogoliubov-de Gennes (BdG) theory, 
which is valid at any temperature,  
while the GL theory is valid 
only just below the critical temperature 
and at the large distance scale. 
In Subsection \ref{sec:BdG}, the BdG equations 
are discussed in the background of a non-Abelian vortex, 
in order to study structures of vortices 
far from the critical temperature and/or 
core structures of the vortices.
Fermions trapped inside the core of the vortex are found.
Inside the vortex core, 
the $SU(2)_{\rm C+F}$ symmetry remains, so that 
localized fermions are classified by the representations 
of $SU(2)_{\rm C+F}$.
Triplet Majorana fermion zero modes, 
which are localized and normalizable,  
and a singlet Majorana fermion zero mode, 
which is localized but non-normalizable,  
are found \cite{Yasui:2010yw}. 
In Subsection \ref{sec:1+1_effective_theory}, 
we construct the low-energy effective theory 
of the localized gapless fermions propagating along the vortex string \cite{Yasui:2010yw}. 
The chemical potential dependence of the 
velocity of gapless modes is obtained. 
In Subsection \ref{sec:supercurrent}, 
we show the absence of supercurrent by Majorana 
fermion zero modes as a new result of this paper, 
unlike the case of Witten's superconducting 
fermionic strings.
In Subsection \ref{sec:index}, 
we study the index theorem for the fermion zero modes 
in the background of a non-Abelian vortex 
\cite{Fujiwara:2011za}.  
We calculate the topological and analytical indices 
and find the agreement between them. 
In Subsection \ref{sec:topological}, 
the characterization of color superconductors as 
a topological superconductor 
is discussed \cite{Nishida:2010wr}. 
Topological superconductors are one of the topics of considerable
interest in condensed matter physics these days.

In Section \ref{sec:na-statistics}, 
as a novel application of Majorana fermion zero modes 
of a non-Abelian vortex, 
we study exchange statistics of non-Abelian vortices in 
$d=2+1$ dimensions. Zero-mode Majorana fermions inside vortices lead to
a new kind of non-Abelian anyons \cite{Yasui:2010yh,Hirono:2012ad}.

In Section \ref{sec:global}, topological objects 
associated with chiral symmetry breaking, 
axial domain walls, non-Abelian and Abelian axial vortices (strings) and Skyrmions, are discussed. 
In Subsection \ref{sec:dw_chiral}, 
axial domain walls are discussed. 
In the chiral limit with massless quarks, 
the instanton-induced potential term 
allows a fractional axial domain wall 
with 1/3 $U(1)_{\rm A}$ winding. 
In the presence of quark masses, 
this domain wall cannot exist alone, 
and only 
an integer axial domain wall, 
made of a set of three axial domain walls with the unit 
$U(1)_{\rm A}$ winding,  
is allowed as a composite wall.  
In Subsection \ref{sec:global_vortex}, 
 Abelian and non-Abelian axial vortices 
are discussed 
in the absence of quark masses and instanton effects   \cite{Balachandran:2001qn,Balachandran:2002je,
Nitta:2007dp,Eto:2009wu,Nakano:2007dq}. 
These vortices are also accompanied by 
${\mathbb C}P^2$ orientational zero modes, 
but these modes are non-normalizable 
unlike those of non-Abelian semi-superfluid vortices. 
In the presence of 
the instanton induced potential 
in the chiral limit, 
each non-Abelian axial vortex 
is attached by one axial domain wall, 
while   
each Abelian axial vortex is attached by three 
axial domain walls \cite{Balachandran:2002je}
and decays into three non-Abelian axial vortices 
each of which is attached by an axial domain wall. 
In the presence of quark masses, 
an Abelian axial vortex is attached by
a composite domain wall consisting of 
three axial domain walls. 
In Subsection \ref{sec:wall-decay}, 
we discuss decays of 
axial domain walls through quantum tunneling.
In Subsection \ref{sec:anomaly}, we discuss 
effects of axial anomaly on axial vortices and axial 
domain walls. 
The axial current flows along 
$U(1)_{\rm B}$ vortices, 
and the electric magnetic 
current flows along axial vortices, 
producing large magnetic fields perpendicular to 
axial domain walls in the presence of 
the external magnetic field. 
In Subsection \ref{sec:skyrmion}, 
Skyrmions as qualitons are discussed \cite{Hong:1999dk}.

In Section \ref{sec:other}, 
we review the topological solitons in the phases other
than the CFL phase.
The CFL phase is the ground state  at high densities
where all three
flavors can be treated as massless. If one lowers the density gradually,
the strange quark mass should be taken into account. 
Then it is expected that kaons form a condensation in addition to the
CFL condensates. 
This is called the CFL+K phase \cite{ Bedaque:2001je}.
If the density is further decreased, only the light flavors ($u$ and
$d$) contribute to the condensate, which is called the 2SC phase \cite{Bailin:1983bm}.
In Sec.~\ref{sec:2SC}, we discuss the domain walls and color magnetic
fluxes in the 2SC phase.
In Sec.~\ref{sec:CFLK}, we review the superconducting strings, vortons, domain
walls, and drum vortons in the CFL+K phase.

Section \ref{sec:conclusion} is devoted to 
a summary and discussions. 
A lot of future problems including possible 
applications to neutron star physics are 
addressed.

In Appendix \ref{sec:susy}, we summarize 
the properties of non-Abelian vortices in supersymmetric QCD 
as a comparison.
In Appendix \ref{sec:toric}, we give a brief review of 
the toric diagram 
which is useful to understand the ${\mathbb C}P^2$ zero modes 
of non-Abelian vortices. 
In Appendix \ref{sec:deriv_LET}, we give a detailed derivation of 
the low-energy effective action of 
 the ${\mathbb C}P^2$ zero modes, 
which was not explained in detail in the original papers.
In Appendix \ref{sec:appendix_dual}, 
we give the derivation of the dual Lagrangian for phonons and gluons,
which describes the interaction of vortices with those particles. 
In Appendix \ref{sec:appendix-fermion}, 
we give a detailed derivation of 
the fermion zero modes, which were not explained 
in the original paper.

\bigskip
This review paper is based on our papers on 
non-Abelian semi-superfluid vortices 
\cite{Nakano:2007dr,
Nakano:2008dc,
Eto:2009kg,
Eto:2009bh,
Eto:2009tr,
Yasui:2010yw,
Yasui:2010yh,
Hirono:2010gq,
Eto:2011mk,
Fujiwara:2011za,
Hirono:2012ad,
Hirono:2012ki,
Yasui:2012zb,
Vinci:2012mc,
Cipriani:2012hr},
non-Abelian axial vortices and axial domain walls 
\cite{Nitta:2007dp,Eto:2009wu,Nakano:2007dq,Eto:2013bxa}, 
superfluid vortices \cite{Kobayashi:2013gba}, 
and other developments, 
but this paper also contains several new results 
in Subsections \ref{sec:Abelian-vortex-decay}, 
\ref{sec:lattice}, \ref{sec:ori-vor-vor-int}, \ref{sec:int-mesons}, 
\ref{sec:supercurrent} and \ref{sec:domain-wall-massive}
and detailed calculations, which were not 
given in the original papers, 
in Appendices \ref{sec:deriv_LET} and \ref{sec:appendix-fermion}. 

%% file: gl-v9.tex
\section{
Low-energy effective theories for high density QCD}
\label{sec:GL}

In this section, we give a brief review of effective descriptions of the color-flavor locking (CFL) phase in high density QCD, i.e., 
the Ginzburg-Landau (GL) theory 
which
is valid around the critical temperature $T \sim T_{\rm c}$,
and the effective theories of massless 
Nambu-Goldstone bosons, 
which are valid at zero temperature $T=0$.
In Sec.~\ref{sec:gl_subsec}, 
we first explain the Ginzburg-Landau (GL) theory, 
and take into account 
the finite strange quark mass and 
the electromagnetic interaction.
We then extend the GL theory 
to incorporate the time dependence of the fields.
In Sec.~\ref{sec:eff_NG}, 
we first briefly review the effective field theories 
for $U(1)_{\rm B}$ phonons at zero temperature, 
and 
we introduce the effective theory for chiral symmetry breaking, 
i.e., the chiral Lagrangian describing
the (pseudo) Nambu-Goldstone bosons, 
the CFL pions, and the $\eta'$ meson.


\subsection{Ginzburg-Landau theories}
\label{sec:gl_subsec}

\subsubsection{The CFL phase}
\label{sec:cfl}

Let us start with giving a brief review of 
the GL free energy
\cite{Iida:2000ha,Iida:2001pg,Giannakis:2001wz}, 
which is a low-energy effective theory for the CFL phase in the color superconductivity of the massless three-flavor QCD 
at sufficiently high baryonic densities.
We first ignore the masses of the quarks, 
which are taken into account in the subsequent sections.

The order parameters are 
the diquark condensates $\Phi_{\rm L,R}$ defined by
\beq
(\Phi_{\rm L})_a^A \sim \epsilon_{abc} \epsilon_{ABC}
\langle (q_{\rm L})_b^B C (q_{\rm L})_c^C \rangle,
\label{eq:cond_L}
\\
(\Phi_{\rm R})_a^i \sim \epsilon_{abc} \epsilon_{ABC}
\langle (q_{\rm R})_b^B C (q_{\rm R})_c^C \rangle,
\label{eq:cond_R}
\eeq
where $q_{\rm L,R}$ stand for the left (right) handed quarks with the indices $a,b,c = 1,2,3 = r,g,b$ being those of colors while $A,B,C=1,2,3 = u,d,s$ 
are those of flavors, 
and $C$ is the charge conjugation operator.
The diquark Cooper pairs are induced by an attractive interaction via one-gluon exchange in
the $s$-wave color antisymmetric channel. The flavor indices must also be anti-symmetric because of
Pauli's principle. 
The most generic form of the GL free energy for uniform matter is
\beq
\Omega_0 &=&  a_{0{\rm L}}\Tr[\Phi_{\rm L}^\dagger \Phi_{\rm L}] + a_{0{\rm R}}\Tr[\Phi_{\rm R}^\dagger \Phi_{\rm R}] \non
&+& b_{1{\rm L}}\left(\Tr[\Phi_{\rm L}^\dagger \Phi_{\rm L}]\right)^2 + b_{1{\rm R}}\left(\Tr[\Phi_{\rm R}^\dagger \Phi_{\rm R}]\right)^2
+ b_{2{\rm L}}\Tr[(\Phi_{\rm L}^\dagger \Phi_{\rm L})^2] + b_{2{\rm R}}\Tr[(\Phi_{\rm R}^\dagger \Phi_{\rm R})^2]\non
&+& b_3\Tr[\Phi_{\rm L}^\dagger \Phi_{\rm L} \Phi_{\rm R}^\dagger \Phi_{\rm R}] + b_4\Tr[\Phi_{\rm L}^\dagger \Phi_{\rm L}]\Tr[ \Phi_{\rm R}^\dagger \Phi_{\rm R}].
\label{eq:free_ene}
\eeq
The color symmetry $\SU(3)_{\rm C}$, the flavor symmetry $\SU(3)_{\rm L} \times \SU(3)_{\rm R}$, the $\U(1)_{\rm B}$ symmetry,
and the $\U(1)_{\rm A}$ symmetry act on $\Phi_{\rm L}$ and $\Phi_{\rm R}$ as
\beq
\Phi_{\rm L} \to e^{i\theta_{\rm B}+i\theta_{\rm A}} U_{\rm C} \Phi_{\rm L} U_{\rm L},\quad
\Phi_{\rm R} \to e^{i\theta_{\rm B}-i\theta_{\rm A}} U_{\rm C} \Phi_{\rm R} U_{\rm R},
\label{eq:full_sym_phi}
\eeq
with $e^{i\theta_{\rm B}} \in \U(1)_{\rm B}$, $e^{i\theta_{\rm A}} \in \U(1)_{\rm A}$, $U_{\rm C} \in \SU(3)_{\rm C}$, and $U_{\rm L,R} \in \SU(3)_{\rm L,R}$.
Hereafter, we omit the last two terms since they can be neglected in the high baryon density region
\footnote{
The coefficients $b_3$ and $b_4$ in (\ref{eq:free_ene}) come from two loop diagrams in which
bubble graphs are connected by the gluon propagator; see Ref.~\cite{Matsuura:PhD} for details.
}.
Although the interactions induced by the perturbative gluon exchanges do not distinguish 
the positive diquark condensates from the negative one, the
non-perturbative instanton effects do. 
It was found that the state with positive parity is favored compared
to the one with negative parity as a ground state \cite{Alford:1998mk,Rapp:1997zu}.
We then choose the positive-parity state in what follows,
\beq
\Phi_{\rm L}=-\Phi_{\rm R} \equiv \Phi.
\label{eq:positive_parity}
\eeq

For later convenience, let us explicitly show the structure of indices of the $3 \times 3$ matrix field $(\Phi)_a{}^A$
\beq
\Phi = \left(
\begin{array}{ccc}
\Phi_{gb}{}^{ds} & \Phi_{gb}{}^{su} & \Phi_{gb}{}^{ud} \\
\Phi_{br}{}^{ds} & \Phi_{br}{}^{su} & \Phi_{br}{}^{ud} \\
\Phi_{rg}{}^{ds} & \Phi_{rg}{}^{su} & \Phi_{rg}{}^{ud} 
\end{array}
\right),
\label{eq:phi_structure}
\eeq
with the color indices $a,b,c=1,2,3=gb, br,rg$ and
the flavor indices $A,B,C=ds, su,ud$.
Let us next give a GL free energy for the inhomogeneous condensate $\Phi$, which
includes gradient energies and gluon fields to $\Omega_0$ given in Eq.~(\ref{eq:free_ene}),
\beq
\Omega \!&=&\! 
\Tr\left[ {1\over 4 \lambda_3} F_{ij}^2 + {\varepsilon_3 \over 2} F_{0i}^2 
+ K_3\D_i \Phi^\dagger \D_i \Phi \right] 
+ V, \non
V\!&=&\! \alpha \Tr\left(\Phi^\dagger \Phi \right)
+ \beta_1 \left[\Tr(\Phi^\dagger\Phi)\right]^2 
+\beta_2 \Tr \left[(\Phi^\dagger\Phi)^2\right] + \frac{3\alpha^2}{4(\beta_1+3\beta_2)},
\label{eq:gl}
\eeq
where $\lambda_{3}$ is a magnetic
permeability, and $\varepsilon_{3}$ is a dielectric constant for gluons, 
$i,j=1,2,3$ are indices for space coordinates, 
and the covariant derivative and the field strength of gluons are
defined by 
\beq
\D_\mu \Phi &=& \p_\mu \Phi - i g_{\rm s} A^a_\mu T^a \Phi, 
\label{eq:covariant_deri}\\ 
F_{\mu \nu} &=& \partial_{\mu}A_{\nu}
-\partial_{\nu}A_{\mu}-ig_{\rm s}[A_{\mu},A_{\nu}]. 
\label{eq:FS_gluon}
\eeq
Here, $\mu,\nu$ are indices for spacetime coordinates and
$g_{\rm s}$ stands for the $\SU(3)_{\rm C}$ coupling constant.
The coupling constants 
$\alpha, \beta_{1,2}, K_{0,3}$ are obtained from the weak-coupling calculations, which are valid 
at a sufficiently high density as \cite{Iida:2000ha,Giannakis:2001wz}
\beq
\alpha &=& 4 N(\mu) \log \frac{T}{T_{\rm c}}, \\
\beta_1 &=& \beta_2 = \frac{7\zeta(3)}{8(\pi T_{\rm c})^2}\, N(\mu)\equiv \beta, \\
K_3 &=&  \frac{7\zeta(3)}{12(\pi T_{\rm c})^2}N(\mu), \\
\lambda_0 &=& \epsilon_0 = 1,\quad
\lambda_3 = \epsilon_3 = 1,
\label{eq:weak}
\eeq
where $\mu$ stands for the quark chemical potential.
We have introduced the density of state $N(\mu)$ at the Fermi surface
\beq
N(\mu) = \frac{\mu^2}{2\pi^2}.
\label{eq:N_mu}
\eeq

The Lagrangian (\ref{eq:gl}) has the same symmetry as QCD 
except for the chiral symmetry,
which is spontaneously broken to the diagonal one, 
reflecting the fact that
the positive parity state is favored in the ground state as Eq.~(\ref{eq:positive_parity}). 
When only $\Phi_{\rm L}$ ($\Phi_{\rm R}$) condenses, the color $\SU(3)_{\rm C}$ and the flavor $\SU(3)_{\rm L}$ ($\SU(3)_{\rm R}$) 
are spontaneously broken down to their diagonal subgroup $\SU(3)_{\rm C+L}$ ($\SU(3)_{\rm C+R}$).
The ground state can be found by minimizing the potential energy in
Eq.~(\ref{eq:gl}), 
\beq
\label{eq:ground}
\Phi = {\rm diag}(\Delta_{\rm CFL}, \Delta_{\rm CFL}, \Delta_{\rm CFL}),\quad
\Delta_{\rm CFL} \equiv \sqrt{-\frac{\alpha}{8\beta}},
\eeq
up to a color and flavor rotation. 
This is so-called the color flavor locking (CFL) phase of the QCD.
The chiral symmetry is spontaneously
broken as $\SU(3)_{\rm L} \times SU(3)_{\rm R} \to SU(3)_{\rm L+R}$ as explained above. 
Since the condensate matrix $\Phi$ is proportional to the identity matrix,  the diagonal symmetry of the
color $\SU(3)_{\rm C}$ and the flavor $\SU(3)_{\rm L+R}$ remains as the genuine symmetry in the CFL phase.

Let us see the detailed structure of the spontaneous symmetry breaking in CFL.
Firstly, recall the actions of the symmetries on the order parameter $\Phi$,
\beq
\Phi \rightarrow e^{i\theta_{\rm B}} U_{\rm C} \Phi U_{\rm F},
\label{eq:sym_action}
\eeq
with $e^{i\theta_{\rm B}} \in \U(1)_{\rm B}$, $U_{\rm C} \in \SU(3)_{\rm C}$, and $U_{\rm F} \in \SU(3)_{\rm L+R}$.
There is a redundancy in the action of the symmetries, and the actual symmetry group is given by
\beq
 G  =
    \frac{\SU(3)_{\rm C} \times \SU(3)_{\rm F} \times \U(1)_{\rm B}}
   {(\mathbb{Z}_3)_{\rm C+B} \times (\mathbb{Z}_3)_{\rm F+B}},
\label{eq:sym_G}
\eeq
where the discrete groups are defined by
\beq
(\mathbb{Z}_3)_{\rm C+B} :\left(\omega^k{\bf 1}_3,{\bf 1}_3,\omega^{-k}\right)\in \SU(3)_{\rm C} \times \SU(3)_{\rm F} \times \U(1)_{\rm B},
\label{eq:Z3_C+B}\\
(\mathbb{Z}_3)_{\rm F+B} :\left({\bf 1}_3,\omega^k{\bf 1}_3,\omega^{-k}\right)\in \SU(3)_{\rm C} \times \SU(3)_{\rm F} \times \U(1)_{\rm B},
\label{eq:Z3_F+B}
\eeq
with $k=0,1,2$ and $\omega$ being
\beq
\omega \equiv e^{2\pi i/3}.
\label{eq:omega}
\eeq
Note that the discrete groups can be rearranged as
\beq
(\mathbb{Z}_3)_{\rm C+B} \times (\mathbb{Z}_3)_{\rm F+B} \simeq
(\mathbb{Z}_3)_{\rm C+F} \times (\mathbb{Z}_3)_{\rm C-F+B}
\label{eq:Z3_decomp}
\eeq
with
\beq
(\mathbb{Z}_3)_{\rm C+F} &:&
\left(\omega^k{\bf 1}_3,\omega^{-k}{\bf 1}_3,1\right)\in \SU(3)_{\rm C} \times \SU(3)_{\rm F} \times \U(1)_{\rm B},
\label{eq:Z3_c+F}\\
(\mathbb{Z}_3)_{\rm C-F+B} &:&
\left({\omega^k\bf 1}_3,\omega^{k}{\bf 1}_3,\omega^{-2k}\right)\in \SU(3)_{\rm C} \times \SU(3)_{\rm F} \times \U(1)_{\rm B}.
\label{eq:Z3_c-F+B}
\eeq
In the ground state, the full symmetry group $G$ is spontaneously broken down to 
\beq
H = \frac{\SU(3)_{\rm C+F}\times (\mathbb{Z}_3)_{\rm C-F+B}}{(\mathbb{Z}_3)_{\rm C+B} \times (\mathbb{Z}_3)_{\rm F+B}} 
\simeq \frac{\SU(3)_{\rm C+F}}{(\mathbb{Z}_3)_{\rm C+F}}.
\label{eq:H}
\eeq
Thus, we find the order parameter space of the ground state as
\beq
 {G \over H} \simeq \frac{\SU(3)_{\rm C-F}\times \U(1)_{\rm B}}{(\mathbb{Z}_3)_{\rm C-F+B}} 
\simeq \U(3)_{\rm C-F+B}.
\label{eq:sym_break_pattern_CFL}
\eeq
This $\U(3)$ manifold is parametrized by 9 would-be Nambu-Goldstone (NG) modes, among which
8 are eaten by the gluons via the Higgs mechanism and only one massless scalar field (referred to
as the $H$ boson) associated with the $\U(1)_{\rm B}$ symmetry breaking remains in the physical spectrum\footnote{
To be precise, the $\U(1)_{\rm B}$ is broken to $\mathbb Z_2$ which
flips the signs of $L$ and $R$ quarks ($q_L \to -q_L$ and
$q_R \to - q_R$). This ${\mathbb Z}_2$ cannot be described in the GL theory. 
}.

Including the broken $U(1)_{\rm A}$ symmetry and the chiral symmetry, the full order parameter space (OPS)
in the CFL phase is given by
\beq
 {\cal M}_{\rm CFL} = \U(3)_{\rm C-F+B} \times 
U(3)_{\rm L-R+A}.
\eeq
The GL Lagrangian (\ref{eq:gl}) describes only the first part, 
while the second part describing
the NG modes associated with the broken $\U(1)_{\rm A}$ and the chiral symmetry 
is discussed in Sec.~\ref{sec:CFL_meson}.

The spectrum of the GL theory (\ref{eq:gl}) can be found by expanding the order parameter $\Phi$
around the ground state given in Eq.~(\ref{eq:ground})  as
\beq
\Phi(x) = \Delta_{\rm CFL}{\bf 1}_3 + \frac{\phi_1(x) + i \varphi(x)}{\sqrt{2}}{\bf 1}_3
+ \frac{\phi_8^a(x) + i \zeta^a(x)}{\sqrt{2}}T^a.
\label{eq:fluctuation}
\eeq
The fields proportional to ${\bf 1}_3$ and $T^a$ belong to the singlet 
(${\bf 1}$) and adjoint 
(${\bf 8}$) representations of the CFL symmetry given in Eq.~(\ref{eq:H}).
The small fluctuations $\zeta^a(x)$ are absorbed by the gluons and 
$\varphi(x)$ remains the massless $\U(1)_{\rm B}$ NG mode (phonon). 
The masses of the gluons, $\phi_1$ and $\phi_8^a$ are given 
respectively as follows: 
\beq
m_{\rm g}^2 = 2\lambda_3 g_{\rm s}^2 \Delta_{\rm CFL}^2 K_3, \quad
m_{1}^2 = - \frac{2\alpha}{K_3}, \quad
m_{8}^2 = \frac{4\beta \Delta_{\rm CFL}^2}{K_3}.
\label{eq:mass_CFL}
\eeq
From this together with Eq.~(\ref{eq:weak}), we find the following relation
\beq
\label{eq:spectra}
m_{\rm g} \sim \sqrt{\lambda_3} g_{\rm s} \mu, \quad m_1 = 2m_8 \sim 2\Delta_{\rm CFL}.
\eeq
Note that $g_{\rm s}\mu \gg \Delta_{\rm CFL}$ at the high density limit, so that we have 
\beq
\kappa_{1,8} \equiv \frac{m_{1,8}}{m_{\rm g}} \ll 1.
\label{eq:kappa}
\eeq
This implies that the CFL phase is in the type I superconductor \cite{Giannakis:2003am}.\footnote{
This does not mean that a state with vortices is unstable in the CFL phase. 
NG bosons for the $\U(1)_{\rm B}$ symmetry breaking induce repulsive force between vortices, which stabilizes the multi-vortex state.
}

Note that the lightest mode in the GL Lagrangian (\ref{eq:gl}) is the NG mode.
As explained above, this exists because of the 
 $\U(1)_{\rm B}$ symmetry  spontaneously broken by the condensations, 
implying that the CFL phase is the $\U(1)_{\rm B}$ superfluid.
Namely, the CFL phase of the color superconductivity of high density QCD has 
the peculiarity of the coexistence of superfluidity and superconductivity.

\subsubsection{Including strange quark mass}
\label{sec:cfl_strange}

So far, we have considered the CFL phase at the asymptotically high density where
all the quark masses $m_{u,d,s}$ are negligible 
compared to the baryon chemical potential $\mu$.
In this subsection, let us consider 
the effect of the finite non-zero strange quark mass  while
the masses of the u and d quarks are kept at zero,
\beq
0\simeq m_{\rm u,d} < m_{\rm s} \ll \mu.
\label{eq:mass_hierachy}
\eeq
The effects of non-zero quark masses become important 
at smaller baryon chemical potentials. 
It was found \cite{Iida:2004cj} that the non-zero quark mass 
together with the $\beta$-equilibrium and the electric charge neutrality changes the CFL phase
to the modified CFL (mCFL) phase where the color-flavor locking symmetry is further broken as
\beq
\SU(3)_{\rm C+L+R} \to \U(1)^2.
\label{eq:sym_mCFL}
\eeq
The important difference between the CFL and mCFL phases is that the quark chemical potentials
$\mu_{\rm u},\mu_{\rm d}, \mu_{\rm s}$ take different values. Hence, there appear difference between the Fermi
momenta and the gaps of the di-quark condensation take different values as
\cite{Iida:2004cj}
\beq
\Delta_{\rm ud} > \Delta_{\rm ds} > \Delta_{\rm us}.
\label{eq:gap_inequality}
\eeq
This is responsible for the symmetry breaking in Eq.~(\ref{eq:sym_mCFL}).
The 
correction to quadratic order
to the GL potential in Eq.~(\ref{eq:gl}) was obtained 
as \cite{Iida:2003cc,Iida:2004cj}
\beq
\delta V &=&  \frac{2}{3}\varepsilon\ \tr\left[\Phi^\dagger\Phi\right] + \varepsilon\ \tr\left[\Phi^\dagger X_3\Phi\right],
\label{eq:GL_modification}\\
\varepsilon &=& N(\mu)\frac{m_{\rm s}^2}{\mu^2} \log\frac{\mu}{T_{\rm c}},
\label{eq:def_epsilon}\\
X_3 &=& {\rm diag}\left(0,\frac{1}{2},-\frac{1}{2}\right).
\label{eq:def_X3}
\eeq

The first term in Eq.~(\ref{eq:GL_modification}) can be absorbed into the definition of $\alpha$ as
\beq
\alpha' \equiv \alpha + \frac{2}{3}\varepsilon.
\label{eq:modif_alpha}
\eeq
If we ignore the second term in Eq.~(\ref{eq:GL_modification}), all the results in the previous sections
are still valid under the understanding of the replacement $\alpha$ with $\alpha'$ in all equations.
For instance, $\Delta_{\rm CFL}$ is replaced with
\beq
 \Delta_{\varepsilon} \equiv
 \sqrt{-\alpha' \over 8\beta}.
 \label{eq:delta_epsilon}
\eeq
Therefore, an essential difference from the massless case is given rise to by the second term 
in Eq.~(\ref{eq:GL_modification}).
Since the term is sufficiently small if $m_{\rm s} \ll \mu$, we will treat it as a perturbation in Sec.~\ref{sec:strange-quark}.

\subsubsection{Including electromagnetic interactions}
\label{sec:cfl_em}

Let us next include the electromagnetic interaction that is realized as a $\U(1)$ action from the right-hand side on $\Phi$ as
\beq
\Phi \to \Phi e^{i e T^{\rm EM}},\quad 
T^{\rm EM} = \sqrt{\frac{2}{3}}\,T^8 = {\rm diag}\left(-\frac{2}{3},\frac{1}{3},\frac{1}{3}\right).
\label{eq:gene_EM}
\eeq
The covariant derivative of $\Phi$ is changed from Eq.~(\ref{eq:covariant_deri}) to
\beq
\D_\mu \Phi &=& \p_\mu \Phi - i g_{\rm s} A^a_\mu T^a \Phi - i e A^{\rm EM}_\mu  \Phi T^{\rm EM}, 
\label{eq:covariant_deri_em}
\eeq
and the field strength is 
\beq
F^{\rm EM}_{\mu\nu} &=& \p_\mu A_\nu^{\rm EM} - \p_\nu A_\mu^{\rm EM}.
\label{eq:FS_EM}
\eeq
The GL thermodynamic potential is modified as
\beq
\Omega = 
\Tr\left[ {1\over 4 \lambda_3} F_{ij}^2 
+ {\varepsilon_3 \over 2} F_{0i}^2 
+ K_3\D_i \Phi^\dagger \D_i \Phi \right] 
+ {1\over 4 \lambda_0} (F^{\rm EM}_{ij}){}^2 
+ {\varepsilon_0 \over 2} (F^{\rm EM}_{0i}){}^2 
+ V,
\eeq
where $V$ is the same as the one in Eq.~(\ref{eq:gl}) and
$\lambda_{0}$ 
and $\varepsilon_0$ are 
the magnetic permeability 
and the dielectric constant 
for the electromagnetic field.

There is mixing between $A_i^{\rm EM}$ and $A_i^8$ in the CFL phase,
since the electric charge is proportional to $T_8$ as shown in Eq.~(\ref{eq:gene_EM}) \cite{Alford:1998mk}.
Let us first rescale the gauge fields by
\beq
\mathcal A_i^a = \frac{1}{\sqrt{\lambda_3}} A_i^a,\quad
\mathcal A_i^{\rm EM} = \frac{1}{\sqrt{\lambda_0}} A_i^{\rm EM}.
\eeq
This redefinition changes the kinetic terms of the electromagnetic fields and the gluons into the canonical
forms as $- \frac{1}{4}\Tr[(\mathcal F_{ij})^2] - \frac{1}{4} (\mathcal F_{ij}^{\rm EM})^2$.
Then, the covariant derivatives are rewritten as
\beq
\D_i \Phi &=& \p_i \Phi - i \sqrt{\lambda_3} g_{\rm s} \mathcal A_i^a T^a \Phi
- i \sqrt{\lambda_0} e \Phi \mathcal A_i^{\rm EM}T^{\rm EM}.
\eeq

When the order parameter $\Phi$ is diagonal  (in particular in the ground state), 
the covariant 
derivative acting on the $\Phi$ can be rewritten as
\beq
\D_i \Phi &=& \p_i \Phi - i \sqrt{\lambda_3} g_{\rm s} \mathcal A^{\tilde a}_i T^{\tilde a} \Phi 
- i \sqrt{\lambda_0} g_{\rm M} {\mathcal A}_i^{\rm M} T^{\rm EM} \Phi , 
\label{eq:covariant_deri_mix}\\
g_{\rm M} &\equiv& \sqrt{e^2 + \frac{3}{2}\frac{\lambda_3}{\lambda_0}g_{\rm s}^2}, \label{eq:g_mix}\\
T^{\rm M} &\equiv& T^{\rm EM} \label{eq:TM}
\eeq
with $\tilde a = 1,2,\cdots,7$, 
in terms of rotated gauge fields defined by 
\beq
 {\mathcal A}_i^{\rm M} \equiv \cos \chi {\mathcal A}_i^{\rm EM} + \sin \chi {\mathcal A}_i^8,
\label{eq:mix_gluon}\\
 {\mathcal A}_i^{0} \equiv - \sin \chi {\mathcal A}_i^{\rm EM} + \cos \chi {\mathcal A}_i^8,
\label{eq:mix_photon}
\eeq
with the mixing angle defined by
\beq
\tan \chi = \frac{\sqrt{\frac{3}{2}\frac{\lambda_3}{\lambda_0}} g_{\rm s}}{e}.
\label{eq:tan_zeta}
\eeq
We have rewritten $T^{\rm EM}$ by $T^{\rm M}$ in order to remember that it acts 
on $\Phi$ from its left while $T^{\rm EM}$ acts on its right.
One can see from Eq.~(\ref{eq:covariant_deri_mix}) that ${\mathcal A}_i^{\rm M}$ couples to $\Phi$
and so is massive in the ground state [see Eq.~(\ref{eq:massive}), below], 
while ${\mathcal A}_i^{0}$ decouples from $\Phi$ and so is massless. 
However, one should note that 
this rotation is valid only when $\Phi$ is diagonal, 
and it is meaningless for general $\Phi$.

The ground state of the theory is unchanged with respect to the case 
without the electromagnetic interaction:
\begin{equation}
\left<\Phi_{a}^{\ A}\right>= \Delta_{\rm CFL} \delta_{a}^{\ A}\,.
\label{eq:vaccgaug}
\end{equation}
Because of the non-trivial mixing, the masses of gluons are modified 
in the CFL phase as 
\begin{eqnarray}
 m^{2}_{{\rm g},\tilde a} 
=2 \lambda_3 g_{\rm s}^{2}\Delta_{\rm CFL}^{2}K_{3} ,\quad
 m^{2}_{{\rm g,M}} 
=\frac{4}{3}\lambda_0 g_{\rm M}^{2}\Delta_{\rm CFL}^{2}K_{3}\,, \label{eq:massive}
\end{eqnarray}
where $\tilde a$ runs from 1 to 7 and $m_{{\rm g,M}}$ is the mass of the gluon corresponding to 
the generator $T^{8}$.

One of the important effects of introducing electromagnetic interactions is the explicit breaking of the $SU(3)_{\rm F}$ flavor symmetry, even if we still consider all quarks to be massless. 
Introducing electromagnetic interactions is equivalent to gauging 
the $U(1)$ subgroup 
generated by $T^{\rm EM}$ inside the $SU(3)_{\rm F}$ flavor symmetry. 
Therefore, the flavor $SU(3)$ is explicitly broken as 
\begin{equation}
SU(3)_{\rm F}\stackrel{T^{\textsc{em}}}{\longrightarrow }SU(2)_{\rm F}\times U(1)_{\textsc{em}}\,,  \label{eq:em-breaking}
\end{equation}
where $SU(2)_{\rm F}$ is a subgroup of $SU(3)_{\rm F}$ 
commuting with $T_{\rm EM}$.
The full set of symmetries of the CFL phase of QCD with electromagnetic interactions is thus given by the following (apart from the chiral symmetry):
\begin{equation}
G=U(1)_{\rm B}\times U(1)_{\textsc{em}}\times SU(3)_{\rm C}\times SU(2)_{\rm F}\, .
\label{eq:symmetrygaug}
\end{equation}
Apart from the unbroken gauge $U(1)$ symmetry, the CFL ground state  Eq.~(\ref{eq:vaccgaug}) has the following diagonal color-flavor symmetry
\begin{equation}
H_{\textsc{em}} = SU(2)_{\rm C+F}\, .\label{eq:SU(2)}
\end{equation}
This reduced symmetry is crucial to understanding the property of 
non-Abelian vortices with electromagnetic coupling.

\subsubsection{Time-dependent Ginzburg-Landau theory\label{sec:tdgl}}

In this section, let us discuss the time-dependent Ginzburg-Landau Lagrangian (TDGL) 
for the CFL phase \cite{Abuki:2006dv}.
The time-independent part of the TDGL is the same as Eq.~(\ref{eq:gl}).
Then, the TDGL is given by 
\beq
&& \Lag = 
\Tr\left[-{\varepsilon_3\over 2} F_{0i}F^{0i} - {1\over 4 \lambda_3} F_{ij}F^{ij} \right]
- {\varepsilon_0\over 2} (F^{\rm EM})_{0i}(F^{{\rm EM}})^{0i} 
- {1\over 4 \lambda_0} (F^{\rm EM})_{ij}(F^{{\rm EM}})^{ij}
\non
&&\hs{5} + \Tr\left[
2 i \gamma K_0 (\D_0  \Phi^\dagger  \Phi - \Phi^\dagger \D_0 \Phi ) 
+ K_0 \D_0\Phi^\dagger \D^0\Phi
+ K_3\D_i \Phi^\dagger \D^i\Phi  \right] 
 - V,
\label{eq:tdgl}
\eeq
with 
\beq 
  K_0 = 3K_3.
\eeq
$K_0$ has not been calculated in the literature, but can be derived
following the same procedure as in Refs.~\cite{Abrahams:1966zz,SadeMelo:1993zz}.

At zero temperature, there is no dissipation. 
However,  the dissipation is present at finite temperature, which cannot be incorporated in the Lagrangian. 
Instead, the Euler-Lagrange equation 
can be modified as
\beq
 K_{\rm D} {\del \Phi \over \del t} =  {\delta {\cal L} \over \delta \Phi^*}  \label{eq:dissipation}
\eeq
with a real GL parameter 
$K_{\rm D} = K_{\rm D}(T,\mu)$, which has not been 
calculated from microscopic theory.
Formally, this term can be obtained by 
replacing $2 i K_0 \gamma \del/\del t$ by 
$(2 i K_0 \gamma - K_{\rm D} ) \del/\del t$ in 
the Euler-Lagrange equation of the Lagrangian in  Eq.~(\ref{eq:tdgl}).

Let us construct 
the $U(1)_{\rm B}$ Noether current:
\beq
&& j_0^{\rm B} = 
\gamma K_0 \Tr (\Phi^\dagger \Phi) 
+ 2 i K_0 \Tr \left[ \D_0  \Phi^\dagger  \Phi - \Phi^\dagger \D_0 \Phi\right] ,\non
&&  j^{\rm B}_i = 2 i K_3 \Tr ({\cal D}_i\Phi^\dagger \Phi - \Phi^\dagger {\cal D}_i \Phi). \label{eq:U(1)Bcurrent} 
\eeq
The Noether current 
satisfies the continuity equation: 
\beq
 \del_0  j^{\rm B}_0 - \del_i j^{\rm B}_i = 0.
\eeq 
The Noether charge defined by 
\beq
   Q^{\rm B} = \int d^3 x j^{\rm B}_0 
\eeq
is a conserved quantity. 

When we ignore the second derivative with respect to 
the time coordinate, the charge density 
is reduced as 
\beq
  j_0^{\rm B} \sim 
\gamma K_0 \Tr \Phi^\dagger \Phi 
\equiv \rho, \label{eq:density}
\eeq 
where $\rho$ is called the superfluid density. 
Then, the superfluid velocity 
is defined by dividing the spatial components 
of the Noether current 
by the superfluid density:
\beq
 J^{\rm B}_i = {2 i K_3 \over \gamma K_0}
{\Tr ({\cal D}_i\Phi^\dagger \Phi - \Phi^\dagger {\cal D}_i \Phi)\over \Tr (\Phi^\dagger \Phi)}. \label{eq:U(1)Bsuperflow} 
\eeq
The vorticity is defined by the rotation 
of the superfluid velocity
\beq
 \omega_i = \1{2} \epsilon_{ijk} \del_j J^{\rm B}_k .
\eeq

The conjugate momentum of $\Phi$ is 
\beq
\Pi_\Phi = K_0 \left(\D_0\Phi^\dagger - 2 i \gamma \Phi^\dagger \right),
\eeq
and the Hamiltonian is given by
\beq
\mathcal H &=& 
\Tr\left[{\varepsilon_3\over 2} (F_{0i})^2 + {1\over 4 \lambda_3} (F_{ij})^2 
+ K_0 \left| \D_0\Phi\right|^2
+ K_3\left|\D_i\Phi\right|^2  \right]\non
&& + {\varepsilon_0\over 2} (F^{\rm EM}_{0i})^2 + {1\over 4 \lambda_0} (F_{ij}^{\rm EM})^2 +  V.
\eeq

\subsection{Effective theories for light fields at zero temperature} \label{sec:eff_NG}

Thus far, we have explained the GL theory, 
which is valid around the transition temperature
$T \sim T_{\rm c}$. 
In this section, we introduce the effective Lagrangians 
for Nambu-Goldstone modes associated 
with spontaneous symmetry breaking, 
which are valid around the zero temperature $T \sim 0$.
We study the low-energy effective theory 
for $U(1)_{\rm B}$ phonons 
in Sec.~\ref{sec:eff_th_phonon}
and the chiral Lagrangian describing 
the CFL mesons and the $\eta'$ meson in Sec.~\ref{sec:CFL_meson}. 
In general, 
the effective theory for Nambu-Goldstone modes 
can be constructed by nonlinear realizations 
up to decay constants \cite{Coleman:1969sm,Callan:1969sn}.  
At low baryon densities, the decay constant of the chiral Lagrangian is determined experimentally.
On the other hand, at high baryon densities,   
the decay constant can be microscopically calculated from QCD, as explained below.

\subsubsection{Effective theory for 
$U(1)_{\rm B}$ phonons at zero temperature}
\label{sec:eff_th_phonon}

In the CFL phase the baryonic $U(1)_{\rm B}$ symmetry is spontaneously broken, so that
a corresponding massless Nambu-Goldstone boson $\varphi_{\rm B}$ appears. It will play important roles
since it is deeply related to superfluidity. 
In this section we give an brief review of low energy effective theories at high density and zero temperature
\cite{Casalbuoni:1999wu,Son:1999cm,Son:2000tu,Son:2002zn}
of the $U(1)_{\rm B}$ Nambu-Goldstone mode.

Let us first take the phase of the diquark condensate $qq$ as the Nambu-Goldstone field.
An effective action takes the form
\beq
\Lag_{\rm B} = 12 f_{\rm B}^2 \left[ \left(\p_0\varphi_{\rm B} + 2\mu\right)^2 - v_{\rm B}^2 (\p_i\varphi_{\rm B})^2 \right],
\label{eq:u1B_eff_theory_1}
\eeq
where the coefficients were calculated \cite{Son:1999cm,Son:2000tu} as
\beq
f_{\rm B}^2 = \frac{3\mu^2}{8\pi^2},\quad
v_{\rm B}^2 = \frac{1}{3},
\eeq
and $\mu$ is the quark chemical potential.
Note that the time derivative is not $\p_0$ but $\p_0 + 2\mu$. This is needed for the effective theory to have
the same symmetry as QCD.  This can be understood as follows.
The Lagrangian of QCD in the medium is given by
\beq
\mathcal L_{\rm QCD} = \bar q (i \gamma_\nu\p^\nu) q + \mu q^\dagger q = \bar q (i\gamma^\nu D_\nu) q,
\eeq
where we introduced a covariant derivative $D_\nu = \p_\nu + i B_\nu$ with
a fictitious (spurion) gauge field $B_\mu$ which corresponds to the baryon current \cite{Forbes:2001gj}. 
$B_\mu$ will be set to $B_\mu = (\mu,0,0,0)$ in the end.
The QCD Lagrangian is invariant under the $U(1)_{\rm B}$ rotation
\beq
q \to e^{i\theta_{\rm B}} q,\quad
\varphi_{\rm B} \to \varphi_{\rm B} + 2 \theta_{\rm B},\quad
B_\nu \to B_\nu - \p_\nu \theta_{\rm B}.
\eeq
We now require the low energy effective field to have this symmetry; one then reaches the Lagrangian
(\ref{eq:u1B_eff_theory_1}).

There is another derivation of the effective theory in gauge invariant fashion \cite{Son:2002zn} by taking phase of the gauge invariant
operator $qqqqqq$ as the $U(1)_{\rm B}$ Nambu-Goldstone field. In Ref.~\cite{Son:2002zn},
it was found that the effective field theory takes the form
\beq
\Lag_{\rm B}' = \frac{N_{\rm C}N_{\rm F}}{12\pi^2} \left[
\left(\p_0\varphi_{\rm B} - 6 \mu\right)^2 - (\p_i\varphi_{\rm B})^2\right]^2.
\label{eq:u1B_eff_theory_2}
\eeq
Expanding this Lagrangian and taking terms to the quadratic order in the derivative with $N_{\rm C} = N_{\rm F} = 3$, 
one reproduces the same Lagrangian as Eq.~(\ref{eq:u1B_eff_theory_1}).

Although we have ignored the amplitude (Higgs) modes in this section,
they have been taken into account in Ref.~\cite{Anglani:2011cw} and interactions
between the $U(1)_{\rm B}$ mode and the amplitude modes have been studied.

\subsubsection{Chiral Lagrangian of the CFL mesons: pions and the $\eta'$ meson}
\label{sec:CFL_meson}

In this subsection, we study the chiral Lagrangian describing 
the Nambu-Goldstone modes for  chiral 
symmetry breaking, i.e., 
the CFL mesons and the $\eta'$ meson. 
So far,  we have not considered the spontaneous symmetry breaking of 
the axial $\U(1)_{\rm A}$ symmetry, since it is explicitly broken by 
quantum effects (the chiral anomaly).
Instantons induce vertices that break the $U(1)_{\rm A}$ symmetry,
since  instantons flip the chirality of fermions.
The instanton effects becomes arbitrarily small at the high
density limit,
so that the chiral anomaly is suppressed.  Therefore, at high density, the spontaneous broken 
$\U(1)_{\rm A}$ symmetry 
introduces a light meson, namely the $\eta'$ meson.
Accordingly, there appears an order parameter corresponding to this.
Including the axial $\U(1)_{\rm A}$ symmetry, 
the flavor symmetry  
$SU(3)_{\rm L} \times SU(3)_{\rm R} \times U(1)_{\rm A}$ of QCD
acts on the gauge invariant 
\beq
\Sigma \equiv \Phi_{\rm L}^\dagger \Phi_{\rm R},
\label{eq:sigma0}
\eeq
as 
\beq
  \Sigma \to 
g_{\rm L}^\dagger \Sigma g_{\rm R} e^{2 i \theta_{\rm A}}, 
\quad 
(g_{\rm L}, g_{\rm R}, e^{2 i \theta_{\rm A}}) \in
\SU(3)_{\rm L} \times \SU(3)_{\rm R} \times \U(1)_{\rm A}. 
\label{eq:sigma-trans}
\eeq
While discrete groups are not well discussed in the literature, 
here we discuss them in more detail.
There is a redundancy of discrete groups in this action, 
and the actual flavor symmetry is 
\beq
 G_{\rm F} 
 ={SU(3)_{\rm L} \times SU(3)_{\rm R} \times U(1)_{\rm A} \over
  ({\mathbb Z}_3)_{\rm L+A} \times  ({\mathbb Z}_3)_{\rm R+A}
}
\simeq {SU(3)_{\rm L} \times SU(3)_{\rm R} \times U(1)_{\rm A} \over
  ({\mathbb Z}_3)_{\rm L+R} \times  ({\mathbb Z}_3)_{\rm L-R+A}
},
  \label{eq:GF}
\eeq
where the redundant discrete groups are 
\beq
(\mathbb{Z}_3)_{\rm L+A} &:& 
\left(\omega^k{\bf 1}_3,{\bf 1}_3,\omega^{-k}\right)
 \in \SU(3)_{\rm L} \times \SU(3)_{\rm R} \times \U(1)_{\rm A},
\label{eq:Z3_L+A}\\
(\mathbb{Z}_3)_{\rm R+A} &:& 
\left({\bf 1}_3,\omega^{k}{\bf 1}_3,\omega^{-k}\right)
 \in \SU(3)_{\rm L} \times \SU(3)_{\rm R} \times \U(1)_{\rm A},
\label{eq:Z3_R+A}\\
(\mathbb{Z}_3)_{\rm L+R} &:& 
\left(\omega^k{\bf 1}_3,\omega^{-k}{\bf 1}_3,1\right)
 \in \SU(3)_{\rm L} \times \SU(3)_{\rm R} \times \U(1)_{\rm A},
\label{eq:Z3_L+R}\\
(\mathbb{Z}_3)_{\rm L-R+A} &:& 
\left({\omega^k\bf 1}_3,\omega^{k}{\bf 1}_3,\omega^{-2k}\right)
 \in \SU(3)_{\rm L} \times \SU(3)_{\rm R} \times \U(1)_{\rm A},
\label{eq:Z3_L-R+A}  
\eeq
with $\omega$ in Eq.~(\ref{eq:omega}). Here the suffix ${\rm A}$ always implies 
$U(1)_{\rm A}$. 
With the same group structures as Eqs.~(\ref{eq:H}) and (\ref{eq:sym_break_pattern_CFL}),
the unbroken symmetry in the ground state $\Sigma \sim {\bf 1}_3$ is 
\beq
 H_{\rm F} 
= 
 {SU(3)_{\rm L+R} \times ({\mathbb Z}_3)_{\rm L-R+A}\over 
({\mathbb Z}_3)_{\rm L+A} \times  ({\mathbb Z}_3)_{\rm R+A}}
\simeq {SU(3)_{\rm L+R} \over ({\mathbb Z}_3)_{\rm L+R}}
  \label{eq:HF}
\eeq
and the order parameter space is found to be 
\beq
{G_{\rm F} \over H_{\rm F}}
\simeq  {SU(3)_{\rm L-R} \times U(1)_{\rm A}
\over ({\mathbb Z}_3)_{\rm L-R+A}}
\simeq U(3)_{\rm L-R+A}.  \label{eq:GFHF}
\eeq

The low-energy effective field theory for the $U(3)_{\rm L-R+A}$  NG modes, 
the chiral Lagrangian at high baryon density and at zero temperature, was first obtained in Ref.~\cite{Casalbuoni:1999wu}.
Let us first restrict the generic field $\Sigma$ 
in Eq.~(\ref{eq:sigma0}) 
by the constraints \footnote{
Note that the GL theory can correctly describe physics at slightly lower temperatures than
the transition temperature. Only there does expansion in terms of the gap $\Delta_{\rm CFL}$ work well because
$\Delta_{\rm CFL}$ is sufficiently small and can be treated as an expansion parameter.
On the other hand, the chiral Lagrangian is an effective theory including only the derivative of massless zero modes;
it can describe physics at zero temperature even if it is not close to a transition temperature. 
}
\beq
\Sigma = \Phi_{\rm L}^\dagger \Phi_{\rm R},\quad 
\Phi_{\rm L}\Phi_{\rm L}^\dagger =  \Phi_{\rm R} \Phi_{\rm R}^\dagger =\mathbf 1_3. \label{eq:sigma}
\eeq
Furthermore, we decompose $\Sigma$ into $\SU(3)$ and $\U(1)$ parts as
\beq
\Sigma = \tilde \Sigma e^{i\varphi_{\rm A}},\quad
\tilde \Sigma =  \Phi_{\rm L}^\dagger  \Phi_{\rm R}.
\eeq
Then, the chiral Lagrangian is given by \cite{Casalbuoni:1999wu}
\beq
\Lag &=& \frac{f_\pi^2}{4}\Tr \left[\nabla_0\tilde \Sigma\nabla_0\tilde \Sigma^\dagger - v_\pi^2 \p_i\tilde \Sigma\p_i\tilde \Sigma^\dagger\right]
+ \frac{3f^2}{4}\left[\p_0V\p_0V^* - v_{\eta'}^2 \p_iV \p_iV^*\right] \\
&+& \left( A \Tr\left[M\tilde \Sigma^\dagger\right] V^* + {\rm h.c.}\right) \non
&+& \left( 
B_1 \Tr[ M\tilde \Sigma^\dagger]\Tr[M\tilde \Sigma^\dagger]V
+ B_2 \Tr [ M\tilde \Sigma^\dagger M\tilde \Sigma^\dagger] V
+ B_3 \Tr [ M\tilde \Sigma^\dagger]\Tr[M^\dagger\tilde \Sigma] + {\rm h.c.}
\right) + \cdots, \nonumber
\label{eq:chiral_lag}
\eeq
where we have omitted the electromagnetic interaction.
Here $M={\rm diag}(m_{\rm u},m_{\rm d},m_{\rm s})$ is a mass matrix, $\tilde \Sigma$ stands for the octet chiral field, 
and $V$ is the axial $U(1)_{\rm A}$ field
\beq
\tilde \Sigma = \exp\left(i\frac{\pi^a T^a}{f_\pi}\right),\quad
V = e^{i\varphi_{\rm A}} = \exp\left(i\frac{2\eta'}{\sqrt{6}f}\right),
\eeq
with $f_\pi$ and $f$ being the octet and singlet decay constants.
The covariant derivative of the time derivative for $\tilde \Sigma$ is defined by \cite{Bedaque:2001je}
\beq
\nabla_0 \tilde \Sigma = \p_0 \tilde \Sigma + i \left(\frac{MM^\dagger}{2\mu}\right) \tilde \Sigma - i \tilde \Sigma \left(\frac{M^\dagger M}{2\mu}\right).
\eeq
The decay constants and speed of mesons are obtained as \cite{Son:1999cm,Son:2000tu} 
\beq
f_\pi^2 = \frac{21-8\log2}{18} \, \frac{\mu^2}{2\pi^2}, \quad
f^2 = \frac{3}{4}\, \frac{\mu^2}{2\pi^2},\quad
v_\pi^2 = v_{\eta'}^2 = \frac{1}{3}.
\label{eq:decay_const_speed_of_meson}
\eeq
The mechanism of chiral symmetry breaking in the high baryon density is very different
from that at low densities. Indeed, the dominant contribution is not  the usual $\left<\bar\psi\psi\right>$
but a four fermion operator
$\left<(\bar\psi \psi)^2\right> \sim \left<\psi\psi\right>^2$. Thus, dominant contributions from the quark mass terms
appear as quadratic terms of $M$ in the chiral Lagrangian, namely the terms 
proportional to $B_i$. These coefficients are obtained as \cite{Son:1999cm,Son:2000tu,Schafer:2001za}
\beq
B_1 = - B_2 = \frac{3\Delta_{\rm CFL}^2}{4\pi^2},\quad B_3 = 0,
\eeq
with $\Delta_{\rm CFL}$ being the gap in the quasi-particle spectrum.
The terms linear in $M$ appear when $\left<\bar\psi\psi\right>$ is not zero. 
Although they vanish at the asymptotically high density limit, they appear through non-perturbative effects, instantons
\cite{Schafer:1999fe,Manuel:2000wm,Son:2000fh,Son:2001jm,Schafer:2002ty}. The coefficient $A$ is calculated
by Ref.~\cite{Schafer:2002ty} to the one-instanton level as
\beq
A = C_N \frac{8\pi^4}{3}\frac{\Gamma(6)}{3^6}\left[\frac{3\sqrt2\pi}{g_{\rm s}}\Delta_{\rm CFL}\left(\frac{\mu^2}{2\pi^2}\right)\right]^2
\left(\frac{8\pi^2}{g_{\rm s}^2}\right)^6\left(\frac{\Lambda}{\mu}\right)^{12}\Lambda^{-3},
\eeq
with $C_N = 0.466 \exp(-1.679 N_{\rm C})1.34^{N_{\rm F}}/(N_{\rm C}-1)!(N_{\rm C}-2)!$.
Note that $A$ is related to the chiral condensate by $A= - \left<\bar\psi\psi\right>/2$ \cite{Schafer:2002ty}.

From the chiral Lagrangian (\ref{eq:chiral_lag}), one can read the masses of meson spectra.
For instance, when all the quark masses are equal, $M = m {\bf 1}_3$, setting $\Sigma={\bf 1}_3$, we have
\beq
\Lag_{\eta'} 
&=& \frac{1}{2}(\p_0\eta')^2 - \frac{v_{\eta'}^2}{2}(\p_i \eta')^2 - V_{\rm 1-inst}(\eta'), \\
V_{\rm 1-inst}(\eta') &=& \left( \frac{2Am}{f^2} + \frac{4m^2B}{f^2}\right) (\eta')^2 + \cdots.
\eeq
Thus the mass of $\eta'$ is
\beq
m_{\eta'}^2\big|_{\rm 1-inst} = \frac{4A}{f^2}m + \frac{8B}{f^2}m^2.
\eeq
Therefore, in the chiral limit, $\eta'$ is massless at the one-instanton level \cite{Schafer:2002ty}.
The $\eta'$-mass in the chiral limit is generated by the two-instanton contribution. 
Although it is hard to evaluate it, the potential for $\eta'$ can be easily read as \cite{Schafer:2002ty}
\beq
V_{\rm sym.}(\varphi_{\rm A}) = -6mA \cos \varphi_{\rm A} - 12 m^2 B \cos \varphi_{\rm A} - 2C \cos 3\varphi_{\rm A},
\label{eq:V_mass_instanton}
\eeq
where we have set the $\theta$ angle of QCD to zero and $B_1=-B_2=B$.
The first term is the one-instanton contribution and the last term is the two-instanton contribution.
We leave $C$ as a free parameter in this work. ``sym.'' in the subscript indicates the flavor symmetric limit
$m_{\rm u} = m_{\rm d} = m_{\rm s} = m$.

Before closing this section, let us show the original derivation of the above chiral Lagrangian 
by Ref.~\cite{Casalbuoni:1999wu}. Let $\Phi_{\rm L}$ and $\Phi_{\rm R}$ be  coset fields for
$SU(3)_{\rm C} \times SU(3)_{\rm L}$ and $SU(3)_{\rm C} \times SU(3)_{\rm R}$, respectively:
\beq
\Phi_{\rm L}\Phi_{\rm L}^\dagger =  \Phi_{\rm R} \Phi_{\rm R}^\dagger =\mathbf 1_3,\quad
\det \Phi_{\rm L} = \det \Phi_{\rm R} = 1.
\eeq
Since an effective theory of QCD should have the same symmetry as QCD, the effective Lagrangian
derived in Ref.~\cite{Casalbuoni:1999wu} is of the form
\beq
\mathcal L &=& - \frac{f_{\pi}^2}{4}\Tr\left[
\left(\Phi_{\rm R}\p_0 \Phi_{\rm R}^\dagger - \Phi_{\rm L}\p_0 \Phi_{\rm L}^\dagger\right)^2
- v_\pi^2\left(\Phi_{\rm R}\p_i \Phi_{\rm R}^\dagger - \Phi_{\rm L}\p_i \Phi_{\rm L}^\dagger\right)^2
\right] \non
&-& C_\alpha \frac{f_\pi^2}{4} \Tr\left[
\left(\Phi_{\rm R}\p_i \Phi_{\rm R}^\dagger + \Phi_{\rm L}\p_i \Phi_{\rm L}^\dagger + 2 i g_{\rm s}A_0\right)^2
- v_\pi^2 \left(\Phi_{\rm R}\p_i \Phi_{\rm R}^\dagger + \Phi_{\rm L}\p_i \Phi_{\rm L}^\dagger + 2 i g_{\rm s}A_i\right)^2
\right]\non
&-& \frac{3f^2}{4} \left(|\p_0V|^2 - v_{\eta'}^2 |\p_iV|^2\right) - \frac{1}{4}\Tr\left[F_{\mu\nu}F^{\mu\nu}\right].
\label{eq:chiral_CG}
\eeq
Since we are interested in the low-energy dynamics, 
we integrate out the gluons $A_\mu$, which are heavy.
This can be done by neglecting the kinetic term of the gluons.
Then, in this Lagrangian, the gluon is just an auxiliary field and can be
eliminated by its equation of motion.
After eliminating the gluon, we are left with the chiral Lagrangian
\beq
\mathcal L &=& - \frac{f_{\pi}^2}{4}\Tr\left[
\left(\Phi_{\rm R}\p_0 \Phi_{\rm R}^\dagger - \Phi_{\rm L}\p_0 \Phi_{\rm L}^\dagger\right)^2
- v_\pi^2\left(\Phi_{\rm R}\p_i \Phi_{\rm R}^\dagger - \Phi_{\rm L}\p_i \Phi_{\rm L}^\dagger\right)^2
\right] \non
&-& \frac{3f^2}{4} \left(|\p_0V|^2 - v_{\eta'}^2 |\p_iV|^2\right)\non
&=& 
\frac{f_\pi^2}{4}\Tr\left[
\p_0\Sigma\p_0\Sigma^\dagger - v_\pi^2 \p_i\Sigma\p_i\Sigma^\dag \right]
- \frac{3f^2}{4} \left(|\p_0V|^2 - v_{\eta'}^2 |\p_iV|^2\right).
\eeq

Based on the formulations above, we discuss the topological solitons 
associated with the chiral symmetry breaking in Sec.~\ref{sec:global}.

%% file: vortices-v9.tex
\section{Vortices}\label{sec:vortices}

In this section we explain various kinds of 
vortices in the CFL phase. 
While we use the Ginzburg-Landau theory valid 
around the transition temperature $T\sim T_{\rm c}$ 
to study vortices,  
the chiral Lagrangian before integrating out gluon 
fields, Eq.~(\ref{eq:chiral_CG}), valid at 
zero temperature $T\sim 0$
can alternatively be used  
if we study vortices with singular cores 
at large distances.
In Sec.~\ref{sec:Abelian-vortices}, 
we discuss Abelian vortices, i.e., 
$\U(1)_{\rm B}$ superfluid vortices, and 
$\U(1)_{\rm A}$ vortices or axion strings. 
We also discuss non-topological color-magnetic flux tubes 
in Sec.~\ref{sec:nt-cmft}. 
In Sec.~\ref{sec:NA-vortices}, 
we introduce non-Abelian vortices, sometimes called semi-superfluid vortices. 
A peculiar feature of non-Abelian vortices, the existence of 
internal orientational zero modes (collective coordinates), is explained in detail.

\subsection{Abelian vortices}\label{sec:Abelian-vortices}

In this subsection, we give two concrete examples of Abelian vortices in the CFL phase.
The first one is Abelian $U(1)_{\rm B}$ global vortices \cite{Forbes:2001gj,Iida:2002ev}, which are topologically stable, while
the second one is 
$U(1)_{\rm A}$ vortices or axion strings
\cite{Forbes:2001gj}.
Since the authors in Ref.~\cite{Forbes:2001gj}
used the effective theory at $T\sim 0$ 
describing only massless particles,  
reviewed in Sec.~\ref{sec:eff_th_phonon}, 
the core structures of vortices cannot be described 
in the absence of massive particles. 
Here, we use the GL theory that can 
deal with the core structures of vortices.
We also study color magnetic fluxes \cite{Iida:2004if}, 
which are non-topological and decay immediately. 

\subsubsection{$\U(1)_{\rm B}$ superfluid vortices}\label{sec:U(1)B}

Let us discuss 
$\U(1)_{\rm B}$ superfluid vortices.
The phases of diquark condensates can be parametrized as
\beq
\Phi_{\rm L} = e^{i\theta_{\rm A} +i\theta_{\rm B}}\Delta_{\rm CFL} {\bf 1}_3 ,\quad
\Phi_{\rm R} = e^{-i\theta_{\rm A} +i\theta_{\rm B}}\Delta_{\rm CFL} {\bf 1}_3 .
\label{eq:phases_condensate}
\eeq
The phases $\ph_{\rm A}$ and $\ph_\B$ are the 
phases of 
the $\U(1)_{\rm A}$ and $\U(1)_{\rm B}$ symmetries, respectively.
Both phase symmetries $\U(1)_{\rm B}$ and  $\U(1)_{\rm A}$ 
are spontaneously broken when the diquarks are condensed,
while the $\U(1)_{\rm A}$ is also  broken explicitly by the chiral
anomaly, namely the instanton effects.
The spontaneous breaking of $\U(1)_{\rm B}$ and $U(1)_{\rm A}$ gives 
rise to $U(1)_{\rm B}$ and $U(1)_{\rm A}$ global vortices, respectively.

The $\U(1)_{\rm B}$ global vortices can be easily 
found in Eq.~(\ref{eq:gl}) by
requiring the condensate matrix to be proportional to the unit matrix 
$\Phi(x) = \phi(x) {\bf 1}_3/\sqrt 3$. 
Then, the Lagrangian is reduced to the following linear sigma model for the complex scalar field $\phi(x)$:
\beq
\Lag_{\U(1)_{\rm B}} = K_0 \left( |\p_0\phi|^2 - v_{\rm B}^2 |\p_i\phi|^2\right) - \left[
\alpha  |\phi|^2 + \frac{4\beta}{3} |\phi|^4 \right] + \frac{3\alpha^2}{16\beta},\quad
v_{\rm B} \equiv \sqrt{\frac{K_3}{K_0}} = \frac{1}{\sqrt{3}}.
\label{eq:gl_U(1)B}
\eeq
The effective field $\phi$ develops the VEV $\left<\phi\right> = \sqrt{3} \Delta_{\rm CFL}$
in the ground state, so that the $\U(1)_{\rm B}$ 
is spontaneously broken. 
Since the first homotopy group is non-trivial,
\beq
 \pi_1[U(1)_{\rm B}] \simeq {\mathbb Z},
\eeq
there exist topologically stable $k$ vortices in general with $k \in {\mathbb Z}$.

\begin{figure}[ht]
\begin{center}
\includegraphics[height=5cm]{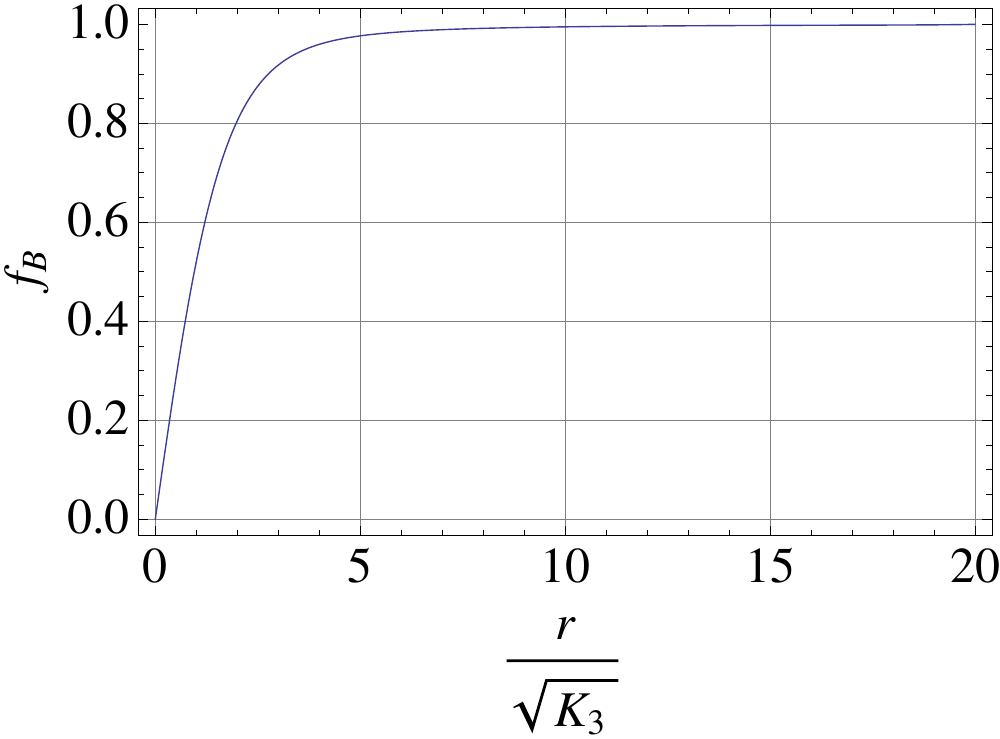}
\caption{
A profile function $f_{\rm B}(r)$ of a minimally winding ($k=1$) $U(1)_{\rm B}$ vortex. The parameters are
$\alpha = -1$ and $\beta = 2$ for simplicity.
}
\label{fig:fB}
\end{center}
\end{figure}

We make a standard ansatz for the axially symmetric $k$ vortex strings, which are infinitely 
long straight lines, say,  along the $x_3$ axis:
\beq
\phi (r,\theta)= \sqrt{3}\Delta_{\rm CFL} f_{\rm B}(r) e^{ik\theta},\quad 
\label{eq:sol_U(1)B}
\eeq
with the cylindrical coordinates $(r,\theta,z)$, i.e.,  
$x^1 + i x^2 \equiv r e^{i\theta}$.  
A numerically obtained profile function $f_{\rm B}$ is shown in Fig.~\ref{fig:fB}.

The circulation of the superfluid velocity 
in Eq.~(\ref{eq:U(1)Bsuperflow}) 
obeys the famous 
Onsager-Feynman quantization
\beq 
 c_B =  \oint d x^i J^{\rm B}_i 
= 2 \pi {4K_3 \over \gamma K_0}  k
\label{eq:circulation-U1}
\eeq
with the vortex number $k \in  \pi_1[U(1)_{\rm B}]$.

The Hamiltonian density with respect to the amplitude function $f_{\rm B}(r)$ is given by
\beq
{\mathcal H}_{\U(1)_{\rm B}} = 3\Delta_{\rm CFL}^3\left(
K_3 f_{\rm B}'{}^2 + \left(\frac{K_3k^2}{r^2} + \alpha \right) f_{\rm B}^2 + 4\beta\Delta_{\rm CFL}^2f_{\rm B}^4\right).
\label{eq:Hamiltonian_U(1)B}
\eeq
The $k$ static global string solution is obtained by solving the following equation of motion:
\beq
K_3 \left(f_{\rm B}'' + \frac{f_{\rm B}'}{r}\right) - \frac{k^2K_3}{r^2} f_{\rm B} - \left(\alpha + 8\beta \Delta_{\rm CFL}^2 f_{\rm B}^2\right)f_{\rm B} = 0,
\label{eq:eom_fB}
\eeq
with the boundary conditions $f_{\rm B}(0) = 0$ and $f_{\rm B}(\infty) \to 1$.
The tension of the string logarithmically diverges as
\beq
T_{\U(1)_{\rm B}} = 6\pi \Delta_{\rm CFL}^2 K_3 k^2 \log \frac{L}{\xi} + \cdots,
\label{eq:ene_U(1)B}
\eeq
where $\cdots$ stands for a finite contribution to the tension, 
$L$ is an IR cutoff scale representing the system size and $\xi \sim m_1^{-1}$ is a UV cutoff representing the size of the vortex core. 

Asymptotic solutions of a minimally winding vortex are given by
\beq
f_{\rm B} &=& c_1 (r/\sqrt{K_3}) + \cdots ,\qquad\qquad r \to 0,\\
f_{\rm B} &=&  1 - \frac{2}{-\alpha (r/\sqrt{K_3})^2} + \cdots,\qquad r \to \infty.
\eeq
Here the constant $c_1$ is known to be $c_1 = 0.58\cdots$ \cite{Vilenkin:1994}.

Note that if one takes quantum anomalies into account, an axial $U(1)_{\rm A}$ current runs inside the
$U(1)_{\rm B}$ vortex \cite{Son:2004tq}. This phenomenon is explained in Sec.~\ref{sec:anomaly}.

\subsubsection{$\U(1)_{\rm A}$ axial vortices \label{sec:U(1)A}}

There exists another vortex string in the CFL phase that is associated with the broken $\U(1)_{\rm A}$ symmetry at high density limit.
It is topologically stable because of non-trivial homotopy 
\cite{Forbes:2001gj,Iida:2002ev}
\beq
 \pi_1[ U(1)_{\rm A}] \simeq \mathbb{Z}.
\eeq 

Note that the instanton-induced potential given in Eq.~(\ref{eq:V_mass_instanton}) explicitly breaks $\U(1)_{\rm A}$, so that
a $\U(1)_{\rm A}$ vortex cannot exist alone but is 
always accompanied by sine-Gordon domain walls
at an intermediate density. 
This is explained in Sec.~\ref{sec:global}.

\subsection{Non-topological color-magnetic fluxes}
\label{sec:nt-cmft}

Here, we discuss color-magnetic flux tubes \cite{Iida:2004if} 
that are non-topological 
because of trivial homotopy group\footnote{
More precisely, gauge symmetry is 
spontaneously broken as 
$SU(3)_{\rm C} \times U(1)_{\rm EM} \to U(1)_0$, 
and the order parameter for gauge symmetry is 
$[SU(3)_{\rm C} \times U(1)_{\rm EM}] /U(1)_0 \simeq SU(3)$.
The first homotopy group is trivial for this $SU(3)$.
} 
\beq 
 \pi_1[SU(3)_{\rm C}] = 0
\eeq
and consequently are unstable.
A color magnetic flux is generated by one generator
of the $SU(3)_{\rm C}$ color group. 
Among all possible generators, 
$T_8$ gives the lightest fluxes, 
because it can be mixed with the electromagnetic 
$U(1)_{\rm EM}$ acting on $\Phi$ from its right.
The corresponding massive gauge field is 
given in Eq.~(\ref{eq:mix_gluon}).

To construct the vortex, we make the ansatz that all the off-diagonal elements of $\Ph$ are zero and denote
\beq
\Phi(x) = {\rm diag} \left(\phi_1(x),\frac{\phi_2(x)}{\sqrt2},\frac{\phi_2(x)}{\sqrt 2}\right).
\label{eq:ansatz:condense_mix}
\eeq
Then, the Hamiltonian density is reduced as
\beq
\tilde {\mathcal H} &=& \frac{1}{4 }(\mathcal F^{\rm M}_{ij})^2 + |\D_i \phi_1|^2 + |\D_i \phi_2|^2 + \tilde V,
\label{eq:hamiltonian_mix}\\
\tilde V &=& \alpha \left(|\phi_1|^2 + |\phi_2|^2\right) + \beta \left(|\phi_1|^2 + |\phi_2|^2\right)^2 +
\beta \left(|\phi_1|^4 + \frac{|\phi_2|^4}{2}\right),
\label{eq:pot_mix}
\eeq
with 
\beq
\D_i \phi_1 &=& \p_i \phi_1+ i \frac{2g_{\rm M}}{3} \mathcal A_i^{\rm M} \phi_1,
\label{eq:cov_deri_mix1}\\
\D_i \phi_2 &=& \p_i \phi_2 - i \frac{g_{\rm M}}{3} \mathcal A_i^{\rm M} \phi_2,
\label{eq:cov_deri_mix2}\\
\mathcal F^{\rm M}_{ij} &=& \p_i \mathcal A_j^{\rm M} - \p_j \mathcal A_i^{\rm M},
\label{eq:FS_mix_em}
\eeq
and
$g_{\rm M} = \sqrt{e^2 + 3\lambda_3g_{\rm s}^2/2\lambda_0}$ as defined in Eq.~(\ref{eq:g_mix}).
In the ground state, the fields develop the VEV as $\left<\phi_1\right> = \Delta_{\rm CFL}$ and
$\left<\phi_2\right> = \sqrt{2} \Delta_{\rm CFL}$, 
so that the $A^{\rm M}_\mu$ 
in Eq.~(\ref{eq:mix_gluon})  is massive 
and $A^0_\mu$ in Eq.~(\ref{eq:mix_photon}) remains massless, 
as discussed before.

Now, we are ready to construct the magnetic flux tube by making the axially symmetric ansatz by
\beq
\phi_1(r,\theta) = f_1(r)e^{-2i\theta},\quad
\phi_2(r,\theta) = f_2(r) e^{i\theta},\quad
A_i^{\rm M} (r,\theta) = -\frac{3 \epsilon_{ij}}{\tilde g} \frac{x^j}{r^2}\chi(r),
\label{eq:ansatz_em_vor}
\eeq
with the boundary condition
\beq
f_1(0) = f_2(0) = \chi(0) = 0,\quad
f_1(\infty) = f_2(\infty) = \chi(\infty) = 1.
\label{eq:bc_em_vor}
\eeq
The magnetic flux that this vortex carries 
can be calculated as 
\beq
\int d^2x\ \mathcal F^{\rm M}_{12} = \frac{3}{g_{\rm M}} \int d^2x\ \frac{\chi'}{r} = \frac{6\pi}{g_{\rm M}}.
\label{eq:em_flux}
\eeq
This solution was studied in Ref.~\cite{Iida:2004if}.
The flux in Eq.~(\ref{eq:em_flux}) is quantized but 
the solution is unstable and decays by turning on the other 
components of gauge fields. 

For later convenience,
let us set $\lambda_0 = \lambda_3 = 1$ and decompose the flux in Eq.~(\ref{eq:em_flux}) 
into gluon $\SU(3)_{\rm C}$ and electromagnetic 
$\U(1)_{\rm EM}$ parts as
\beq
F_{12}^8 = F^{\rm M}_{12} \sin \zeta,\quad
F^{\rm EM}_{12} = F^{\rm M}_{12} \cos \zeta,
\label{eq:FSs_mix}
\eeq
and take the limit $\zeta = \pi/2$, namely the limit where the electromagnetic interaction is ignored.
Then we find that this flux tube has the pure color-magnetic flux 
\beq
\Phi^8_{\rm A} = \int d^2x\ F_{12}^8 = \frac{4\sqrt{3}\pi}{g_{\rm s}}.
\label{eq:flux_abelian}
\eeq
Note that $F_{12}^8$ is not a gauge invariant quantity but $\tr[(F_{12})^2] = (F_{12}^8)^2$ is gauge invariant.

We should mention the stability of this flux tube. Since we chose one $\U(1)$ gauge orbit inside
the $\SU(3)$ group by hand, this flux tube is a kind of a trivial embedding of the $\U(1)$ local vortex
in a non-Abelian gauge theory. Of course, there are no reasons for such vortices to be stable,
because they are not protected by topology. In order to clarify the stability, one should work out
the small fluctuation analysis expanded around this solution.

Unstable color-flux tubes were also studied in the 2SC phase 
\cite{Alford:2010qf,Sedrakian:2013rr}.
Color-magnetic flux tubes in quark-gluon plasma 
were studied in Ref.~\cite{Liao:2007mj}, 
where they are claimed to be metastable.

\subsection{Non-Abelian vortices: 
stable color-magnetic flux tubes}\label{sec:NA-vortices}

In this section, we study 
non-Abelian vortices in the CFL phase
in the GL theory. 
While the use of the GL theory is limited near the
transition temperature where the gap $\Delta_{\rm CFL}$ is sufficiently small,
 the existence of non-Abelian vortices is not restricted 
to that region. 
One can also construct these vortices 
by using the chiral Lagrangian 
(\ref{eq:chiral_CG}) before integrating out the gluons, 
although the chiral Lagrangian cannot 
 describe the short distance structure at the center of the vortex, 
and the vortex configurations are singular at their core 
since we restricted $|\det \Phi_{\rm L}| = |\det\Phi_{\rm R}|=1$. 
We also study vortices in the BdG equation in Sec.~\ref{sec:fermion}, 
which is not restricted to any region.

\subsubsection{Minimal (M$_1$) non-Abelian vortices}\label{sec:M1}

In this subsection, 
we explain the so-called non-Abelian vortices, which are 
the most stable vortices in the CFL phase \cite{Balachandran:2005ev}. 
For simplicity, here we omit the mixing of the color and electric charges by simply setting $e=0$, 
which we take into account in Sec.~\ref{sec:elemag}.  
We consider only static configurations.
Therefore, we study solutions to the Euler-Lagrange equations associated with the Lagrangian
\beq
\Lag = 
\tr\!\!\left[- K_3\D_i \Phi^\dagger \D_i \Phi
- {F_{ij}^2 \over 4 \lambda_3}  \right]
-  \alpha \Tr\!\!\left(\Phi^\dagger \Phi \right)
- \beta \left(\left[\Tr(\Phi^\dagger\Phi)\right]^2 
- \Tr\!\!\left[(\Phi^\dagger\Phi)^2\right]\right)
+ \frac{3\alpha^2}{16\beta},
\label{eq:lag_NA_vor}
\eeq
where we discarded the terms irrelevant for the discussions below. 
In the rest of this section, we will set $\lambda_3 = 1$.
In order to get a static straight vortex string along the $x_3$-axis, we make the following ansatz:
\beq
\Phi(r,\theta) &=& \Delta_{\rm CFL} \left(
\begin{array}{ccc}
e^{i\theta} f(r) & 0 & 0 \\
0 & g(r) & 0 \\
0 & 0 & g(r)
\end{array}
\right),\label{eq:ansatz_NA_vortex}\\
A_i(r,\theta) &=& \frac{\epsilon_{ij}x^j}{g_{\rm s}r^2} \left(1-h(r)\right) \left(
\begin{array}{ccc}
-\frac{2}{3} & 0 & 0\\
0 & \frac{1}{3} & 0\\
0 & 0 & \frac{1}{3}
\end{array}
\right).\label{eq:ansatz_NA_vortex2}
\eeq
Similarly, we can take 
\beq
\Phi(r,\theta) &=& \Delta_{\rm CFL} \left(
\begin{array}{ccc}
g(r) & 0 & 0 \\
0 & e^{i\theta} f(r) & 0 \\
0 & 0 & g(r)
\end{array}
\right),\label{eq:ansatz_NA_vortex-b1}\\
A_i(r,\theta) &=& \frac{\epsilon_{ij}x^j}{g_{\rm s}r^2} \left(1-h(r)\right) \left(
\begin{array}{ccc}
\frac{1}{3} & 0 & 0\\
0 & -\frac{2}{3} & 0\\
0 & 0 & \frac{1}{3}
\end{array}
\right), \label{eq:ansatz_NA_vortex-b2}
\eeq
or 
\beq
\Phi(r,\theta) &=& \Delta_{\rm CFL} \left(
\begin{array}{ccc}
 g(r) & 0 & 0 \\
0 & g(r) & 0 \\
0 & 0 & e^{i\theta} f(r)
\end{array}
\right),\label{eq:ansatz_NA_vortex-c1}\\
A_i(r,\theta) &=& \frac{\epsilon_{ij}x^j}{g_{\rm s}r^2} \left(1-h(r)\right) \left(
\begin{array}{ccc}
\frac{1}{3} & 0 & 0\\
0 & \frac{1}{3} & 0\\
0 & 0 & -\frac{2}{3}
\end{array}
\right).\label{eq:ansatz_NA_vortex2-c2}
\eeq
These vortices were first found by 
Balachandran, Digal, and Matsuuta 
in Ref.~\cite{Balachandran:2005ev}.
They call them semi-superfluid 
non-Abelian vortices. 
For either case,
the boundary conditions for the profile functions $f,g,h$ at spacial infinity should be determined
in such a way that the configurations reach the ground state. Therefore, we impose the boundary
condition
\beq
f(\infty) = 1,\quad g(\infty) = 1,\quad h(\infty) = 0,
\label{eq:bc_NAvor_1}
\eeq
where the last condition implies that we are in the pure gauge at infinity.
We also need to fix the values of the fields at the origin,
which ensures that the solutions are regular there:
\beq
f(0) = 0,\quad g'(0) = 0,\quad h(0) = 1.
\label{eq:bc_NAvor_2}
\eeq

The decomposition of the $U(3)$ action 
in Eq.~(\ref{eq:ansatz_NA_vortex}) 
to 
the $U(1)_{\rm B}$ and $SU(3)_{\rm C}$ actions 
can be found as 
\begin{eqnarray} 
\Phi  = \Delta_{\rm CFL} 
\left(
\begin{array}{ccc}
e^{i\theta} f(r) & & \\
& g(r) & \\
& & g(r)
\end{array}
\right)
= \Delta_{\rm CFL} e^{\frac{i\theta}{3}} \left(
\begin{array}{ccc}
e^{\frac{2i\theta}{3}} f(r)& &\\
& e^{-\frac{i\theta}{3}} g(r) &\\
& & e^{-\frac{i\theta}{3}} g(r) 
\end{array}
\right) .
\label{eq:asym_NA_vor_phi0}
\end{eqnarray}
At the end $\theta=2\pi$ of the closed loop 
in the order parameter space,  
the $U(1)_{\rm B}$ contribution becomes 
$\omega = \exp (2\pi i /3)$. 
This factor is canceled by the $SU(3)_{\rm C}$ part 
$\diag  (\omega^2,\omega^{-1},\omega^{-1})
= \omega^{-1} {\bf 1}_3$.  
Only the $U(1)_{\rm B}$ part with fractional winding 
cannot make a closed loop, while a contribution from the $SU(3)_{\rm C}$ part makes it possible to 
have a closed loop. 
One observes that the presence of 
the center group 
$\mathbb Z_3$ of the $SU(3)_{\rm C}$ group is 
essential. 
This is the same for the other two configurations. 

Let us discuss the shape of the profile functions $f,g,h$. 
The corresponding Euler-Lagrange equations are given by
\beq
&&\left[ \triangle - \frac{(2h+1)^2}{9r^2}-\frac{m_1^2}{6}\left(f^2+2g^2-3\right) - \frac{m_8^2}{3}(f^2-g^2)\right] f = 0,
\label{eq:1}\\
&&\left[ \triangle - \frac{(h-1)^2}{9r^2}-\frac{m_1^2}{6}\left(f^2+2g^2-3\right) - \frac{m_8^2}{6}(f^2-g^2)\right] g = 0,
\label{eq:2}\\
&&h'' - \frac{h'}{r} - \frac{m_{\rm g}^2}{3}\left( g^2(h-1) + f^2(2h+1)\right) = 0.
\label{eq:3}
\eeq
Here the masses $m_1$, $m_8$ and $m_{\rm g}$ are given in Eq.~(\ref{eq:mass_CFL}).

To get the vortex configurations, we should solve the above 
coupled three ordinary differential equations. 
Although it is impossible to solve them analytically, 
we can solve them numerically. 

Before demonstrating the numerical solutions, let us examine the asymptotic behavior of the profile functions
that can be analytically solvable. 
From these asymptotics, 
we find several peculiar properties of the non-Abelian vortices in the CFL phase. 
To this end, we consider small fluctuations around the asymptotic values $(f,g,h) = (1,1,0)$ and define
\beq
\delta F(r) = (f(r)+2g(r)) - 3,\quad
\delta G(r) = f(r)-g(r)-0,\quad
\delta h(r) = h(r)-0.
\label{eq:fluct_NA_vor}
\eeq 
$\delta F(r)$ is the fluctuation of the trace part of $\Phi$ and $\delta G(r)$ is that for the traceless part proportional to $T_8$ as
\beq
\Phi = \Delta_{\rm CFL} {\bf 1}_3 
+ \Delta_{\rm CFL}  
\left(
\begin{array}{ccc}
\frac{1}{3} & & \\
& \frac{1}{3} & \\
& & \frac{1}{3}
\end{array}
\right) \delta F (x) 
+ \Delta_{\rm CFL} 
\left(
\begin{array}{ccc}
\frac{2}{3} & & \\
& -\frac{1}{3} & \\
& & -\frac{1}{3}
\end{array}
\right) \delta G (r) + \cdots.
\label{eq:fluct_NA_vor2}
\eeq
The linearized field equations for the fluctuations are given by
\beq
\left(\triangle - m_1^2 - \frac{1}{9r^2}\right) \delta F = \frac{1}{3r^2},\label{eq:asym_1}\\
\left(\triangle - m_8^2 - \frac{1}{9r^2}\right) \delta G = \frac{2}{3r^2}\delta h,\label{eq:asym_2}\\
\delta h'' - \frac{\delta h'}{r} - m_{\rm g}^2\delta h = \frac{2}{3}m_{\rm g}^2 \delta G.\label{eq:asym_3}
\eeq
An approximate solution of  the first equation is \cite{Eto:2009kg}
\beq
\delta F = q_1 \sqrt{\frac{\pi}{2m_1 r}}\, e^{-m_1 r}  - \frac{1}{3m_1^2 r^2} + O\left((m_1r)^{-4}\right).
\label{eq:asym_F}
\eeq
The first term is much smaller than the others which is usually neglected as in the case of the $U(1)$ global vortex.
The dominant terms decrease polynomially, which is a common feature of global vortices. 
At high baryon density where $m_{\rm g} \gg m_{1,8}$, the solutions of Eqs.~(\ref{eq:asym_2}) and (\ref{eq:asym_3})
are given by
\beq
\delta G = q_8 \sqrt{\frac{\pi}{2m_8r}}\,e^{-m_8 r},\quad
\delta h = - \frac{2}{3} \frac{m_{\rm g}^2}{m_{\rm g}^2 - m_8^2} \delta G.
\label{eq:asym_G}
\eeq
Here $q_{1,8}$ are some constants that should be determined numerically.
This behavior of $\delta h$ is counterintuitive because the gluon has the magnetic mass $m_{\rm g}$ by
the Higgs mechanism in the CFL phase, 
so naively one expects $\delta h \sim e^{-m_{\rm g} r}$.
Since the asymptotic behaviors of the vortex string are deeply related to the inter-vortex forces,
we expect that the inter-vortex forces of the non-Abelian vortices in CFL are quite different from those
in conventional metallic superconductors.\footnote{
The exponential tails of the famous Abrikosov-Nielsen-Olesen vortices \cite{Abrikosov:1956sx,Nielsen:1973cs}
in a conventional Abelian-Higgs model
are like $\exp(-m_{H}r)$ for the massive Higgs field $H$ with mass $m_H$ and 
$\exp(-m_e r)$ for the massive Abelian gauge boson with mass $m_e$.
The layer structures are exchanged for $m_e > m_H$ (type II) and $m_e < m_H$ (type I).
However, for the strong type II region $(m_e > 2m_H)$, the tail of the gauge boson becomes
$\exp(-2m_Hr)$ \cite{Plohr:1981cy,Perivolaropoulos:1993uj,Eto:2009wq}. The width of the gauge field cannot become 
smaller than half of that 
$1/m_H$ of the scalar field.
}
In the opposite case $m_{\rm g} \ll m_8$, the asymptotic behaviors are changed from Eq.~(\ref{eq:asym_G}) as
\beq
\delta G = - \frac{2q_{\rm g}}{3} \frac{1}{(m_8^2-m_{\rm g}^2)r^2}\sqrt{\frac{\pi}{2m_8r}}\, e^{-m_8 r},\quad
\delta h = q_{\rm g} \sqrt{\frac{\pi m_{\rm g} r}{2}}\, e^{-m_{\rm g}r}.
\eeq
In this case the asymptotic behaviors are governed by $m_{\rm g}$, which is smaller than $m_8$.
The coefficients $q_8$ and $q_{\rm g}$ depend on the masses $m_1$, $m_8$, and $m_{\rm g}$ and 
can be determined numerically \cite{Eto:2009kg}.

\begin{figure}[ht]
\begin{center}
\begin{minipage}[b]{0.32\linewidth}
\centering
\includegraphics[width=\textwidth]{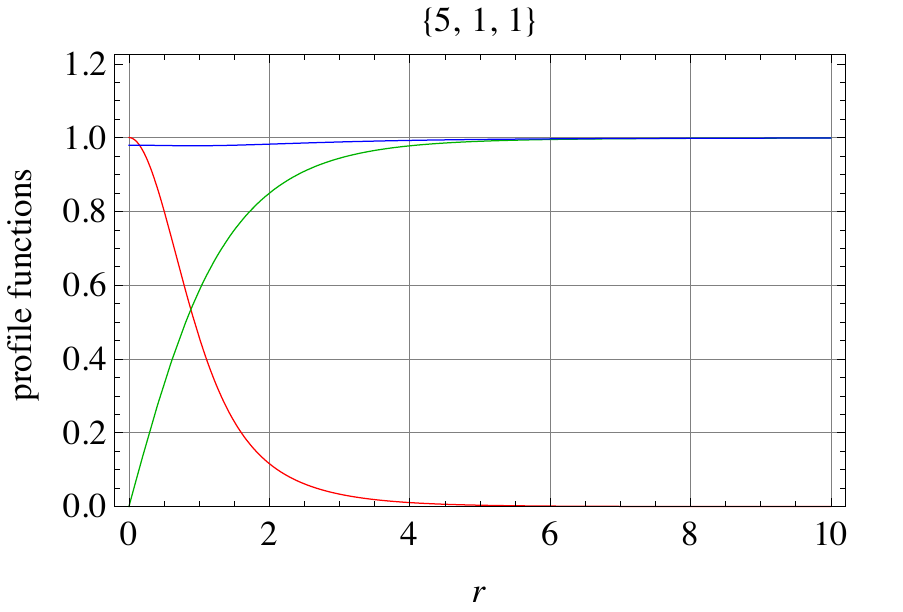}
\end{minipage}
\begin{minipage}[b]{0.32\linewidth}
\centering
\includegraphics[width=\textwidth]{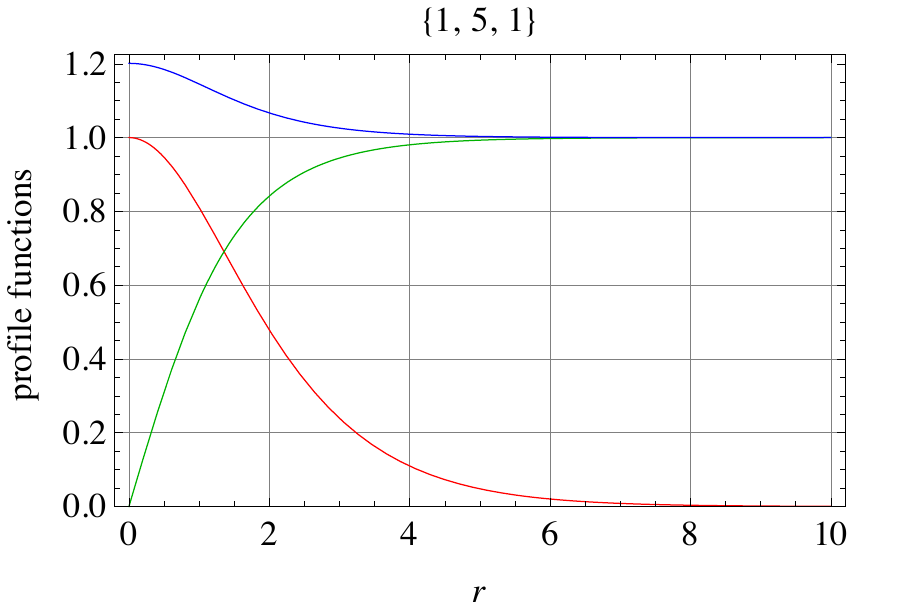}
\end{minipage}
\begin{minipage}[b]{0.32\linewidth}
\centering
\includegraphics[width=\textwidth]{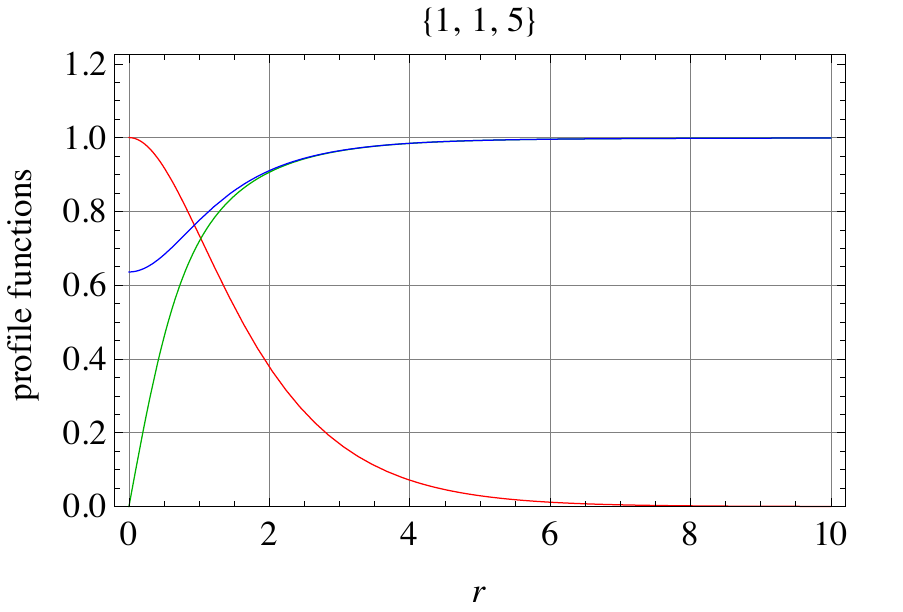}
\end{minipage}
\caption{The non-Abelian vortex profile functions $\{f(r),g(r),h(r)\}=\{\text{green},\text{blue},\text{red}\}$.
The mass parameters are set to be $\{m_{\rm g},m_1,m_8\} = \{5,1,1\}, \{1,5,1\}, \{1,1,5\}$. $g(r)$ is almost flat when
$m_1 \simeq m_8$.}
\label{fig:profiles}
\end{center}
\end{figure}

\begin{figure}[ht]
\begin{center}
\begin{minipage}[b]{0.65\linewidth}
\centering
\includegraphics[width=\textwidth]{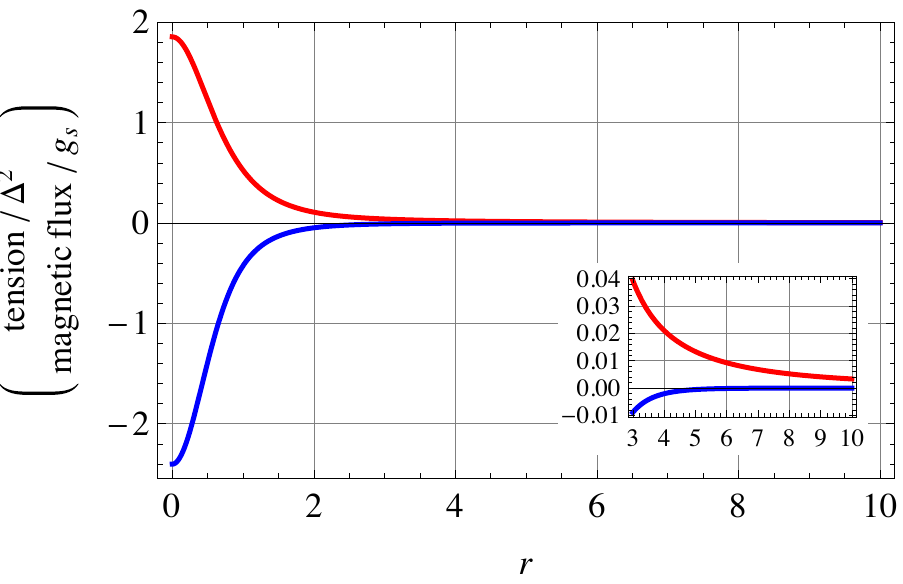}
\end{minipage}
\caption{Distributions of the energy density (red) and the non-Abelian magnetic flux (blue) for the minimally winding non-Abelian vortex. The parameters are chosen tto be $(m_{\rm g},m_1,m_8) = (10,2,1)$. While the energy density has a long tail, the magnetic flux whose transverse size is of order $m_8^{-1}$ 
exponentially converges to 0.}
\label{fig:basic_sol}
\end{center}
\end{figure}

The full numerical solutions for Eqs.~(\ref{eq:1}), (\ref{eq:2}), and (\ref{eq:3}) were obtained in Ref.~\cite{Eto:2009kg}.
It was found there that the shapes of the profile functions depend on the mass parameters
$m_{1,8,g}$. In particular, the value of $g(0)$ at the origin is quite sensitive to the ratio $m_1/m_8$.
Roughly speaking, $g(0)$ becomes larger than 1 for $m_1 > m_8$, and is smaller than 1 for $m_1<m_8$,
see Fig.~\ref{fig:profiles}.
Since $f(r)-g(r)$ plays the role of an order parameter for the breaking of $\SU(3)_{\rm C+F}$, information on 
the profile functions $f(r)$ and $g(r)$ is important. For the high baryon density region, we have $m_1 = 2 m_8 \ll m_{\rm g}$.
Therefore, the non-Abelian vortex in the CFL phase has $g(0)$ greater than 1, as shown in the right panel of Fig.\,\ref{fig:profiles}.
The profile functions $f(r),g(r),h(r)$, the energy, and the color-magnetic flux densities
for a particular choice of parameters $(m_1,m_8,m_{\rm g}) = (2,1,10)$ are
shown in Fig.\,\ref{fig:basic_sol}.
Note that $m_{\rm g}$ is much greater than $m_{1,8}$, nevertheless the color-magnetic flux has almost same width
as the scalar density distribution. One reason to explain this counterintuitive behavior is the asymptotic behavior
of the profile functions that we explained above. 
The asymptotic tail of the gluon is $\sim e^{-m_8 r}$, which 
depends only  on $m_8$ when
$m_8 < m_{\rm g}$. Therefore, the width of the flux density is not of order $m_{\rm g}^{-1} \sim 10$ but is of order $m_8^{-1}\sim 1$.

Let us compare the non-Abelian vortex explained in this section with the Abelian vortices in Sec.~\ref{sec:Abelian-vortices}.
The $U(1)_{\rm B}$ vortex is characterized by the first homotopy group $\pi_1[U(1)_{\rm B}]$.
As can be seen from Eq.~(\ref{eq:sol_U(1)B}), the $\U(1)_{\rm B}$ vortex has an integer winding number.
Therefore, this vortex is called an integer vortex.
The non-Abelian vortex is also characterized by  $\pi_1[\U(1)_{\rm B}]$, but its winding number takes the value
of a fractional number. 
The minimal winding number is quantized by $1/3$. 
The vortices with fractional winding
numbers are the so-called fractional vortices. 
To understand this, let us write the asymptotic behavior of $\Phi$ given in
Eq.~(\ref{eq:asym_NA_vor_phi0}) at $r \to \infty$ as
\begin{eqnarray} 
\Phi \sim \Delta_{\rm CFL} e^{\frac{i\theta}{3}} {\bf 1}_3.  
\label{eq:asym_NA_vor_phi} 
\end{eqnarray}
Here ``$\sim$'' stands for the equivalence under the $\SU(3)_{\rm C}$ gauge transformation.
This explicitly shows that the minimally winding non-Abelian vortex has 
only 1/3 winding number in the $\U(1)_{\rm B}$ space.
Consequently, 
the circulation of the superfluid velocity 
in Eq.~(\ref{eq:U(1)Bsuperflow}) 
obeys a fractional  
Onsager-Feynman quantization 
\beq 
 c_B =  \oint d x^i J^{\rm B}_i = {2 \pi \over 3}  {4K_3 \over \gamma K_0}  k
\label{eq:circulation-NA}
\eeq
with the vortex number $k \in  \pi_1[U(3)_{\rm C-F+B}]$. 
This is 1/3 of the circulation (\ref{eq:circulation-U1}) of a $U(1)_{\rm B}$ vortex.\footnote{
The rational Onsager-Feynman quantizations
are known for 
superfluid $^3$He \cite{PhysRevLett.55.1184,Salomaa:1987zz,volovik2009universe}, 
chiral $p$-wave superconductors 
\cite{PhysRevLett.55.1184,Salomaa:1987zz,volovik2009universe,
PhysRevLett.86.268,Jang14012011}, 
and spinor BECs \cite{Ho:1998zz,JPSJ.67.1822,Semenoff:2006vv,Kobayashi:2008pk},  
while irrational Onsager-Feynman quantizations
are known for 
multi-gap or multi-component superconductors \cite{Babaev:2001hv,Smiseth:2004na,Babaev:2004rm,Babaev:2007,
0295-5075-80-1-17002,JPSJ.70.2844,PhysRevLett.88.017002,Nitta:2010yf} 
and multi-component (non-spinor) BECs \cite{PhysRevLett.88.180403,Kasamatsu:2005,Eto:2011wp,Cipriani:2013nya,Cipriani:2013wia,Eto:2012rc,Eto:2013spa,Nitta:2013eaa}.
}

Finding the minimally winding solution is very important to 
find which configuration is the most stable in the CFL phase.
Since the non-Abelian vortex is a global vortex, 
its tension consists of two parts: a logarithmically divergent part and 
a finite part.
\begin{equation}
\mathcal T_{{\rm M}_1}=\mathcal T_{{\rm div;M}_1}
+\mathcal T_{{\rm fin;M}_1}.
\end{equation}
A dominant contribution to the tension (logarithmically divergent) comes 
from the kinetic term as
\beq
\mathcal T_{{\rm div;M}_1} \simeq K_3\int d^2x\ \tr \D_i\Phi (\D_i\Phi)^\dagger = \frac{1}{9} \times 6\pi \Delta_{\rm CFL}^2 K_3 \log \frac{L}{\xi}.
\label{eq:ene_NA}
\eeq
This should be compared with the tension of the $\U(1)_{\rm B}$ integer vortex given in Eq.~(\ref{eq:ene_U(1)B}).
Since the minimal winding number of the non-Abelian vortex is $1/3$, 
the tension of the non-Abelian vortex is $1/3^2 = 1/9$ times as large as
that of the $\U(1)_{\rm B}$ integer vortex. 
Thus, we conclude that the non-Abelian vortex is the most stable 
string configuration in the CFL phase.
In Sec.~\ref{sec:Abelian-vortex-decay}, 
we discuss how a $\U(1)_{\rm B}$ integer vortex
decays into non-Abelian vortices.

\begin{figure}[htbp]
\begin{center}
\includegraphics[width=8cm]{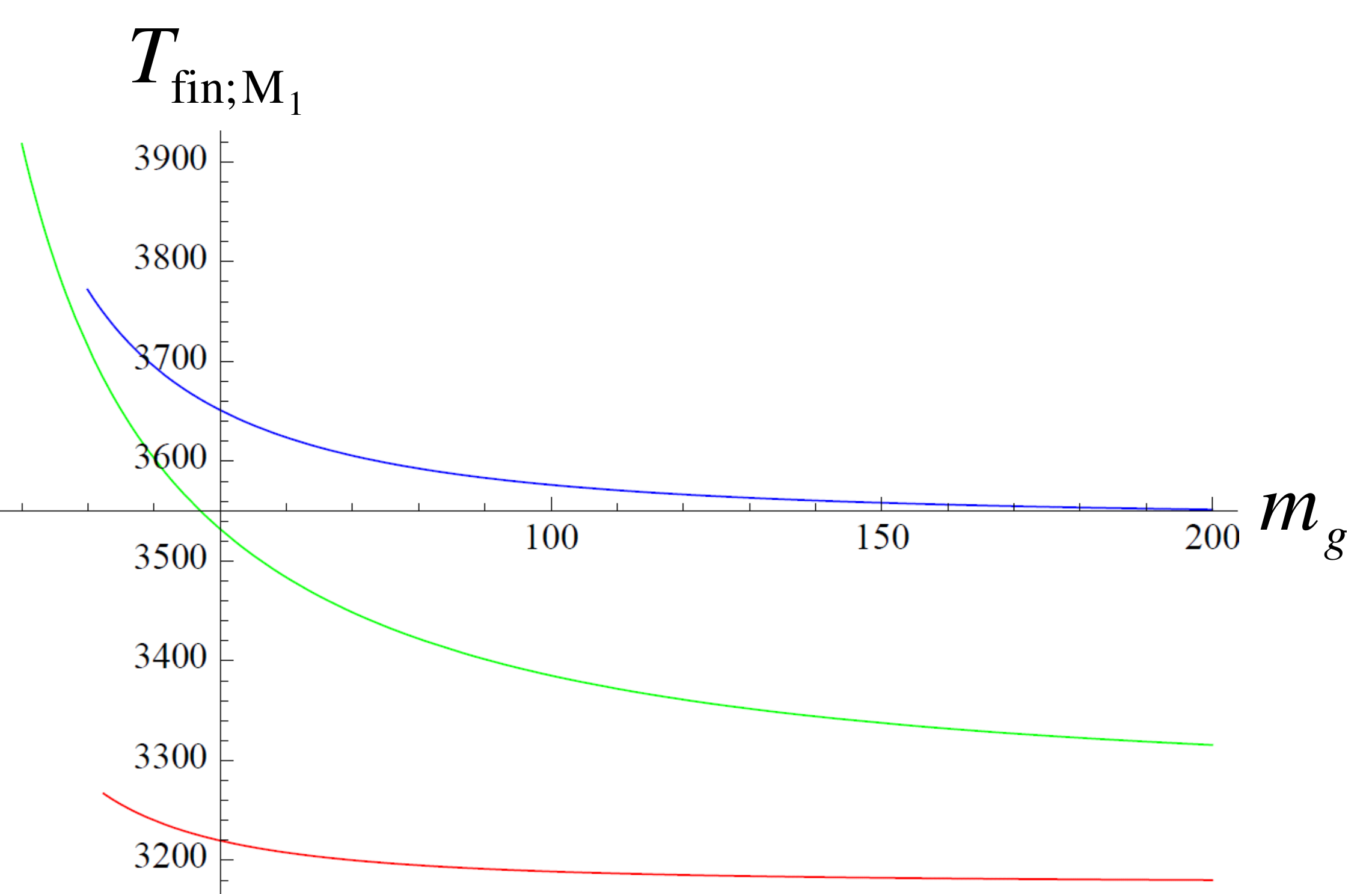}
\caption{The monotonic decreasing of the finite contribution to the tension of the vortex as function of the gauge coupling $g_{\rm s}$.
The plot is against the mass of the gauge bosons $m_{\rm g}$. 
The blue line corresponds to the ``realistic'' parameters: $\mu\sim 500$ MeV, $\Lambda\sim200$ MeV, $T_{\rm c}\sim 10$ MeV, $T\sim 0.9 T_{\rm c}$, which correspond to $\Delta_{\textsc{cfl}}=7$ MeV, $K_{1}=9$, $m_1=34$ MeV, $m_8=17$ MeV. 
The red and green lines correspond respectively to $(m_1,m_8)=(50$ MeV, $10$ MeV) and $(m_1,m_8)=(10$ MeV, $50$ MeV). The logarithmically divergent tension has been cut-off at a distance $L=0.4$ MeV$^{-1}$. Notice that the choice of $L$ is arbitrary. A different value would correspond to a logarithmic shift in the total tension, but the monotonic decrease of the tension would be the same, since it is only given by the dependence on $g_{\rm s}$ of the finite term $\mathcal T_{\rm fin;M_1}.$}
\label{fig:tension}
\end{center}
\end{figure}

Let us next consider the finite contribution to the vortex tension.
We discuss the dependence of the energy on the coupling constant $g_{\rm s}$. As shown in Eq.~(\ref{eq:ene_NA}) the logarithmically divergent contribution to the tension does not depend on the gauge coupling. However, finite corrections to the energy do depend  in general on the gauge coupling and on the various coefficients in the potential energy. 
These finite contributions have been numerically calculated 
as the behavior during a change in $g_{\rm s}$ in Ref.~\cite{Vinci:2012mc}.
Fig.~\ref{fig:tension} shows that the tension decreases monotonically with increasing gauge coupling. 
This behavior is generic and independent of the values of scalar masses. 
It can be intuitively understood by noticing the $1/g_{\rm s}^{2}$ dependence in the kinetic terms for the gauge potentials. 

Let us next compare the color-magnetic fluxes of the non-Abelian vortices and non-topological color fluxes discussed in Sec.~\ref{sec:nt-cmft}.
As shown in Eq.~(\ref{eq:flux_abelian}), the flux of the 8th direction in the $\SU(3)_c$ space of
the Abelian vortex (in the limit of $\zeta = \pi/2$) is $4\sqrt{3}\pi/g_{\rm s}$. With the ansatz (\ref{eq:ansatz_NA_vortex2}),
the amount of flux of the non-Abelian vortex can be easily obtained as
\beq
A_i = - \sqrt{\frac{2}{3}} \frac{\epsilon_{ij}x^j}{g_{\rm s}r^2}(1-h)T_8 \quad \to \quad 
\Phi_{\rm NA}^8 = \int d^2x~ F_{12}^8 = \frac{1}{3} \times \Phi_{\rm A}^8.
\label{eq:color_flux_NA}
\eeq
Thus, with respect to the color-magnetic flux, the non-Abelian vortex is the minimal configuration in the CFL phase.

\subsubsection{Non-minimal (M$_2$) non-Abelian vortices}\label{sec:M2}

A similar but slightly different vortex configuration from 
the minimal (M$_1$) vortices, called M$_2$ vortices, 
was also found in Ref.~\cite{Balachandran:2005ev}.
The ansatz for the M$_2$ vortex is quite similar to that in Eqs.~(\ref{eq:ansatz_NA_vortex}) 
and (\ref{eq:ansatz_NA_vortex2}) as
\beq
\Phi(r,\theta) &=& \Delta_{\rm CFL} \left(
\begin{array}{ccc}
q(r) & 0 & 0 \\
0 & e^{i\theta}p(r) & 0 \\
0 & 0 & e^{i\theta}p(r)
\end{array}
\right),\label{eq:ansatz_M2}\\
A_i(r,\theta) &=& -\frac{\epsilon_{ij}x^j}{g_{\rm s}r^2} \left(1-h(r)\right) \left(
\begin{array}{ccc}
-\frac{2}{3} & 0 & 0\\
0 & \frac{1}{3} & 0\\
0 & 0 & \frac{1}{3}
\end{array}
\right).\label{eq:ansatz_M2_2}
\eeq
The ansatz for $\Phi$ can be rewritten as
\beq
\Phi(r,\theta) = \Delta_{\rm CFL}  e^{i\frac{2\theta}{3}}
\left(
\begin{array}{ccc}
e^{-i\frac{2\theta}{3}} & 0 & 0 \\
0 & e^{i\frac{\theta}{3}} & 0 \\
0 & 0 & e^{i\frac{\theta}{3}}
\end{array}
\right)
\left(
\begin{array}{ccc}
q(r) & 0 & 0 \\
0 & p(r) & 0 \\
0 & 0 & p(r)
\end{array}
\right).
\label{eq:decompose_M2}
\eeq
This shows that an M$_2$ vortex has the 2/3 winding in the $\U(1)_{\rm B}$ space, which is twice as large as
that of an M$_1$ vortex. 
For the rest, the ansatz goes into the
$\SU(3)_{\rm C}$ orbit; more explicitly, $S^1 \subset \SU(3)_{\rm C}$ which is generated by $T_8$. 
In comparison with the M$_1$ vortex
given in Eqs.~(\ref{eq:ansatz_NA_vortex}) 
and (\ref{eq:ansatz_NA_vortex2}), 
one finds that the circulation of the M$_2$ vortex goes 
in the opposite direction in the $\SU(3)_{\rm C}$ orbit. 
Namely, the color of the M$_1$ 
vortex in Eq.~(\ref{eq:ansatz_NA_vortex}) is $\bar r = gb$, 
while that of the M$_2$ vortex in Eq.~(\ref{eq:decompose_M2}) is 
$r = \overline{gb}$.
Like the M$_1$ vortex, the tension of the M$_2$ vortex consists of a divergent part and a finite part.
The divergent part is given by
\beq
{\mathcal T}_{{\rm div;M}_2} = \frac{4}{9} \times 6\pi^2 \Delta_{\rm CFL}^2 K_3 \log \frac{L}{\xi}.
\eeq
This is four times as large as the divergent part of the tension of the M$_1$ vortex.
This implies that an M$_2$ vortex with a red flux 
decays into two M$_1$ vortices with green and blue colors
with opposite directions, 
as illustrated in Fig.~\ref{fig:split}(b).

On the other hand, the color-magnetic flux contributions to the tension are almost the same 
as those of the M$_1$ vortex. 
This is because the differences  are just $\bar r$ or $r$.

Notice that an M$_1$ vortex with the winding number 2 has the same divergent tension
as $\mathcal T_{{\rm div;M}_2}$ because of
\beq
\Phi(r,\theta) = \Delta_{\rm CFL}  e^{i\frac{2\theta}{3}}
\left(
\begin{array}{ccc}
e^{i\frac{4\theta}{3}} & 0 & 0 \\
0 & e^{-i\frac{2\theta}{3}} & 0 \\
0 & 0 & e^{-i\frac{2\theta}{3}}
\end{array}
\right)
\left(
\begin{array}{ccc}
f_2(r) & 0 & 0 \\
0 & g_2(r) & 0 \\
0 & 0 & g_2(r)
\end{array}
\right).
\label{eq:decompose_M1x2}
\eeq
For the finite contribution of the color-magnetic flux to the energy, the M$_2$ vortex has lower cost
since the M$_1$ vortex with  winding number 2 has 
a color-magnetic flux twice as great as that of an M$_2$ vortex.

\subsubsection{Orientational zero modes 
of non-Abelian vortices}\label{sec:orientational}

Minimal (M$_1$) non-Abelian vortices

The vortices in the CFL phase are called non-Abelian vortices because they carry color magnetic fluxes, 
or equivalently the color gauge $SU(3)_{\rm C}$ transformation participate in the loop in the order 
parameter space.
The three configurations in Eqs.~(\ref{eq:ansatz_NA_vortex})--(\ref{eq:ansatz_NA_vortex-b2}) carry 
corresponding color magnetic fluxes.
Since these configurations cannot be transformed into each other by the gauge transformation, these are all
physically different vortices. 
However, note that the above three vortices 
do not exhaust
all possible configurations. 
Indeed, there exists a continuous family of 
an infinite number of vortex configurations.
To see it, let us decompose the ansatz 
Eqs.~(\ref{eq:ansatz_NA_vortex}) and  (\ref{eq:ansatz_NA_vortex2})
as
\beq
\Phi = \Delta_{\rm CFL} {\rm diag}(f e^{i\theta},g,g) \sim \Delta_{\rm CFL} e^{\frac{i\theta}{3}} \left( \frac{F}{3}{\bf 1}_3 + \sqrt{\frac{2}{3}}GT_8\right),
\label{eq:NA_decomp}
\eeq
where we have introduced 
\beq
 F \equiv f + 2g, \quad G \equiv f - g
\eeq 
and have transformed the configuration by a certain 
$\SU(3)_{\rm C}$ gauge symmetry. 
Continuously degenerate configurations can be obtained by the color-flavor rotation as
\beq
\Phi \to \Delta_{\rm CFL} e^{\frac{i\theta}{3}} 
 \left( \frac{F}{3}{\bf 1}_3 + \sqrt{\frac{2}{3}}GU T_8 U^\dagger \right),\quad U \in \SU(3)_{\rm C+L+R}.
\eeq
The three configurations in Eqs.~(\ref{eq:ansatz_NA_vortex})--(\ref{eq:ansatz_NA_vortex-b2}) are 
particular diagonal configurations. 
It is now obvious that the color-flavor $\SU(3)_{\rm C+L+R}$ symmetry of the ground state in the CFL phase is {\it spontaneously}
broken in the presence of a non-Abelian vortex because of 
the term proportional to $T_8$ in Eq.~(\ref{eq:NA_decomp}). 
In the vortex core, there remains 
the unbroken symmetry 
\beq
  K = \U(1)_{\rm C+L+R} \times \SU(2)_{\rm C+L+R}
\eeq 
where 
$U(1)_{\rm C+L+R}$ is the Abelian subgroup generated by $T_8$ and $SU(2)_{\rm C+L+R}$ is a $SU(2)$ subgroup that
commutes with $U(1)_{\rm C+L+R}$.
Therefore, there appear, 
 in the {\it vicinity} of the vortex,
Nambu-Goldstone (NG) modes 
associated with this spontaneous symmetry breaking: 
\beq
{H_{\rm CFL} \over K} = 
\frac{\SU(3)_{\rm C+L+R}}{\U(1)_{\rm C+L+R} \times \SU(2)_{\rm C+L+R}} \simeq \mathbb{C}P^2.
\label{eq:quotient}
\eeq
This space is known as 
the 2D complex projective space.
These modes are called the orientational moduli 
(collective coordinates) of the non-Abelian vortex. 
The existence of these modes was first pointed out 
in Ref.~\cite{Nakano:2007dr} in the context of dense QCD, 
but it was known before in the context 
of supersymmetric QCD, as summarized in Appendix \ref{sec:susy}. 
Points on the $\mathbb{C}P^2$ manifold correspond to 
the different fluxes. 
One can move  from point to point by $\SU(3)_{\rm C+L+R}$.

The order parameter of the breaking of $\SU(3)_{\rm C+L+R}$ is $|G| = |f-g|$. Since both $f$ and $g$ asymptotically reach $1$ at the boundary,
the color-flavor locked symmetry is not broken far from the non-Abelian vortex. On the other hand, as can be seen in Fig.~\ref{fig:basic_sol},
$|G|$ takes a maximum value at the center of the non-Abelian vortex. Hence, we expect that the orientational NG modes are localized 
on the non-Abelian vortex. 
In Sec.~\ref{sec:LEEA}, we prove that the orientational modes are in fact normalized 
and propagate along the non-Abelian vortex as gapless excitations. 
We construct the low-energy effective theory of these modes 
on the vortex world-volume. 

Non-minimal (M$_2$) non-Abelian vortices

Let us next consider the orientation of the M$_2$ vortex.
Going away from the M$_2$ vortex, one asymptotically reaches the ground state where 
the $\SU(3)_{\rm C+L+R}$ color-flavor symmetry holds. On the other hand, since 
the condensation field $\Phi$ at the center of the M$_2$ vortex is
\beq
\Phi \propto {\rm diag}(1,0,0),
\label{eq:center_M2}
\eeq
the color-flavor symmetry which does not change this is the same as that for the M1 vortex.
Namely, $\U(1)_{\rm C+L+R} \times \SU(2)_{\rm C+L+R}$.
There the M$_2$ vortex also has the orientational zero modes
\beq
{H_{\rm CFL} \over K} = 
\frac{\SU(3)_{\rm C+L+R}}{\U(1)_{\rm C+L+R} \times \SU(2)_{\rm C+L+R}} \simeq \mathbb{C}P^2.
\label{eq:quotient_M2}
\eeq

However, 
the symmetry restored at the center of an M$_2$ vortex is not
$\U(1)_{\rm C+L+R} \times \SU(2)_{\rm C+L+R}$ but is enlarged to
$\U(1)_{\rm C+L+R} \times \SU(2)_{\rm L+R} \times \SU(2)_{\rm C}$.
The restoration of the gauge symmetry $\SU(2)_{\rm C}$ is in sharp contrast to the M$_1$ vortex, for which
no gauge symmetries are recovered.

%% file: dynamics-v9.tex
\section{Dynamics of vortices} \label{sec:dynamics}

The dynamics of non-Abelian semisuperfluid vortices are 
generally similar to those of 
vortices in superfluids and atomic BEC, 
which have been studied extensively \cite{0953-8984-13-12-201,RevModPhys.81.647,
Kasamatsu2009351}. 
(From a field theoretical point of view, 
see Refs.~\cite{0951-7715-11-5-006,mantontopological}.)
In Sec.~\ref{sec:translational}, we construct the effective 
field theory of translational zero modes, known as Kelvin modes, 
and study dynamics of 
a single vortex string in terms of the low-energy effective theory.
In Sec.~\ref{sec:intervortex-force},
the intervortex force between two non-Abelian vortices is  derived 
and dynamics of two vortices and a vortex ring is summarized. 
In Sec.~\ref{sec:Abelian-vortex-decay},
we discuss the decaying process of a $U(1)_{\rm B}$ Abelian superfluid vortex (M$_2$ vortex) into a set of three (two) non-Abelian vortices. 
The creation of vortices and formation of a colorful lattice of 
non-Abelian vortices under rotation are discussed in Sec.~\ref{sec:lattice}.
In Sec.~\ref{sec:relativistic-superfluid}, 
we discuss the relation between relativistic strings 
in relativistic scalar field theories and superfluid vortices.

\subsection{The translational zero modes (Kelvin modes)}\label{sec:translational}
\subsubsection{The effective theory of translational zero modes}
In this subsection, we study dynamics of vortices 
in terms of the effective field theory. 
Here, we concentrate on translational zero modes, 
which are common for $U(1)_{\rm }$ Abelian vortices 
and non-Abelian vortices, so that we calculate in 
$U(1)_{\rm }$ Abelian vortices. 
The dynamics of orientational zero modes 
is studied in the next section. 

The existence of a vortex spontaneously breaks 
the translational symmetry. 
When a vortex is placed along the $z$-axis, 
it breaks two translational symmetries in the $x$-$y$ plane.
When the vortex string fluctuates,  
the configuration can be written as
\begin{align}
\Phi = f(\bar{r}) e^{i (\bar{\theta} + \alpha)}, \quad
\bar{r} = \sqrt{(x - X)^2 + (y - Y)^2}, \quad
\bar{\theta} = \tan^{-1}\left(\frac{y - Y}{x - X}\right), \label{eq:Kelvin-ansatz}
\end{align} 
where  $X = X(t,z)$ and $Y = Y(t,z)$ denote the position
of the vortex in the $x$-$y$ plane which is a field 
in the vortex world-sheet, 
and $(\bar r, \bar \theta)$ are the polar coordinates from the vortex center. 
Inserting Eq.~\eqref{eq:Kelvin-ansatz} into Eq.~\eqref{eq:tdgl}, 
and 
integrating the Lagrangian density over the $x$-$y$ plane, 
we obtain the effective theory \cite{Kobayashi:2013gba}
\beq
{\cal L}_{\rm eff} = - {\cal T} 
+ 4 \pi \gamma K_0 (Y \del_t X - X \del_t Y ) 
- {{\cal T}\over 2} [(\del_z X)^2 +(\del_z Y)^2],
\label{eq:eff-th-Kelvin}
\eeq
up to the quadratic order of $X$ and $Y$ 
and the leading order in derivatives. 
Here, ${\cal T}$ is the tension of the vortex, 
given in Eq.~\eqref{eq:ene_U(1)B} for 
a $U(1)_{\rm B}$ vortex and 
Eq.~\eqref{eq:ene_NA} for a non-Abelian vortex.
For a vortex in $d=2+1$, the second term 
can be found in Refs.~\cite{0951-7715-11-5-006,mantontopological}. 
The first and third terms in Eq.~(\ref{eq:eff-th-Kelvin}) 
are consistent with the spatial part of the Nambu-Goto action 
\cite{Nambu:1974zg,Goto:1971ce} at this order:  
\beq
 {\cal L}_{\rm NG} = - {\cal T} \sqrt {1 - (\del_z X)^2 - (\del_z Y)^2}.
\eeq

\begin{figure}[ht]
\begin{center}
\begin{minipage}[b]{0.5\linewidth}
\centering
\includegraphics[width=\textwidth]{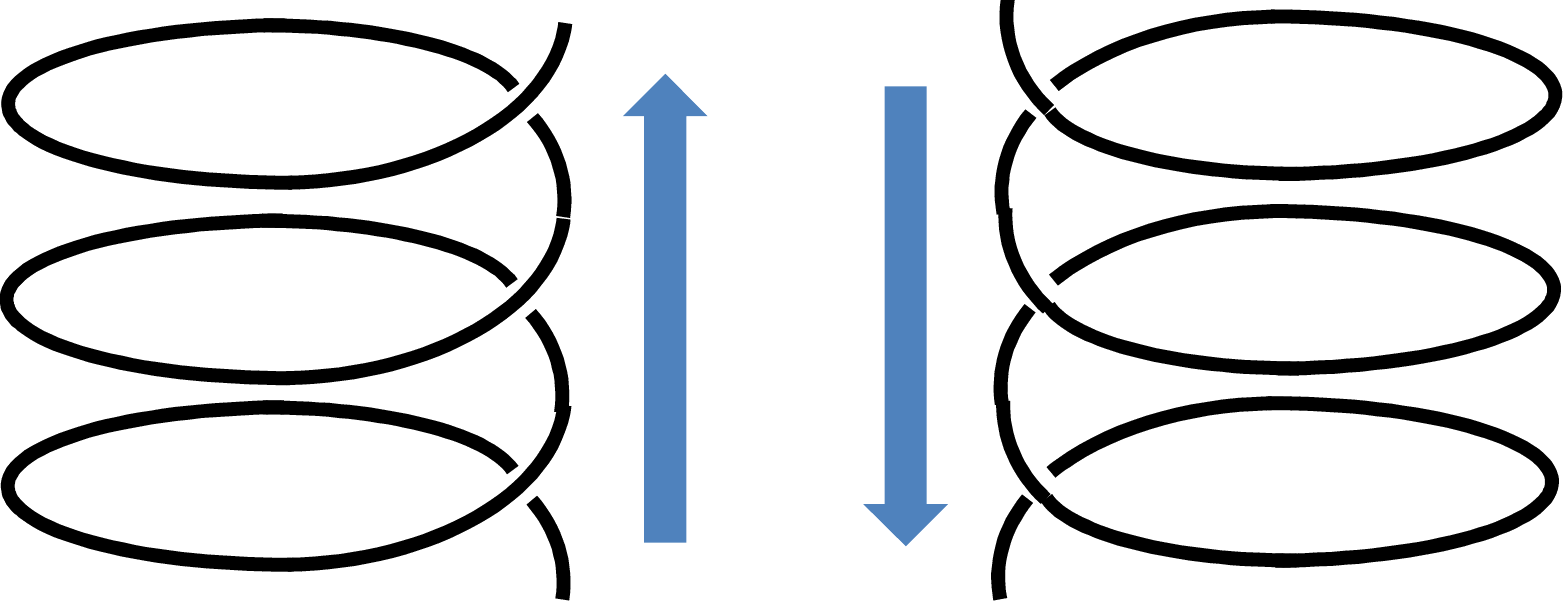}
\end{minipage}
\caption{Kelvin waves. 
The Kelvin waves in Eqs.~(\ref{eq:kelvin-wave1}) 
and (\ref{eq:kelvin-wave2}) 
propagate in definite directions indicated by the arrows.
\label{fig:kelvin-wave}}
\end{center}
\end{figure}

Let us remark on the effective Lagrangian in Eq.~(\ref{eq:eff-th-Kelvin}). 
Because the time derivative is at the first order in the Lagrangian, 
$X$ and $Y$ are not independent fields; 
rather, they are momentum conjugate to each other: 
\beq 
 P_X = {\del {\cal L}\over \del (\del_t X)} \sim Y, \quad 
 P_Y = {\del {\cal L}\over \del (\del_t Y)} \sim X.
\eeq 
Therefore, although two translational symmetries are broken, 
there appears only one independent gapless Nambu-Goldstone mode,  known a Kelvin mode or Kelvon if quantized.
There are two typical solutions of the effective 
Lagrangian (\ref{eq:eff-th-Kelvin}):
\beq
& X_1 = A \cos(k_z z - \omega t + \delta_1), \quad
Y_1= A \sin(k_z z - \omega t + \delta_1),  \label{eq:kelvin-wave1}\\
& X_2 = A \sin(k_z z + \omega t + \delta_2), \quad
Y_2= A \cos(k_z z + \omega t + \delta_2),  \label{eq:kelvin-wave2}
\eeq
where $\delta_1$ and $\delta_2$ are arbitrary constants.
The first and second solutions show the clockwise and counterclockwise spiral Kelvin waves 
propagating along the vortex string in opposite directions,  
as illustrated in Fig.~\ref{fig:kelvin-wave}.
The Kelvin waves have ``chirality": 
the (counter)clockwise waves can propagate 
only in one (the other) direction, 
which is well known in superfluids.

Since $X$ and $Y$ are not independent 
but are momentum conjugate to each other, 
there is only one dispersion relation
\beq
\omega = { {\cal T} \over 4 \pi \gamma K_0}  k_z^2 , 
\eeq
which is quadratic. 
The Kelvon is an example of type-II Nambu-Goldstone modes.\footnote{
\label{footnote:NG-modes}
When a continuous symmetry is spontaneously broken in 
{\it relativistic} field theories, there appear 
as many Nambu-Goldstone modes as the number of broken generators (for internal symmetries), but it is not the case 
for {\it non-relativistic} field theories. 
The type-I and II Nambu-Goldstone modes are defined to 
be gapless modes with linear and quadratic dispersion relations, respectively. 
For spontaneously broken {\it internal} symmetries, 
it has been proved in Refs.~\cite{Watanabe:2012hr,Hidaka:2012ym} 
that one type-I Nambu-Goldstone mode corresponds to one broken generator and one type-II Nambu-Goldstone mode corresponds to two broken generators,
showing the saturation of the equality of 
the Nielsen-Chadha inequality \cite{Nielsen:1975hm}, 
$N_{\rm I} + 2 N_{\rm II} \geq N_{\rm BG}$,  
where $N_{\rm I}$, $N_{\rm II}$ and $N_{\rm BG}$ 
are the numbers of 
type-I and   
type-II Nambu-Goldstone modes, 
and spontaneously broken generators, respectively.
However, 
there is no general statement for space-time symmetries 
for either relativistic or non-relativistic theories. 
The presence of a vortex breaks the rotational and translational 
symmetries but the former do not give 
independent Nambu-Goldstone modes 
\cite{Ivanov:1975zq} 
(see also Ref.~\cite{Clark:2002bh}),
since a rotation can be reproduced by 
infinitesimal local translations \cite{Low:2001bw}. 
Then, the equality of the Nielsen-Chadha inequality is saturated for both relativistic and non-relativistic case 
if one counts only translational modes 
\cite{Kobayashi:2013gba}.  
See also Ref.~\cite{Nitta:2013mj} 
for discussion of Nambu-Goldstone modes for 
space-time symmetry in the presence of a vortex. 
} 

Defining $\psi \equiv X+iY$, 
the effective Lagrangian in Eq.~(\ref{eq:eff-th-Kelvin}) 
can be rewritten as 
\beq
{\cal L}_{\rm eff.} = - {\cal T} 
+ 2 \pi i \gamma K_0 (\psi^* \del_t \psi - \psi \del_t \psi^*)
- {{\cal T}\over 2} |\del_z \psi|^2 .
\label{eq:eff-th-Kelvin2}
\eeq

The two comments are addressed here.
The translational modes are exact Nambu-Goldstone modes 
only for an infinite system size. For a finite system, 
a gap as a correction is present in 
the effective theory \cite{Kobayashi:2013gba}:
\beq
 {\cal L}_{\rm gap} = {\pi \over R^2} (X^2 + Y^2),
\eeq
which is ``tachyonic.'' 
However, this does not imply 
the instability in non-relativistic cases. 
Instead, the chirality is broken because of this term :
Kelvin waves with wave length longer than 
some critical length propagate in 
a direction  opposite to that of modes with shorter lengths,
which is contrary to conventional understanding. 

When one includes higher order terms as the next-leading order, 
 one obtains the nonlinear term $V \sim |\psi|^4$ as the potential 
in the effective Lagrangian.  
The Eular-Lagrange equation of the 
total Lagrangian is a nonlinear Schr\"{o}dinger equation, 
which is integrable and and admits soliton solutions. 
These solitons describe  
nonlinear waves propagating along the vortex string, 
known as 
Hasimoto solitons \cite{hasimoto1972soliton}.

\subsubsection{Magnus and inertial forces}
\label{sec:magnus_inert}
Here, we restrict ourselves to 2+1 dimensions. 
The translational zero modes $X$ and $Y$ are free fields 
in the leading order. 
Let us consider multiple vortices 
with the position $(X_i, Y_i)$ of the $i$th vortex.  
The effective Lagrangian of interacting vortices 
can be written as 
\beq
{\cal L}_{\rm eff}  
= 
\sum_i \left[
- {\cal T} 
+ 4 \pi \gamma K_0 (Y_i \del_t X_i - X_i \del_t Y_i ) 
\right]
- E_{\rm int} (X_i,Y_i) 
\label{eq:eff-th-multi-vortex}
\eeq
where $E_{\rm int}$ is the interaction energy among 
vortices,  
which is calculated in the next subsection.
The interaction of vortices with 
an external superfluid velocity 
can be introduced by the Galilei transformation 
\beq
 (X',Y') = (X,Y) - (J^{\rm B, ext}_x,J^{\rm B, ext}_y) t
\eeq
with an external superfluid velocity  $J^{\rm B, ext}_{x,y}$: 
\beq
{\cal L}_{\rm eff}  
&=& 
\sum_i \left[
- {\cal T} 
+ 4 \pi \gamma K_0 (Y_i \del_t X_i - X_i \del_t Y_i ) 
\right]
\non
&& 
- E_{\rm int} (X_i,Y_i) 
- 4 \pi \gamma K_0 \sum_i (Y_i J^{\rm B, ext}_x -X_i J^{\rm B, ext}_y ).
\label{eq:eff-th-multi-vortex2}
\eeq 
The equations of motion in the presence of contributions 
from other vortices and an external flow 
can be written as 
\beq
&& 4 \pi \gamma K_0 {\del X^i \over \del t} 
= + {\del E_{\rm int} \over \del Y^i}  + 4 \pi \gamma K_0 J^{\rm B, ext}_x \non
&& 4 \pi \gamma K_0 {\del Y^i \over \del t}  
= - {\del E_{\rm int} \over \del X^i} +  4 \pi \gamma K_0 J^{\rm B, ext}_y.
\label{eq:eom-XY0}
\eeq
The first terms on the right-hand sides are called the inertial force.

\begin{figure}[ht]
\begin{center}
\begin{minipage}[b]{0.5\linewidth}
\centering
\includegraphics[width=\textwidth]{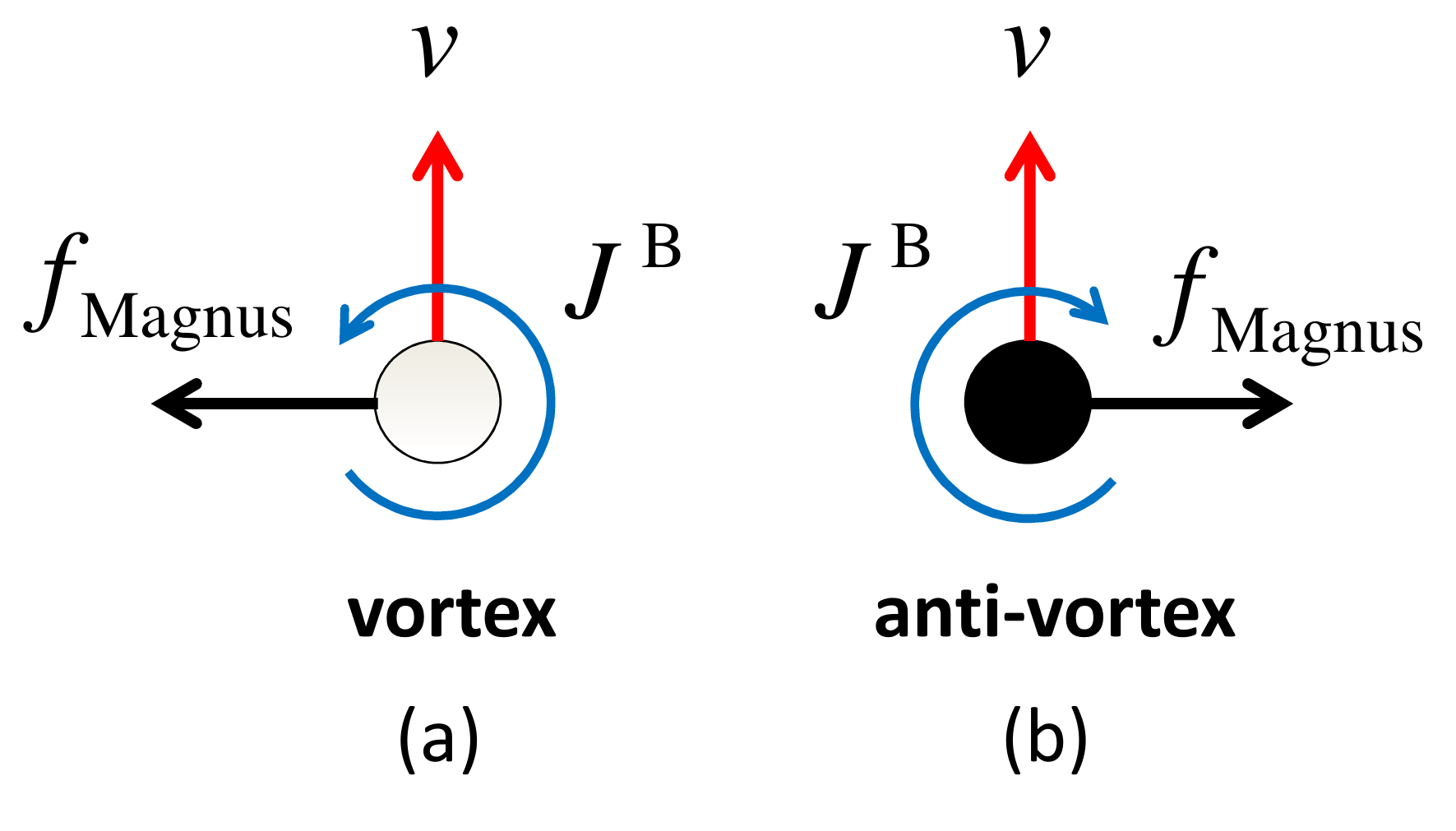}
\end{minipage}
\caption{Magnus force. 
$v$ is the velocity of a moving vortex or anti-vortex. 
$f_{\rm Magnus}$ is the Magnus force. 
The trajectories of the vortex and anti-vortex are bent 
into the direction of  $f_{\rm Magnus}$, 
as a curve ball.
\label{fig:magnus}}
\end{center}
\end{figure}

The dissipation is present 
at finite temperature
as the term proportional to $K_{\rm D}$  
in Eq.~(\ref{eq:dissipation}).  
It is known that the contributions from the dissipation 
can be taken into account in  
the equations of motion in Eq.~(\ref{eq:eom-XY0}) as 
\cite{0953-8984-13-12-201,RevModPhys.81.647,
Kasamatsu2009351} 
\beq
&& 4 \pi \gamma K_0 {\del X^i \over \del t} 
= + {\del E_{\rm int} \over \del Y^i}  
+ 4 \pi \gamma K_0 J^{\rm B, ext}_x 
- K_{\rm D} {\del E_{\rm int} \over \del X^i} - K_{\rm D}  J^{\rm B, ext}_y ,  \non
&& 4 \pi \gamma K_0 {\del Y^i \over \del t}  
= - {\del E_{\rm int} \over \del X^i} +  4 \pi \gamma K_0J^{\rm B, ext}_y
- K_{\rm D} {\del E_{\rm int} \over \del Y^i} + K_{\rm D}  J^{\rm B, ext}_x .
\label{eq:eom-XY}
\eeq 
The third and fourth terms on the right hand sides are 
called the Magnus forces, 
both proportional to the coefficient $K_{\rm D}$ 
of the dissipation term 
in Eq.~(\ref{eq:dissipation}). 
In particular, the third term is a contribution 
from the other vortices and the fourth term 
is a contribution from the external flow. 
Note that the Magnus forces exist only when 
the dissipation $K_{\rm D}$ is present at finite temperature.  
When a vortex moves with velocity $v$, there is an external flow in the rest frame of the vortex. 
Therefore, the vortex feels the Magnus force from the external flow as illustrated in Fig.~\ref{fig:magnus}.
The trajectories of the vortex and anti-vortex are bent 
into the directions of  $f_{\rm Magnus}$, 
as a curve ball in fluid dynamics. 

\subsection{Interaction between non-Abelian vortices}\label{sec:intervortex-force}

In this subsection, we study the interaction between two vortices with general orientation 
in the internal space at large distance 
\cite{Nakano:2007dr,Nakano:2008dc}.

\subsubsection{Intervortex force}
We first take one vortex 
as a reference $\Phi_0$, 
\begin{eqnarray}
 \Phi_0  = {\rm diag}(e^{i\theta}f,g, g).\label{eq:reference}
\end{eqnarray} 
Then, the other vortex  $\Phi$ with general orientation 
in ${\mathbb C}P^2$ relative to the reference vortex 
should be obtained 
by $SU(3)_{\rm C+F}$ transformation to $\Phi_0$. 
When we consider two vortices only,  
only an $SU(2)_{\rm C+F} \left( \subset SU(3)_{\rm C+F} \right)$ rotation 
is enough to be considered for relative orientation to $\Phi_0$
without loss of generality:
\begin{eqnarray}
 \Phi 
 = \Phi_0 U_{\rm F}
 =\left(\begin{array}{cc}
   \left(\begin{array}{cc}
     e^{i\theta} f & 0 \\
                 0 & g \\  
   \end{array} \right)
  u_{\rm F}^{-1} & 0 \\
               0 & g  
 \end{array} \right)
= 
 \left(\begin{array}{cc}
  \left(\begin{array}{cc}
   e^{i\theta} a f & e^{i\theta} b f \\
            -b^* g & a^* g \\  
  \end{array} \right) & 0  \\
                    0 & g 
\\
 \end{array} \right),
  \label{phigene}
\end{eqnarray}
where 
$u_{\rm F} \equiv 
\left(\begin{array}{cc}
a^* & -b \\
b^* & a \\  
\end{array} \right)$
(with $|a|^2 + |b|^2 = 1$) is an element of $SU(2)_{\rm F}$. 
This corresponds to a ${\mathbb C}P^1$ submanifold in the whole ${\mathbb C}P^2$. 
Any color gauge transformation keeps 
the physical situation unchanged if they are regular. 
Here we implement a {\it twisted} color transformation 
of $SU(2)_{\rm C}$, given by
\begin{equation}\label{twist1}
u_{\rm C}(\theta, r)=
\left(\begin{array}{cc}
a^* & -b e^{i \theta F(r) }  \\
b^* e^{-i \theta F(r)}  & a 
\end{array} \right) 
\end{equation}
with $F(r)$ being an arbitrary regular function 
with boundary conditions  $F(0)=0$ and $F(\infty)=1$. 
The former condition has been imposed 
to make the transformation regular at the center of string. 
This is possible because 
$\pi_1 [SU(2)_{\rm C}] = 0$.
The upper left 
$2 \times 2$ minor matrix of 
$\Phi$ in Eq.~(\ref{phigene}) is transformed to 
\begin{eqnarray}\label{phi2twist1}
u_{\rm C}(r, \theta)
\left(\begin{array}{cc}
e^{i\theta} a f & e^{i\theta} b f \\
-b^* g & a^* g \\  
\end{array} \right)
&=&
\left(\begin{array}{cc}
  |a|^2 f e^{i \theta}+|b|^2 g e^{i\theta F} 
& a^* b \left[-e^{i\theta F}+f e^{i\theta} \right] \\
a b^* \left[-1+f e^{i(1-F)\theta} \right] 
&|a|^2 g+|b|^2 f e^{i(1-F)\theta }   
 \end{array} \right) \nonumber \\
&\simeq&  
\left(\begin{array}{cc}
 e^{i \theta} & 0 \\
 0 & 1 
 \end{array} \right) 
\quad \mbox{for} \quad r \gg m_1^{-1},m_8^{-1},m_{\rm g}^{-1}. 
\end{eqnarray}
This result means that $\Phi \simeq \Phi_0$ 
for the large distance $r \gg m_1^{-1},m_8^{-1},m_{\rm g}^{-1}$. 
Also, the fully opposite orientation can be obtained 
by another color gauge transformation, 
\begin{equation}\label{twist2}
\left( 
\begin{array}{cc}
e^{-i \theta F(r)}  & 0 \\
0 & e^{i \theta F(r)} 
 \end{array} 
\right)
u_{\rm C}
\left( 
\begin{array}{cc}
f e^{i \theta}  & 0 \\
0 & g 
 \end{array} 
\right)u_{\rm F}^{-1} 
\simeq 
\left( 
\begin{array}{cc}
1  & 0 \\
0 &  e^{i \theta}
 \end{array} 
\right) 
\quad \mbox{for} \quad {r \gg m_1^{-1},m_8^{-1},m_{\rm g}^{-1}}. 
\end{equation}
Note that one cannot change the topological number 
by use of this kind of regular gauge transformation. 
We have thus seen that spatial infinity of the string configurations 
is the same and does not depend on the orientational zero modes.
This is just a consequence of the fact that the orientational zero modes are normalizable.

All semi-superfluid vortices with general orientation 
are equivalent to each other away from the core. 
In other words, 
vortices rotated by the flavor $SU(2)_{\rm F}$ revert to 
the reference vortex given by Eq.~(\ref{eq:reference}) 
via the color gauge transformation at much longer distances than the coherence length. 
This fact simplifies the problem of the static long range force 
between two strings significantly significantly.

\begin{figure}
\begin{center}
\includegraphics[width=0.4\linewidth]{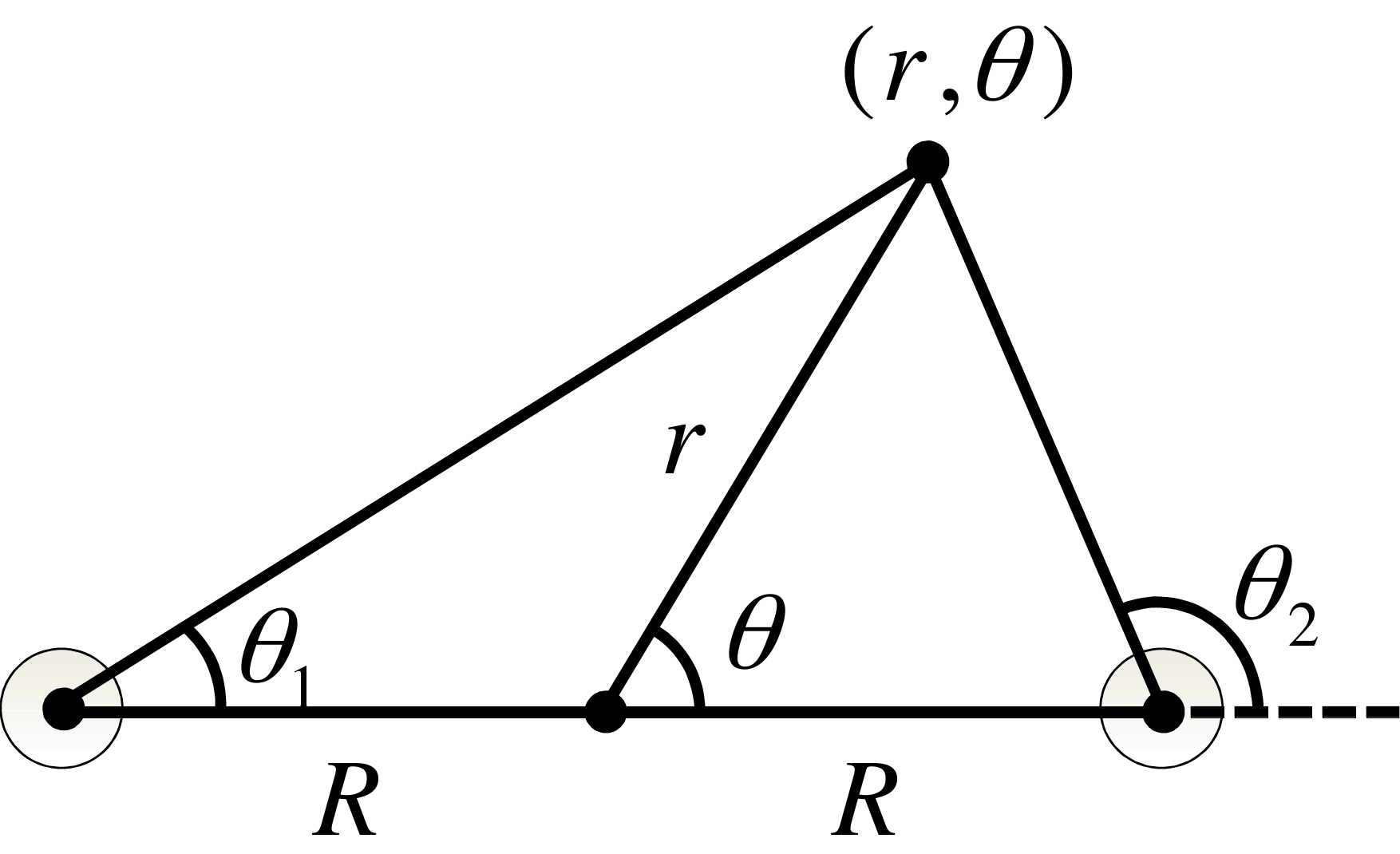}
\end{center}
\caption{\label{fig:interaction} 
A configuration of two semi-superfluid non-Abelian vortices 
with interval $2R$ 
in the polar coordinates $(r, \theta)$.
$r_{1,2}$ is the distance from the vortices $\Phi_{1,2}$, 
and $\theta_{1,2}$ is the angle around them. }
\end{figure}

Here, we consider the interaction between arbitrary two vortices $\Phi_1$ and $\Phi_2$.
placed at $(r, \theta)=(R,\pi)$ and $(R,0)$ 
in parallel along the $z$-axis; see Fig.~\ref{fig:interaction}. 
We eventually decide on  
the expression of a long range static force between two strings, 
which is valid if the strings are sufficiently separated. 
The interval between strings is much larger than both 
the coherence length 
and the penetration depth: 
$R \gg m_{1}^{-1}, m_{8}^{-1},m_{\rm g}^{-1}$. 
The first vortex $\Phi_{1}$ 
is approximated everywhere by the asymptotic profile (\ref{eq:reference}): 
$\Phi_{1}={\rm diag}\left( e^{i\theta_{1}}, 1, 1 \right)$. 
The second string $\Phi_{2}$ has the profile (\ref{phigene}) 
with general orientation relative to $\Phi_1$. 
At the large distance of  interest, however, 
it is equivalent to the reference string configuration (\ref{eq:reference}): 
$\Phi_{2} \simeq {\rm diag}\left( e^{i\theta_{2}}, 1, 1 \right)$. 
$\Phi_{1,2}$ becomes an anti-string by changing the signs of $\theta_{1,2}$.

The total profile of the two string system is given 
by the Abrikosov ansatz: 
\beq
 \Phi_{\rm tot}=\Phi_1\Phi_2, \quad
 A^{\theta}_{\rm tot}=A^{\theta}_1+A^{\theta}_2.
\eeq
The first ansatz does not depend on the ordering of the matrices 
because the second vortex transforms to diagonal at large distances  
as shown in Sec.~3.
$A^{\theta}_{1,2}$ is the gauge field configuration 
accompanied with the single string system of $\Phi_{1,2}$. 
For an anti-vortex, $A^{\theta}_{1,2}$ changes the signs.

In order to obtain the static force between them, 
we first calculate the interaction energy density of the two string system, 
which is obtained by subtracting two individual string energy 
densities  
from the total configuration energy density; 
\begin{eqnarray}\label{F}
{\cal E}_{\rm int}(r,\theta,R)
&\simeq&  \Tr
\left( |D \Phi_{\rm tot}|^2 -|D \Phi_1|^2
-|D \Phi_2|^2  \right) \nonumber \\
&=& \pm
\frac{2}{3}
\left[ 
\frac{-R^2 + r^2}
{R^4 + r^4 -2 R^2 r^2 \cos (2 \theta)}
\right], 
\end{eqnarray}
where we have neglected the potentials $V(\Phi_{\rm tot})=V(\Phi_1)=V(\Phi_2)=0$ 
and the field strength $F_{ij}^aF^{aij}=0$ at large distance \cite{Perivolaropoulos:1991du}. 
Here and below, the upper (lower) sign indicates the quantity 
for the vortex-vortex (vortex-anti-vortex) configuration.

The tension, the energy of the string per unit length,  
is obtained for $R \ll L$ by integrating the energy density
over the $x$-$y$ plane as
\begin{eqnarray}\label{E}
E_{\rm int}(R,L)= \pm
\int _{0}^{L} dr 
\int _{0}^{2 \pi} d\theta  r {\cal E}_{\rm int}(r,\theta,R)
= \pm \frac{2 \pi}{3} 
\left[ -\log 4 -2 \log R  + \log \left(R^2 + L^2 \right)  \right],
\label{eq:int-energy}
\end{eqnarray}
where the system size $L$ has been 
introduced as the IR cutoff.  
The force between the two vortices is obtained by 
differentiating $E$ by the interval:
\begin{eqnarray}\label{eq:force}
 f(a,L)
 = {\mp} 
 \frac{\partial E_{\rm int}}{2 \partial R}
 =  \pm \frac{2 \pi}{3} 
\left(\frac{1}{R} - \frac{R}{R^2 + L^2}  \right) 
\simeq  \pm \frac{ 2 \pi }{3 R},
\end{eqnarray}
where the last expression is for $L \to \infty$.
We can see that the force is repulsive (attractive) 
for the vortex-vortex (vortex-anti-vortex) configuration. 
The overall factors $1/3$ in Eqs.~(\ref{F})--(\ref{eq:force}) are attributed 
to the fact 
that the tension of the fundamental non-Abelian vortex 
is reduced by $1/3$ compared to the usual Abelian vortex, 
then leading to $1/3$ erosion in magnitude of the force.

Note that the result does not depend on whether the superconductivity 
is of type I or II. 
This has an important meaning in the case of color superconductivity
since, although the perturbation theory indicates 
the color superconductivity is of type I for the whole density regime
\cite{Giannakis:2003am},
the most fundamental strings, semi-superfluid strings, 
can be stable at any density regime where the CFL phase is realized.
This result also implies that 
global $U(1)_{\rm B}$ superfluid vortices 
\cite{Forbes:2001gj,Iida:2002ev} 
$\Phi \simeq {\rm diag} (e^{i \theta},e^{i \theta},e^{i \theta})$ 
studied in Sec.~\ref{sec:U(1)B}
as well as the M$_2$ vortices  \cite{Balachandran:2005ev} 
$\Phi \simeq {\rm diag} (1,e^{-i \theta},e^{-i \theta})
\simeq (e^{2i \theta},1,1)$ 
studied in Sec.~\ref{sec:M2} are both 
unstable to decay into 
$3$ or 2 semi-superfluid vortices, respectively,  
as discussed in Sec.~\ref{sec:Abelian-vortex-decay}.

This contrasts with the case of global non-Abelian vortices  \cite{Nakano:2007dq} discussed in Sec.~\ref{sec:global}, 
where the $U(1)$ Abelian string is marginally unstable, 
$i.e.$, 
no force exists between two strings with opposite orientations.

The force between two vortices at a short distance 
remains an important problem. 
In this regard, the intervortex force at arbitrary 
distance was calculated in a related model 
in which $U(1)_{\rm B}$ is gauged \cite{Auzzi:2007wj}.

\subsubsection{Dynamics of two vortices 
and a vortex ring}  \label{sec:dynamics-two}

\begin{figure}[ht]
\begin{center}
\centering
\begin{tabular}{cc}
\includegraphics[width=0.5\linewidth]{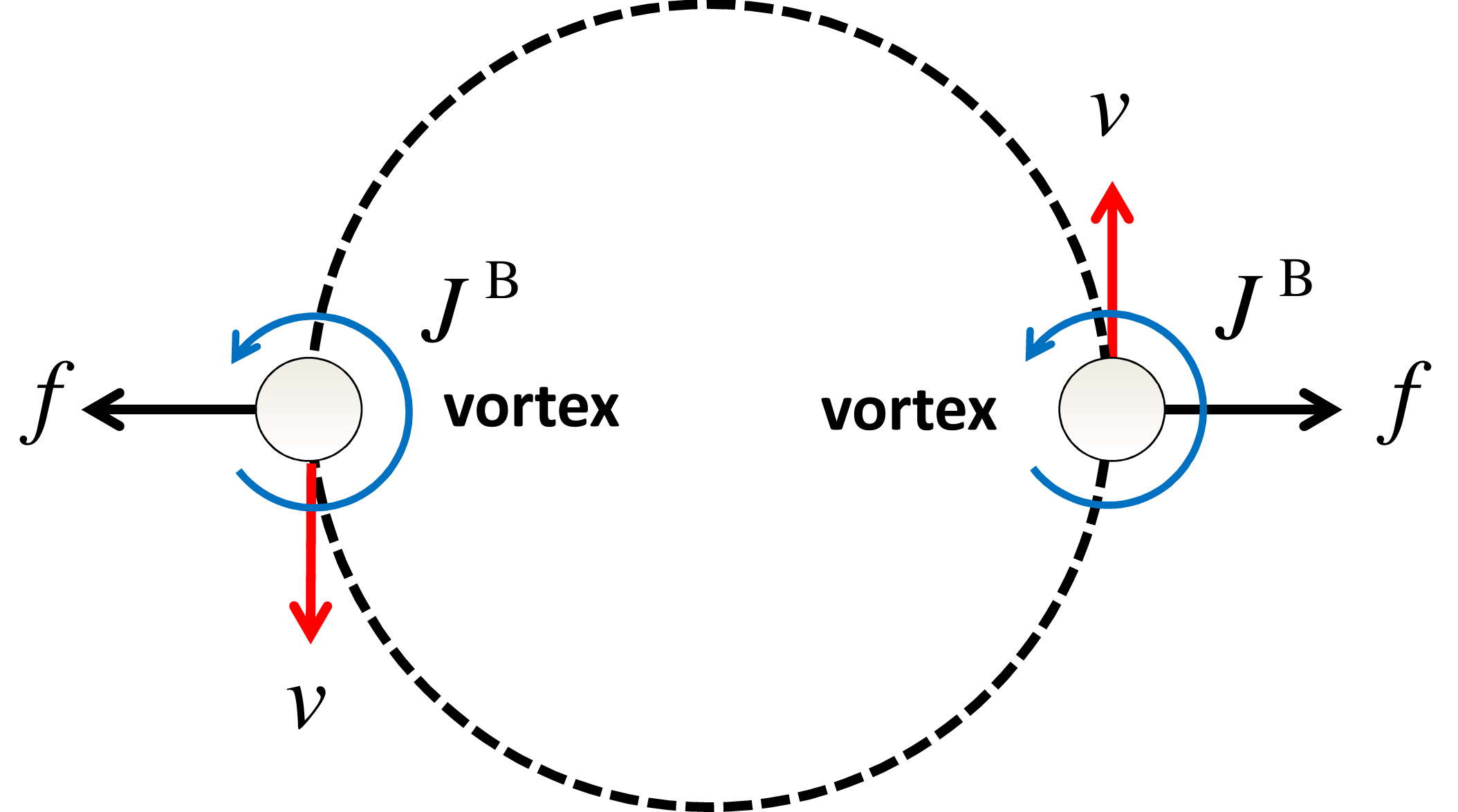}
&\hspace{1cm}
\includegraphics[width=0.32\linewidth]{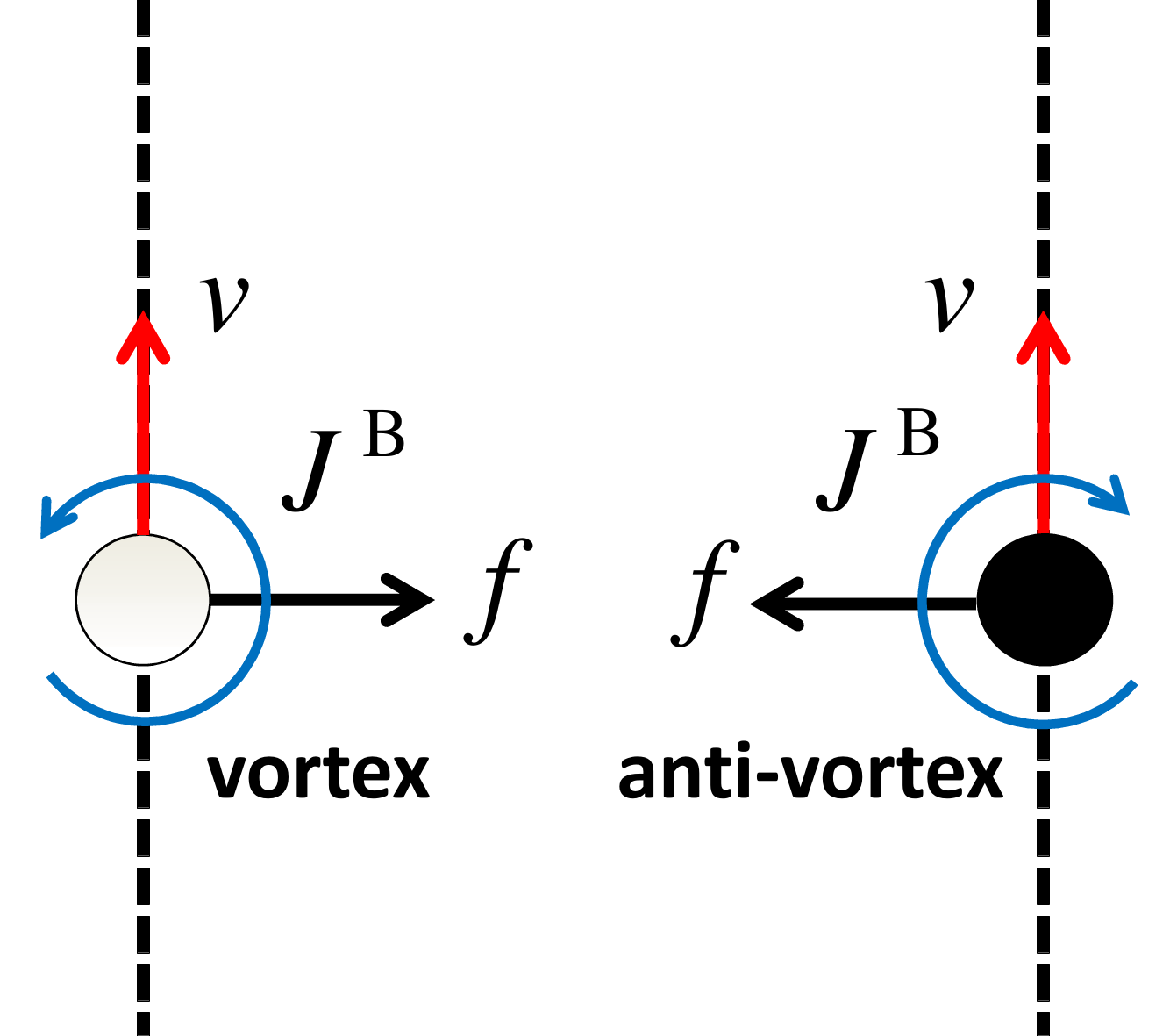}\\
(a) &(b)
\end{tabular}
\caption{
Dynamics of two vortices.
(a) Two vortices repelling each other rotate around each other 
by the inertial force represented by 
$f$ in Eq.~(\ref{eq:force}). 
The velocity is faster as the radius of the circular orbit is smaller.
(b) A vortex and an anti-vortex attracting each other 
move parallel  by the inertial force. 
The velocity is faster as the distance between them is shorter. 
\label{fig:inertial}}
\end{center}
\end{figure}

We denote positions of two vortices by 
$(X_1,Y_1)$ and $(X_2,Y_2)$.  
In the absence of the dissipation ($K_{\rm D}=0$) 
at zero temperature, 
the equations of motion contain 
only the inertial force terms on the right-hand sides 
of Eq.~(\ref{eq:eom-XY}).
Then, the two vortices separated by the interval $2R$ 
rotate around each other,  
as in Fig.~\ref{fig:inertial}(a). 
\beq
 X_1 + i Y_1 = - X_2 - iY_2  =  R \exp (i t /R^2),
\label{eq:vortex-vortex}
\eeq
where we set the coefficients to be one.
On the other hand, 
a vortex and an anti-vortex move parallel to each other 
due to the inertial force as in Fig.~\ref{fig:inertial}(b):
\beq
 X_1 + i Y_1 = - X_2 + i Y_2  =  R + i (1/R) t  
\label{eq:vortex-anti-vortex}
\eeq
where we set the coefficients to be one.
The velocity of the pair is faster when the distance 
between them is smaller. 
\begin{figure}[ht]
\begin{center}
\centering
\includegraphics[width=0.5\linewidth]{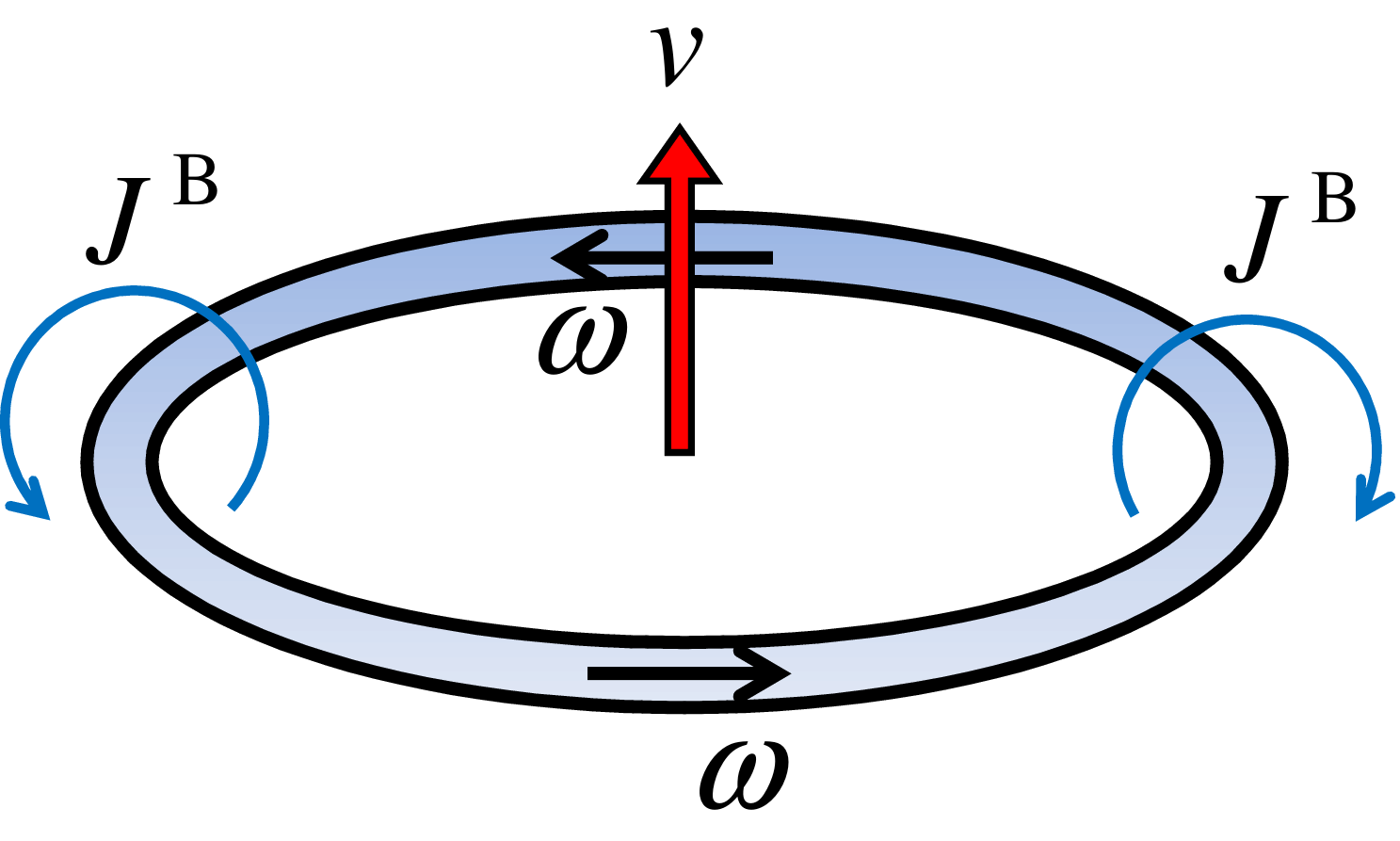}
\caption{A vortex ring. 
It moves in the direction perpendicular to the ring. 
The velocity is faster as the size of the ring is smaller.
The red, black and white arrows denote 
the velocity,  vorticity and flow, respectively. 
\label{fig:vortex-ring}}
\end{center}
\end{figure}

Let us mention what happens 
when the dissipation term is present 
($K_{\rm D} \neq 0$) at finite temperature.
For a pair of vortices, the radius of the circle orbit 
gradually increases because of the Magnus force 
from each other, the second terms 
of Eq.~(\ref{eq:eom-XY}). 
For a vortex and an anti-vortex, 
the distance between them 
gradually decreases due to the Magnus force 
from each other, 
and eventually they collide and annihilate each other. 

The situation of  a pair of a vortex and 
an anti-vortex in $d=2+1$ dimensions 
can be extended to a vortex ring 
in 3+1 dimensions,  
as illustrated in Fig.~\ref{fig:vortex-ring}. 
The vortex ring moves by the inertial force 
and is stable in the absence of the dissipation 
at zero temperature. 
In the presence of the dissipation at finite temperature, 
the radius of the ring decreases in time 
due to the Magnus force 
and eventually it decays. 
As the radius becomes smaller, 
the velocity becomes faster 
according to Eq.~(\ref{eq:vortex-anti-vortex}).


\subsection{Decays of $U(1)_{\rm B}$ vortices and 
non-minimal M$_2$ non-Abelian vortices }
\label{sec:Abelian-vortex-decay}

\begin{figure}[h]
\begin{center}
\includegraphics[width=0.6\linewidth,keepaspectratio]
{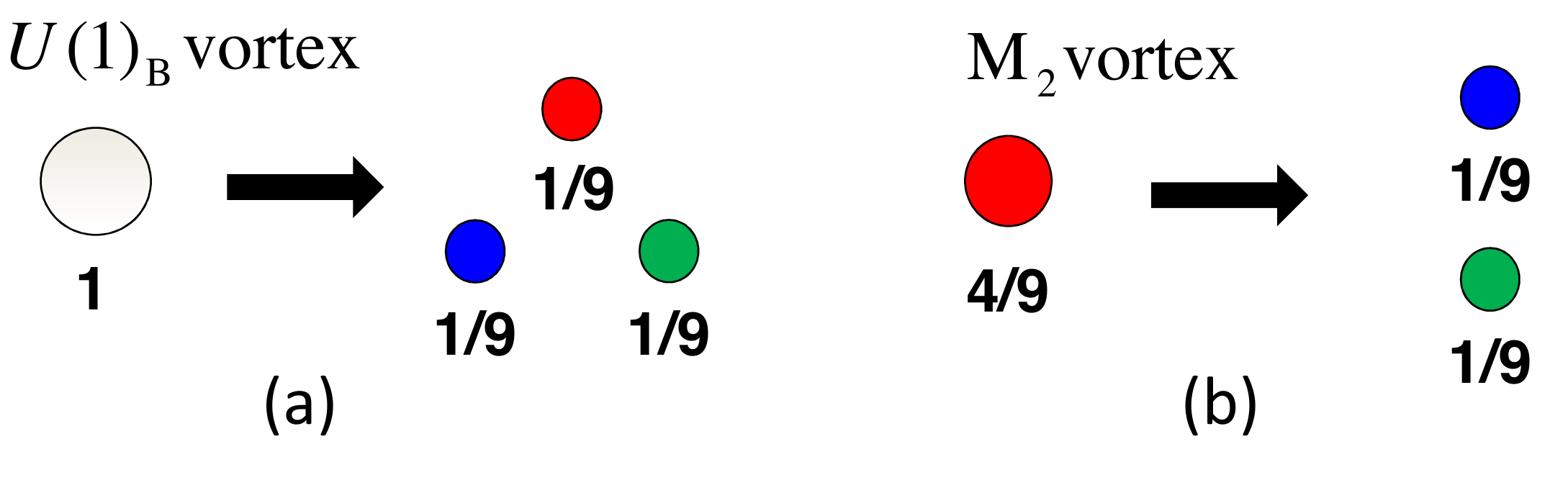}
\end{center}
\caption{\label{fig:split} 
Decays of a $U(1)_{\rm B}$ superfluid vortex 
and an M$_2$ vortex.
(a) A global $U(1)_{\rm B}$ superfluid vortex decays 
into a set of three non-Abelian vortices with 
the total magnetic flux canceled out. 
(b) An M$_2$ vortex with a red magnetic flux 
decays into two M$_1$ vortices with 
green and blue magnetic fluxes directed 
in the opposite direction.
The numbers represent the energy ratio  
when all of them are infinitely separated.
}
\end{figure}

The logarithmically divergent part of the tension of 
non-Abelian vortices is given in Eq.~(\ref{eq:ene_NA}),
while the tension of a $\U(1)_{\rm B}$ integer vortex is 
given in Eq.~(\ref{eq:ene_U(1)B}).
Therefore, 
the tension of a non-Abelian vortex is $1/3^2 = 1/9$ times as large as the tension of a $\U(1)_{\rm B}$ vortex. 
Since the total energy of the constituent three 
non-Abelian vortices is $3 \times 1/9 = 1/3$ 
of that of the $U(1)_{\rm B}$ vortex 
when they are infinitely separated, 
we conclude that the $\U(1)_{\rm B}$ vortex breaks up into 
three non-Abelian vortices with three different color fluxes, say red, blue, and green color fluxes, with total fluxes canceled out, 
as illustrated in Fig.~\ref{fig:split}(a): 
\beq
\Phi &=& \Delta_{\rm CFL} {\rm diag.}\,(e^{i\theta},e^{i\theta},e^{i\theta}) \non
 &\to& \Delta_{\rm CFL} {\rm diag.}\,(e^{i\theta_1},1,1) \times 
 {\rm diag.}\,(1,e^{i\theta_2},1) \times
 {\rm diag.}\,(1,1,e^{i\theta_3}), \label{eq:Abelian-decay}
\eeq
where $\theta_{1,2,3}$ is the angle of polar coordinates 
at each vortex.
In fact, 
in Sec.~\ref{sec:intervortex-force}, 
we have seen that the interaction between two non-Abelian vortices is repulsive at large separation. 
There remains the possibility of metastability of 
the $U(1)_{\rm B}$ vortices, 
since each vortex in Eq.~(\ref{eq:Abelian-decay}) must carry 
a color magnetic flux that contributes finite energy corrections 
to the tension.  
However, they decay with perturbations. 
After the decay, these three vortices rotate as in 
Fig.~\ref{fig:inertial}, because of the inertial force. 
Without dissipation at zero temperature, 
they rotate forever,
while the radius increases in the presence of the dissipation 
at a finite temperature. 

In the same way, a non-minimal M$_2$ vortex also decays 
into two non-Abelian vortices as in Fig.~\ref{fig:split}(b). 
Let us suppose an M$_2$ vortex carries 
a red magnetic flux; 
it then decays into two M$_1$ vortices with 
green and blue magnetic fluxes directed 
in the opposite direction.

\begin{figure}
\begin{center}
\includegraphics[width=0.4\linewidth,keepaspectratio]{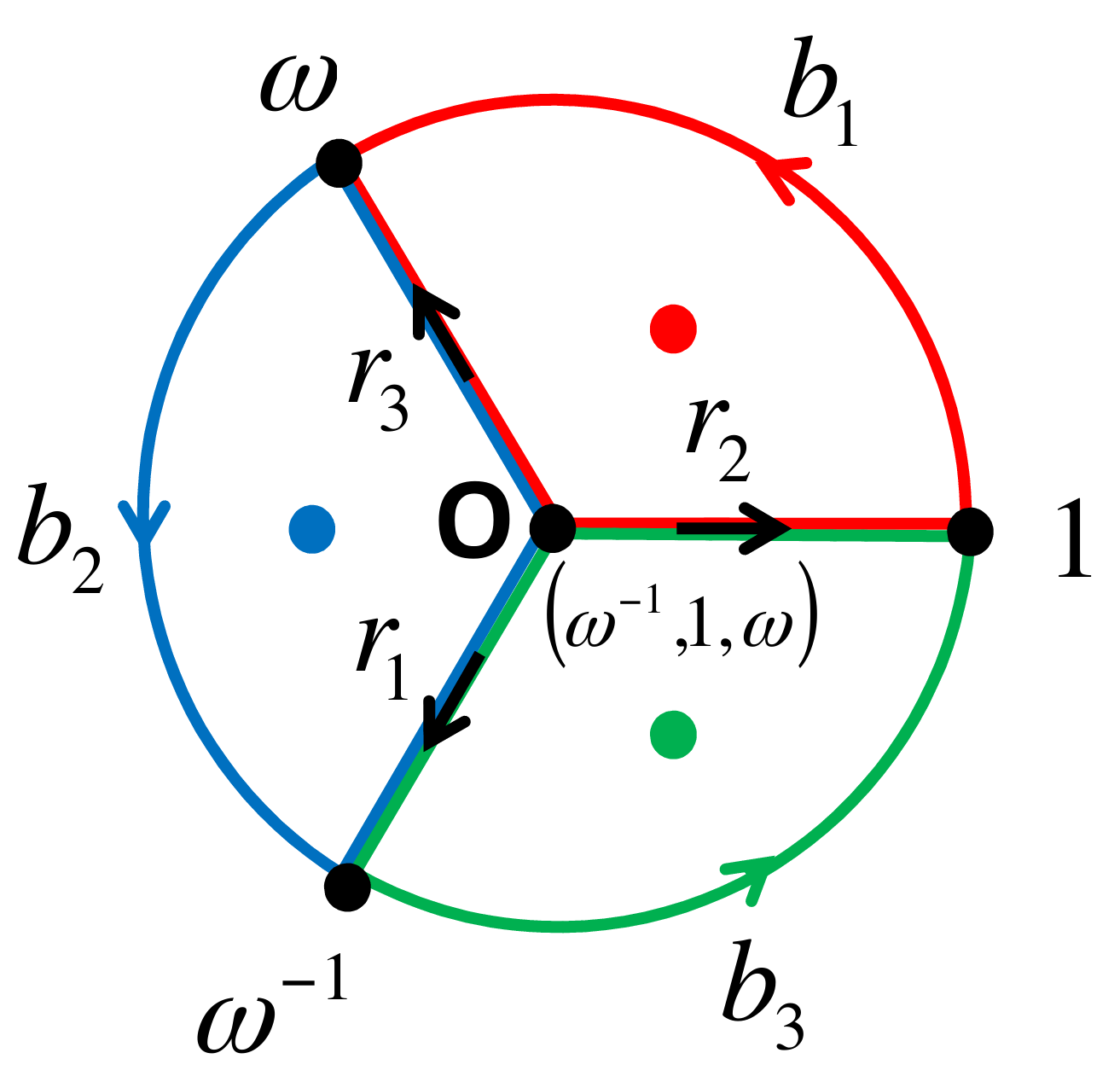}
\end{center}
\caption{\label{fig:vortex-decay} 
Detailed configurations of 
decays of a $U(1)_{\rm B}$ superfluid vortex.  
The $U(1)_{\rm B}$  vortex initially located at the origin O decays 
int to three non-Abelian vortices, denoted by 
the red, green, blue dots. 
The $b_1$, $b_2$ and $b_3$ are the paths with angles $2\pi/3$ 
at the boundary at spatial infinity, 
and $r_1$, $r_2$, $r_3$ denote the paths from the origin O to spatial 
infinities. 
For $b_1$, $b_2$ and $b_3$, 
the $U(1)_{\rm A}$ phase is rotated by 
$\exp [i \theta \diag (1,1,1)]$ with the angle $\theta$ 
of the polar coordinates. 
}
\end{figure}
Next, let us show a detailed configuration of the decaying process 
of a $U(1)_{\rm B}$ Abelian vortex 
in Fig.~\ref{fig:vortex-decay}.
The Abelian vortex initially located at the origin O decays 
into three non-Abelian vortices, denoted by 
the red, green and blue dots, carrying corresponding color fluxes.
The red, blue and green non-Abelian vortices 
are encircled by the paths
\beq
 b_1 - r_3 + r_2, \quad 
 b_2 - r_1 + r_3, \quad 
 b_3 - r_2 + r_1 \label{eq:paths}
\eeq
respectively.
At the boundary of spatial infinity, 
the $U(1)_{\rm B}$ phase is rotated by 
$\exp [i \theta \diag (1,1,1)]$ with the angle $\theta$ 
of the polar coordinates from the origin O.  
Therefore,  
the $U(1)_{\rm B}$ phase is rotated 
by $2\pi/3$ along each of the paths 
$b_1$, $b_2$ and $b_3$. 
Let us suppose that the three paths in Eq.~(\ref{eq:paths}) 
enclose the three configurations in 
Eqs.~(\ref{eq:ansatz_NA_vortex}), (\ref{eq:ansatz_NA_vortex-b1}), and (\ref{eq:ansatz_NA_vortex-c1}),
 respectively. 
Then, we find that 
the {\it color gauge} transformations $g(r) \in SU(3)_{\rm C}$ occur 
along the paths $r_1$, $r_2$ and $r_3$ as 
\beq 
r_1: && g(r) 
= \exp [i u(r) \diag(0,-1,1)]
=\bigg\{\begin{array}{c}
  \diag (1,1,1), \quad r=0 \cr 
  \diag (1,\omega^{-1},\omega) , \quad r=\infty
\end{array},   \non
r_2: && g(r) 
= \exp [i u(r) \diag(1,0,-1)]
=\bigg\{\begin{array}{c}
  \diag (1,1,1), \quad r=0 \cr 
  \diag (\omega,1,\omega^{-1}) , \quad r=\infty
\end{array}, \label{eq:path-decay}\\
r_3: && 
g(r) = \exp [i u(r) \diag(-1,1,0)]
=\bigg\{\begin{array}{c}
  \diag (1,1,1), \quad r=0 \cr 
  \diag (\omega^{-1},\omega,1) , \quad r=\infty
\end{array}, \nonumber
\eeq
respectively, 
with a monotonically increasing function $u(r)$ 
with the boundary conditions 
$u(r=0)=0$ and $u(r=\infty)=2\pi/3$.\footnote{
This route for the decomposition of an integer vortex into 
three fractional vortices is exactly the same as 
that of three-gap superconductors
\cite{Nitta:2010yf} 
and three-component Bose-Einstein condensates 
\cite{Eto:2012rc,Cipriani:2013wia,Nitta:2013eaa}.
} 
We find that the origin O is consistently given by 
\beq
 \Phi = \Delta_{\rm CFL} \diag (\omega^{-1},1,\omega).
\eeq 
From a symmetry, permutations of each component 
are equally possible.
The configuration of Fig.~\ref{fig:vortex-decay}) 
is a $U(1)_{\rm B}$ vortex  
encircled by the path $b_1+b_2+b_3$ corresponding 
to unit circulation 
at large distance.
However, at short distance, it is separated into 
a set of three non-Abelian vortices, 
 each of which is encircled by the paths in Eq.~(\ref{eq:paths}). 
The paths $r_i$ contribute to 
color magnetic fluxes and each $b_i$ corresponds 
to 1/3 quantized circulation.

\subsection{Colorful vortex lattices under rotation}\label{sec:lattice}

\subsubsection{Vortex formation and vortex lattices  
as a response to rotation in conventional superfluids
}
Here we consider the response of the CFL matter to rotation. 
Let us first recall what happens if one rotates an ordinary superfluid.
Suppose that one has a superfluid in a vessel; 
let us rotate the vessel
with angular velocity $\Omega$.
The ground state of the system can be determined by minimizing the free
energy. 
In a rotating system, the free energy is modified as 
$
 F' = F - \bm \Omega \cdot \bm L
$, where $\bm L$ is the angular momentum vector. 
At low temperatures, the entropy term can be neglected and we just have
to minimize $H' = H  - \bm \Omega \cdot \bm L$, where $H$ and $H'$ are
the Hamiltonian in the rest and rotating frames.

The time evolution of a rotating system is generated by $H'$. Let us
first recall the reason for this using a simple example, a non-relativistic point
particle in a rotating frame. 
The Lagrangian of a point particle with mass $m$ is written as 
$
 L = m \bm v^2/2,
$
where $\bm v$ is the velocity in the rest frame.
The conjugate momentum and the Hamiltonian are given by 
\begin{equation}
 \bm p = \frac{\p L}{\p \bm v} = m \bm v,
\quad  H = \bm p \cdot \bm v - L = \frac{m {\bm v}^2}{2}.
\end{equation}
Now let us move to the description in the rotating frame at an angular
velocity $\bm \Omega$. 
Let $\bm v'$ be the 
velocity of the particle in the rotating system. 
The two velocities are related by 
\begin{equation}
\bm v = \bm v' +  {\bm \Omega} \times \bm r.
\label{eq:cood-rel}
\end{equation}
Since the Lagrangian mechanics is covariant under general coordinate
transformation, we can switch to the rotating frame just by substituting
$\bm v'$ into $\bm v$, 
\begin{equation}
 L = \frac{m}{2}
\left(\bm v' +   \bm \Omega \times \bm r 
\right)^2,
\end{equation}
The conjugate momentum and Hamiltonian in the rotating system are given by 
\begin{equation}
 \bm p' = \frac{\p L}{\p \bm v'} = m (\bm v' +  \bm \Omega \times \bm r )
  \left( = \bm p\right) , 
\quad 
H' =\bm p' \cdot \bm v' - L 
= \frac{\bm p'^2}{2m}  - \bm \Omega \cdot \left(\bm r \times \bm p'
 \right) ,
\end{equation}
where we have used the cyclic property of the cross product, 
$
\bm p' \cdot \left( \bm \Omega \times \bm r \right)
=
\bm \Omega \cdot \left(\bm r \times \bm p'
 \right) 
$.
Noting that $\bm p' = \bm p$, we can rewrite the Hamiltonian in the
rotating frame as 
\begin{equation}
\begin{split}
  H'
&= \frac{\bm p'^2}{2m}  - \bm \Omega \cdot \left(\bm r \times \bm p'
 \right) \\
&=  \frac{\bm p^2}{2m}  - \bm \Omega \cdot \left(\bm r \times \bm p
 \right) \\
&= H - \bm \Omega \cdot \bm L, 
\label{eq:rotating-hamiltonian}
\end{split}
\end{equation}
which connects the Hamiltonians in the rest and rotating frames.
The discussion above can be straightforwardly extended to many-body
systems, as long as the interaction potential is invariant under rotation.

Now let us discuss the response of a superfluid to rotation. 
We consider the situation where a cylinder 
with radius $R$ is filled with a superfluid  
and there is one vortex in the center of the vessel.
Then the superfluid velocity is given by $ \bm v = \frac{n}{2 \pi r}
\bm e_{\theta}$ with $n$ the winding number.
The energy of a vortex per unit length is 
\begin{equation}
E = \int d^2x \frac{1}{2}\rho \, \bm v^2 
= \frac{\rho n^2}{4 \pi} {\rm log} \left(\frac{R}{a}\right),
\end{equation}
where $\rho$ is the superfluid density and $a$ is the core radius ($R
\gg a$).
The angular velocity per unit length is written as 
\begin{equation}
 L = \int d^2x \ \left( \bm r \times (\rho \bm v) \right)_z
 = \int_a^R 2 \pi r \cdot r \rho |\bm v| \simeq \frac{1}{2} n \rho R^2,
\end{equation}
where the term proportional to $a^2$ is neglected.
Thus, the energy in the rotating system is given by
\begin{equation}
 E' = 
\frac{\rho n^2}{4 \pi} {\rm log} \left(\frac{R}{a}\right)
-  \frac{1}{2} \Omega n \rho R^2.
\end{equation}
If this energy is less than $0$, one vortex state is favored compared to
the state without a vortex.
We can define the critical angular velocity $\Omega_{\rm c}$,
\begin{equation}
\Omega_{\rm c} \equiv \frac{n}{2 \pi R} {\rm log} \left(\frac{R}{a} \right).
\end{equation}
Thus, if $\Omega > \Omega_{\rm c}$, the state with the vortex is more favorable than the trivial state.

If one further increases the rotational speed, multiple vortices 
are generated along the rotational axis.
They all have the same winding number, so the intervortex force 
is repulsive. All the vortices repel each other, resulting in the
formation of a vortex lattice.
In the end, the vortex lattice co-rotates with the vessel,
which means that the superfluid velocity at the edge coincides with the
speed of the wall. Then the circulation $\Gamma$ can be calculated as $2 \pi
R \cdot R \Omega$. This should be equal to the sum of the circulations of all
the vortices inside the vessel, $\Gamma = n N$, where $N$ is the total
number of vortices.
Therefore, the total number of vortices 
\begin{equation}
 N = \frac{2S \Omega}{n},
\end{equation}
where $S$ is the cross section of the vessel.
In the case of an ordinary superfluid, the circulation of each vortex is equal to one, $n=1$.

Let us see the creation of vortices from another point of view.
We here consider a BEC and denote the condensate by a complex
scalar field $\Phi(x)$.
In a rotating frame, the gradient term of the energy functional 
can be written as (see Eq.~(\ref{eq:rotating-hamiltonian}))
\beq 
 E_{\rm grad} = 
 \Phi^\ast 
\left( \frac{ \hat {\bf p}^2}{2m} - {\bf \Omega} \cdot {\bf \hat L} 
\right)\Phi
    = \1{2m} |[\del_i - im ({\bf r} \times {\bf \Omega})_i]\Phi|^2 
       -\1{2}m r^2 \Omega^2 |\Phi|^2  \label{eq:rotating}
\eeq
where $\hat L_i = i \epsilon_{ijk}r_j \del_k$ and $\hat p_i = \p_i$, and 
the equality is meant up to a total derivative.
The combination $D_i =\del_i - i m({\bf r} \times {\bf \Omega})_i$ 
can be seen as a covariant derivative on the field $\Phi$.
Then $A_i \equiv m \,  {\bf r} \times {\bf \Omega}$ can be seen as a gauge
field and this system looks like a charged field under a constant
magnetic field $2{\bf \Omega}$.
Therefore, just like a type II superconductor under an external magnetic
field, 
vortices come into a rotating superfluid. 
Let us make a comment on the trapping potential. 
One should add an external potential to trap the condensate, 
such as $V_{\rm trap} = \1{2} \omega^2 r^2$ for BECs. 
This term and the last term in Eq.~(\ref{eq:rotating}) 
can be combined as 
$ -\1{2} r^2 \Omega^2 |\Phi|^2+ V_{\rm trap} = \1{2} (\omega^2  - \Omega^2) r^2 |\Phi|^2$. 
When the rotation speed $\Omega$ is less than $\omega$, 
the condensates can be trapped.

\subsubsection{Colorful vortex lattices}
Then, what happens if one rotates CFL matter? 
We can repeat the energetical argument. When the vessel is large enough,
the energy of a vortex is dominated by the superfluid velocity part and
the contribution from the color flux is negligible, 
Thus, what is modified in the argument above is that $n$ should be equal
to one third, $n = 1/3$.
For realistic neutron stars, the number of vortices can be estimated as 
\begin{equation}
 N_v \simeq 1.9 \times 10^{19 } 
\left(\frac{ 1 \,{\rm ms }}{ P_{\rm rot}} \right)
\left(\frac{\mu/3}{300  \,{\rm MeV}}  \right)
\left( \frac{R}{10  \,{\rm km} } \right)^2,
\label{eq:rot-vortex-num}
\end{equation}
where $P_{\rm rot}$ is the rotational period, $\mu$ is the baryon
chemical potential, and the parameters are
normalized by typical values.
The corresponding intervortex spacing is given by 
\begin{equation}
 \ell \equiv 
\left(\frac{\pi R^2}{N_v}\right)^{1/2} 
\simeq
4.0 \times 10^{-6}~ {\rm m} 
\left(\frac{ P_{\rm rot}}{ 1 \, {\rm ms}} \right)^{1/2}
\left(\frac{300 \, {\rm MeV}}{\mu/3}  \right)^{1/2}.
\end{equation}
Since the intervortex spacing is far larger than the size of a vortex
core, which is given by inverse gluon/meson masses,  
gluons and mesons would not affect the force between two vortices. 
The intervortex force is dominated by the exchange of $U(1)_{\rm B}$ phonons.
This justifies the treatment above, in which we have only considered the
contribution of $U(1)_{\rm B}$ circulations.


In the discussion using the free energy, we can only determine the
ground state of the system. 
The dynamical process of vortex generation can be nontrivial, especially
for non-Abelian vortices.
Basically, as one gradually increases the speed of rotation, vortices
enter one by one from the edge of the superfluid. However, in the case
of the non-Abelian vortices, it has a color flux and 
one vortex cannot be created because of the color conservation. 
A vortex with unit $U(1)_{\rm B}$ winding number does not have a color
flux, but it is energetically unstable. 
So, one plausible idea is that a $U(1)_{\rm B}$ vortex is created first; it then decays into three non-Abelian vortices,  
as discussed in Sec.~\ref{sec:Abelian-vortex-decay}.
Such two-fold vortex generation is seen in the simulation of rotating
three-component BEC \cite{Cipriani:2013wia}, in which
integer-quantized vortices are created first and then decay into fractional vortices.

\begin{figure}[htbp]
\begin{center}
  \includegraphics[width=90mm]{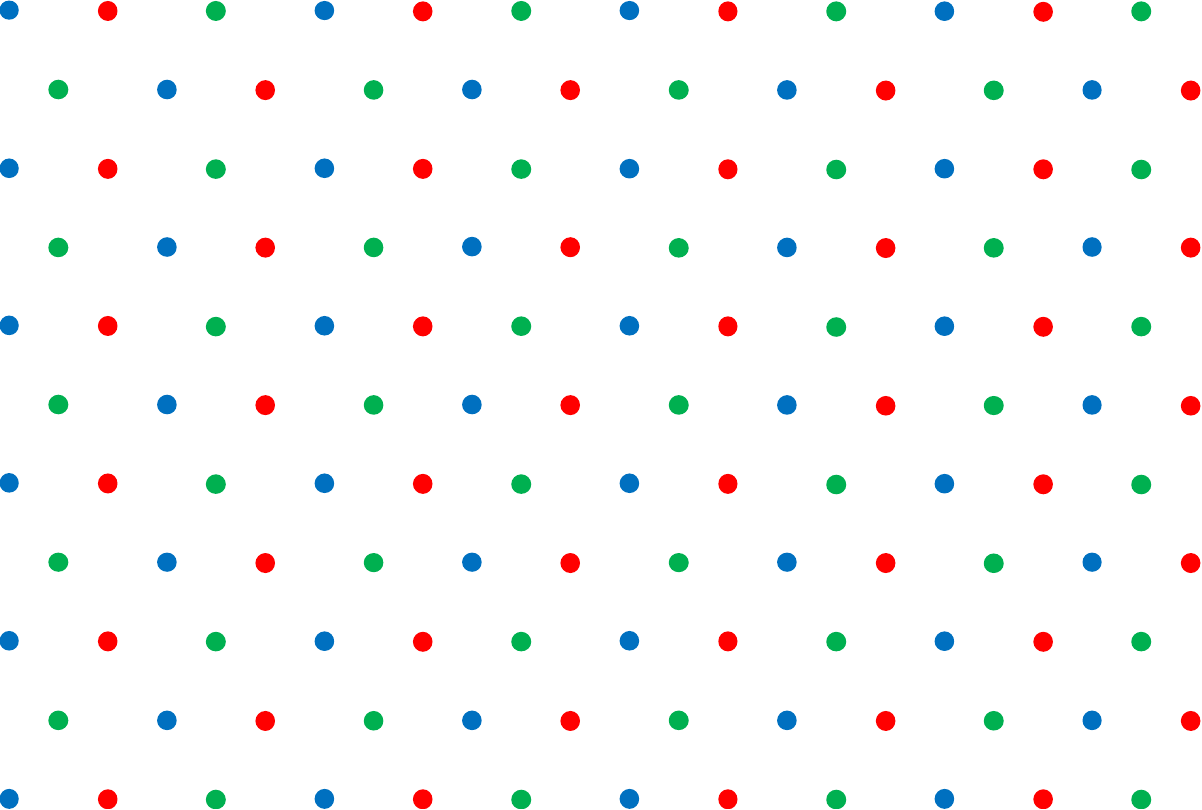}\\
(a) Ordered colorful vortex lattice\\
\vspace{5mm}
  \includegraphics[width=90mm]{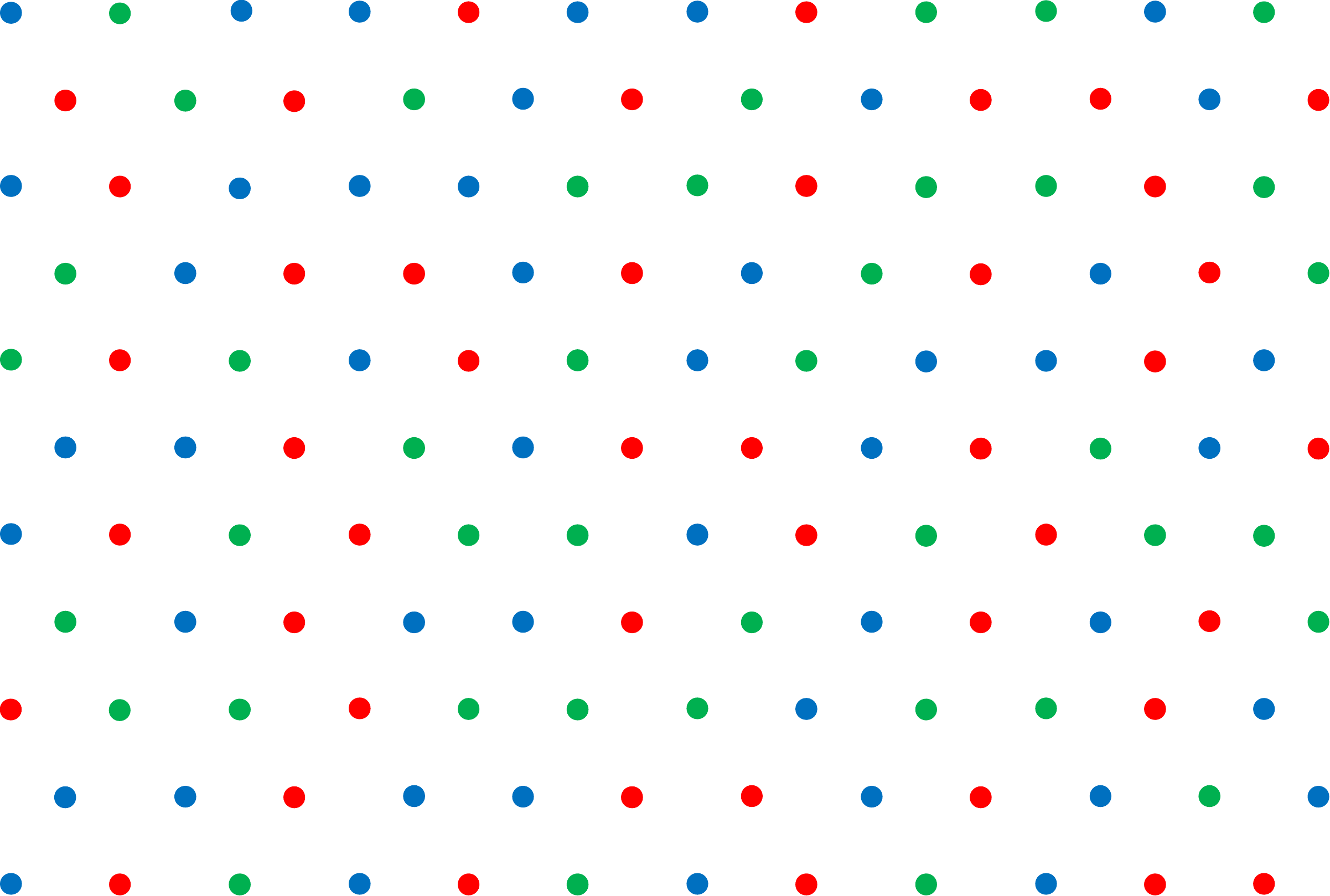}\\
(b) Disordered colorful vortex lattice
\end{center}
 \caption{
Colorful vortex lattices in a rotating frame. 
Colors are 
(a) ordered and (b) disordered. 
In both cases, vortices with ignoring color constitute 
an Abrikosov lattice, and total color fluxes are canceled out. 
(a) The color magnetic fluxes of each color constitute 
of an Abrikosov lattice.
(b) The color magnetic fluxes of each color {\it do not} 
constitute an Abrikosov lattice.
}
\label{fig:lattice}
\end{figure}

Another feature of rotating CFL matter is that the lattice is
``colorful'' in the sense that each vortex carries a color flux.  
There can be various patterns of the configurations of 
the colored vortices, with vanishing total color fluxes.
In Fig.~\ref{fig:lattice}(a), we show a possibility for an ``ordered" colorful vortex lattice.
In this case, the color magnetic fluxes of each color constitute an Abrikosov lattice, 
and the total configuration itself neglecting 
colors is also in the form of an Abrikosov lattice. 
In general, even when the total configuration 
neglecting colors is in the form of an Abrikosov lattice, 
there may exist a vortex lattice 
that is ``disordered" as in Fig.~\ref{fig:lattice}(b), 
in the sense that 
each colors is placed at random 
with total colors canceled out. 
Even ordered, there can be a defect of color orderings. 

In the simulation of a three-component BEC, 
ordered lattices have always been 
obtained as mentioned above \cite{Cipriani:2013wia}. 
However, in this simulation, one has introduced a very small perturbation 
to split one integer vortex into three fractional vortices 
after the integer vortex lattice is formed. 
In this sense, there could be a defect of color ordering 
in general if several perturbations were introduced. 

Gravitational waves from vortex lattices in the CFL phase 
(and the 2SC phase)
were studied in Ref.~\cite{Glampedakis:2012qp} 
and compared with neutron star  observations.
However, vortex lattices of unstable color magnetic fluxes 
were assumed, so the calculation should be revised 
by considering colorful vortex lattices.

\subsubsection{Tkachenko modes}

A Tkachenko mode is a gapless excitation 
with a quadratic dispersion in the presence of a vortex lattice in a rotating system.
This mode corresponds to deformation of the lattice 
and the phonon does not appear independently; 
see Ref.~\cite{Sonin:1987zz} for a review of 
the Tkachenko modes.
If one ignores the centrifugal force, 
there are approximate translational symmetries 
in the system, 
which are spontaneously broken by the lattice, 
resulting in an approximate 
type-II Nambu-Goldstone mode identified 
with the Tkachenko mode \cite{Watanabe:2013iia}. 

The Tkachenko mode should also exist in a vortex
lattice of non-Abelian vortices, if the CFL phase is realized inside neutron stars. 
The characteristic timescale of this oscillation is
consistent with the quasiperiodic oscillation, which is of the order of
$100-1000$ days, observed in the pulsars PSR B0531+21 and PSR B1828-11 \cite{Noronha:2007qf, Shahabasyan:2009zz}.
It is an open question whether there is a color effect 
for the Tkachenko mode of a colorful vortex lattice.

\subsection{Superfluid vortices from relativistic strings}
\label{sec:relativistic-superfluid}

A direct relation between a relativistic global vortex in medium and a non-relativistic superfluid vortex was
discussed in Refs.~\cite{Lund:1976ze,Davis:1989gn} and later also in the CFL phase \cite{Forbes:2001gj}. 
When a relativistic global vortex is put into an uniform medium that violates the Lorentz invariance,
it acquires a non-zero angular momentum and behaves as if it is a non-relativistic superfluid vortex \cite{Davis:1989gn}.
To see this, let us consider the low energy effective theory for the $U(1)_{\rm B}$ Nambu-Goldstone mode
at high density and zero temperature in 
Sec.~\ref{sec:eff_th_phonon}, 
\beq
\Lag = f_{\rm B}^2 \left[(\p_0\varphi_B + 2\mu)^2 - v_{\rm B}^2 (\p_i\varphi_{\rm B})^2\right].
\eeq
Note that the time derivative $\p_0\varphi_{\rm B}$ is replaced by $\p_0\varphi_B + 2\mu$ in order for 
the effective theory to have the same symmetry as the microscopic theory, QCD.
Because of this peculiar dependence of $\mu$, the order parameter gets the following time rotation:
\beq
e^{i\varphi_{\rm B}} \sim e^{-2i\mu t}.
\label{eq:time_depend_phase}
\eeq
As a result, a global $U(1)_{\rm B}$ vortex acquires a non-zero angular momentum \cite{Forbes:2001gj}.
From the time dependence in Eq.~(\ref{eq:time_depend_phase}), one can easily see that
a relativistic global string in medium has non-zero angular momentum \cite{Forbes:2001gj}: 
\beq
M_{12} = K_0 \int d^2x \left( 
x_1 \p_2\phi^*\p_0\phi -x_2 \p_1\phi^*\p_0\phi
\right) +{\rm c.c.} \sim k \mu K_0.
\label{eq:ang_mom_U(1)B}
\eeq
Furthermore, by interpreting a dual of the Kalb-Ramond field strength $H^{\nu\lambda\rho}$ (completely antisymmetric tensor)
as the superfluid velocity $\p_\mu \varphi_{\rm B}$ \cite{Lund:1976ze,Davis:1989gn},
\beq
2 \mu\, \delta_{0\kappa} = \p_\kappa\theta_{\rm B} \sim \epsilon_{\kappa\nu\rho\lambda}H^{\nu\lambda\rho},
\eeq
one can reproduce interaction similar to the inertial force of superfluid vortices,  
which is one of the peculiar properties of 
superfluid vortices; see Sec.~\ref{sec:dynamics-two}.
For instance, the stability of a moving vortex 
ring was shown in  Refs.~\cite{Davis:1989gn,PhysRevD.40.4033}.

%% file: eff-th-v9.tex
\section{Dynamics of orientational zero modes}\label{sec:LEEA}

In this section, we review the dynamics of the 
orientational zero modes of non-Abelian vortices.
In Sec. \ref{sec:orientational_eff} we explain the low-energy effective theory for the orientational zero modes 
on a non-Abelian vortex world-sheet, which is a basis throughout this section.
We include the effects of the strange quark mass in Sec. \ref{sec:strange-quark} and discuss the instability of 
non-Abelian vortices. The effects of electromagnetic interaction are 
taken 
into account
in Sec. \ref{sec:elemag}. 
In Sec. \ref{sec:monopoles}, we discuss quantum monopoles on a non-Abelian vortex. The monopoles form
a bound state, which can be thought of as bounded kinks that appear on the vortex world-sheet $\mathbb CP^2$
theory when we take quantum effects into account. 
Application to the quark-hadron duality is 
also discussed.
Yang-Mills instantons trapped inside a non-Abelian vortex are discussed in Sec. \ref{sec:instanton-in-vortex}.

\subsection{Low-energy effective theory of orientational zero modes}\label{sec:orientational_eff}

As is shown in Sec.~\ref{sec:orientational}, the non-Abelian vortex has the internal orientational moduli 
$\mathbb{C}P^2$ associated with the spontaneous symmetry breaking of the color-flavor locking symmetry.
The orientational moduli are  massless modes that propagate along the host non-Abelian vortex.
In this subsection, we study 
the propagation of the massless orientational modes on a straight non-Abelian vortex. We assume that the vortex is on the $z$-axis.

In the rest of this subsection, we omit all the massive modes and concentrate on the orientational zero modes
of $\mathbb{C}P^2$. Let us first identify the orientational zero modes in the background fields $\Phi$ and $A_\mu$.
We start with a particular non-Abelian vortex solution,
\beq
\Phi^\star(x,y) = \Delta_{\rm CFL} e^{\frac{i\theta}{3}} \left(\frac{F(r)}{3}{\bf 1}_3 + G(r)\sqrt{\frac{2}{3}}~T_8\right),\quad
A_i^\star(x,y) = - \frac{1}{g_{\rm s}}\frac{\epsilon_{ij}x^j}{r^2} h(r)\sqrt{\frac{2}{3}}~T_8.
\label{eq:NA_sol_ref}
\eeq
Then the general solution can be obtained by acting the color-flavor locked symmetry on them as
\beq
\Phi(x,y) &=& U\Phi^\star U^\dagger = \Delta_{\rm CFL} e^{\frac{i\theta}{3}} \left(\frac{F(r)}{3}{\bf 1}_3 - G(r)\left<\phi\phi^\dagger\right>\right),
\label{eq:sol_ori0}\\
A_i(x,y) &=& UA_i^\star U^\dagger = \frac{1}{g_{\rm s}}\frac{\epsilon_{ij}x^j}{r^2} h(r) \left<\phi\phi^\dagger\right>.
\label{eq:sol_ori}
\eeq
Here $\left<\mathcal O\right>$ is the traceless part of a square matrix $\mathcal O$ and
we introduced a complex 3-component vector $\phi$ 
that satisfies the following relation:
\beq
- \sqrt{\frac{2}{3}}~UT_8U^\dagger  = \phi\phi^\dagger - \frac{1}{3}{\bf 1}_3 \equiv \left<\phi\phi^\dagger\right>.
\label{eq:identify_orientation}
\eeq
Taking the trace of this definition gives a constraint
\beq
\phi^\dagger \phi = 1.
\label{eq:const_phi}
\eeq
Furthermore, the phase of $\phi$ is redundant by definition.
Thus, we find that $\phi$ represents the homogeneous coordinates of $\mathbb{C}P^2$.

Now we promote the moduli parameters $\phi$ to the fields depending on the coordinates $t,z$
of the vortex world-sheet by using the so-called moduli space approximation (first introduced by Manton
for BPS monopoles \cite{Manton:1981mp,mantontopological}). 
Since we are interested in 
the low-energy
dynamics of the orientational moduli fields,
we restrict ourself to the study of the dynamics with a typical energy scale that is much lower than the mass scales of
the original theory, namely $m_1$, $m_8$, and $m_{\rm g}$. 
Namely, we consider the situation where
\beq
\left|\partial_\alpha \phi(z,t)\right| \ll \min\left\{m_1,m_8,m_{\rm g}\right\},\quad (\alpha = z,t).
\label{eq:low_condition}
\eeq
Omitting the higher derivative terms (massive modes) with respect to $z$ and $t$, the low-energy effective theory
for the orientational moduli field can be obtained by plugging $\Phi(x,y;\phi(z,t))$ and $A_i(x,y;\phi(z,t))$
into the full Lagrangian (\ref{eq:gl}). 
Note that $\Phi(x,y;\phi(z,t))$ and $A_i(x,y;\phi(z,t))$ depend on the full spacetime coordinates, 
which are obtained by replacing the moduli parameter $\phi$ with the moduli field $\phi(z,t)$ in Eq.~(\ref{eq:sol_ori}).
In order to construct the Lagrangian of the effective theory, we also
have to determine the $x^\alpha$ dependence of the gauge fields
$A_\alpha$, which are exactly zeros in the background solution.
For this purpose, following Ref.~\cite{Shifman:2004dr}, we make an ansatz
\beq
A_\alpha = \frac{i\rho_\alpha(r)}{g_{\rm s}} \left[\left<\phi\phi^\dagger\right>,\partial_\alpha\left<\phi\phi^\dagger\right>\right],
\label{eq:ansatz_SY}
\eeq
where $\rho_\alpha(r)$ ($\alpha=z,t$) are unknown functions of
the radial coordinate $r$ at this stage.

\begin{figure}[ht]
\begin{center}
\begin{minipage}[b]{0.32\linewidth}
\centering
\includegraphics[width=\textwidth]{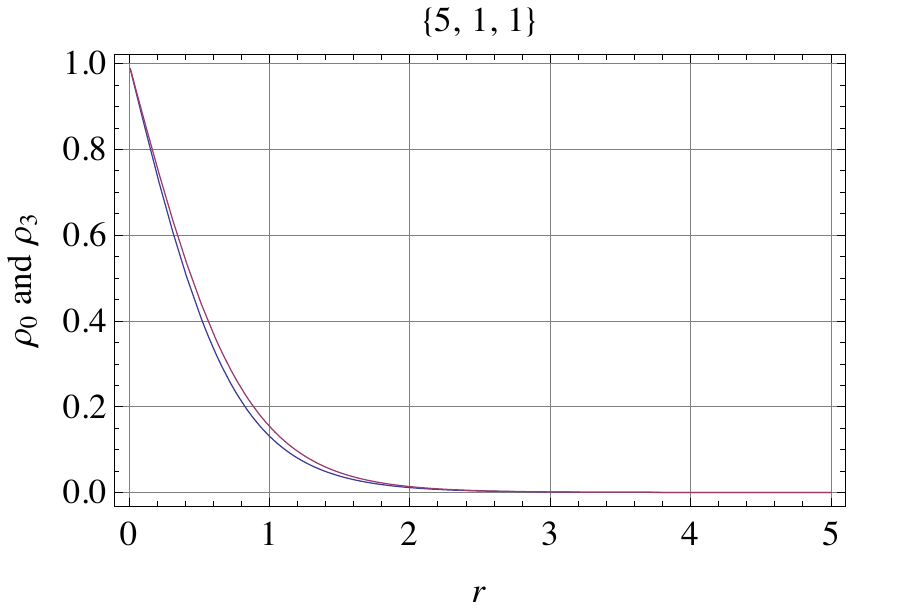}
\end{minipage}
\begin{minipage}[b]{0.32\linewidth}
\centering
\includegraphics[width=\textwidth]{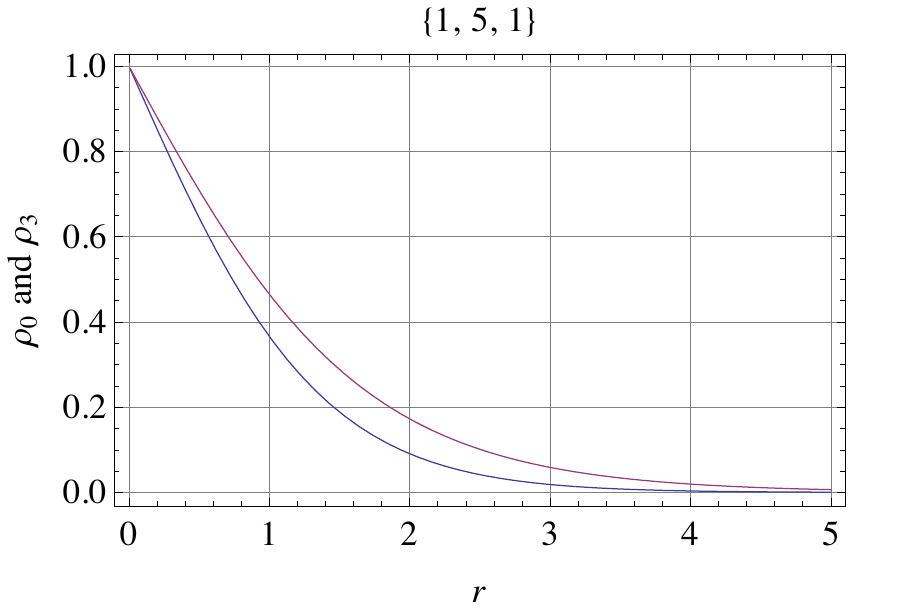}
\end{minipage}
\begin{minipage}[b]{0.32\linewidth}
\centering
\includegraphics[width=\textwidth]{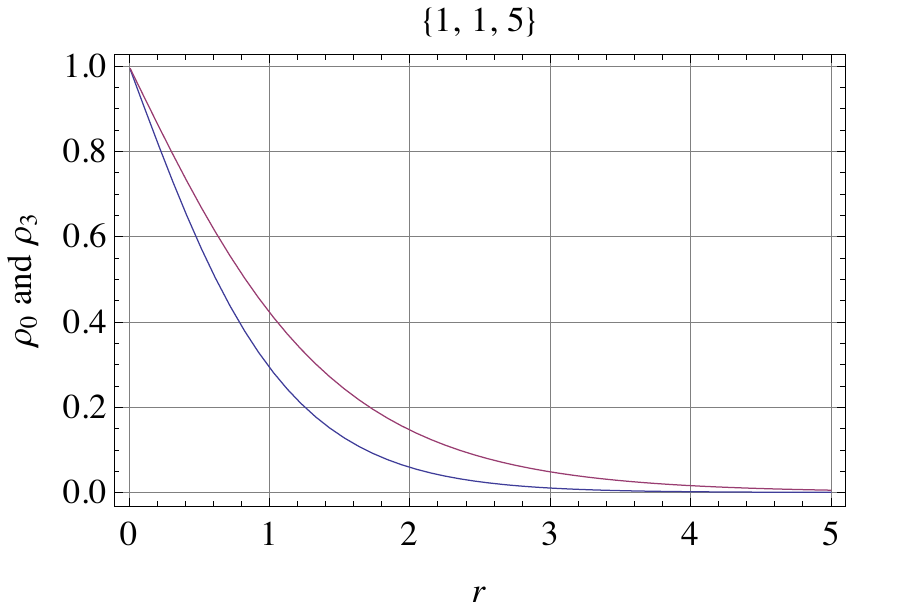}
\end{minipage}\\
\begin{minipage}[b]{0.32\linewidth}
\centering
\includegraphics[width=\textwidth]{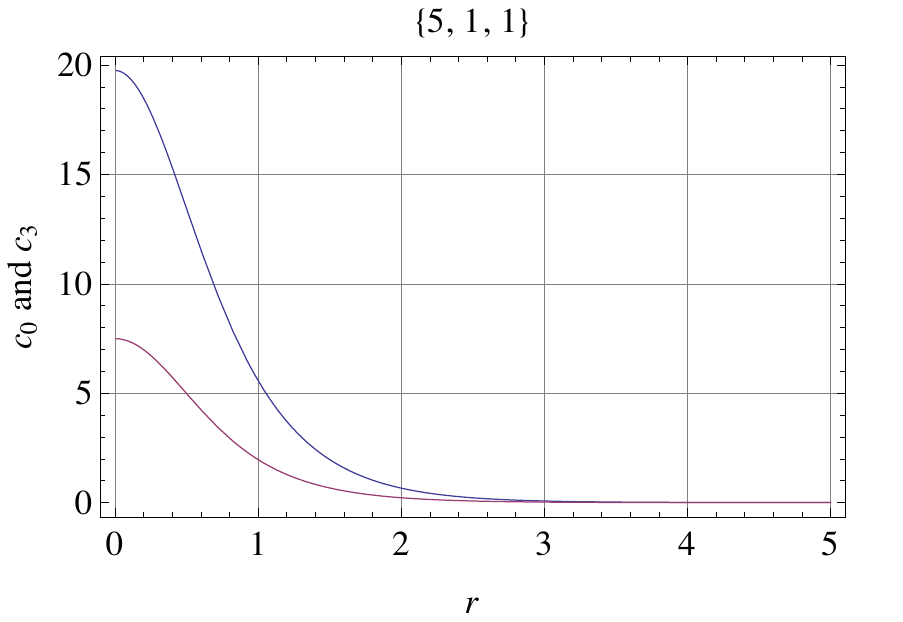}
\end{minipage}
\begin{minipage}[b]{0.32\linewidth}
\centering
\includegraphics[width=\textwidth]{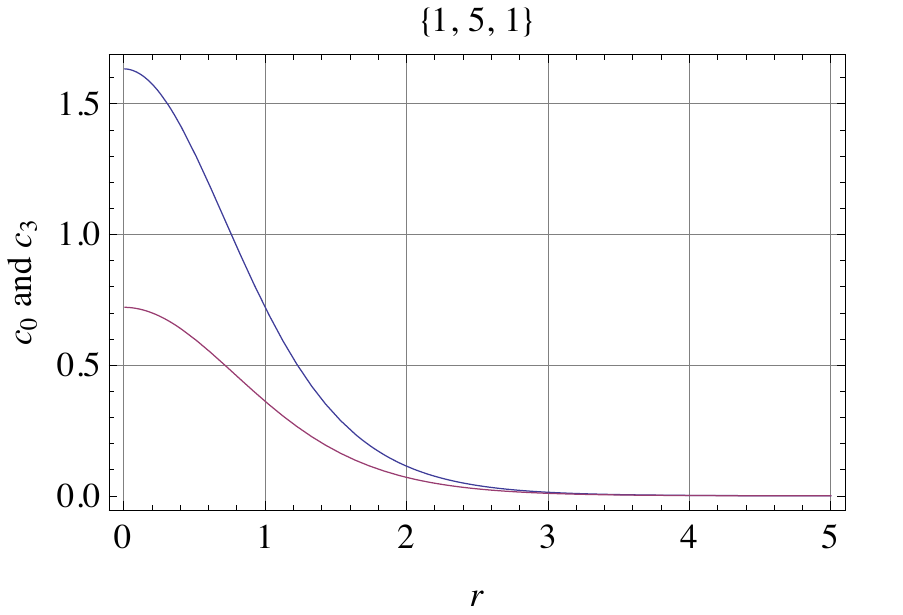}
\end{minipage}
\begin{minipage}[b]{0.32\linewidth}
\centering
\includegraphics[width=\textwidth]{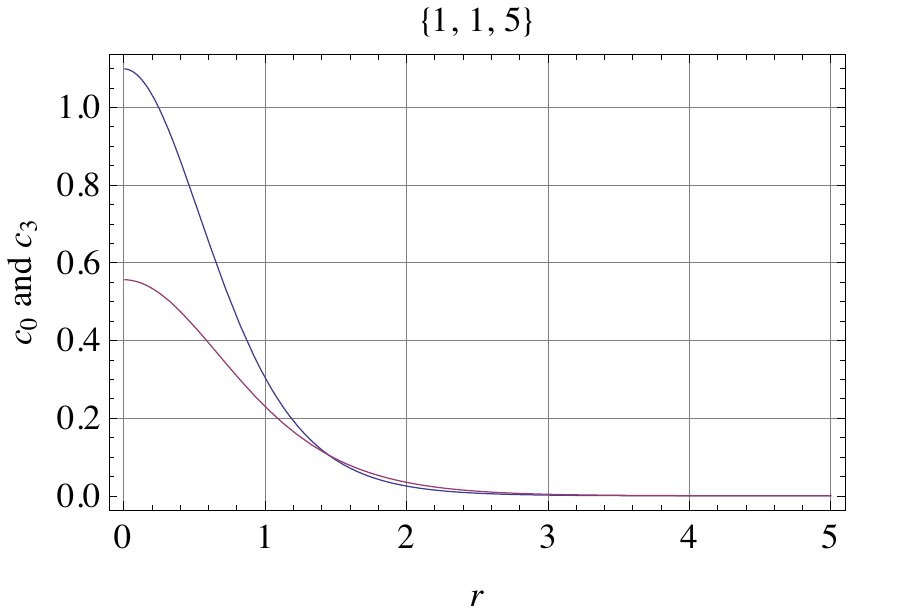}
\end{minipage}
\caption{The numerical solutions of $\rho_0(r)$ (blue) and $\rho_3(r)$ (purple) for $\{m_{\rm g},m_1,m_8\} = \{1,5,1\}, \{5,1,1\}, \{1,1,5\}$.
The K\"ahler class densities $c_0(r)$ (blue) and $c_3(r)$ (purple) are also shown in the 2nd line.
}
\label{fig:kahler_class}
\end{center}
\end{figure}

Substituting all the fields $\Phi$ and $A_\mu$ into Eq.~(\ref{eq:gl}), we are led to the effective Lagrangian
that consists of two parts,
\beq
\Lag_{\rm eff} = \Lag_{\rm eff}^{(0)} + \Lag_{\rm eff}^{(3)},
\label{eq:eff_lag}
\eeq
with
\beq
\Lag_{\rm eff}^{(0)} &=& \int dxdy\ \tr \left[ - \frac{\varepsilon_3}{2}F_{0i}F^{0i} + K_0 \D_0\Phi^\dagger \D^0\Phi\right],
\label{eq:eff0}\\
\Lag_{\rm eff}^{(3)} &=& \int dxdy\ \tr \left[ - \frac{1}{2\lambda_3}F_{3i}F^{3i} + K_3 \D_3\Phi^\dagger \D^3\Phi\right].
\label{eq:eff3}
\eeq
The Lagrangian of the ${\mathbb C}P^2$ model 
can be uniquely written 
up to the overall factors of the decay constants 
as
\beq
\Lag_{\rm eff}^{(\alpha)} &=& C^\alpha \Lag_{\mathbb{C}P^2}^{(\alpha)},
\label{eq:eff_lag_coeff}\\
\Lag_{\mathbb{C}P^2}^{(\alpha)} &=& \p^\alpha \phi^\dagger \p_\alpha \phi 
+ \left(\phi^\dagger\p^\alpha\phi\right) \left(\phi^\dagger\p_\alpha\phi\right),
\label{eq:lag_cp2}
\eeq
where no summation is taken for $\alpha$.
What we have to determine is the coefficient $C^\alpha$ in Eq.~(\ref{eq:eff_lag_coeff}).
After manipulating straightforward but complicated calculations given in Appendix \ref{sec:deriv_LET}, 
we find that the coefficients are expressed as
\beq
C^0 \!\!\!\!&=&\!\!\!\! \frac{4\pi\alpha_3}{\lambda_3 g_{\rm s}^2} \int\!\! dr\frac{r}{2}\left[
\rho_0'{}^2 + \frac{h^2}{r^2}(1-\rho_0)^2
+ \frac{\beta_3m_{\rm g}^2}{\alpha_3}\left(
(1-\rho_0)(f-g)^2+\frac{f^2+g^2}{2}\rho_0^2\right)
\right],\label{eq:C0}\\
C^3 \!\!\!\!&=&\!\!\!\! \frac{4\pi}{\lambda_3 g_{\rm s}^2} \int\!\! dr\frac{r}{2}\left[
\rho_3'{}^2 + \frac{h^2}{r^2}(1-\rho_3)^2
+ m_{\rm g}^2\left(
(1-\rho_0)(f-g)^2+\frac{f^2+g^2}{2}\rho_3^2\right)
\right],
\label{eq:C3}
\eeq
with $\alpha_3 \equiv \varepsilon_3\lambda_3$ and $\beta_3 \equiv K_0/K_3$.
The coefficients $C^{0,3}$ should be determined in such a way that the energy is minimized.
To this end, we regard the coefficients as ``Lagrangian'' for the undetermined scalar fields $\rho_\alpha$.
Namely, we should solve the Euler-Lagrange equations
\beq
\rho_0'' + \frac{\rho_0'}{r} + (1-\rho_0) \frac{h^2}{r^2} - \frac{\beta_3m_{\rm g}^2}{2\alpha_3}\left[
(f^2+g^2)\rho_0 - (f-g)^2\right] =0,
\label{eq:rho0}\\
\rho_3'' + \frac{\rho_3'}{r} + (1-\rho_3) \frac{h^2}{r^2} - \frac{m_{\rm g}^2}{2}\left[
(f^2+g^2)\rho_3 - (f-g)^2\right] =0,
\label{eq:rho3}
\eeq
with the boundary conditions
\beq
\rho_{0,3}(0) = 1,\qquad
\rho_{0,3}(\infty) = 0.
\label{eq:bc_rho}
\eeq
Numerical solutions of $\rho_0(r)$, $\rho_3(r)$, and the K\"ahler class densities $c_0(r)$ and $c_3(r)$ 
($C_i = (4\pi/g_{\rm s}^2)\int dr\, rc_i(r)$) with $\lambda_i = \varepsilon_i = 1$ ($i=0,3$) for the mass choices 
$\{m_1,m_8,m_{\rm g}\} = \{1,5,1\}, \{5,1,1\}, \{1,1,5\}$ are shown in Fig.~\ref{fig:kahler_class}.
The corresponding background solutions are shown in Fig.~\ref{fig:profiles}.

Let us roughly estimate $C^{0,3}$.
To this end, we first recall the boundary condition $f(\infty) = g(\infty) = 1$ and $h(\infty) = 0$.
The profile functions $f(r),g(r)$, and $h(r)$ approach the above boundary values at $r \sim \Delta_{\rm CFL}^{-1}$
which is the typical inverse mass of the background solution; see Eqs.~(\ref{eq:asym_F}) and (\ref{eq:asym_G}).
Reflecting this property, the function $\rho_{0,3}$ also approaches its boundary values $\rho_{0,3}(\infty) = 0$
around $r \sim \Delta_{\rm CFL}^{-1}$. Because of $\rho_{0,3}(0) = 1$, the values of the square brackets of the integrands of
$C^{0,3}$ at the origin are $(f(0)^2+g(0)^2)\beta_3 m_{\rm g}^2/2\alpha_3 \sim
\beta_3 m_{\rm g}^2/2\alpha_3$ and
$(f(0)^2+g(0)^2) m_{\rm g}^2/2 \sim
m_{\rm g}^2/2$, respectively. Therefore, one can estimate $C^{0,3}$ as
\beq
C^0 \sim \frac{\beta_3 m_{\rm g}^2}{\lambda_3 g_{\rm s}^2 \Delta_{\rm CFL}^2} \sim \frac{\beta_3\mu^2}{\Delta_{\rm CFL}^2},\quad
C^3 \sim \frac{m_{\rm g}^2}{\lambda_3 g_{\rm s}^2 \Delta_{\rm CFL}^2} \sim \frac{\mu^2}{\Delta_{\rm CFL}^2}.\label{eq:C0C3}
\eeq
Note that $\lambda_3$ dependence disappears in the above expression.
From this result, we can also estimate the velocity of the $\mathbb CP^2$ modes propagating along
the vortex string as
\beq
v^2 = \frac{C^3}{C^0} \sim \beta_3^{-1} = \frac{1}{3}.
\eeq
This should be taken with some care due to the uncertainty of the numerical solution; 
it is just a rough parametrical estimation. Indeed, since $C^{0,3}$ depends
on the chemical potential $\mu$, $v^2$ should depend on $\mu$ as well. The
precise $\mu$ dependence of the velocity of the $\mathbb CP^2$ mode has not yet been determined
in the literature.

Let us make a comment on the types of Nambu-Goldstone modes; see footnote \ref{footnote:NG-modes}. 
While the Kelvin mode associated with translational symmetries is a type II Nambu-Goldstone mode 
with a quadratic dispersion, 
the  $\mathbb CP^2$ orientational zero modes 
found in this section are type I 
Nambu-Goldstone modes with a linear dissipation. 
Both of them are localized Nambu-Goldstone modes 
in the presence of the vortex background,
unlike conventional  Nambu-Goldstone modes 
in the ground state. 

\subsection{Effects of strange quark mass}
\label{sec:strange-quark}

Here we consider effects of strange quark mass on the non-Abelian vortices. 
As addressed in Sec.~\ref{sec:cfl_strange}, we can treat the effects as a perturbation in the
high density region $\mu \gg m_{\rm s}$. After absorbing the small term $\varepsilon \propto m_{\rm s}^2$
given in Eq.~(\ref{eq:def_epsilon}) into $\alpha$ as Eq.~(\ref{eq:modif_alpha}),
the background non-Abelian vortex solution, then, is given by just solving the same equations as
Eqs.~(\ref{eq:1})--(\ref{eq:3}) with $\alpha \to \alpha'$.
Furthermore, the leading order terms in the low-energy effective
Lagrangian for the orientational moduli fields are the same as those given in Eqs.~(\ref{eq:eff_lag}),
(\ref{eq:eff_lag_coeff}), and (\ref{eq:lag_cp2}) with the replacement $\alpha \to \alpha'$.

\begin{figure}
\begin{center}
\includegraphics[height=7cm]{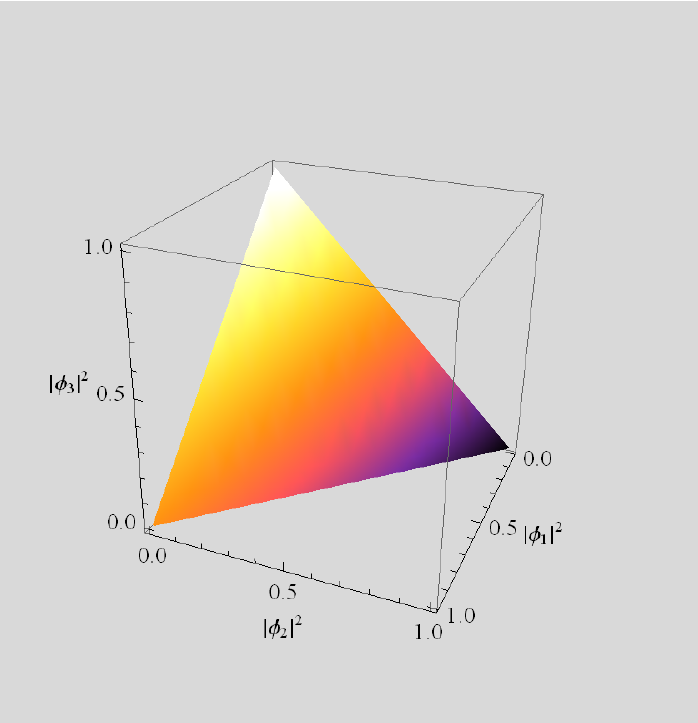}
\includegraphics[height=7cm]{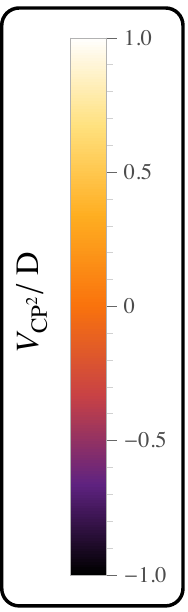}
\label{fig:cp2_pot}
\caption{Contour plot of the effective potential for the $\mathbb{C}P^2$ modes in the $|\phi_1|^2$-$|\phi_2|^2$-$|\phi_3|^2$
space. The colors represent the values of the potential.}
\end{center}
\end{figure}

Let us now turn to find a contribution of the second term in Eq.~(\ref{eq:GL_modification}) in the low-energy
effective Lagrangian. 
By substituting Eq.~(\ref{eq:sol_ori}) into 
the second term in Eq.~(\ref{eq:GL_modification}), we get the following potential term in the $\mathbb{C}P^2$
nonlinear sigma model:
\beq
V_{\mathbb{C}P^2} &=& D \left(|\phi_3|^2 - |\phi_2|^2\right),
\label{eq:pot_eff}\\
D&\equiv& \pi \varepsilon \Delta_{\varepsilon}^2  \int^\infty_0 dr~r\left(g^2-f^2\right),
\label{eq:D}
\eeq
where $(\phi_1,\phi_2,\phi_3)$ is the homogeneous coordinate of $\mathbb{C}P^2$ which satisfies the
constraint $|\phi_1|^2 + |\phi_2|^2 + |\phi_3|^2 = 1$.
Note that $\Delta_{\varepsilon}$, $f$, and $g$ should be obtained after replacing $\alpha$ with $\alpha'$, as in Eq.~(\ref{eq:delta_epsilon}).
Note also that $D$ is positive and finite because $g-f$ is positive for everything over $r$ and gets exponentially
smaller on going away from the vortex; see Eq.~(\ref{eq:asym_G}).
In the limit $\varepsilon \to 0$, the low-energy effective theory is the massless $\mathbb{C}P^2$ nonlinear
sigma model where all the points on $\mathbb{C}P^2$ are degenerate. This is because non-Abelian vortices with
different orientations have the same tensions. Once non-zero $\varepsilon$ is turned on, almost all the points
of $\mathbb{C}P^2$ are lifted and only one special point $(\phi_1,\phi_2,\phi_3) = (0,1,0)$ remains as the 
global minimum of the effective potential; see Fig.~\ref{fig:cp2_pot}.
This means that the non-Abelian vortex with the specific orientation with color ``blue-red" and flavor ``s-u" is energetically 
favored,
\beq
\Phi_{su} \to {\rm diag}(\Delta_{ds},\ \Delta_{su} e^{i\theta},\ \Delta_{ud}),\qquad {\rm as}\quad r \to \infty.
\label{eq:su_vor}
\eeq
This matches the fact that the pairing gap $\Delta_{su}$ is smaller than $\Delta_{ds},\Delta_{ud}$ as 
in Eq.~(\ref{eq:gap_inequality}) so that the vortex whose string tension is proportional to $\Delta_{su}$ is 
more easily created than others. 
As shown below, the details of the dynamics suggest that 
all configurations except for the $(0,1,0)$ vortex
are no longer stable, including 
the $(1,0,0)$ and $(0,0,1)$ vortices.

Let us estimate the lifetime of unstable vortices. For simplicity, we consider the decay from the $(1,0,0)$ vortex 
to the $(0,1,0)$ vortex (from the left-bottom vertex to the right-bottom vertex of the triangle in Fig.~\ref{fig:cp2_pot}).
Since the effective potential is lifted for $|\phi_3|\neq0$ direction, it is reasonable to set $\phi_3=0$ from the beginning.
Then we are left inside the $\mathbb{C}P^1$ submanifold in $\mathbb{C}P^2$.
It is useful to introduce an inhomogeneous coordinate $u(t) \in \mathbb{C}$ by
\beq
\phi_1 = \frac{1}{\sqrt{1+|u|^2}},\quad
\phi_2 = \frac{u}{\sqrt{1+|u|^2}}.
\label{eq:inh_cp1sub}
\eeq
Then the low-energy effective Lagrangian can be rewritten as
\beq
\Lag_{\mathbb{C}P^1} = C^0 \frac{|\dot u|^2}{(1+|u|^2)^2} + D \frac{|u|^2}{(1+|u|^2)}.
\label{eq:lag_cp1sub}
\eeq
The typical timescale appearing with this Lagrangian is
\beq
\tau = \sqrt{\frac{C^0K_0}{D}}.
\label{eq:typical_scale}
\eeq
In principle, we can numerically calculate $\tau$ for each $\mu$. Instead, here, we provide
a simple analytical estimation. Since the profile function $f(r)$ 
for the winding pairing gap ($g$, $h$, and $\rho$) increases (decreases) with a typical scale 
$r \sim \Delta_{\varepsilon}^{-1}$ for $m_{\rm g} \gg m_{1,8}$ as shown in Eqs.~(\ref{eq:asym_F}) and (\ref{eq:asym_G}),
we find from Eq.~(\ref{eq:C0})
\beq
C^0 \sim \frac{m_{\rm g}^2}{g_{\rm s}^2\Delta_{\varepsilon}^2\lambda_3} \sim \left(\frac{\mu}{\Delta_{\varepsilon}}\right)^2 \frac{1}{\lambda_3}.
\label{eq:rough_c0}
\eeq
Furthermore, $D$ is estimated from Eq.~(\ref{eq:D}) as
\beq
D \sim \varepsilon \sim m_{\rm s}^2\log\frac{\mu}{\Delta_{\varepsilon}}.
\label{eq:rough_D_estimate}
\eeq
Thus, the typical timescale of the rollover from the $(1,0,0)$ vortex to the $(0,1,0)$ vortex is
estimated by
\beq
\tau \sim \frac{1}{m_{\rm s}\sqrt{\lambda_3}}\left(\frac{\mu}{\Delta_{\varepsilon}}\right)^2 \left(\log\frac{\mu}{\Delta_{\varepsilon}}\right)^{\frac{1}{2}}.
\label{eq:lifetime}
\eeq
In the limit $m_{\rm s} \to 0$, $\tau \to \infty$ as anticipated.

It is interesting to investigate the possible astrophysical implications of 
the above results. When the core of a neutron star cools down below the 
critical temperature of the CFL phase, a network of non-Abelian vortices 
will be formed by the Kibble-Zureck mechanism. Remarkably, the extrapolation of 
our formula (\ref{eq:lifetime}) to the intermediate density regime 
relevant to the cores of neutron stars ($\mu \sim 500$ MeV) with 
$\Delta_{\varepsilon} \sim 10$ MeV and $m_{\rm s} \simeq 150$ MeV suggests that all the 
vortices decay radically with a lifetime of order $\tau \sim 10^{-21}$ 
s (assuming $\lambda_3 \sim 1$). Although this result should be taken with some care due to the 
uncertainty of the numerical factor in Eq.~(\ref{eq:lifetime}), it is 
reasonable to expect that only one type of non-Abelian vortex, 
which corresponds to the point $(0,1,0)$ in the $\mathbb{C}P^2$ space, 
survives as a response to the rotation of neutron stars in reality. 
The other decaying non-Abelian vortices will emit NG bosons, quarks, 
gluons, or photons during thermal evolution of neutron stars.

Note that, since we are interested in a realistic situation like that inside the core of a neutron star, 
we have imposed $\beta$-equilibrium and electric charge neutrality conditions for deriving
the $m_{\rm s}$-correction to the effective potential in Eq.~(\ref{eq:GL_modification}).
Let us see how the effective Lagrangian changes when we remove the conditions.
The direct $m_{\rm s}$ correction was obtained in Refs.~\cite{Iida:2003cc,Iida:2004cj} as
\beq
\delta V &=& \frac{2}{3}\varepsilon \Tr\left[\Phi^\dagger \Phi\right] + \varepsilon \Tr\left[\Phi^\dagger X_8 \Phi\right],\\
X_8 &=& {\rm diag}\left(\frac{1}{3},\frac{1}{3},-\frac{2}{3}\right).
\eeq
This leads to the effective potential on the $\mathbb CP^2$ manifold
\beq
V_{\mathbb CP^2}' = \frac{2}{3}D \left(2|\phi_3|^2 - |\phi_1|^2 - |\phi_2|^2\right) = \frac{2}{3}D \left(3|\phi_3|^2 - 1\right).
\eeq
In this case the ground state of the potential is the $\mathbb CP^1$ sub-manifold defined by $|\phi_1|^2 + |\phi_2|^2 = 1$.

\subsection{Effects of electromagnetic fields}\label{sec:elemag}

\subsubsection{The effects on the electromagnetic coupling 
on non-Abelian vortices}

We have neglected the effects on the coupling of the CFL matter 
to 
the electromagnetic fields $A_{\mu}^{\rm EM}$
thus far.  
As discussed in Sec.~\ref{sec:cfl_em}, 
the electromagnetic symmetry is embedded into 
the flavor $SU(3)_{\rm F}$ symmetry, 
and it mixes with one of the color gauge fields 
in the CFL ground state.  
In this subsection, we discuss the electromagnetic 
coupling to non-Abelian vortices 
following Ref.~\cite{Vinci:2012mc}.
There are two main consequences of the electromagnetic coupling to 
non-Abelian vortices.
One is a gauging of $U(1)$ symmetry in the ${\mathbb C}P^2$ zero modes, 
i.e., the low energy theory becomes a $U(1)$ gauged  
${\mathbb C}P^2$ model. 
The other is explicit breaking of $SU(3)_{\rm F}$ flavor symmetry, 
which induces the effective potential in the low energy effective theory. 
A physical consequence of the former is discussed later in 
Sec.~\ref{sec:polalizer}. 
Here, we concentrate on the latter effect.

The $SU(3)_{\rm F}$ flavor symmetry is explicitly broken as in Eq.~(\ref{eq:em-breaking}) 
by the electromagnetic coupling.  
Consequently, all non-Abelian vortices are not transformed 
by the $SU(3)_{\rm C+F}$ action once the electromagnetic coupling is 
taken into account. 
Only the $SU(2)_{\rm C+F} \times U(1)_{\rm EM}$ 
subgroup remains exact as in Eq.~(\ref{eq:SU(2)}). 
In order to classify all possible configurations, 
we consider 
a closed loop in the order parameter space generated 
by the gauge/baryon group, which is given by the following transformation on the order parameter, 
\begin{eqnarray}
&\left<\Phi(\infty,\theta)\right> 
= e^{i \theta_{\rm B}(\theta)} e^{i \gamma^{a}(\theta)T^{a}}\left<\Phi(\infty,0)\right>  \,  e^{i \alpha(\theta) T^{\textsc{EM}}} 
\end{eqnarray}
where 
$\theta$ ($0 \leq \theta < 2\pi$) 
is the angle coordinate of the space and 
$\theta_{\rm B}$, $\gamma^{a}$ and $\alpha$ are monotonically increasing functions.
Since $\Phi$ must be single valued, one obtains a relation at 
$\theta = 2\pi$:
\begin{eqnarray}
 & e^{i \gamma^{a}(2 \pi)T^{a}}  =e^{-i \theta_{\rm B}(2 \pi)} e^{-i \alpha(2 \pi) T^{\textsc{EM}}}\,.
 \label{eq:loops}
\end{eqnarray}
This is an equation for the possible symmetry transformations giving a closed loop, or, equivalently, a possible vortex configuration. Notice that existence and stability of vortices related to various solutions of the above equation can in principle only be inferred by a direct study of equations of motion. It is possible to determine all the solutions of Eq.~(\ref{eq:loops}), e.g. using an explicit parametrization of the elements of $SU(3)$. In what follows we will analyze three types of solutions that cannot be related by $SU(2)$ color-flavor transformations: 
1)  ``Balachandran-Digal-Matsuura (BDM)'' case \cite{Balachandran:2005ev}, 
2) ``$\mathbb CP^{1}$'' case, and 
3)  ``Pure color'' case.

1) ``BDM'' case

The first possibility is a closed loop generated by $T^{8}$ in  $SU(3)$ and the electromagnetic $T^{\textsc{EM}}$ alone:
\begin{eqnarray}
& e^{i \gamma^{8}(\theta) T^{8}} =e^{-i \theta_{\rm B}(\theta)} e^{-i \,\alpha(\theta)T^{\textsc{EM}}}.
\end{eqnarray}
By putting $\theta = 2\pi$, we obtain the relation
\beq
 {\gamma^{8}(2\pi) \over \sqrt6}+ {\alpha(2\pi)\over 3} =-{2\pi \over 3}, \quad \theta_{\rm B}(2\pi)={2\pi \over 3}\,.&
\eeq
 The equation above determines the  phases $\gamma^{8}$ and $\alpha$ only up to a linear combination. This is a consequence of the fact that the two generators $T^{8}$ and $T^{\textsc{EM}}$ are proportional and indistinguishable from each other on diagonal configurations. The configuration above is invariant under  $SU(2)$ color-flavor transformations (singlet). 
Balachandran, Digal, and Matsuura considered this case only \cite{Balachandran:2005ev}.

{2)  ``$\mathbb CP^{1}$'' case}

The second possibility is a closed loop generated by winding in the $SU(3)$ group around the $T^{3}$ direction too, in addition to $T^{8}$ and $T^{\textsc{EM}}$:
\beq
& e^{i (\gamma^{3}(\theta) T^{3}+\gamma^{8}(\theta) T^{8})} =e^{-i \theta_{\rm B}(\theta)} e^{-i \,\alpha(\theta)T^{\textsc{EM}}}.
\eeq
By putting $\theta = 2\pi$, we obtain the relation
\beq
 {\gamma^{8}(2\pi)\over \sqrt6}+ {\alpha(2\pi)\over 3} 
= {\pi \over 3}, \quad 
\gamma^{3}(2\pi)=\pm {\sqrt2} \pi,   \quad \theta_{\rm B}(2\pi)={\pi\over 3}.
\eeq
The configuration is not preserved by color-flavor transformations, and  we can generate a orientational moduli space using  $SU(2)$ color-flavor transformations. In fact, we have the most general configuration of this type in the form:
\begin{eqnarray}\label{eq:CFModuli}
\gamma^{3}(2\pi)T^{3}\rightarrow \gamma^{b}(2\pi)T^{b}\, ,
\quad |\gamma^{b}|= {\sqrt2} \pi, \quad b=1,2,3\,,
\end{eqnarray}
where $\rightarrow$ denotes a replacement; the $T^{a}$ above are the generators of $SU(3)$ that commute with $T^{8}$ and form an $SU(2)$ subgroup.
Since the continuous family generated by the $SU(2)$ group 
is characterized by a $\mathbb CP^{1}$ submanifold inside 
the whole $\mathbb CP^{2}$ space, 
we call them  ``$\mathbb CP^{1}$'' vortices.

{3)  ``Pure color'' case}

In terms of the vector $ |\gamma^{a}|$ introduced above, the two previous cases correspond to $ |\gamma^{a}|=0$ and $ |\gamma^{a}|={\sqrt2}\pi$. These are the only two cases for which a non-trivial electromagnetic phase is allowed. In all the other cases, we must have:

\beq
e^{i \gamma^{a}(\theta)T^{a}}=e^{-i\theta_{\rm B}(\theta)}.
\eeq
By putting $\theta = 2\pi$, we obtain the relation
\beq
 \theta_{\rm B}(2\pi) = {2 \pi \over 3},\quad \alpha(2\pi)=0\,.
\eeq
We call this the ``pure color'' case because it does not involve  electromagnetic transformations, and is thus equivalent to the case of 
pure color vortices studied in the previous subsections 
without the electromagnetic coupling. Notice that this case spans a whole $\mathbb CP^{2}$, while $SU(2)$ color-flavor transformations generate only  $\mathbb CP^{1}$ orbits (apart from the case where only $\gamma^{8}$ is non-zero, which is invariant).

We stress again that the cases listed above are just a consequence of boundary conditions when we search for closed loops at spatial infinity. 
Moreover, all the cases are topologically equivalent 
in the absence of the electromagnetic coupling.

\subsubsection{BDM vortex}

The ``BDM case'' corresponds to the vortex studied by Balachandran, Digal and Matsuura in Ref.~\cite{Balachandran:2005ev}. Since we only need $T^{8}$ to generate the correct winding for this vortex, the idea is to restrict the action~(\ref{eq:gl}) to include only the gauge fields $A^{8}$ and $A^{\rm EM}$, by keeping all the other gauge fields to zero. Formally the action reduces to that of a $U(1)\times U(1)$ gauge theory, which we can then express in terms of the massless and massive combinations in Eqs.~\eqref{eq:mix_gluon} and \eqref{eq:mix_photon}:
\begin{eqnarray}
SU(3)_{\rm C}\times U(1)_{\textsc{EM}}\rightarrow U(1)_{8}\times U(1)_{\textsc{EM}}\simeq U(1)_{0}\times U(1)_{\rm M}\,,
\end{eqnarray}
where the arrow means that we truncate the model to its Abelian subalgebra.   
The Lagrangian reads
\begin{align}
\mathcal L &=  \Tr\left[-\frac14\frac32 F^{0}_{ij}F^{0ij}\right] 	
- \Tr\left[\frac14\frac32 F^{\rm M}_{ij}F^{{\rm M}ij}+ K_{1} {\cal D}_{i} \Phi^{\dagger}{\cal D}^{i} \Phi 	
-\beta_{2}(\Phi^{\dagger}\Phi)^{2}+m^{2}\Phi^{\dagger}\Phi  \right]  \nonumber \\
&
-\beta_{1}(\Tr[\Phi^{\dagger}\Phi])^{2}-\frac{3m^{4}}{4(3\beta_{1}+\beta_{2})},   \label{eq:diaglagBDM}
\end{align}
with 
\beq 
{\cal D}_{i}=\p_{i}-ig_{\rm M}A^{\rm M}_i T^{\rm M},
\eeq 
and $g_{\rm M}$ and $T^{\rm M}$ in 
Eqs.~(\ref{eq:g_mix}) and (\ref{eq:TM}), respectively.
This form of the Lagrangian
is applicable only for diagonal configurations for the order parameter $\Phi$. In the action above, the massless field decouples completely, as expected, while the massive field $A^{\rm M}_i$ couples to $\Phi$ in the standard way. The BDM vortex is constructed analogously to Sec.~\ref{sec:NA-vortices} with the only difference from the un-coupled case being the new coupling constant $g_{\rm M}$ instead of $g_{\rm s}$:
\begin{align}
& \Phi(r,\theta)_{\textsc{bdm}}= \Delta_{\textsc{cfl}}
\begin{pmatrix}
 e^{i\theta}f(r) &  0 & 0  \\
0  &  g(r) &  0 \\
 0 & 0  & g(r)   
\end{pmatrix} , \nonumber \\ 
& A_{i}^{\rm M}T^{\rm M}= \frac1{g_{\rm M}}
 \frac{\epsilon_{ij} x^{j}}{r^{2}}[1-h(r)] \, T^{\rm M}
, \quad A_{i}^{0}=0 \,.
\label{eq:BDMvortex}
\end{align}
As shown in Fig.~\ref{fig:tension}, the tension of the vortex decreases monotonically with the gauge coupling. Since we have $g_{\rm M}>g_{\rm s}$, the BDM vortex has a smaller tension as compared to the corresponding vortex in the un-coupled case. 

\subsubsection{$\mathbb CP^{1}$ vortex}

The $\mathbb CP^{1}$ case was 
considered for the first time in Ref.~\cite{Vinci:2012mc}. 
We have identified a vortex configuration that is a solution of the equations of motion that corresponds to these new boundary conditions. To see this, we notice that, similarly to the previous case, we can consistently restrict the action to include only the gauge fields corresponding to generators commuting with $T^{8}$. This corresponds to formally reduce to the case
\begin{align}
SU(3)_{\rm C}\times U(1)_{\textsc{EM}}
\rightarrow SU(2)_{\rm C}\times U(1)_{8}\times U(1)_{\textsc{EM}}\simeq SU(2)_{\rm C}\times U(1)_{0}\times U(1)_{\rm M}\,,
\end{align}
similarly to what we have done for the BDM case. 
The truncated Lagrangian now is as follows:
\begin{align}
	\mathcal L &= \Tr\left[-\frac14\frac32 F^{0}_{ij}F^{0ij}\right]
- \Tr\left[\frac14 \frac32 F^{\rm M}_{ij}F^{{\rm M}ij} -\frac14 F^{b}_{ij}F^{bij} 	+K_{1} {\cal D}_{i} \Phi^{\dagger}{\cal D}^{i} \Phi -\beta_{2}(\Phi^{\dagger}\Phi)^{2}+m^{2}\Phi^{\dagger}\Phi  \right] \non
& -\beta_{1}(\Tr[\Phi^{\dagger}\Phi])^{2}
-\frac{3m^{4}}{4(3\beta_{1}+\beta_{2})}\,,
	\label{eq:diaglagCP1}
\end{align}
where
\beq
	& {\cal D}_{i}=\p_{i}-ig_{\rm M}A^{\rm M}T^{\rm M}-i g_{\rm s}A^{b}T^{b} \, , \quad b=1,2,3 \, ; \quad [T^{b},T^{8}]=0 \,.
\eeq
Here, the index $b$ is relative to the $SU(2)_{\rm C}$ factor.
As before, the massless combination decouples completely, and we can set it to zero. 
Among ${\mathbb C}P^1$ vortices, 
diagonal configurations are 
${\mathbb C}P^{1}_{+}$ and 
${\mathbb C}P^{1}_{-}$ vortices 
given 
in the second and third configurations in
Eqs.~(\ref{eq:ansatz_NA_vortex-b1})-(\ref{eq:ansatz_NA_vortex2-c2}).
One of the simplest vortex configurations
 of the type considered here 
has the following diagonal form:
\begin{align}
	& \Phi(r,\theta)_{\mathbb CP^{1}_{+}}= \Delta_{\textsc{cfl}}
\left(
\begin{array}{ccc}
 g_{1}(r)  &  0 & 0  \\
0  &  e^{i\theta}f(r)  &  0 \\
 0 & 0  & g_{2}(r)   
\end{array}
\right) \, , \nonumber \\ 
	&  A_{i}^{\rm M}T^{\rm M}=-\frac12\frac1{g_{\rm M}}\frac{\epsilon_{ij} x^{j}}{r^{2}}[1-h(r)] T^{\rm M}, \nonumber \\
	& A_{i}^{3}T^{3}=\frac{1}{\sqrt2}\frac1g_{\rm s}\frac{\epsilon_{ij} x^{j}}{r^{2}}[1-l(r)]\,T^{3}\,. \label{eq:CP1ansatz}
\end{align}
As explained in Eq.~(\ref{eq:CFModuli}), we can generate a full $\mathbb CP^{1}$ of solutions by applying $SU(2)_{\rm C+F}$ rotations to the configuration above. Once the ansatz \eqref{eq:CP1ansatz} is inserted into the equations of motion, we obtain the equations given in the Appendix 
of Ref.~\cite{Vinci:2012mc}. We have numerically solved these equations, and determined the energy of the vortex configuration. The tension, as schematically shown in Fig.~\ref{fig:ToricTens}, is higher for the $\mathbb CP^{1}$ vortices than for the BDM vortex. This result can be intuitively understood if we recall the observation of the previous section, where we noticed that the tension of a color vortex decreases monotonically with the gauge coupling. The $\mathbb CP^{1}$ vortex is built with both $g_{\rm s}$ and $g_{\rm M}$ couplings; thus, the interactions depending on $g_{\rm s}<g_{\rm M}$ contribute to increase the tension with respect to a vortex built exclusively from $g_{\rm M}$.

Since the $\mathbb CP^{1}$ vortex is a solution of the equations of motion, it is a critical configuration for the energy density functional. 
It must be a local minimum, and thus a metastable configuration.

\subsubsection{Pure color vortex}

\begin{figure}[htbp]
\begin{center}
\includegraphics[width=0.5\linewidth,keepaspectratio]
{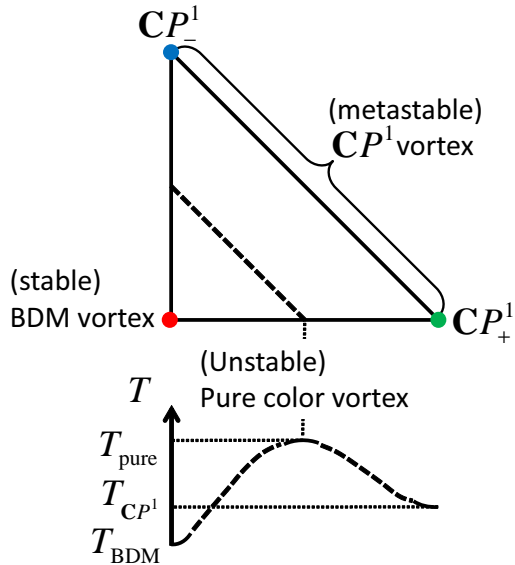}
	\caption{The three different special vortex solutions. The BDM vortex sits in the bottom left corner of the toric diagram and has the lowest tension. The other two diagonal vortices are located at the remaining corners and are connected by color-flavor transformations. All the upper right edge of the diagram represents the whole family of vortex configurations related by this $SU(2)_{\rm C+F}$ transformations, namely a $\mathbb{C}P^{1}$ moduli space. These vortices have a slightly larger tension compared to the BDM configuration. The dashed line in the middle of the diagram corresponds to vortices not winding along the $T^{8}$ direction in the color group. These vortices have the largest tension and do not involve the electromagnetic gauge field. 
}
	\label{fig:ToricTens}
\end{center}
\end{figure}

Let us now come to the third case, the ``pure color'' case. This situation is realized when the boundary conditions are the same as the case without electromagnetic coupling. This means that, at infinity, only the non-Abelian gauge fields give non-trivial winding, and thus non-vanishing fluxes, while the electromagnetic gauge field is zero everywhere. However, this situation is generically not compatible with the coupled equations of motion. An easy way to see this is the fact that boundary conditions in the pure color case imply a non-zero flux for the massless field $A_{i}^{0}$. Since $A_{i}^{0}$ is massless and unbroken, 
there is no topology, compatible with the equations of motion, stabilizing and confining the flux.
We thus expect that no solutions exist in general for the pure case. However, a special configuration that we call the ``pure color vortex'' is an exception to this statement. This configuration is defined as that corresponding to the following boundary conditions:
\begin{eqnarray}
&e^{i \gamma^{a}(2\pi)T^{a}}=e^{-i\theta_{\rm B}(2\pi)}\,; \quad \gamma^{8}\equiv 0\,. &
\end{eqnarray}
Notice that the conditions above are perfectly consistent. We can take any configuration of $\gamma^{a}$ phases and set $\gamma^{8}$ to zero using a gauge(-flavor) transformation. This means that we can consistently set:
\begin{eqnarray}
A_{i}^{8}=A_{i}^{\rm EM}=0, \quad \Rightarrow \quad A_{i}^{\rm M}=A_{i}^{0}=0\,.
\end{eqnarray}
In fact, $A_{i}^{\rm EM}=0$ means that for the pure color vortex we can consistently restrict the action 
by simply dropping all the terms involving $A_{i}^{\rm EM}$. As a consequence, the pure color vortex is exactly the same configuration as we get in the un-coupled case. Moreover, it satisfies the full equations of motion of the coupled case, because of the consistent restriction. Notice that the dependence of the electromagnetic gauge coupling also disappears completely from the restricted action. This means that the tension of the pure color vortex is also the same as the tension of the un-coupled vortices, involving only $g_{\rm s}$. As represented in Fig.~\ref{fig:ToricTens}, the tension of this vortex is larger than both the BDM and the $\mathbb CP^{1}$ vortices.

Because of this very fact we are led to conclude that the pure vortex is in fact a stationary point for the energy functional, but it corresponds to the global maximum. Both the BDM and the $\mathbb CP^{1}$ vortices are, on the other hand, local minima, with the BDM vortex being the absolute minimum. The $\mathbb CP^{1}$ vortices are, however, metastable, and if long-lived can play a crucial role in the physics  of the CFL superconducting phase together with the BDM vortex. The whole situation is schematically summarized in Fig.~\ref{fig:ToricTens}. The numerical values of the tensions are compared as 
\begin{eqnarray}
&\mathcal T_{\text{pure}}-\mathcal T_{\textsc{bdm}}=0.0176 \,{\rm MeV}^{2}, \quad 
 \mathcal T_{\text{pure}}- \mathcal T_{\mathbb CP^{1}}= 0.0044 \,{\rm MeV}^{2} \, ,
 \label{eq:tensions}
\end{eqnarray}
for the same ``realistic'' choice of parameters made in Fig.~\ref{fig:tension}, where in addition we have chosen $m_{\rm g}=92$ MeV and a value for the electromagnetic coupling constant of $e^{2}=1/137$. Notice that the expressions shown above do not depend on the infrared regulator $L$. The diverging parts of the tensions in Eq.~\eqref{eq:ene_NA} 
are equal and cancel out in the differences.

\subsubsection{Magnetic fluxes}

There are two main differences between color magnetic flux tubes with and without electromagnetic coupling. The first one, as we have already examined, is the lifting of the moduli space $\mathbb CP^{2}$ to leave  the stable BDM vortex and the family of  metastable degenerate $\mathbb CP^{1}$ vortices. The second is the fact that coupled vortices now carry a non-trivial electromagnetic flux. This is given by the fact that coupled vortices are made of the massive field $A_{i}^{\rm M}$, which is in turn a linear combination of color and electromagnetic gauge fields. 

\begin{figure}[htbp]
\begin{center}
\includegraphics[width=0.20\linewidth,keepaspectratio]{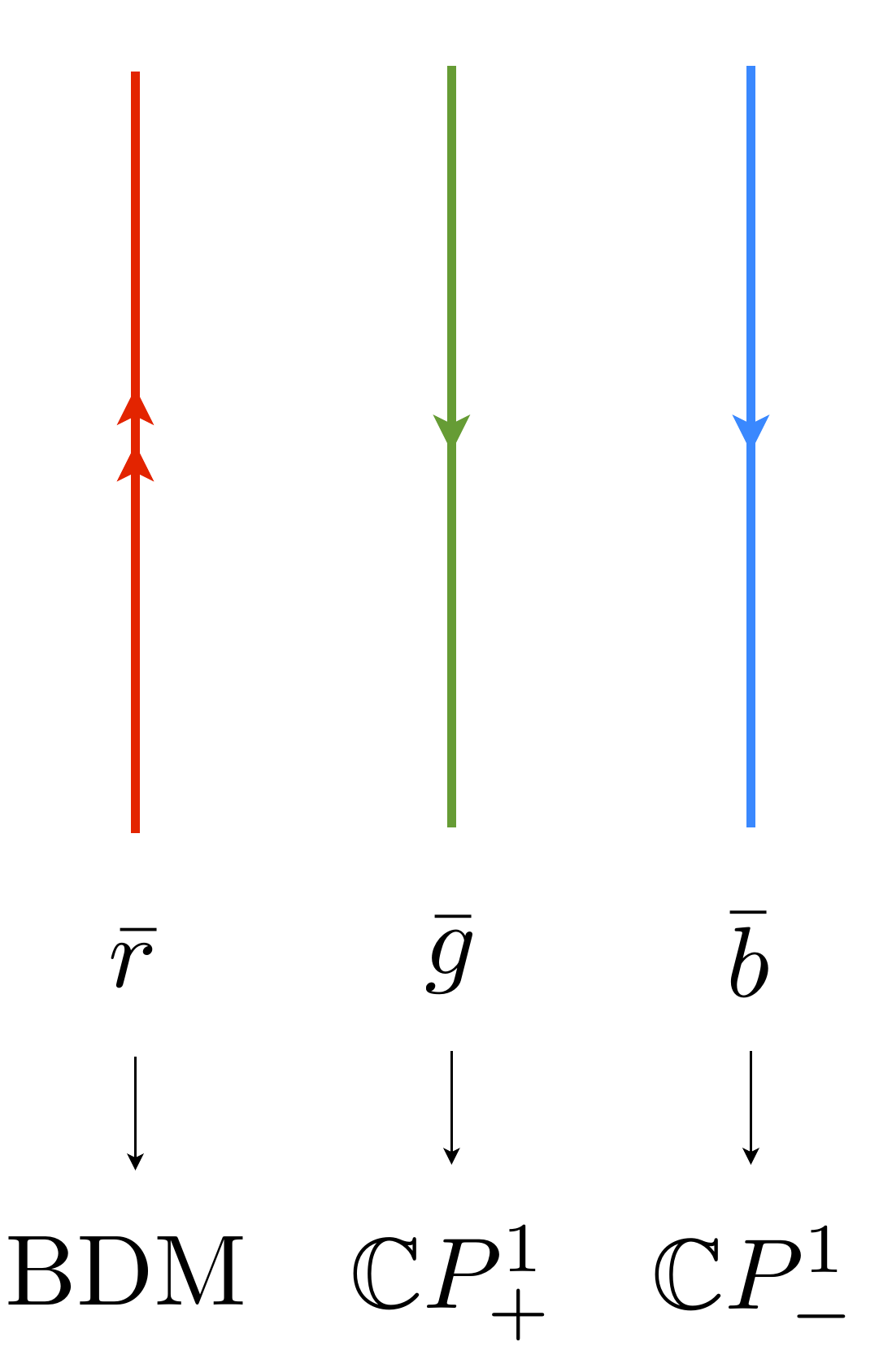}
	\caption{The three different types of vortices we consider. The BDM vortex is labeled as the anti-red $\bar{r}$ vortex, as explained in the text, and carries twice the electromagnetic flux of the other solutions, the anti-blue $\bar{b}$ and anti-green $\bar{g}$ vortices.}
	\label{fig:VortColors}
\end{center}
\end{figure}

The BDM vortex carries a quantized $A_{i}^{\rm M}$ flux:
\begin{eqnarray}
& \displaystyle \Phi^{\rm M}_{\textsc{bdm}} 
= \oint \vec A^{\rm M}\cdot  d\vec l=\frac{2 \pi}{g_{\rm M}}\,.
 \end{eqnarray}
Because of the mixing in the ground state, this means the following non-quantized fluxes for the color and electromagnetic fields:
\begin{align}
& \Phi^{8}_{\textsc{bdm}}=\sqrt\frac{2}{3}\frac{1}{1+\delta^{2}}\frac{2 \pi}{g_{\rm s}},\quad \Phi^{\textsc{EM}}_{\textsc{bdm}}=\frac{\delta^{2}}{1+\delta^{2}}\frac{2 \pi}{e}\, , \quad \delta^{2} \equiv \frac23 \frac{e^{2}}{g_{\rm s}^{2}} \, .\label{eq:fluxes}
 \end{align}
 The fluxes of the $\mathbb CP^{1}$ vortices can be similarly determined; see Fig.~\ref{fig:VortColors}:
\begin{align}
&\Phi^{\rm M}_{\mathbb CP^{1}}=-\frac12 \Phi^{\rm M}_{\textsc{bdm}} \, 
 \quad \Rightarrow \quad   
\Phi^{8}_{\mathbb CP^{1}}=-\frac12 \Phi^{8}_{\textsc{bdm}},\quad 
\Phi^{\textsc{EM}}_{\mathbb CP^{1}}=-\frac12 \Phi^{\textsc{EM}}_{\textsc{bdm}}, \nonumber \\
 &\Phi^{3}_{\mathbb CP^{1}_{+}} \, =  \,  \oint \vec A^{3}\cdot  d\vec l=\frac{1}{\sqrt2}\frac{2 \pi}{g_{\rm s}}\,,\quad \Phi^{3}_{\mathbb CP^{1}_{-}}=-\Phi^{3}_{\mathbb CP^{1}_{+}}\,.
 \end{align}
 Moreover, the quantized circulations of the BDM  and $\mathbb CP^{1}$ vortices are the same as that of the usual un-coupled vortices, and are equal to $c_{B}\sim 1/3$ 
in Eq.~(\ref{eq:circulation-NA}).

Notice that the expressions determined above imply that a color-neutral bound state of vortices necessarily also carries no electromagnetic flux, and vice versa. In particular, in the un-coupled case we need at least a bound state of three vortices, ($\bar r, \bar g, \bar b$) to obtain a color-less state, which is nothing but the $U(1)_{\rm B}$ vortex. In the coupled case, this ``minimal'' color-less bound state is obtained with the combination (${\rm BDM},\, \mathbb CP^{1}_{+},\,\mathbb CP^{1}_{-}$). The bound state carries an integer circulation $c_{B} \sim 1$ in Eq.~(\ref{eq:circulation-U1}).

\subsubsection{Quantum mechanical decay of the ${\mathbb C}P^1$ vortices}
As we have seen, the ${\mathbb C}P^1$ vortices are classically metastable because of the potential barrier 
provided by the presence of the pure color vortex. 
However, here we show that the ${\mathbb C}P^1$ vortices are quantum mechanically unstable, and they can decay to the BDM vortex by quantum tunneling. 
We estimate this decay probability.

As discussed in the previous subsections, we do not expect static solutions of the equations of motion apart from the BDM, the $\mathbb CP^{1}$, and the pure color vortices. However, we can try to define an effective potential interpolating between the three types of solutions. At generic orientations, the boundary conditions are the same as those for the un-coupled vortices. In fact, the un-coupled vortex evaluated on the coupled action Eq.~(\ref{eq:gl}) exactly gives the same tension as that of un-coupled vortices. This value of the tension is an upper bound for vortex configurations that have a fixed boundary condition corresponding to a generic point in $\mathbb CP^{2}$. The tension of the configuration that really minimizes the energy, for that fixed boundary conditions, defines an ``effective'' potential on the  $\mathbb{C}P^{2}$ moduli space induced by the electromagnetic interactions. Moreover, since the tension of vortices is mainly modified by the contribution of the mixed $A_{i}^{\rm M}$, which couples with a larger gauge coupling, we expect the qualitative behavior of the effective potential to be of the form represented in Fig.~\ref{fig:ToricTens}.

This qualitative picture is enough to address some important features of coupled color-magnetic flux tubes. The fact that the potential has more than one local minimum allows for the existence of kinks interpolating between the two vortices corresponding to the various (meta)stable configurations; these kinks are interpreted as confined monopoles.  Moreover, the presence of kinks, and the higher tension of  $\mathbb CP^{1}$ vortices  with respect to the BDM vortex, implies a decay rate of the former vortices into the latter through quantum tunneling. This tunneling proceeds by enucleation of kink/anti-kinks pairs along the vortex \cite{Preskill:1992ck,Shifman:2002yi,Voloshin:1985id}  and it is similar to the quantum decay of a false vacuum in 1+1 dimensions \cite{Voloshin:1985id}.  An analogous situation arises  for pure color vortices \cite{Eto:2011mk,Gorsky:2011hd}, where  the potential is generated by quantum non-perturbative effects of the  $\mathbb CP^{2}$ non-linear sigma model.

 The decay rate can be roughly estimated in our case by following the arguments of Refs.~\cite{Preskill:1992ck,Voloshin:1985id}. The enucleation of a couple of kinks costs an energy of order $M_{\rm kink}$. Moreover, they are created at a critical distance $L_{\rm crit}$ such that the energy cost for the pair production is balanced by the energy gain due to the presence of an intermediate vortex with smaller tension: $L_{\rm crit} \Delta \mathcal T \sim M_{\rm kink}$. The decay probability rate per unit length is thus:
\begin{equation}
P\sim e^{-M^{2}_{\rm kink}/ \Delta \mathcal T}\,.
\end{equation}

We now apply the formula above to our specific case. The mass of the kinks can be estimated as being of the order of the square root of the height of the potential times the ``size'' $\beta$ of the moduli space: 
\begin{equation}
M_{\rm kink}\sim \beta \sqrt{\mathcal T_{\text{pure}}-\mathcal T_{\mathbb CP^{1}}}\,.
\end{equation}
The quantity $\beta$ has been evaluated analytically and numerically in  Refs.~\cite{Eto:2011mk,Gorsky:2011hd}:
\begin{equation}
\beta \sim K_{1}^{2}/\beta_{1}\sim \mu^{2}/T_{\rm c}^{2},
\end{equation}
and it turns out to be large for our ``realistic'' regime, $\beta\sim2500$. The tension difference is given by $\Delta \mathcal T=\mathcal T_{\text{pure}}-\mathcal T_{\textsc{bdm}}$. We have already reported numerical estimates of the  quantities above in Eq.~(\ref{eq:tensions}), for a special value of the couplings.
The decay probability is then:
\begin{equation}
P\sim e^{-\beta\,R},\quad R\equiv\frac{\mathcal T_{\text{pure}}-\mathcal T_{\mathbb CP^{1}}}{\mathcal T_{\text{pure}}-\mathcal T_{\textsc{bdm}}}
\label{eq:Ratio}
\end{equation}

\begin{figure}[htbp]
\begin{center}
\begin{tabular}{cc}
\includegraphics[height=4.5cm]{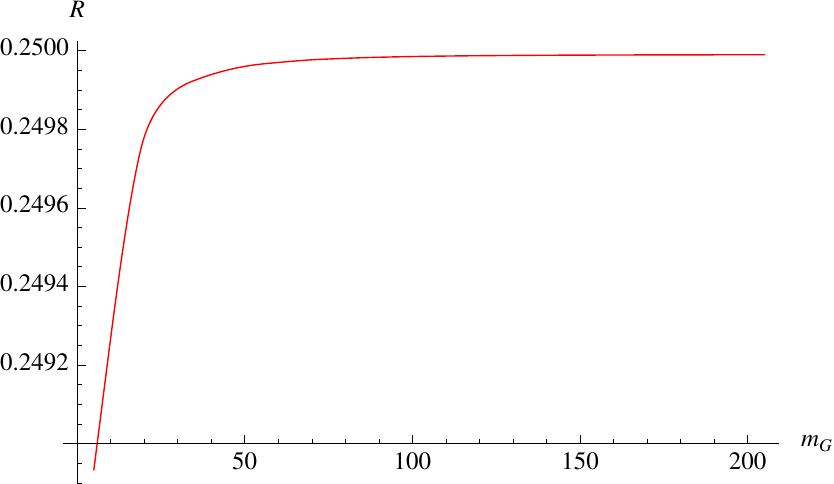} & 
\includegraphics[height=4.5cm]{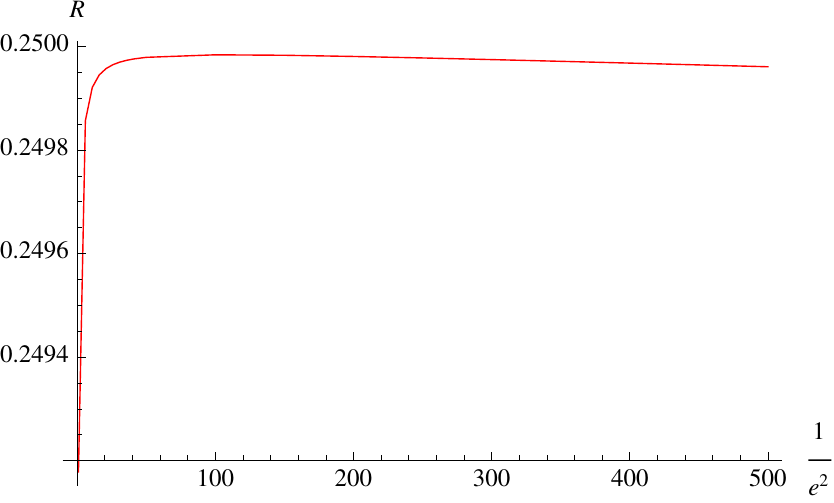} 
\end{tabular}
	\caption{The numerical value of the ratio $R$ of Eq.~(\ref{eq:Ratio}). In the left panel, we plot it as a function of the mass $m_{\rm g}$, fixing the value of the electromagnetic gauge coupling to its realistic value $e^{2}=1/137$. In the right panel, we plot the same quantity as a function of $e$, fixing the value of the mass $m_{\rm g}$ to a typical value: $m_{\rm g}=92$ MeV. Notice that the value of the ratio $R$ is always of order 1 for the wide range of physical parameter values chosen.}
	\label{fig:Ratio}
	\end{center}
\end{figure}

Substituting the numerical values of Eq.~(\ref{eq:tensions}), we obtain $R\sim0.25$. We have also studied the dependence of this decay probability in terms of a more general set of values of the gauge couplings. The numerical results are shown in Fig.~\ref{fig:Ratio}. In the left panel  we have plotted the quantity $R$ as a function of the mass $m_{\rm g}$, which depends on the gauge coupling $g_{\rm s}$, where we have set $e^{2}=1/137$.  In the right panel, we show the same quantity as a function of the electromagnetic coupling $e$, where we have set $m_{\rm g}=92$ MeV. We see that $R$ has a very mild dependence on the value of both $m_{\rm g}$ and $e$. As shown in Fig.~\ref{fig:Ratio}, $R$ is always a quantity of order 1.  We have limited ourselves to considering large values of the gauge bosons ($m_{\rm g}\gtrsim 10$ MeV) and   small values of the electromagnetic gauge coupling ($e\lesssim1$), as expected in realistic settings in the CFL phase. We thus see that, for the range of values of the couplings considered, the ratio $R$ is never small enough to compensate the large ``moduli space'' factor $\beta$. We thus estimate the probability of decay of  $\mathbb CP^{1}$ vortices in BDM vortices to be exponentially small in realistic settings. 

 Notice that the estimate made above is fully justified only in the case in which the size of the kink is negligible as compared to the critical length $L_{\rm crit}$, while  in our case the two sizes are comparable. This estimate is very rough, and has to be corrected including possible Coulomb interactions between the kink/anti-kink pair. However, it should correctly capture the order of magnitude of the decay probability. 
It is an open problem to give a more precise estimate of the decay probability  of  $\mathbb CP^{1}$ vortices in the CFL with electromagnetic coupling and vanishing quark masses. However, as we shall see in the next section, the effects of a non-vanishing strange quark mass overshadow the effects of the potential induced by electromagnetic interactions in more realistic settings.

\subsubsection{Comparison with other potential terms}
\label{sec:stability}

In this subsection, we compare the potential generated semi-classically by the electromagnetic interactions with the other potentials, {\it i.e.}, 
the quantum mechanically induced potential and 
the potential induced by the strange quark mass.

First, the quantum potentials in 
Eqs.~(\ref{eq:ene_N_minima}) and (\ref{eq:mass_gap}) are 
 exponentially small and negligible as compared to the semiclassical potential induced by electromagnetic interactions.

Second, the potential induced by electromagnetic interactions is comparable with the effects of the strange quark mass 
discussed in Sec.~\ref{sec:strange-quark} 
only when the strange quark mass is very negligible as compared to the chemical potential ($\epsilon\ll m^{2}$). 
These densities are not realistic in, e.g., neutron star cores. 
We then conclude that  $\mathbb CP^{1}_{+}$
 is the most relevant in realistic densities, 
for instance  in the inner core of neutron stars.
This is interesting because 
Balachandran, Digal, and Matsuura concluded that 
the BDM vortex is the most stable 
\cite{Balachandran:2005ev}.

\subsection{Quantum monopoles and the quark-hadron duality}\label{sec:monopoles}

Understanding the confinement of quarks and gluons is one of the most important question in QCD.
One plausible scenario is the dual superconducting picture of the QCD vacuum \cite{Nambu:1974zg,Zichichi:1976uh,Mandelstam:1974pi}:
Assuming the condensation of putative magnetic monopoles, the color electric flux is squeezed and
the quarks would be confined.
Although this dual picture can explain a number of properties in the QCD vacuum (see, e.g., \cite{Suganuma:1993ps,Kondo:1997pc})
and is shown to be realized in the $\mathcal N=2$ supersymmetric QCD \cite{Seiberg:1994rs,Seiberg:1994aj}, the condensation
or even the existence of magnetic monopoles have not been justified in real QCD without 
dramatic assumptions \cite{Greensite:2003bk}.
If the magnetic monopoles indeed exist in real QCD, it is natural to expect that they would also appear
in QCD at finite temperature $T$ and finite quark chemical potential $\mu$. 
In the quark-gluon plasma phase at high $T$, a number of instances for evidence of the existence 
of monopoles were suggested in the model calculation in conjunction with the lattice QCD simulations
(for reviews, see Refs.~\cite{Chernodub:2008vn,Shuryak:2008eq}).

In this subsection, we ask whether the monopole exists or not in QCD at large $\mu$.
It is indeed an ideal situation to investigate this question because of the following two reasons:
Firstly, the physics is under theoretical control in this regime because the QCD coupling constant
$g_s$ is weak according to the asymptotic freedom. Secondly, the ground state is the most symmetric
three-flavor CFL phase in the color superconductivity.
By making use of these two advantages, we will show that mesonic bound states of confined
monopoles appear inside the non-Abelian vortices in the CFL phase.

We start with solving $\mathbb{C}P^{N_{\rm C}-1}$ model in $1+1$ dimensions, 
taking the quantum effects into account. The case with $N_{\rm C} =3$ corresponds to the 
low energy effective theory on the non-Abelian vortex. Unfortunately, the solution to the $\mathbb{C}P^2$
nonlinear sigma model is not known  so far. Thus, we solve the model to leading order of
$1/N_{\rm C}$ expansion following \cite{D'Adda:1978kp,Witten:1978bc}. Owing to the qualitative similarity of the solutions
to the $\mathbb{C}P^1$ and $\mathbb{C}P^{N_{\rm C}-1}$ models, the solution to the $\mathbb{C}P^2$
model should be well approximated by taking $N_{\rm C}=3$ at the end.

Let us first rescale the variables as
\beq
t \to \sqrt{C^0} t,\quad
z \to \sqrt{C^3} z,\quad
\phi \to \sqrt{C^0C^3} \phi.
\label{eq:rescaling}
\eeq
Then the low energy effective Lagrangian given in Eqs.~(\ref{eq:eff_lag})--(\ref{eq:lag_cp2}) can be
rewritten in the following form:
\beq
S = \int d^2x\ \left[\left|(\p_\alpha - iA_\alpha) \phi \right|^2 - \sigma \left(|\phi|^2 - \frac{N_{\rm C}}{3}\sqrt{C^0C^3}\right)\right].
\label{eq:cp2_Q_act}
\eeq 
Here we have performed the Hubbard-Stratonovich transformation by introducing 
the auxiliary gauge field $A_\alpha$ and also the Lagrange multiplier $\sigma$.
After the rescaling, the partition function has the standard normalization as
\beq
Z = \int \left[d\phi d\phi^\dagger d\sigma dA_\alpha\right]e^{iS}.
\label{eq:partition_func}
\eeq
Integrating out $\phi$ and $\phi^\dagger$, one obtains 
\beq
Z = \int\left[d\sigma A_\alpha\right] \exp
\left\{ - N_{\rm C} \tr \log\left[ - \left(\p_\alpha + i A_\alpha\right)^2 - \sigma \right]
+ \frac{iN_{\rm C}}{3}\sqrt{C^0C^3}\int d^2x~\sigma
\right\}.
\label{eq:partition_func2}
\eeq
Following \cite{Witten:1978bc}, we solve this in a stationary phase approximation. From the ``Lorentz'' symmetry,
the saddle point should be a point with $A_\alpha = 0$ and $\sigma=$ constant.
Then, we vary the partition function in terms of $\sigma$ leading to the following gap equation
\beq
i\frac{\sqrt{C^0C^3}}{3}+
\int^{\Lambda = \Delta_{\rm CFL}} \frac{d^2k}{(2\pi)^2} \frac{1}{k^2-\sigma + i\epsilon}  = 0.
\label{eq:gap_eq}
\eeq
Note, here, that we have introduced the cutoff scale $\Lambda = \Delta_{\rm CFL}$ which is not
a dynamical cutoff of the $\mathbb{C}P^2$ model but is a physical cutoff $\Delta_{\rm CFL}$ 
below which the description in the low energy effective theory does make sense. 
After integration, we get
\beq
M^2 \equiv \sigma = \Delta_{\rm CFL}^2 e^{-4\pi \frac{\sqrt{C^0C^3}}{3}} \sim \Delta_{\rm CFL}^2 e^{-\gamma \left(\frac{\mu}{\Delta_{\rm CFL}}\right)^2} > 0,
\label{eq:mass_gap}
\eeq
where $\gamma$ is a certain positive constant and we have used $C^0 \sim C^3 \sim \left(\mu/\Delta_{\rm CFL}\right)^2$ 
under the assumption $\lambda_3 \sim \varepsilon_3 \sim 1$.
Looking back to the original action Eq.~(\ref{eq:cp2_Q_act}), one sees that a positive expectation value
of $\sigma$ is a mass of $\phi$ and $\phi^\dagger$.

Let us next consider fluctuations of $\sigma$ and $A_\alpha$ around the saddle point.
It was shown \cite{Witten:1978bc} that higher order terms in $\sigma$ and $A_\alpha$ are suppressed in the large-$N_{\rm C}$ limit;
only the quadratic terms are relevant. It turns out that the relevant Feynman diagrams are the 
propagator of $A_\alpha$ at the one-loop level \cite{Witten:1978bc}. Thus the effective world-sheet theory including the quantum effects
to leading order of $1/N_{\rm C}$ is 
\beq
\Lag_{\rm eff}^{\rm quant} = \left|(\p_\alpha - iA_\alpha)\phi\right|^2 - M^2 |\phi|^2 - \frac{N_{\rm C}}{48\pi M^2}F_{\alpha\beta}^2.
\label{eq:lag_quantum}
\eeq
By rescaling $A_\alpha$ so that the kinetic terms of $A_\alpha$ are canonically normalized,
\beq
A_\alpha \to \sqrt{\frac{12\pi M^2}{N_{\rm C}}}~A_\alpha,
\label{eq:rescale_A}
\eeq
the final form of the quantum effective theory reads
\beq
\Lag_{\rm eff}^{\rm quant} = \left|(\p_\alpha - ieA_\alpha)\phi\right|^2 - M^2 |\phi|^2 - \frac{1}{4}F_{\alpha\beta}^2,
\quad e\equiv \sqrt{\frac{12\pi}{N_{\rm C}}}~M.
\label{eq:lag_eff_q}
\eeq
In summary, taking the quantum effects into account, the auxiliary gauge fields become dynamical and
$\phi$ acquires the $U(1)$ charge $e$ and 
a non-zero mass, which is consistent with the Coleman-Mermin-Wagner theorem in $1+1$ dimensions
\cite{Coleman:1973ci,PhysRevLett.17.1133}.

\begin{figure}[h]
\begin{center}
\includegraphics[width=10cm]{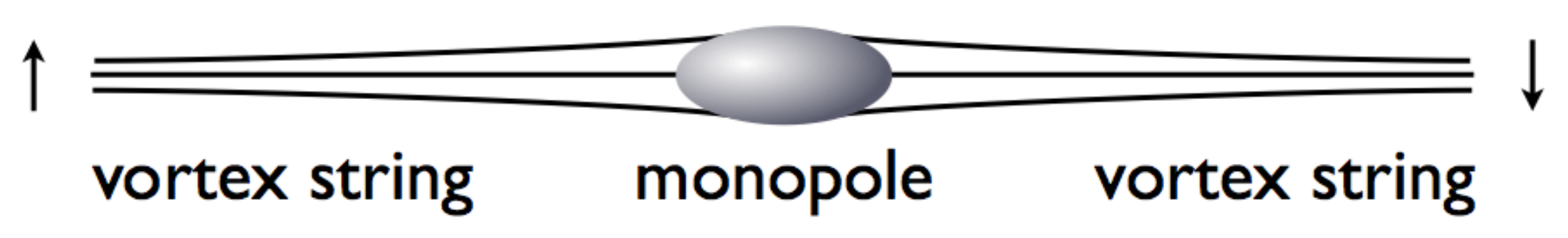}
\caption{A monopole as a kink of the orientational moduli $\phi$ in supersymmetric models.}
\label{fig:conf_mono}
\end{center}
\end{figure}

We are now ready to understand the above quantum phenomena in $1+1$ dimensions from the $3+1$ dimensional viewpoint.
A clue is that the particles $\phi$ and $\phi^\dagger$ in $1+1$ dimensions can be interpreted as topological
solitons, namely a kink and an anti-kink, as was found by Witten \cite{Witten:1978bc}. This might be best seen in a supersymmetric
$\mathbb{C}P^{N_{\rm C}-1}$ model, which has $N_{\rm C}$ degenerate quantum vacua with $\mathbb Z_{N_{\rm C}}$ symmetry.
The kinks interpolate two adjacent vacua among the $N_{\rm C}$ vacua.
Remember that, for us, the $\mathbb{C}P^{N_{\rm C}-1}$ fields $\phi$ and $\phi^\dagger$ are the internal orientational moduli of the
non-Abelian vortex in the CFL phase. Therefore, the kink on the vortex is a particle soliton in $3+1$ dimensions and is
a junction of two different non-Abelian vortices. Thereby, it is nothing but a magnetic monopole in the Higgs phase;
see Fig.~\ref{fig:conf_mono}.
The mass of the monopole can be read from Eq.~(\ref{eq:lag_quantum}) as
\beq
M_{\rm monopole} = M \sim \Delta_{\rm CFL} e^{-\frac{\gamma}{2}\left(\frac{\mu}{\Delta_{\rm CFL}}\right)^2}.
\label{eq:monopole_mass}
\eeq
This is much smaller than the mass $\sim \Delta_{\rm CFL} \log(L/\xi)$ of the non-Abelian vortex of the unit length.
Dealing with the monopole as the soliton in the low energy world-sheet theory is justified by this hierarchy in the mass scales.

\begin{figure}[h]
\begin{center}
\includegraphics[width=15cm]{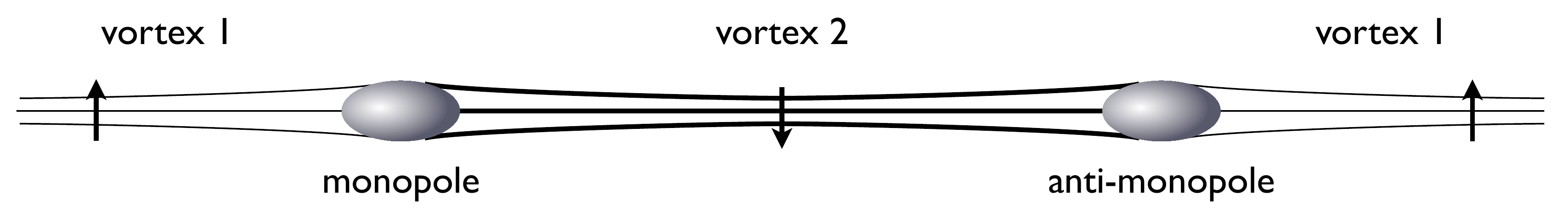}
\caption{A confined monopole and anti-monopole in non-supersymmetric models.}
\label{fig:conf_mono2}
\end{center}
\end{figure}

Though the kinks on the vortex string are deconfined in the supersymmetric $\mathbb{C}P^{N_{\rm C}-1}$ model \cite{Hanany:2004ea,Gorsky:2004ad},
they are confined in the non supersymmetric case. This is because the particles $\phi$ and $\phi^\dagger$ have
charges $\pm e$ for the dynamical $U(1)$ gauge field as in Eq.~(\ref{eq:lag_eff_q}). In one space dimension
the Coulomb potential is linear, so that, even if the charges are small, they are confined.
This is the reason why Witten referred to $\phi$ and $\phi^\dagger$ as ``quarks'' (for us they are magnetic
monopoles).
The linear potential between a monopole at $z=z_1$ and an anti-monopole at $z=z_2$ is given by
\beq
V_{\rm conf} = e^2|z_1 - z_2|.
\label{eq:pot_conf}
\eeq
We illustrate the situation where the monopole and the anti-monopole are confined on the non-Abelian vortex in
Fig.~\ref{fig:conf_mono2}.
The monopole and the anti-monopole are accompanied by semi-infinite long strings (vortex 1 in Fig.~\ref{fig:conf_mono2}).
The tension of vortex 1 is given by
\beq
T_1 = T_{\rm classical} - \frac{N_{\rm C}}{3}M^2 \sim T_{\rm classical} - \frac{N_{\rm C}}{3}\Delta_{\rm CFL}^2 e^{-\gamma\left(\frac{\mu}{\Delta_{\rm CFL}}\right)^2}.
\label{eq:t1}
\eeq
From Eq.~(\ref{eq:pot_conf}) we see that the vortex string (vortex 2 in Fig.~\ref{fig:conf_mono2}) has a tension $T_2$ bigger 
than that of vortex 1 by
\beq
T_2 - T_1 = e^2 = \frac{12\pi}{N_{\rm C}}M^2 \sim \frac{12\pi}{N_{\rm C}} \Delta_{\rm CFL}^2 e^{-\gamma\left(\frac{\mu}{\Delta_{\rm CFL}}\right)^2}.
\label{eq:t2}
\eeq

There is another perspective for understanding of this phenomenon \cite{Gorsky:2004ad}.
The vacuum structure of the non-supersymmetric $\mathbb{C}P^{N_{\rm C}-1}$ model
can be realized by looking at the $\vartheta$ dependence of the theory \cite{Witten:1998uka,Shifman:1998if,Gorsky:2004ad} by adding
the $\vartheta$ term
\beq
\Lag_\vartheta = \frac{\vartheta}{2\pi} \epsilon^{\alpha\beta}\p_\beta A_\alpha,
\label{eq:lag_theta}
\eeq
where $A_\alpha$ is that before rescaling (\ref{eq:rescale_A}).
Recalling that the vacuum energy $E(\vartheta)$ is of order $N_{\rm C}$ in the large $N_{\rm C}$ limit,
$E(\vartheta)$ is expressed as \cite{Witten:1998uka,Gorsky:2004ad}
\beq
E(\vartheta) = N_{\rm C} M^2 f\left(\frac{\vartheta}{N_{\rm C}}\right).
\label{eq:ene_largeN}
\eeq
Here $f(\vartheta)$ is an even function of $\vartheta$ due to the CP symmetry under which
$\vartheta$ transforms as $\vartheta \to -\vartheta$.
Furthermore, $E(\vartheta)$ is a periodic function as
\beq
E(\vartheta + 2\pi) = E(\vartheta).
\label{eq:periodicity_theta}
\eeq
One might suspect that these two conditions are incompatible at first sight.
However, there is a way out: both of them can be satisfied when
$E(\vartheta)$ is a multibranched function as
\beq
E(\vartheta) = N_{\rm C} M^2 \min_k \left\{f\left(\frac{\vartheta + 2\pi (k-1)}{N_{\rm C}}\right)\right\},\quad
k=1,2,\cdots, N_{\rm C}.
\label{eq:ene_theta}
\eeq
Expanding $f(\vartheta) = f_0 + f_2 \vartheta^2 + \cdots$ and considering that higher order terms
in $\vartheta$ are suppressed at large $N_{\rm C}$, the vacuum energy at $\vartheta=0$ is given by
\beq
E_k(0) = - \frac{N_{\rm C}}{3}M^2 + \frac{12\pi}{N_{\rm C}}M^2 (k-1)^2,\quad k=1,2,\cdots,N_{\rm C},
\label{eq:ene_N_minima}
\eeq
where we have set $f_0 = -1/3$ and $f_2 = 12\pi$ in such a way that Eqs.~(\ref{eq:t1}) and (\ref{eq:t2})
are reproduced. Therefore, there exist $N_{\rm C}$ local minima among which
only one $(k=1)$ is a genuine ground state while the others are quasi vacua.
Each local minimum corresponds to a quantum vortex state. 
Namely, the genuine ground state $(k=1)$ is the stable vortex with $T_1 = T_{\rm classical} + E_1(0)$, 
and the quasi vacua with $k \ge 2$ are the metastable string with the tension $T_k = T_{\rm classical} + E_k(0)$.

\begin{figure}[h]
\begin{center}
\includegraphics[width=11cm]{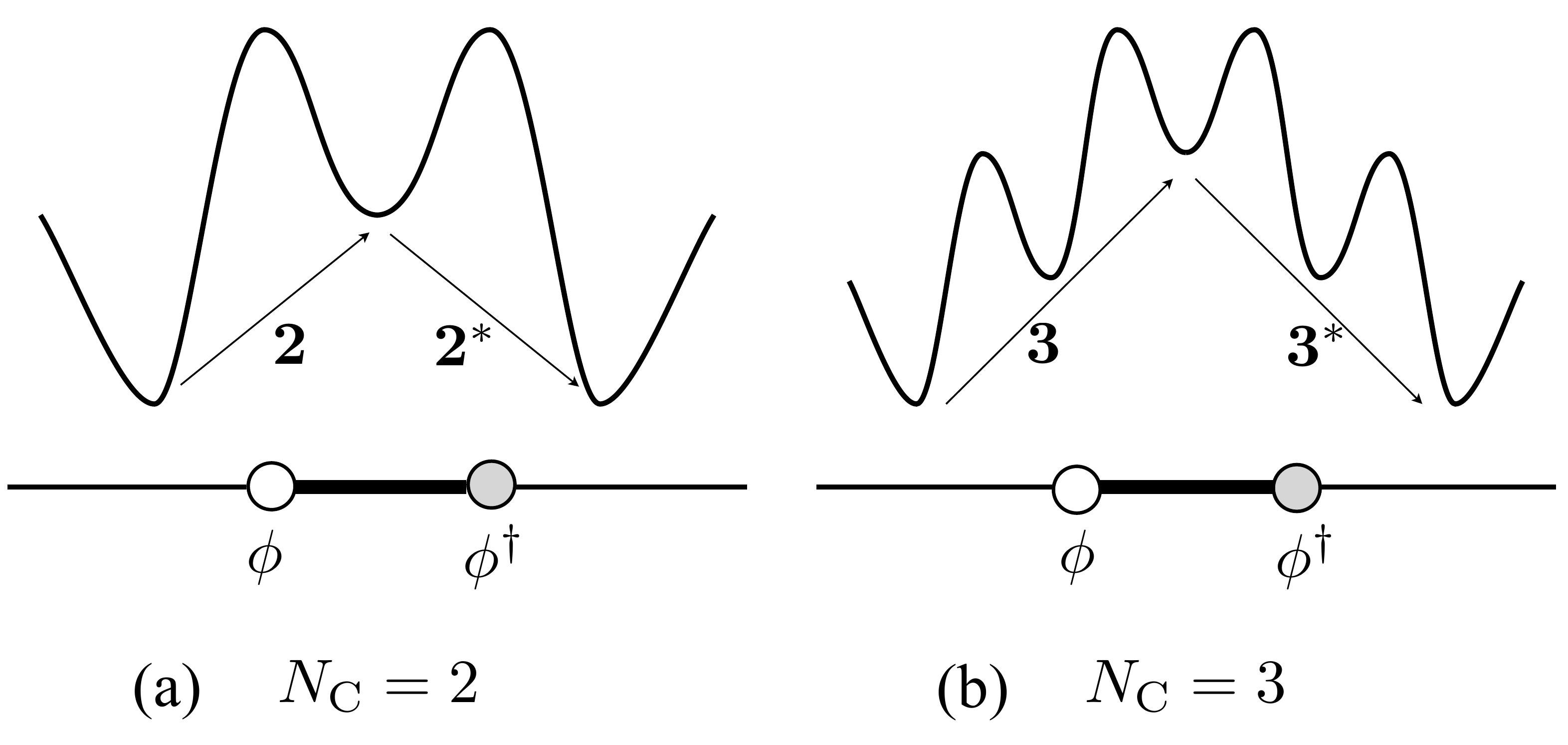}\caption{A schematic illustration of 
the nonperturbative potential and kinks interpolating between 
the ground state and the metastable states, 
in the cases of (a) $N_{\rm C}=2$ and (b) $N_{\rm C}=3$.
Kinks ($\phi$ and $\phi^{\dag}$) can be identified with monopoles 
($M$ and $\bar M$) from the bulk $3+1$ dimensional point of view. 
The total configuration is a bound state of a monopole and 
an antimonopole.
}
\label{fig:monopole-pair}
\end{center}
\end{figure}

It is now natural to interpret $\phi$ and $\phi^\dagger$
as a kink and an antikink interpolating the adjacent local minima on a vortex, 
respectively \cite{Gorsky:2004ad}; taking into account the codimension, 
this bound state neutral to the $\U(1)$ charge can be identified 
as the bound state of a monopole and an antimonopole in terms of the original 
3+1 dimensions, as illustrated in Fig.~\ref{fig:monopole-pair}: 
a monopole and an antimonopole with the mass $M$ 
are confined into the mesonic bound state by the linear potential.
A similar understanding has been demonstrated in Ref.~\cite{Markov:2004mj}
based on a comparison with SUSY QCD.

Let us discuss the representations of the bound state of the monopole and anti-monopole.
The physical degrees of freedom of $\phi$ is $2(N_{\rm C}-1)$ after fixing 
$\U(1)$ gauge symmetry.
For example, the inhomogeneous coordinates are
\beq
\phi = {1 \over \sqrt {1+|b_1|^2}} \left(
\begin{array}{c}
1\\
b_1 
\end{array}
\right),\quad
\phi = {1 \over \sqrt {1+|b_1|^2 + |b_2|^2}}
\left(
\begin{array}{c}
1\\b_1\\b_2
\end{array}
\right),
\label{eq:inhomo_cp1_cp2}
\eeq 
for $N_{\rm C}=2$ and $N_{\rm C}=3$, respectively. 
For $N_{\rm C}=2$, $\phi$ ($\phi^{\dag}$) represents one (anti)kink,
as can be seen in Fig.~\ref{fig:monopole-pair}(a).
Each of them corresponds to one (anti)monopole.
For the case of the CFL phase with $N_{\rm C}=3$, 
one (anti)monopole is a composite state of $N_{\rm C}-1=2$ (anti)kinks, 
each of which has one complex moduli (position and phase), 
as seen in Fig.~\ref{fig:monopole-pair}(b). 
Since the fields $\phi$ and $\phi^{\dag}$ 
transform as  (anti)fundamental
representations under $\SU(N_{\rm C})$, respectively, 
these (anti)monopoles belong to ${\bf N_{\rm C}}$ (${\bf N_{\rm C}}^*$) 
fundamental representations of $\SU(N_{\rm C})$.
The (anti)monopoles are not free particles but  appear as a mesonic bound state
that belongs to 
${\bf N_{\rm C}} \otimes {\bf N_{\rm C}}^* = {\bf 1} \oplus {\bf N_{\rm C}^2-1}$ 
representation. Note that it was shown in Refs.~\cite{Zamolodchikov:1978xm,Zamolodchikov:1992zr} that 
the singlet in this decomposition does not appear in the spectrum 
 in the ${\mathbb C}P^1$ model ($N_{\rm C}=2$).
This was interpreted in Ref.~\cite{Markov:2004mj} as indicating that 
the singlet corresponds to a set of a monopole and an antimonopole 
with opposite charges, which is unstable to decay.
Although there is no such a calculation for $N_{\rm C} \geq 3$, 
we expect that the same holds.

\begin{table*}[t]
\begin{center}
  \begin{tabular}{|l c c|}
    \hline
    Phases & Hadron phase & Color-flavor locked phase \\ \hline \hline
          & Confinement   & Higgs \\
    Quarks & Confined & Condensed \\ 
    {\it Monopoles}  & {\it Condensed?} & {\it Confined} \\
    Coupling constant & Strong  & Weak \\ 
    Order parameters & Chiral condensate $\langle \bar q q \rangle$ 
	& Diquark condensate $\langle qq \rangle$\\ 
    Symmetry & $\SU(3)_{\rm L} \times \SU(3)_{\rm R} \times \U(1)_{\rm B}$ 
	& $\SU(3)_{\rm C} \times \SU(3) _{\rm L} \times \SU(3)_{\rm R} \times \U(1)_{\rm B}$\\
	  & $\rightarrow \SU(3)_{\rm L+R}$ & $\rightarrow \SU(3)_{\rm C+L+R}$\\ 
    Fermions & ${\bf 8}$ baryons & ${\bf 8}$ + ${\bf 1}$ quarks  \\ 
    Vectors & ${\bf 8}$ + ${\bf 1}$ vector mesons  & ${\bf 8}$ gluons  \\
    NG modes & ${\bf 8}$ pions ($\bar q q$)  
	& ${\bf 8}$ + ${\bf 1}$ pions ($\bar q \bar q qq $)  \\
	& $H$ boson & $H$ boson \\ \hline
  \end{tabular}
  \end{center}
  \caption{
    Comparisons of the physics between the hadron phase 
    and the CFL phase in massless three-flavor QCD: symmetry breaking patterns 
	[the $\U(1)_{\rm A}$ and discrete symmetries are suppressed here]
	and the elementary excitations. 
    There is still one missing piece in the table; the properties
    of monopoles in the hadron phase, for which we speculate that the
    condensation of monopoles corresponds to the condensation 
    of quarks in the CFL phase.
    See the text for further explanations.}
  \label{tab:continuity}
  \end{table*}

Next, we would like to discuss the implications of  the color-octet of the magnetic-mesonic bound states.
Because  they live in the CFL phase, they are also flavor-octet under $\SU(3)_{\rm C+L+R}$.
Clearly, these bound states resemble the flavor-octet mesons formed by
quark-antiquark pairs in the hadron phase.
Thus we are naturally led to speculation on the idea of the ``quark-monopole duality": 
the roles played by quarks and monopoles are interchanged between
the hadron phase (at low density) and the CFL phase (at high density).
If this is indeed the case, our results in the CFL phase would imply the condensation 
of monopoles in the hadron phase; namely, it embodies the
dual superconducting scenario for the quark confinement 
in the hadron phase \cite{Nambu:1974zg,Zichichi:1976uh,Mandelstam:1974pi}.

The possible quark-monopole duality may have some relevance to 
``hadron-quark continuity" which is the one-to-one correspondence without any phase transitions
between the hadron phase and the CFL phase conjectured by Sch\"afer and Wilczek \cite{Schafer:1998ef}.
The hadron-quark continuity may be realized 
in the QCD phase structure in the three-flavor limit as explicitly shown 
in \cite{Hatsuda:2006ps,Yamamoto:2007ah}.
In Table~\ref{tab:continuity} we summarize
the correspondence in the quark-monopole duality and the hadron-quark continuity.
A number of pieces of nontrivial evidences that support the hadron-quark continuity have been identified: 
the same symmetry breaking patterns; 
the fact that the confinement phase is indistinguishable from the Higgs phase \cite{Fradkin:1978dv,Banks:1979fi};
the one-to-one correspondence of the elementary excitations such as the baryons,
vector mesons \cite{Hatsuda:2008is}, and pions \cite{Yamamoto:2007ah};
and the equivalence of the form of the partition functions
in a finite volume called the $\epsilon$-regime \cite{Yamamoto:2009ey},
between the hadron phase and the CFL phase.

\begin{figure}[h]
\begin{center}
\includegraphics[width=0.3 \textwidth] {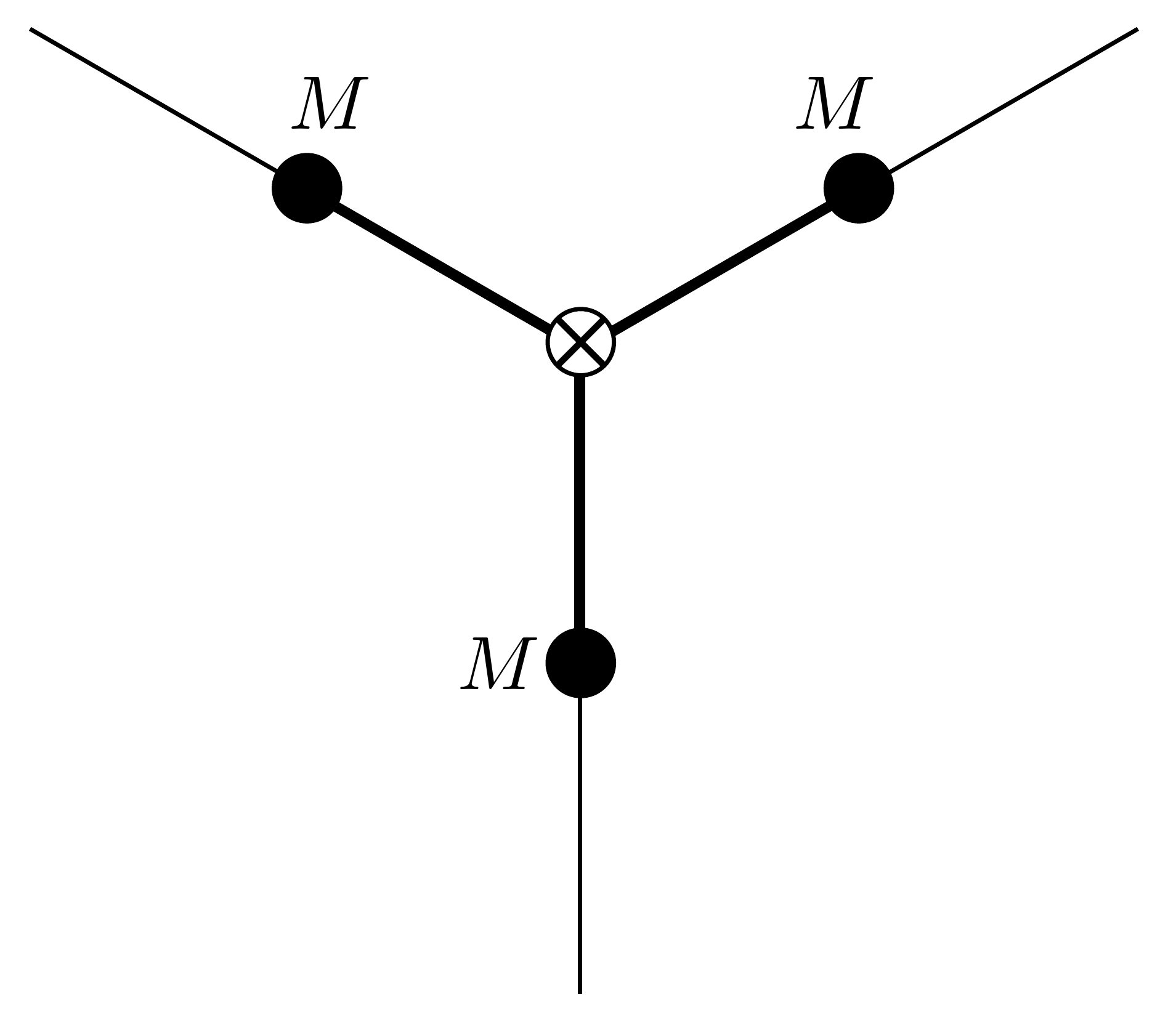}
\caption{Baryonic bound state of three monopoles 
(dark discs $M$ indicate the monopoles). 
They are connected by a junction of  
color magnetic flux tubes. The total color is canceled out and a
$\U(1)_{\rm B}$ vortex represented by $\otimes$ forms at the junction point.
}
\label{fig:baryon}
\end{center}
\end{figure}

So far, we have considered  mesonic bound sates.
What is the counterpart of baryonic states in the CFL phase?
It has been found in Ref.~\cite{Bali:2000gf} by lattice QCD simulations that 
three quarks are connected by a Y-shaped junction of color electric flux tubes.
It is a natural expectation that a junction of three non-Abelian vortices 
with total color fluxes canceled out at the junction point forms; see Fig.~\ref{fig:baryon}.
We note that they carry correct baryon number;
each non-Abelian vortex carries the $\U(1)_{\rm B}$ winding number 1/3, 
and all of them join together to constitute 
one $\U(1)_{\rm B}$ vortex with a $\U(1)_{\rm B}$ winding number one.
However, we have not specified the electromagnetic charges 
of fluxes at this stage because we have ignored 
the electromagnetic coupling of vortices in this section.

Finally, some comments are in order.
The first one is on the relation between classical treatments of the orientational zero modes in other sections and
the quantum treatment explained in this subsection. First of all, in order for the $\mathbb{C}P^2$ nonlinear sigma model
to work as a low energy effective field theory, the energy scale of the orientational modes must be sufficiently smaller than the cutoff scale $\Delta_{\rm CFL}$. Furthermore, 
since the $1+1$ dimensional $\mathbb{C}P^N$ nonlinear
sigma model is asymptotic free, the low energy physics of the low energy effective theory is highly quantum mechanical. 
The typical quantum mass scale is of the quantum monopole mass $M$ given 
in Eq.~(\ref{eq:monopole_mass}). Therefore, the orientational modes can be dealt with classically within the energy scale
smaller than $\Delta_{\rm CFL}$ and larger than $M$ which is very small at the high density limit.

Next, let us make a comment on the effect of the strange quark mass $m_{\rm s}$
with the electric charge neutrality and the $\beta$-equilibrium conditions, 
as is expected in the physical dense matter like inside the neutron stars. 
This situation is considered previously without 
the quantum effects in Sec.~\ref{sec:strange-quark}. Firstly, note that
without the quantum effects there are no monopoles in the CFL phase \cite{Eto:2009tr}.
This is because the effective potential without the quantum effect does not have
any local minima but has only one global minima, as was obtained in Eq.~(\ref{eq:pot_eff}),
so that no meta-stable kinks can form.
Even if we take into account the quantum effects, they are
negligibly-small: the scale of the effective potential $\sim m_{\rm s}^2/g_{\rm s}$ 
is much larger than that induced by the quantum effects $\sim \Delta_{\rm CFL}^2 e^{-\gamma \mu^2/\Delta_{\rm CFL}^2}$
for realistic values of the parameters, $m_{\rm s} \sim 100 \ {\rm MeV}$, 
$\mu \sim 500 \ {\rm MeV}$, and $\Delta _{\rm CFL}\sim 50 \ {\rm MeV}$.
Hence, the confined monopoles will be washed out by $m_{\rm s}$. 
We expect that the notion of the quark-monopole duality 
works well close to the three-flavor limit.
It is also a dynamical question whether the hadron-quark continuity survives 
when one turns on $m_{\rm s}$; there are other candidates for the ground state
at intermediate $\mu$ other than the CFL phase under the stress of $m_{\rm s}$, 
such as the meson condensed phase, 
the crystalline Fulde-Ferrell-Larkin-Ovchinikov phase 
\cite{Fulde:1964zz,larkin:1964zz}, 
the gluon condensed phase, etc \cite{Alford:2007xm}.

\subsection{Yang-Mills instantons trapped inside a non-Abelian vortex}\label{sec:instanton-in-vortex}

In this subsection, we discuss Yang-Mills instantons
\cite{Belavin:1975fg}, 
classified by the third homotopy group of 
the gauge symmetry 
\beq
 \pi_3[SU(3)_{\rm C}] \simeq {\mathbb Z}. \label{eq:YMinst2}
\eeq
The instanton number is $k = (8 \pi^2 /g_{\rm s}^2) \int F \wedge F 
\in \pi_3[SU(3)_{\rm C}]$.

At asymptotic large $\mu$, 
bulk instanton effects 
with the energy $\sim 1/g_{\rm s}^2 \gg 1$ 
are highly suppressed due to the
asymptotic freedom of QCD and the screening of 
Yang-Mills instantons \cite{Schafer:2002ty,Yamamoto:2008zw}. 
As classical solutions, Yang-Mills instantons are unstable against shrinkage 
in the CFL ground state, which can be understood 
from  Derrick's scaling argument \cite{Derrick:1964ww}. 
Instead, Yang-Mills instantons can exist stably inside 
a non-Abelian vortex  \cite{Eto:2004rz,Fujimori:2008ee,Eto:2006pg}.  
They are lumps or sigma model instantons \cite{Polyakov:1975yp} 
supported by the second homotopy group 
\beq
 \pi_2 ({\mathbb C}P^{2}) \simeq {\mathbb Z} 
\label{eq:YMinst}
\eeq 
in the ${\mathbb C}P^{2}$ model 
as the low-energy effective theory of a non-Abelian vortex.

Lump solutions exist inside a ${\mathbb C}P^1$ submanifold 
of  the whole ${\mathbb C}P^{2}$. 
Let $z \in {\mathbb C}$ be 
the complex coordinate made 
of Wick-rotated coordinates of the vortex-world sheet. 
By using the projective coordinate $w$ of ${\mathbb C}P^1$,  
the lump solution can be written as
\beq
 w(z) = \sum_{a=1}^k {\lambda_a \over z - z_a}
\eeq
with the size and phase moduli $\lambda_a \in {\mathbb C}^*$  
and the position moduli $z_a \in {\mathbb C}$ of the $a$-th instanton among $k$ instantons where  
$k \in \pi_2 ({\mathbb C}P^{2}) \simeq {\mathbb Z}$. 
The instanton energy in the vortex world-sheet is 
$C^{0,3} \sim (\mu/\Delta_{\rm CFL})^2 \gg 1/g_{\rm s}^2$ [see Eq.~(\ref{eq:C0C3})],
which is further suppressed more than the bulk instantons, 
consistent with the result in the last subsection. 
The fact that the instanton energy 
$C^{0,3}\sim (\mu/\Delta_{\rm CFL})^2$
inside the vortex is larger than the one $1/g_{\rm s}^2$ 
in the bulk  implies that instantons are repulsive from vortices.

In the case of the SUSY QCD, quantum effects 
in the $d=1+1$ dimensional vortex world-sheet  
can be explained by instanton effects in the original 
$d=3+1$ dimensional theory \cite{Shifman:2004dr,Hanany:2004ea,Shifman:2007ce}.
As in the SUSY QCD, the quantum effects inside the vortex
may be explained by instantons trapped in it, which remains a problem for the future.

%% file: interaction-v9.tex
\section{Interactions of non-Abelian vortices with quasiparticles}\label{sec:int}

In this section, we discuss the interaction of non-Abelian
vortices with quasiparticles in the color-superconducting  medium. 
It is necessary to determine the interaction to discuss physical
phenomena such as scattering or radiation of quasiparticles by vortices.
We can also investigate the interaction between vortices 
using vortex-quasiparticle interaction, since the
intervortex force is mediated by quasiparticles.

In the first subsection, we discuss the interaction of vortices with
phonons, which are the Nambu-Goldstone mode associated with the breaking
of the ${U}(1)_{\rm B}$ symmetry, and gluons. 
In particular, the interaction with gluons is dependent on the orientation
of a vortex. 
This gives rise to an orientation-dependent interaction energy between two
vortices.

In the second subsection, we discuss the interaction of vortices with
CFL mesons. 
The CFL mesons are the Nambu-Goldstone bosons for the breaking of chiral
symmetry. 

In the third subsection, we investigate the interaction of vortices
with photons and its phenomenological consequences.
The orientational zero modes localized on vortices are charged with
respect to ${U}(1)_{\rm EM}$ symmetry. The interaction Lagrangian is 
determined by the symmetry consideration. 
Based on the interaction, we discuss the
scattering of photons off a vortex. 
We also discuss the optical property of a vortex lattice (see Sect.~\ref{sec:lattice}), which is
expected to be formed if CFL matter exists inside the core of a rotating
dense star.
We show that a lattice of vortices serves as a polarizer of photons.

\subsection{Interaction with phonons and gluons}\label{sec:int-phonon-gluon}
Here we discuss the interaction of vortices with gluons and phonons.
For this purpose, we use a method called  dual transformation. 
Dual transformation relates theories that have different Lagrangians
and variables but possess equivalent equations of motion. 
This method is useful in dealing with topological defects, since
topological defects in an original theory are
described as particles in its dual theory. 
After a dual transformation, we can deal with the interaction of
topological defects by the methods of the ordinary field theory.
The action of the dual theory is derived by using the method of path
integration.
For example, let us take phonons in three spatial dimensions, which are described
by a massless scalar field. 
In the dual action, phonons are described by a massless antisymmetric
tensor field $B_{\mu\nu}$ \cite{Kalb:1974yc, Freedman:1980us}.
Antisymmetric tensor fields have been utilized in describing vortices in
superfluids or perfect fluids \cite{Lee:1993ty,hatsuda1994topological, Sato:1994vz}. 
In a dual formulation, 
the field $B_{\mu\nu}$ is introduced via the method of path integration. 
On the other hand, the gluons, which are massive because of the
Higgs mechanism, are described by massive antisymmetric tensor fields in the
dual theory \cite{Seo:1979id}. 

\subsubsection{Dual action and vortex-quasiparticle interaction}
Starting from the time-dependent Ginzburg-Landau effective
Lagrangian in the
CFL phase (\ref{eq:tdgl}), a dual Lagrangian can be derived via the
method of path integration. The derivation is given in
Appendix~(\ref{sec:appendix_dual}). 
In the dual theory, 
gluons and phonons are described by antisymmetric tensors,
$B^a_{\mu\nu}$ and $B^0_{\mu\nu}$, respectively.
The low-energy action for phonons and gluons interacting with vortices
is given by \cite{Hirono:2010gq}
\begin{equation}
 {\mathcal{ S}} = {\mathcal{ S}}_0 + {\mathcal{
S}}_{\rm int},
\label{eq:dual-action}
\end{equation}
where the free part $\mathcal{S}_0$ is defined as 
\begin{equation}
\mathcal{S}_0 = 
 \int d^4 x \left[
  - \frac{1}{12 \tilde{K}_{\mu\nu\sigma}} \left(
					 H^a_{\mu\nu\sigma}H^{a,\mu\nu\sigma}
  + 					 H^0_{\mu\nu\sigma}H^{0,\mu\nu\sigma}
  \right)
- \frac{1}{4} {m_{\rm g}}^2 \left(B^a_{\mu\nu} \right)^2
  \right].
\label{eq:dual-action-free}
\end{equation}
In the above equation,
$
H^a_{\mu\nu\sigma} \equiv 
\p_{\mu} B^a_{\nu\sigma} +  \p_{\nu} B^a_{\sigma\mu} +  \p_{\sigma} B^a_{\mu\nu} 
$ and 
$
H^0_{\mu\nu\sigma} \equiv 
\p_{\mu} B^0_{\nu\sigma} +  \p_{\nu} B^0_{\sigma\mu} +  \p_{\sigma} B^0_{\mu\nu} 
$
are field strength tensors of gluons and phonons, and $m_{\rm g}$ is the
mass of the gluons.
The factor 
$
\tilde{K}_{\mu\nu\sigma} \equiv \epsilon_{\rho\mu\nu\sigma} K^{\rho}
$ comes from the lack of Lorentz invariance, where we have defined
$K_\mu = (K_0, K_3,K_3,K_3)^T$.
The first (second) terms are kinetic term for gluons (phonons). 
The third one is the mass term for gluon fields, which is induced via
the Higgs mechanism.
In the presence of a vortex, 
the mass of gluons is dependent on the distance from the center of
the vortex according to the change of the values of diquark condensates.

The interaction part ${\mathcal{S}}_{\rm int}$ is written as 
\begin{equation}
{\mathcal{S}}_{\rm int}
=- \int d^4 x  \left[2\pi m^0 B^0_{\mu\nu} \omega^{0,\mu\nu}
+ \frac{m_{\rm g}}{g_{\rm s}}  B^a_{\mu\nu} \omega^{a,\mu\nu}\right],
\label{eq:dual-action-int}
\end{equation}
where $\omega^0_{\mu\nu}$ and $\omega^a_{\mu\nu}$ are vorticity tensors,
which depend on the vortex configuration, and 
$m^0$ is a space-dependent function given by the vortex profile functions.
Their specific forms are discussed later.
The vorticity tensors have finite values only around the core of a vortex.
Thus, although gluons and phonons propagate in the four-dimensional spacetime
the interaction is localized around the vortex.

Now we discuss the properties of the interaction (\ref{eq:dual-action-int}).
First, let us look at the interaction of vortices with ${U}(1)_{\B}$ phonons. 
This part is essentially the same as a vortex in a superfluid.
For a general vortex configuration we can write the Abelian component of
the vorticity tensor as 
\begin{equation}
  \left(\omega^0\right)_{\rho\sigma}(x) \equiv \frac{1}{2\pi}\epsilon_{\mu\nu\rho\sigma}
   \p^\nu \p^\mu
  \pi_{\rm MV}(x) 
.
\label{eq:abelian-vorticity}
\end{equation}
where $\pi_{\rm MV}(x)$ is the multivalued part of the phase of the
order parameter fields.
The multivalued part is in general allowed, since it is a phase.
Equation (\ref{eq:abelian-vorticity}) appears to automatically vanish,
but in fact it does not, since the two derivatives do not commute, which
reflects the multivaluedness of $\pi_{\rm MV}(x) $.
For a general vortex configuration, the vorticity tensor can be written as 
\begin{equation}
 (\omega^0)^{\mu\nu}(x) = 
\frac{1}{N_{\rm C}}\int d\tau d \sigma
\frac{\p (X^\mu, X^\nu)}{\p (\tau,  \sigma) } 
\delta^{(4)} (x - X^\mu(\tau, \sigma)),
\end{equation}
where $N_{\rm C}$ is the number of colors ($N_{\rm C}=3$) and  $X^\mu(\tau, \sigma)$ is the the space-time position of 
the vortex parametrized by world-sheet coordinates $\tau$ and $\sigma$.
The interaction of vortices with ${U}(1)_{\B}$ phonons is
rewritten as
\begin{equation}
{\mathcal{S}}^{\rm Ph}_{\rm int} = -\frac{2\pi m^0}{N_{\rm C}} \int d
 \sigma^{\mu\nu} B^0_{\mu\nu}, \label{eq:phonon}
\end{equation}
where 
$d \sigma^{\mu\nu} \equiv \frac{\p (X^\mu,X^\nu)}{\p (\tau,
\sigma)} d \tau d \sigma $ 
 is an area element of the vortex world-sheet. 
The interaction (\ref{eq:phonon}) is a natural generalization of the gauge 
interaction of a point particle,
\begin{equation}
 \mathcal{S} = \int dx^\mu A_\mu.
\end{equation}
The factor $1/N_{\rm C}$, which is equal to the $U(1)_{\rm B}$ winding
number of vortices with the lowest energy, 
reflects the fact that the strength of the interaction is 
proportional to the winding number with respect to ${U}(1)_{\B}$
symmetry.
We also note that ${U}(1)_{\B}$ phonons $B_{\mu\nu}^0$ do not couple to the
orientational zero modes. 
Phonons are blind to the orientation of a vortex.

Next, let us look at the interaction of vortices with 
gluons. 
The non-Abelian vorticity tensor $\omega^a_{\mu\nu}$ is written as 
\begin{multline}
  \omega^a_{\lambda \sigma} = 
\epsilon_{\lambda \sigma \mu \nu} 
\Biggl\{
  \p_\nu 
 \Bigl\{
 -\frac{16}{N_{\rm C}} \gamma(r) 
\left(
  \p_{\mu} \theta + 2N_{\rm C} \gamma \delta_{\mu 0}
\right)
\phi^\+ T^a \phi 
\\
+i\alpha(r) (1+\beta(r))
  \left(
    \phi^\+ T^a \p_\mu \phi -  \p_\mu\phi^\+ T^a \phi  +
  2 \phi^\+ T^a \phi  \p_\mu\phi^\+ \phi
  \right)
  \Bigr\}  
\\
-\frac{4}{N_{\rm C}} \alpha(r) \gamma(r) (1+\beta(r))
 \left(
   \p_{[\mu} \phi^\+ T^a \phi
  + \phi^\+ T^a \p_{[\mu} \phi
 \right)
\left(
\p_{\nu]} \theta + \frac{N_{\rm C}K^\prime_0}{2K_0} \delta_{\nu] 0}
\right)
\\
-  \frac{i}{2}\alpha(r)^2 (1+\beta(r)^2) 
\\
\times  \left[
   \phi^\+ T^a \phi \p_{[ \mu } \phi^\+  \p_{ \nu] }  \phi
   +  \p_{[ \mu } \phi^\+ T^a  \p_{ \nu] } \phi
 + \phi^\+ T^a \p_{[\mu} \phi \p_{\nu]}\phi^\+ \phi
 + \p_{[\mu} \phi^\+ T^a \phi \p_{\nu]}\phi^\+ \phi
  \right]
\Bigr\}.
\label{eq:non-abelian-vorticity}
\end{multline}
where $\alpha(r), \beta(r)$, and $\gamma(r)$ are functions of the
distance from the vortex core and are written in terms of vortex
solutions, and the parameter $\gamma$ is the coefficient of the term with
one time derivative in Eq.~(\ref{eq:tdgl}).
The leading-order part in the deviation of the order parameter from the
ground-state value is given by 
\begin{equation}
 \omega^a_{\lambda \sigma} 
= 
\epsilon_{\lambda \sigma \mu \nu}
  \p^\nu 
  \left[
 -\frac{16}{N_{\rm C}} \gamma(r)
\left\{ 
\p^\mu \theta + 2 N_{\rm C}\gamma \delta^{\mu 0}
\right\}
\phi^\+ T^a \phi
  \right]. 
\end{equation}
As can be seen in Eq.~(\ref{eq:non-abelian-vorticity}), 
gluons actually couple to the orientational zero modes on the vortex.
As a result, gluons are emitted through the interaction 
(\ref{eq:non-abelian-vorticity}) 
when a wave of the $\mathbb{C}P^{2}$ orientational modes propagates 
along a vortex-line.
By using the interaction derived above (\ref{eq:non-abelian-vorticity}), 
we can estimate the amount of radiated gluons from a propagating
 wave in $\mathbb{C}P^{2}$ orientational space.

\subsubsection{Orientation dependence of the vortex-vortex interaction}\label{sec:ori-vor-vor-int}
\begin{figure}[htbp]
\begin{center}
  \includegraphics[width=90mm]{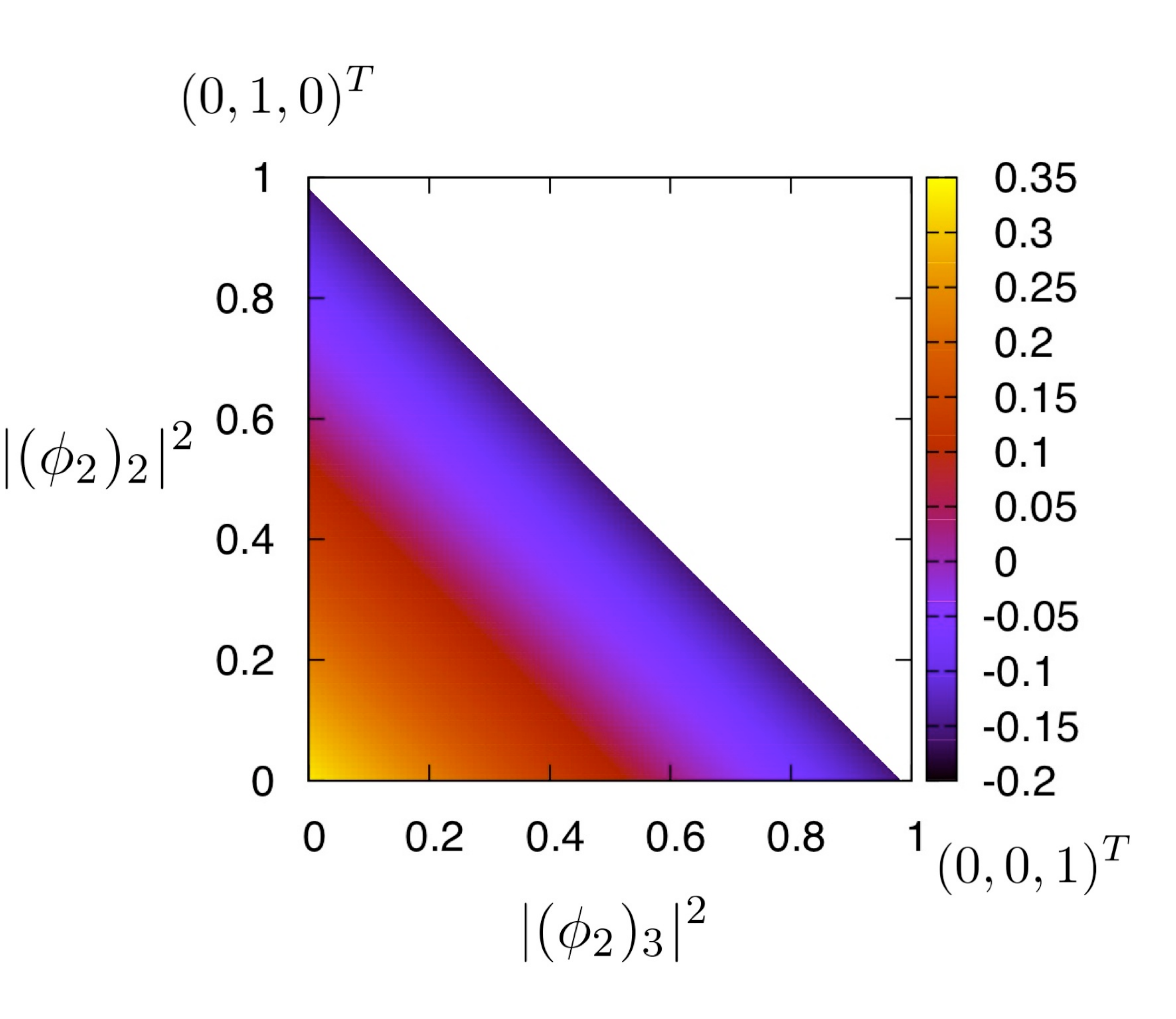}
\end{center}
 \caption{
The value of $G(\phi_1, \phi_2)$ as a function of $|(\phi_2)_2|^2$ and
 $|(\phi_2)_3|^2$.
If the orientation $\phi_2$ is in the red zone the interaction is
 repulsive, while in the blue zone the interaction is attractive.
}
\label{fig:orientation}
\end{figure}
As an application of the dual Lagrangian obtained above, 
let us discuss the orientation
dependence of the interaction energy of two vortices placed in parallel.
We assume that the orientation of each vortex is constant along the
vortex.
The interaction energy due to the gluon exchange is
proportional to 
\begin{equation}
G(\phi_1, \phi_2) \equiv
 \phi^\+_1 T^a \phi_1 ~ 
 \phi^\+_2 T^a \phi_2,
\end{equation}
where $\phi_1$ and $\phi_2$ denote the orientations of the first and
second vortices, respectively.
We have shown in Fig.~\ref{fig:orientation} the value of $G(\phi_1, \phi_2)$ as a function of
$\phi_2$.
We have taken $\phi_1$ as $\phi_1=(1,0,0)^T$ without loss of generality.
Figure \ref{fig:orientation} indicates that, 
if the two orientations are close in the $\mathbb{C}P^2$ space, 
the interaction through gluon exchanges is repulsive, while 
if the orientations are far apart, the interaction is attractive.
This orientation-dependent interaction is expected to be important when
the distance between two vortices is small, {\rm e.g.} when two vortices
cross.

\subsection{Interaction with mesons}\label{sec:int-mesons}

The interactions with phonons and gluons are topological, in the sense
that the interaction term does not involve the metric. 
In contrast, 
the interaction with photons is not topological.
Here, we discuss the interaction with mesons 
studied in Sect.~\ref{sec:CFL_meson}, 
which is also non-topological. 
In this case, one cannot use a dual transformation to obtain the
interaction Lagrangian.

First, let us remind ourselves of the effective action of mesons 
discussed in Sec.~\ref{sec:CFL_meson}.
The gauge invariant 
$\Sigma = \Phi_{\rm L}^\dagger \Phi_{\rm R}$ 
defined in Eq.~\eqref{eq:sigma} transforms under
 the flavor symmetry
$U(1)_{\rm A} \times 
SU(3)_{\rm L} \times SU(3)_{\rm R}$ as
\begin{equation}
 \Sigma 
 \to e^{i \alpha} g_{\rm L}^\dagger \Sigma  g_{\rm R}, \quad
 e^{i \alpha} \in U(1)_{\rm A}, \quad 
 g_{\rm L} \in SU(3)_{\rm L}, \quad 
 g_{\rm R} \in SU(3)_{\rm R}. 
\end{equation}
The chiral symmetry is broken to 
the vector symmetry $SU(3)_{\rm L+R}$ 
with $g_{\rm L} = g_{\rm R}$
in the CFL ground state 
$\Sigma = \Delta_{\rm CFL} {\bf 1}_3$.
The mesons are 
$\Sigma = \Delta_{\rm CFL} ^2 g_{\rm L}^\dagger g_{\rm R} 
= \Delta_{\rm CFL} ^2 g^2(x) 
= \Delta_{\rm CFL} ^2 U(x)$ 
with $g_{\rm L}^\dagger = g_{\rm R} = g(x)$.

In a non-Abelian vortex background, 
$\Phi_{\rm L}= - \Phi_{\rm R} = \diag (f(r)e^{i \theta},g(r),g(r))$, 
the gauge invariant $\Sigma$ takes the form of
\begin{equation}
 \Sigma_v = \diag (f^2(r),g^2(r),g^2(r)) 
\quad [ \to  \diag (0,g_0^2,g_0^2)  \mbox{ at } r=0],
\end{equation}
with a constant $g_0 =g(r=0)$. 
In the presence of the vortex, 
the gauge invariant $\Sigma$ becomes 
\begin{equation}
 \Sigma = \sqrt U^\dagger \Sigma_v \sqrt U
\end{equation}
Then, the chiral Lagrangian can be written as
\begin{equation}
 {\mathcal L} 
= \sum_{\mu} f_{\mu} \tr (\partial_{\mu} \Sigma^\dagger \partial^{\mu} \Sigma) 
= \sum_{\mu} f_{\mu} 
 \tr [2 \alpha_{\rm L} \Sigma_v \alpha_{\rm R} \Sigma_v
- (\alpha_{\rm L}^2 + \alpha_{\rm R}^2) \Sigma_v^2] 
  \label{eq:meson-vortex}
\end{equation}
with the decay constants $f_{\mu}=(1/2)(f_\pi,f_\pi v_\pi)$ 
in 
Eq.~(\ref{eq:decay_const_speed_of_meson}), 
and the left and right Maurer-Cartan forms 
\begin{equation}
 \alpha_{\rm R} \equiv \sqrt U^\dagger \del_{\mu}  \sqrt U , \quad 
 \alpha_{\rm L} \equiv   \sqrt U \del_{\mu}  \sqrt U^\dagger .
\end{equation}
Far apart from the vortex core, 
it reduces to 
\begin{equation}
 {\mathcal L} = - \Delta_{\rm CFL} ^2 \sum_{\mu} K_{\mu} 
\tr (U^\dagger \partial_{\mu} U)^2  
\end{equation}
as expected.
The Lagrangian in Eq.~(\ref{eq:meson-vortex}) 
describes the mesons in the vortex background. 

\subsection{Interaction with electromagnetic fields}
\label{sec:int-em}

Here we discuss the electromagnetic properties of
non-Abelian vortices in the CFL phase, and their phenomenological
consequences.
Although the bulk CFL matter is electromagnetically neutral,
the orientational zero modes are charged, as discussed later.
Thus the vortices interact with photons.
For this purpose we consider the low-energy effective action of
orientational zero modes interacting with photons.
Using the action, we discuss the scattering of photons by a vortex.

In the following analysis, we neglect the mixing of photons and gluons.
The gauge field, $A^\prime_{\mu}$, which remains massless in the CFL phase, 
 is a mixture of the photon $A_{\mu}$ and a gluon part $A^8_\mu$,
$
 A^\prime_{\mu} = -\sin \zeta A_{\mu} + \cos \zeta A^8_\mu.
$
Here, the mixing angle $\zeta$ is given by 
$
\tan \zeta = \sqrt{3}g_{\rm s}/{2e} $\cite{Alford:1998mk}, where $g_{\rm s}$ and $e$ are
the strong and electromagnetic coupling constants.
At accessible densities ($\mu \sim 1 \ {\rm GeV}$), the fraction of the
 photon is given by $\sin \zeta \sim 0.999$,
and so, the massless field $A^\prime_\mu$ consists mostly of the ordinary
 photon and includes a small amount of the gluon.
As a first approximation, we neglect the mixing of the gluon to the
massless field.

We denote the orientational zero modes by a complex three-component vector $\phi \in
\mathbb{C}P^2$, which satisfies $\phi^\+ \phi=1$.
When we neglect the electromagnetic interaction, 
the low-energy effective theory on the vortex that is placed along the
$z$ axis
is described by the
following $\mathbb{C}P^2$ nonlinear sigma model \cite{Eto:2009bh}:
\begin{equation}
\Lcal_{\mathbb{C}P^2} =  \sum_{\alpha = 0, 3} C_\alpha 
\left[
\p^\alpha \phi^\+ \p_\alpha \phi + (\phi^\+ \p^\alpha \phi)  (\phi^\+ \p_\alpha \phi)
\right],
\end{equation}
where the orientational moduli $\phi$ are promoted to dynamical fields, 
and $C_\alpha$ are numerical constants.
Under the color-flavor locked transformation, 
the $\mathbb{C}P^2$ fields $\phi$ transform as 
\begin{equation}
 \phi \rightarrow U \phi,
\end{equation}
with $U \in {SU}(3)_{\rm C+F}$. 

\subsubsection{Coupling of orientation modes with electromagnetic fields}
Now, let us consider the electromagnetic interactions.
The electromagnetic ${U}(1)_{\rm EM }$ group is a subgroup of
the flavor group ${SU}(3)_{\rm F}$. The generator of ${U}(1)_{\rm EM }$ is
$T_8 = \frac{1}{\sqrt{6}}\diag (-2,1,1)$ in our choice of basis.
The effect of the electromagnetic interaction is incorporated by gauging the
corresponding symmetry.
The low-energy effective action on the vortex should be modified 
to the following gauged $\mathbb{C}P^2$ model, 
\begin{equation}
\Lcal_{g\mathbb{C}P^2} = \sum_{\alpha = 0, 3} C_\alpha 
\left[
\D^\alpha \phi^\+ \D_\alpha \phi + (\phi^\+ \D^\alpha \phi)  (\phi^\+ \D_\alpha \phi)
\right],
\end{equation}
where the covariant derivative is defined by 
\begin{equation}
\D_\alpha \phi = \left( \p_\alpha - ie \sqrt{6} A_\alpha T_8  \right) \phi.
\end{equation}
Thus the low-energy behavior is described by the $\mathbb{C}P^2$ modes
localized on the vortex and photons propagating in three-dimensional space.
Hence, the effective action is given by 
\begin{equation}
S =  
\int \left( 
      \frac{\varepsilon_0}{2}  \bm E^2 - \frac{1}{2 \lambda_0 } \bm B^2
 \right)  d^4x
+ 
\int \Lcal_{g \mathbb{C}P^2} dzdt,
\label{eq:cp2-photon-action-0}
\end{equation}
where $\varepsilon_0$ and $\lambda_0$ are the dielectric constant and
permeability of the CFL matter, respectively.
We can formally recover the Lorentz invariance in the kinetic terms of
the photons by the following rescaling:
\begin{equation}
 A'_0 = \sqrt{\varepsilon_0} A_0, \quad  A'_i = \frac{1}{\sqrt{\lambda_0}}
  A_i, \quad t' = v t, 
\end{equation}
where $v \equiv 1/\sqrt{\varepsilon_0 \lambda_0}$.
By further rescaling the parameters as
\begin{equation}
 e'= \sqrt{\lambda_0}e, \quad C'_0 = C_0 v, \quad C'_3= \frac{C_3}{ v},
\end{equation}
we can write the Lagrangian in the following form,
\begin{equation}
v S = -\frac{1}{4} \int  F_{\mu\nu}F^{\mu\nu} d^4x
+ 
\int \Lcal'_{g \mathbb{C}P^2} dzdt,
\label{eq:cp2-photon-action}
\end{equation}
where
\begin{equation}
\Lcal'_{g\mathbb{C}P^2} = \sum_{\alpha = 0, 3} C'_\alpha 
\left[
\D'^\alpha \phi^\+ \D'_\alpha \phi + (\phi^\+ \D'^\alpha \phi)  (\phi^\+ \D'_\alpha \phi)
\right], \quad 
\D'_\alpha \phi = \left( \p_\alpha - ie' \sqrt{6} A_\alpha T_8  \right) \phi.
\end{equation}
In the discussions below, primes are omitted for notational simplicity. 

\subsubsection{Scattering of photons off a vortex}
We can discuss the consequences of the charged degrees of freedom on the
vortex using the low-energy action (\ref{eq:cp2-photon-action}).
For example, let us consider the photon scattering off a vortex.
The equation of motion of the photon field derived from 
the effective action is given by
\begin{equation} 
\begin{split}
\p^\mu F_{\mu\nu}  
&= 
-C_\nu ie \sqrt{6} 
\ \delta(x_\perp) (\delta_{0\nu} + \delta_{3 \nu} ) \\
 &\times 
\bigl\{
\phi^\+ T_8 \D_\nu \phi  
 - 
(\D_\nu \phi)^\+ T_8 \phi
- 2 \phi^\+ \D_\nu \phi \phi^\+ T_8 \phi 
\bigl\} ,
\end{split}
\end{equation}
where
$\delta ( x_\perp)\equiv  \delta(x)\delta(y)$ is the delta function in
the transverse plane.
We consider a situation where 
a linearly polarized photon is normally incident on the vortex.
We assume that the electric field of the photon is parallel to the
vortex.
Then, the problem is $z$-independent and we can set $\theta=\theta(t)$,
$A_t=A_x=A_y=0$, and $A_z = A_z(t,x,y)$. 
The equation of motion is rewritten as 
\begin{equation}
\begin{split}
 (& \p_t^2  - \p_x^2 - \p_y^2) A_z(t,x,y) \\
 & =
12 C_3 e^2 
\left\{
\phi^\+ (T_8)^2 \phi + (\phi^\+ T_8 \phi)^2
\right\}
A_z(t,x,y) \ \delta(x_\perp) \\
&\equiv 
12 C_3 e^2 f(\phi)
A_z(t,x,y) \ \delta(x_\perp), 
\end{split}
\label{eq:gauge-eom}
\end{equation}
where we have defined 
\begin{equation}
f(\phi) \equiv \phi^\+ (T_8)^2 \phi + (\phi^\+ T_8 \phi)^2.
\label{eq:orientation-dep}
\end{equation}
Equation (\ref{eq:gauge-eom}) is the
same as the one discussed by Witten in the
context of superconducting strings \cite{Witten:1984eb}, except for the
orientation-dependent factor, $f(\phi)$.
The cross section per unit length of a vortex, $d\sigma/dz$, is
calculated as 
\begin{equation}
 \frac{d\sigma}{d z}
=\frac{\left( 12 C_3 e^2 f(\phi) \right)^2 \eta^2 } {8 \pi} \lambda
= 288 \pi  \left(C_3 \alpha \eta f(\phi) \right)^2 \lambda,
\end{equation}
where $\lambda$ is the wavelength of the incident photon, $\eta$ is a
numerical factor of order unity, and $\alpha$ is the fine structure constant.
On the other hand, if the electric field of the wave is perpendicular to
the vortex, the photon is not scattered, since current can flow only
along the vortex.

\subsubsection{Vortex lattice as cosmic polarizer}
\label{sec:polalizer}

\begin{figure}[tbp]
 \begin{center}
  \includegraphics[width=90mm]{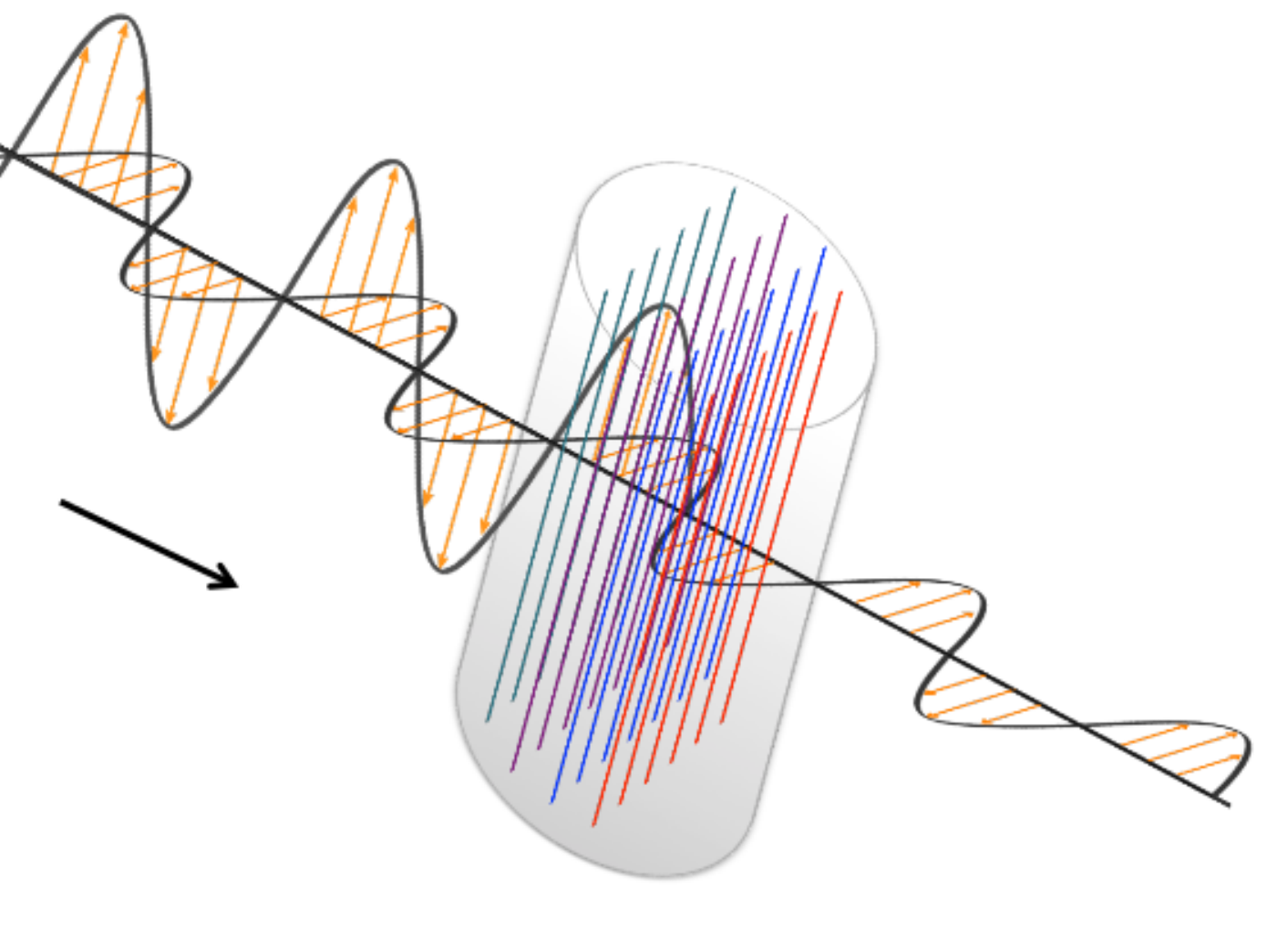}
 \end{center}
 \caption{
Schematic figure of two linearly polarized photons entering a vortex lattice. 
Photons propagate in the direction of the big arrow.
The small arrows indicate the electric field vector.
The waves whose electric fields are parallel to the vortices are
 attenuated inside the lattice, 
while the ones with perpendicular electric fields passes through it.
}
 \label{fig:polarizer}
\end{figure}

The electromagnetic property of vortices can be 
phenomenologically important as it may lead to some observable effects.
As an illustration of such an effect, 
we show that a lattice of vortices works as a polarizer of photons. 
The rotating CFL matter is expected to be threaded with quantum vortices
along the axis of rotation, 
resulting in the formation of a vortex lattice, as discussed in
Sect.~\ref{sec:lattice} 
\cite{Nakano:2007dr,Sedrakian:2008ay,Shahabasyan:2009zz}.
This is basically the same phenomenon as when one rotates atomic superfluids. 
Suppose that a linearly polarized photon is incident on a vortex lattice
as shown in Fig.~\ref{fig:polarizer}.
If the electric field of the photon is parallel to the vortices, 
it induces currents along the vortices,
which results in the attenuation of the photon.
On the other hand, waves with electric fields
perpendicular to the vortices are not affected.
This is exactly what a polarizer does.
A lattice passes electromagnetic waves of a specific polarization and blocks
waves of other polarizations.
This phenomenon, 
resulting from the electromagnetic interaction of vortices, 
may be useful for finding observational evidence of the existence of 
CFL matter.

We consider a situation
where electromagnetic waves of some intensity normally enter the vortex lattice.
We first assume that the electric fields of the waves are parallel to
the vortices.
The fraction of the loss of intensity when the wave passes through the
lattice for distance $dx$ is
\begin{equation}
\left< \frac{d \sigma}{dz} \right>
n_v dx \equiv \frac{dx}{L},
\end{equation}
where $n_v$ is the number of vortices per unit area.
We defined the length $L$ by
\begin{equation}
L \equiv 1/\left( n_v \left< \frac{d\sigma}{dz} \right> \right)=
\ell^2 / \left< \frac{d\sigma}{dz} \right>,
\end{equation}
with the intervortex spacing $\ell$. 
As the cross section depends on the internal state (value of $\phi$) of the vortex,
we have introduced the averaged
scattering cross section $\langle d\sigma/dz \rangle$ over the ensemble
of the vortices.
Let us denote the intensity of waves at a distance $x$ from the surface of the
lattice as $I(x)$.
$I(x)$ satisfies 
\begin{equation}
 \frac{I(x+dx)}{I(x)} = 1-\frac{dx}{L}.
\end{equation}
Therefore, the $x$ dependence of $I(x)$ is characterized by the
following equation
\begin{equation}
 \frac{I'(x)}{I(x)} = -\frac{1}{L}.
\end{equation}
This equation is immediately solved as 
$
 I(x) = I_0 e^{-x/L},
$
where $I_0$ is the initial intensity. 
Hence, the waves are attenuated with the characteristic length $L$.

Let us make a rough estimate of the attenuation length.
The total number of vortices can be estimated, as in
Ref.~\cite{Iida:2002ev}, as 
\begin{equation}
 N_v \simeq 1.9 \times 10^{19 } 
\left(\frac{ 1 {\rm ms }}{ P_{\rm rot}} \right)
\left(\frac{\mu/3}{300 \ {\rm MeV}}  \right)
\left( \frac{R}{10 \ {\rm km} } \right)^2,
\label{eq:num-vortex}
\end{equation}
where $P_{\rm rot}$ is the rotation period, $\mu$ is the baryon
chemical potential and $R$ is the radius of the CFL matter inside dense
stars.
We have normalized these quantities by typical values.
The intervortex spacing is then written as 
\begin{equation}
 \ell \equiv 
\left(\frac{\pi R^2}{N_v}\right)^{1/2} 
\simeq
4.0 \times 10^{-6}~ {\rm m} 
\left(\frac{ P_{\rm rot}}{ 1 \ {\rm ms }} \right)^{1/2}
\left(\frac{300 \ {\rm MeV}}{\mu/3}  \right)^{1/2}.
\end{equation}
Therefore, the characteristic decay length of the electromagnetic waves
is estimated as
\begin{equation}
L =
 \frac{\ell^2}{288 \pi  \left(C_3 \alpha \eta \right)^2 
\left< f(\phi)^2 \right>   \lambda}
\simeq \frac{1.2 \times 10^{-11} \ {\rm m}^2}{\lambda}.
\label{eq:decay-length}
\end{equation}
Here we have determined the value of $f(\phi)$ by considering the effect of a
finite strange quark mass $m_{\rm s}$.
The finite strange quark mass breaks the flavor ${SU}(3)$ symmetry and
gives rise to a potential in the $\mathbb{C}P^2$ space,
as discussed in Ref.~\cite{Eto:2009tr}.
When $m_{\rm s}$ is larger than the typical kinetic energy of the
$\mathbb{C}P^2$ modes,  which is given by the
temperature $T \leq T_{\rm c}\sim 10^1 \ {\rm MeV}$, 
and is small enough that 
the description by the Ginzburg-Landau theory based on chiral
       symmetry is valid, 
the orientation of vortices falls into $ \phi_0^T = (0,1,0)$.
This assumption is valid for the realistic value of $m_{\rm
s} \sim 10^2 \ {\rm MeV}$.  
The orientation dependence of the cross section is encapsulated in the
function $f(\phi)$ defined in Eq.~(\ref{eq:orientation-dep}).
Since $f(\phi_0) = 1/3 \neq 0$, photons interact with a vortex with
this orientation. 
Therefore, we have taken $\langle f(\phi)^2 \rangle = f(\phi_0)^2=1/9$.
We have also taken $\eta =1$, $\mu = 900~{\rm MeV}$ and $\Delta_{\rm CFL} = 100
~{\rm MeV} $, from which 
the values of $C_3$ are determined \cite{Eto:2009tr}.

If we adopt $R \sim 1 \ {\rm km}$ for the radius of the CFL core, 
the condition that the intensity is significantly
decreased within the core is written as $L \leq 1 \ {\rm km}$. 
The condition can also be stated in terms of the wavelength of the photon as
\begin{equation}
\lambda \geq 1.2 \times 10^{-14} \ {\rm m} \equiv \lambda_{\rm c}.
\label{eq:critical-wavelength}
\end{equation}
Hence, a lattice of vortices serves as a wavelength-dependent filter of photons.
It filters out the waves with electric fields parallel to the vortices, 
if the wavelength $\lambda$ is larger than $\lambda_{\rm c}$.
The waves that pass through the lattice are linearly
polarized ones with the direction of their electric fields perpendicular
to the vortices, as shown schematically in Fig.~\ref{fig:polarizer}.

One may wonder why a vortex lattice with a mean vortex distance $\ell$ 
can block photons with wavelengths many orders smaller than $\ell$.
It is true the probability that a photon is scattered during its
propagation for a small distance ({\rm e.g. } $\sim\ell$) is small.
However, while the photon travels through the lattice, the scattering
probability is accumulated and the probability that a photon remains
unscattered decreases exponentially. 
Namely, the small scattering probability is compensated by the large
number of vortices through which a photon passes.
This is why the vortex mean distance and the wavelength of the
attenuated photons can be different.

%% file: boojum-v9.tex
\section{Colorful boojums at a CFL  interface}\label{sec:boojum}

\subsection{What are boojums?}

When vortices cross or end on
an interface between two distinct superfluid phases 
or on a boundary,
they may form interesting structures called
``boojums''
\cite{Mermin,volovik2009universe}.
Boojums are composite topological objects consisting 
of vortices, an interface, and possibly monopoles 
at endpoints of vortices, and 
are topologically characterized by 
in terms of the relative homotopy group 
\cite{volovik2009universe}.
Various types of boojums have been studied  
in nematic liquids \cite{volovik1983topological}, 
superfluids at the edge of a container filled with $^4$He, 
at the A-B phase boundary of $^3$He \cite{Blaauwgeers:2001ev,Hanninen:2003}, 
and in multi-component or spinor Bose-Einstein condensates 
\cite{Takeuchi:2006,Kasamatsu:2010aq, Nitta:2012hy,Kasamatsu:2013lda,Kasamatsu:2013qia,
Borgh:2012es,Borgh:2012sj} in condensed matter physics.
Such boojums are also considered in field theories 
such as nonlinear sigma models \cite{Gauntlett:2000de},
Abelian gauge theories \cite{Shifman:2002jm} 
and non-Abelian gauge theories \cite{Isozumi:2004vg,Eto:2005sw,Auzzi:2005yw,Eto:2008mf}; 
see  Refs.~\cite{Eto:2006pg,Shifman:2007ce,2009supersymmetric} as reviews. 

Here we discuss  
``colorful boojums''  \cite{Cipriani:2012hr}
 appearing at the interface \cite{Giannakis:2003am,Alford:1999pb,Alford:2001zr} 
of a color superconductor.  
It is most likely that the $npe$ phase exists in the core of neutron stars, 
where neutrons and protons are 
superfluid and superconducting, respectively. 
In high density region of nuclear matter, 
a spin triplet (p-wave) pairing of neutrons
is energetically favored compared to an s-wave pairing \cite{Takatsuka:1992ga}.  
The inner region may be characterized by the presence of hyperons or of the CFL phase. 
We consider the interface 
of a hadron phase and a color superconductor  
 \cite{Giannakis:2003am,Alford:1999pb,Alford:2001zr}.\footnote{
The system we described is the simplest configuration in the pure CFL phase, which is realized at very high densities $\mu \gg m_{\rm s}^2/\Delta_{\textsc{cfl}}$. However, in a more realistic setting, the strange quark mass stresses the CFL phase and other phases have to be considered. For example, when the chemical potential is low enough that $m_{\rm s} \gtrsim m_{\rm u,d}^{1/3}\Delta_{\textsc{cfl}}^{2/3}$, kaons  can condense, leading to the so called  CFL-K$^0$ phase \cite{Schafer:2000ew,Bedaque:2001je,Kaplan:2001qk,Bedaque:2001at}, where other kinds of vortices arise \cite{Kaplan:2001hh,Buckley:2002ur}.  However in the high density regime the boojum can be modified only in the vicinity of the interface, while the overall structure of the junction is kept unchanged.
}
Since neutron stars are rotating rapidly and are accompanied by large magnetic fields, 
both superfluid and superconducting vortices exist in the $npe$ phase \cite{baym1969superfluidity}.
Such vortices are expected to explain the pulsar glitch phenomenon \cite{Anderson:1975zze}. 
On the other hand, 
when the matter in the CFL phase is rotating, 
such as in the core of neutron stars, 
superfluid vortices \cite{Forbes:2001gj,Iida:2002ev} 
are inevitably created and expected to constitute a vortex lattice.
However, the simplest superfluid vortex carrying integer circulation is unstable and decay \cite{Nakano:2007dr,Forbes:2001gj,Iida:2002ev} into a set of three non-Abelian vortices, 
each of which carries a color-magnetic flux and 
1/3 quantized circulation  \cite{Balachandran:2005ev,Eto:2009kg}, 
as discussed in Sec.~\ref{sec:Abelian-vortex-decay}.
Non-Abelian vortices constitute 
a colorful vortex lattice in the CFL phase side, 
as discussed in Sec.~\ref{sec:lattice}. 

It was discussed in Ref.~\cite{Cipriani:2012hr} 
that boojum structures are created when  
superfluid and superconducting vortices of the $npe$ phase
penetrate into the CFL core. 
The shape of the neutron boojum was shown to 
split into three color magnetic vortices. 
The colorful boojum is  accompanied by 
confined color-magnetic monopoles of two different kinds, 
Dirac monopoles and surface superconducting currents.

We first study the structure of a colorful boojum 
in the CFL phase side in Sec.~\ref{sec:boojum-CFL} 
followed by the structure in the $npe$ phase 
in Sec.~\ref{sec:boojum-npe}. 

\subsection{Colorful boojums at the CFL phase side}
\label{sec:boojum-CFL}

\subsubsection{The shape of boojums}
When a non-Abelian vortex hits the boundary 
of the CFL phase, 
it cannot go out from the CFL phase 
since it carries the color magnetic flux. 
Since $U(1)_{\rm B}$ vortices do not 
have fluxes, they can go out. 
One $U(1)_{\rm B}$ vortex decays 
into three non-Abelian vortices with 
total color fluxes canceled out, 
as discussed in Sec.~\ref{sec:Abelian-vortex-decay}.
In other words, non-Abelian vortices 
can go out only when three of them meet 
to cancel the total color fluxes as imposed by 
color neutrality.
Such connected points form 
``colorful boojums.''
Including the electromagnetic interactions, 
the only possibility is the formation of a BDM, a $\mathbb{C}P^{1}_{+}$ and  a $\mathbb{C}P^{1}_{-}$ vortex. 
In fact, as explained in Eq.~(\ref{eq:fluxes}), the BDM vortex carries a $U(1)_{\rm EM}$ magnetic flux  	
$\Phi^{\textsc{EM}}_{\textsc{bdm}}=	\frac{\delta^{2}}{1+\delta^{2}}\frac{2 \pi}{e}$ 
with $\delta^{2} = \frac23 \frac{e^{2}}{g_{\rm s}^{2}}$
while the $\mathbb{C}P^{1}_{\pm}$ vortex flux is $\Phi^{\textsc{EM}}_{\mathbb{C}P^{1}_\pm} = - \frac12 \Phi_{\textsc{EM}}^{\textsc{bdm}}$.
Then, $U(1)_{\rm EM}$ magnetic fluxes are canceled out.

In order to present a 
structure of the boojum in the CFL phase side, 
one can model the system as a regular lattices of boojums where the relative separation between vortices in a single boojum is denoted by $y(z)$ as a function of the distance $z$ from the interface. The center of the $i$th boojum at the interface is indicated by $\vec x_{i}$. We use the Nambu-Goto action and approximate the interaction of vortices with that of global parallel vortices, 
given in Eq.~\eqref{eq:int-energy}. 
The energy, per unit length, of the lattice is then:
\begin{align}
E_{\rm tot}(y)& =  3 \mathcal N \mathcal  T\sqrt{1+(dy/dz)^{2}}+V_{\rm int}(y),  \nonumber \\
V_{\rm int}(y)& =  -3 \mathcal N \mathcal  T  \log |y|-9 \mathcal  T \sum_{i>j}\log |\vec x_{i}-\vec x_{j}| \, , 
\end{align}
where $\mathcal N$ is the number of boojums and $\mathcal T$ is the tension of a color-magnetic vortex. 
The first term in the potential above is the interaction energy between vortices in the same boojum while the second term represents the one between boojums, where the shape of the boojum has been neglected. Notice that the first term in the potential has to be regularized in the limit $y\rightarrow 0$. 
\begin{figure}[ht]
\begin{center}
   \includegraphics[width=10cm]{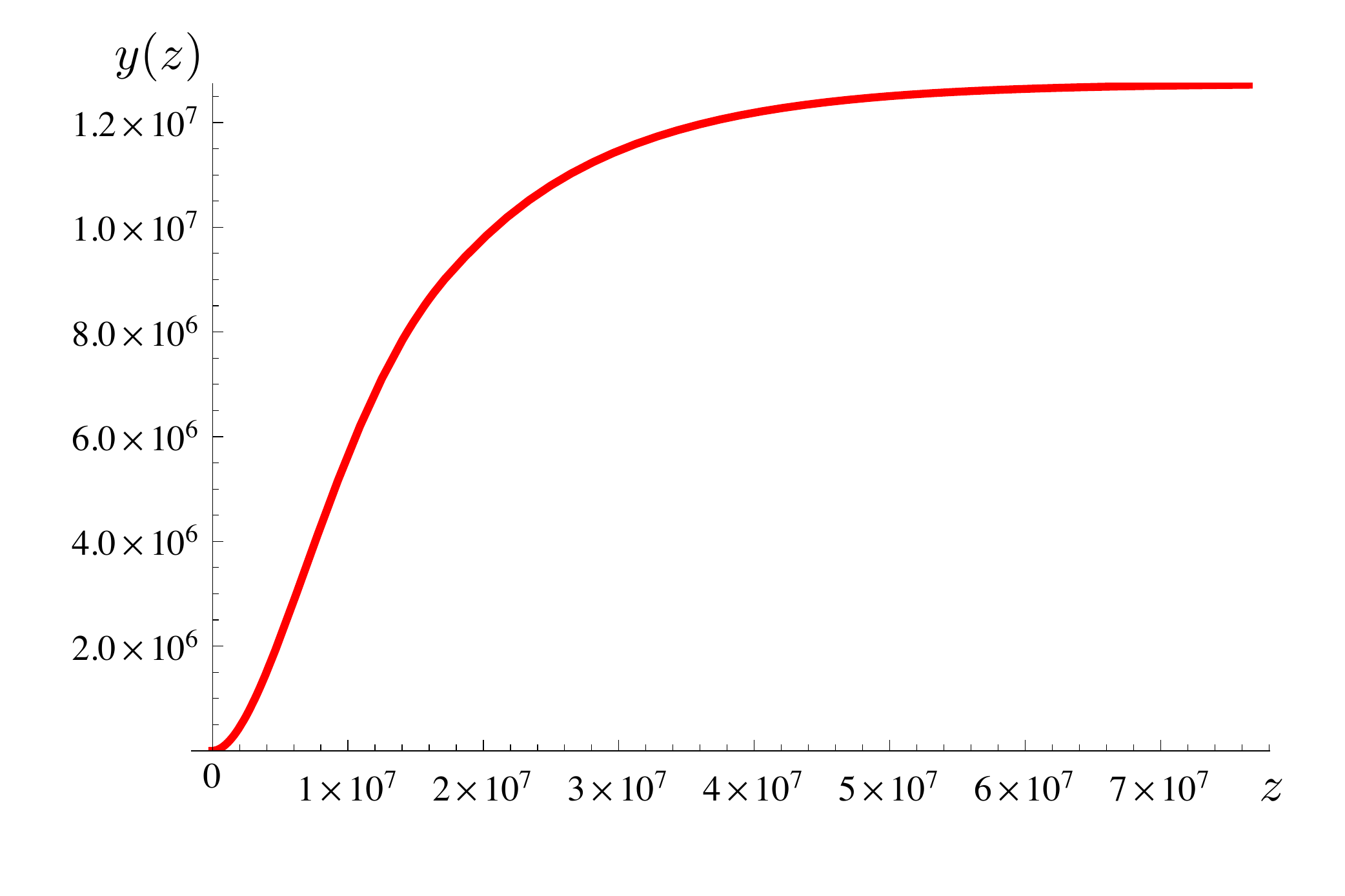}\\
(a)\\
   \includegraphics[width=10cm]{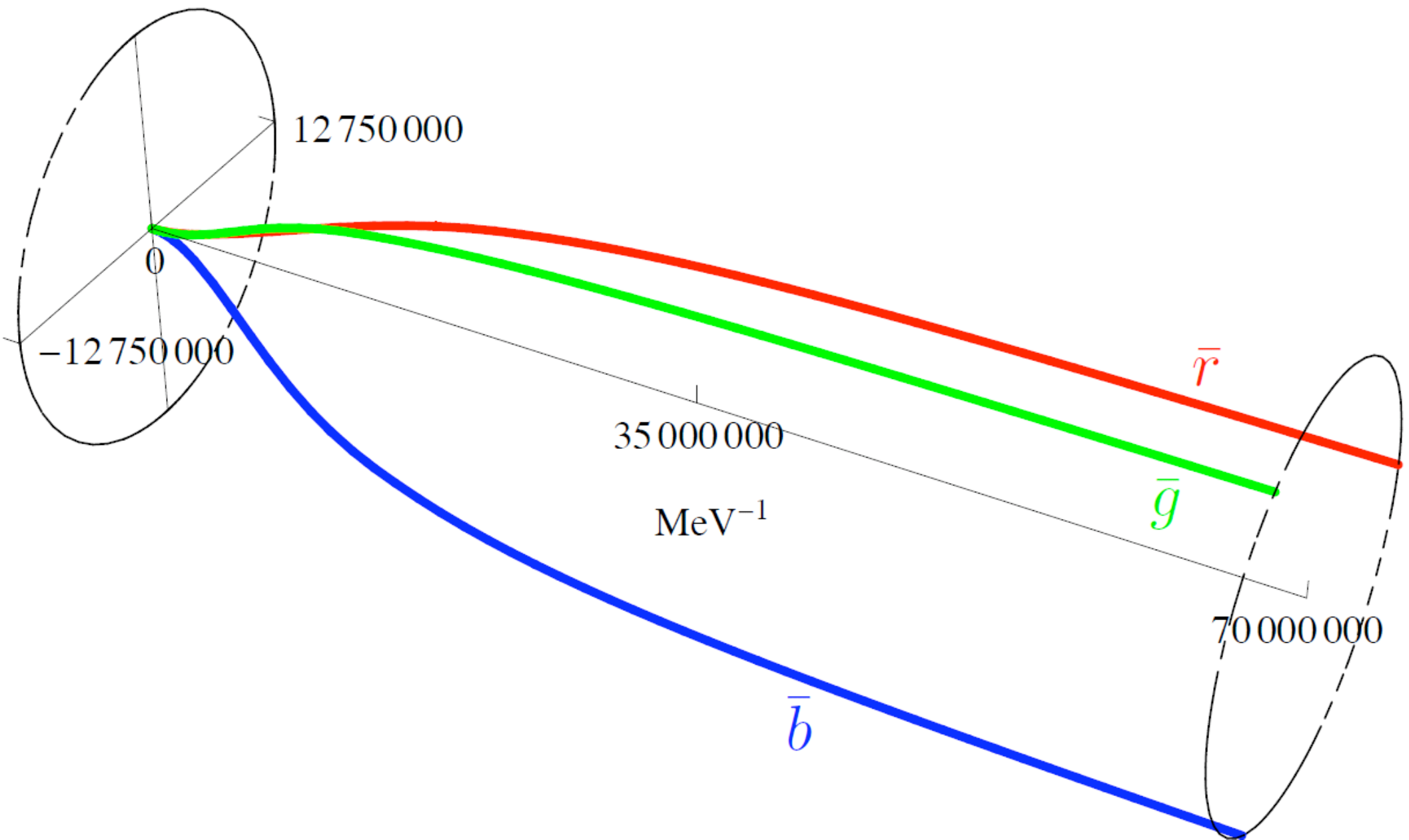}\\
(b)
\end{center}
	\caption{
(a) Transverse shape and 
(b) 
three dimensional shape of a colorful boojum. The $npe$-CFL interface is on the left, where three color vortices are originated from the boojum. We have used typical values of the GL parameters as done, for example, in Ref.~\cite{Vinci:2012mc}. 
In (b), we have ignored the effect of the strange quark mass. 
}
\label{fig:Boojum}
\end{figure}
The numerical shape of the boojum is shown in Fig.~\ref{fig:Boojum},  where we have evaluated the interaction potential for a one dimensional lattice for simplicity. The most important property of the boojum that can be inferred from this very simplified numerical analysis is that the longitudinal  size scales proportionally to the transverse lattice spacing. We have also checked that this property and the shape of the boojum also do not depend on the choice of the regularization of the interaction potential for coincident vortices.

\subsubsection{Formation of 
colorful monopoles with strange quark mass}
At a typical distance $\xi$ from the interface, though, the BDM and the $\mathbb{C}P^{1}_{-}$ solutions will ``decay'' to the $\mathbb{C}P^{1}_{+}$ vortex, due to their instability.  
We can estimate the length scale $\xi$ by referring to the low-energy effective Lagrangian 
\eqref{eq:eff_lag_coeff} with \eqref{eq:lag_cp2}.
Following 
the same steps of Eq.~\eqref{eq:lag_cp1sub}
we obtain
\begin{equation}
	\xi \sim m_{\rm s}^{-1} \left( \frac\mu{\Delta_{\varepsilon}} \right)^{2} \log\left( \frac\mu{\Delta_{\varepsilon}} \right)^{-1/2}  \sim 131 \, {\rm GeV}^{-1} \, ,
\end{equation}
with 
the physical quantities being
	$\mu \sim 1 \, {\rm GeV}, \Delta_{\varepsilon} \sim 100 \, {\rm MeV}, K_{3} = 9$.
This length has to be compared with the thickness $d$ of the interface, which can be seen as a domain wall between the two different phases \cite{Giannakis:2003am}. Using the same values for physical parameters, we get 
\beq 
d \simeq 0.1 \xi.
\eeq
Then the vortices decay at large distances from the interface. 

Since vortices decay into others with different fluxes, each junction corresponds in fact to a monopole. 
Unlike the Dirac monopole at the endpoint of a proton vortex, 
this is a confined {\it color magnetic} monopole attached by vortices from its both sides. 
The monopole connecting $\mathbb{C}P^{1}_{-}$ to $\mathbb{C}P^{1}_{+}$ vortices is a pure color magnetic monopole, because the two vortices have the same $U(1)_{\rm EM}$ magnetic flux but different color-magnetic fluxes; the junction between BDM and $\mathbb{C}P^{1}_{+}$ is instead realized by a color-magnetic and $U(1)_{\rm EM}$ magnetic monopole, because for these vortices both the $U(1)_{\rm EM}$ magnetic and color-magnetic fluxes are different. 
The colorful boojum 
is qualitatively depicted in Fig.~\ref{fig:boojums}.

\subsection{Colorful boojums at the $npe$ phase side}
\label{sec:boojum-npe}

\subsubsection{Matching condition}
The question of what are formed at a boojum point 
outside the CFL phase now arises. 
The CFL phase may be connected 
the kaon condensed (CFL+K) phase, 
2SC phase, or 
hyperon phase.  
However, in a realistic situation for the 
core of neutron stars, 
there is the $npe$ phase.
Here, we discuss what are connected 
to the colorful boojums in the $npe$ phase. 
Note that, in the following discussion, 
we do not assume that 
the CFL phase is directly connected to  the $npe$ phase, 
but that the other phases mentioned above 
can be sandwiched between 
the CFL and the $npe$ phase.

When quarks travel around a vortex, they acquire an Aharanov-Bohm phase in general. Such phases 
have to match across the interface. 
In the CFL phase, the order parameter is 
$\langle qq \rangle$, which behaves 
\beq 
\langle qq \rangle \sim e^{i\theta}  
\mbox{:  $U(1)_{\rm B}$ vortex} \label{eq:qq}
\eeq 
for the endpoint of 
a $U(1)_{\rm B}$ vortex or a triplet of non-Abelian 
semi-superfluid vortices, indicating that the quark fields 
get a phase 
\beq 
 \Delta \theta|_{U(1)_{\rm B}} =  {2\pi \over 2} = \pi
   \label{eq:phase-U(1)B}
\eeq 
corresponding to 
$1/2$ $U(1)_{\rm B}$ winding, 
when they travel around a $U(1)_{\rm B}$ vortex 
or a triplet of  semi-superfluid vortices. 
In particular, all quark fields 
including $u$ and $d$ quarks 
relevant in hadronic matter obtain  
a phase in Eq.~(\ref{eq:phase-U(1)B}) 
around the $U(1)_{\rm B}$ vortex.

The order parameters in the $npe$ phase are 
$\langle nn \rangle \sim \langle (udd)(udd) \rangle$ 
for neutron condensation 
and 
$\langle pp \rangle \sim \langle (uud)(uud) \rangle$ 
for proton condensation.\footnote{
At high density, 
a spin triplet (p-wave) pairing is favored for 
 neutron condensation more than s-wave pairing 
\cite{Takatsuka:1992ga}. 
In this case, half-quantized vortices are possible, 
but here we count neutron vortices as 
integer vortices for simplicity. 
} 
We label the winding numbers of the order parameters 
$\langle nn \rangle$ and $\langle pp\rangle$ 
in the presence of a vortex 
by $(p,q)$.
In the presence of a neutron vortex $(1,0)$, 
these order parameters behave as
\beq 
(1,0): \quad \langle nn \rangle \sim \langle (udd)(udd) \rangle \sim e^{i \theta} ,  \quad 
\langle pp \rangle \sim \langle (uud)(uud) \rangle \sim 1.
\label{eq:nn2}
\eeq
From these behaviors, we find that 
the $u$ and $d$ quark wave functions 
are of the form 
\beq
(1,0): \quad (u, d) \sim (e^{-i (1/6) \theta} ,   e^{ i (2 /6) \theta }) \label{eq:10}
\eeq
in the presence of a neutron vortex.
In the same way, the winding of 
the order parameters in 
the presence of a proton vortex $(0,1)$,
\beq 
(0,1): \quad 
 \langle nn \rangle \sim \langle (udd)(udd) \rangle \sim 1 ,  \quad 
\langle pp \rangle \sim \langle (uud)(uud) \rangle \sim e^{i \theta}
\label{eq:pp2}
\eeq 
leads the wave functions of $u$ and $d$ quarks as
\beq
(0,1): \quad 
(u, d) \sim (e^{i (2/6) \theta} ,   e^{- i (1 /6) \theta }).
 \label{eq:01}
\eeq
From Eqs.~(\ref{eq:10}) and (\ref{eq:01}), 
a composite of neutron and  proton vortices 
has the winding numbers of 
the order parameters and quarks 
\beq 
&& (1,1): \quad
  \langle nn \rangle \sim \langle (udd)(udd) \rangle \sim e^{i \theta} , \quad 
\langle pp \rangle \sim \langle (uud)(uud) \rangle \sim e^{i \theta} , \\
&& (1,1): \quad 
(u, d) \sim (e^{i (1/6) \theta} ,   e^{i (1 /6) \theta }),
\eeq 
respectively. 
From this,  we see that $u$ and $d$ quarks 
get a phase 
\beq 
  \Delta \theta|_{(1,1)} 
= {2\pi \over 6}  = {\pi \over 3}, \label{eq:phase-pn}
\eeq 
corresponding to $1/6$ $U(1)_{\rm B}$ winding,
when they encircle a composite $(1,1)$ of 
neutron and proton 
vortices in the $npe$ phase. 

From Eqs.~(\ref{eq:phase-U(1)B}) and 
(\ref{eq:phase-pn}), 
we have a relation 
\beq
 \Delta \theta|_{(3,3)} (= 3 \Delta \theta|_{(1,1)})
 = \Delta \theta|_{U(1)_{\rm B}}. 
\label{eq:matching}
\eeq
This implies that 
$u$ and $d$ quarks have continuous wave functions 
only when 
three neutron vortices {\it and} three proton vortices  
join at a colorful boojum point 
in the $npe$ phase. 

\begin{figure}[ht]
\begin{center}
	\includegraphics[width=8cm]{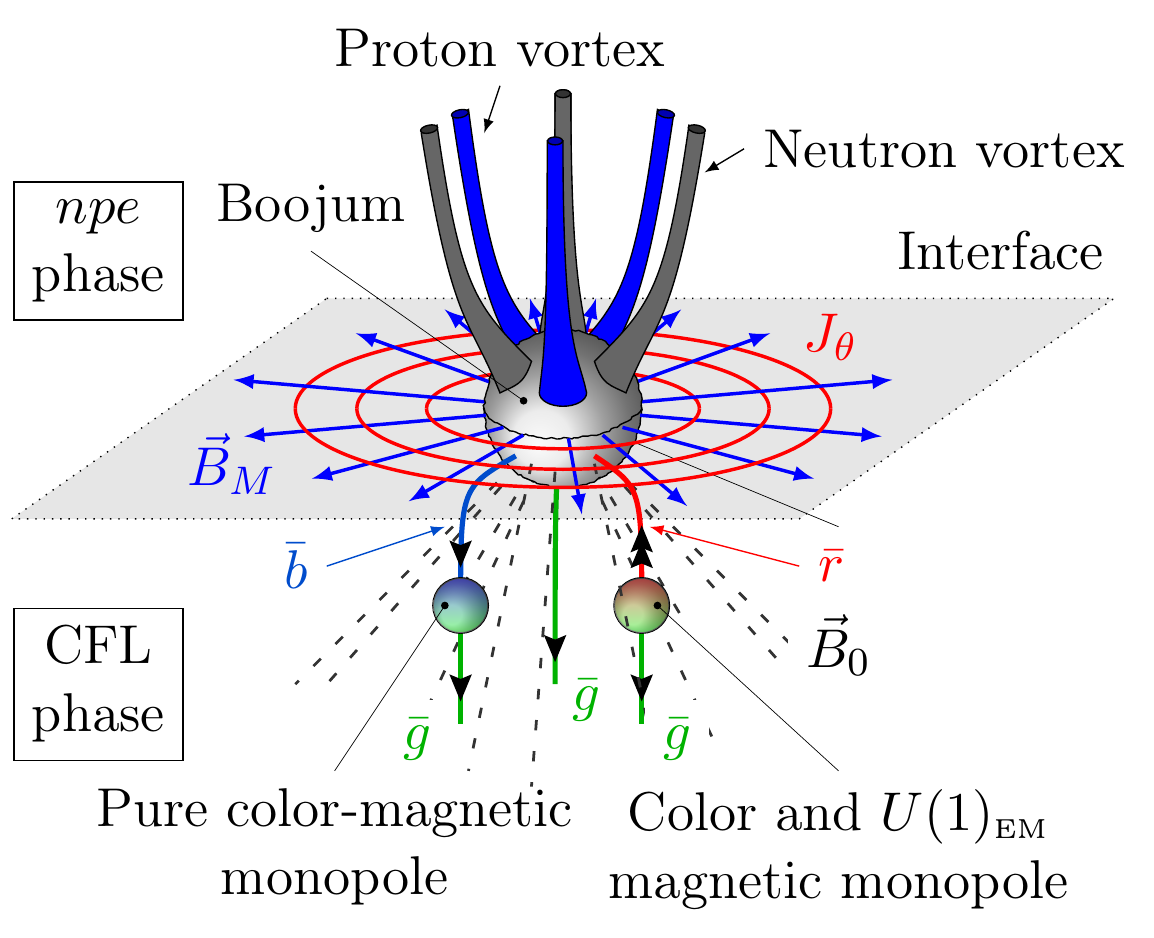}
\end{center}
\caption{
\label{fig:boojums}
Colorful boojum.
Three neutron vortices and three proton vortices end 
on a boojum at the interface in the 
$npe$ phase side, and
the three BDM ($\bar{r}$), $\mathbb{C}P^{1}_{+}$ ($\bar{g}$) and $\mathbb{C}P^{1}_{+}$ ($\bar{b}$) vortices 
end on it in the CFL phase side.
 The black arrows along the three vortices represent $U(1)_{\rm EM}$ magnetic flux. 
The two colorful monopole junctions which 
exist in the presence of the strange quark mass 
are also depicted.  
The pure magnetic fluxes split into a massive $\vec{B}^{\rm M}$ component, which is screened by a surface current and bent along the interface, and a massless $\vec{B}^{0}$ component emanating from the boojum, which looks like a Dirac monopole. 
}
\end{figure}

\subsubsection{Magnetic fluxes}
The proton condensation is electrically charged.
Therefore, the lowest energy configuration of
 a proton vortex is accompanied by a magnetic flux, 
which must be quantized 
\beq
 \Phi_{0}=\pi / e 
\eeq 
as ANO vortices in metallic superconductors.
When this flux penetrate into the CFL phase, 
the $U(1)_{\rm EM}$ magnetic flux is converted into both the fluxes corresponding to the  massive and massless combinations $A^{\rm M}$ and $A^{0}$. This is due to conservation of flux and to the the mixing in Eqs.~\eqref{eq:mix_gluon} and \eqref{eq:mix_photon}, 
respectively. 
The massive combination $A^{\rm M}$ is 
screened by a surface color-magnetic current circulating around the contact point. 
Unlike metallic superconductors, this is completely screened and cannot enter the CFL phase 
even if the flux is larger than the quantized flux of the non-Abelian vortex, 
because the non-Abelian vortex also has to carry the 1/3 quantized circulation. 
A rough estimate of the behavior of the current in proximity of the vortex endpoint can be obtained by using the London equation valid for an ordinary superconductor. Then we obtain 
\beq 
 J_{\theta} \simeq {\Phi_{0} \over 2\pi r},
\eeq 
where $\theta$ and $r$ are the planar polar coordinates centered at the contact point. 
On the other hand, the massless combination $A^{0}$ can spread freely into the CFL phase, being no superconducting currents that can screen it. 
This object looks like a Dirac monopole, as is common in 
boojums in other systems, such as helium superfluids.
This boojum  is qualitatively depicted 
in Fig.~\ref{fig:boojums}.

One may wonder that the formation of vortices under rotation would lead
to the generation of magnetic fields inside neutron stars.
However, it turns out that the magnetic field induced at the colorful
boojums are in fact quite small, and are estimated as 
\begin{equation}
B \sim 
{ N_v \times \Phi_0
\over (10 \, {\rm km})^2} 
= 10^{19} \times 10^{-7} \, \mbox{G cm}^2  \times 10^{-8} \, {\rm m}^2 
\simeq 1 \, {\rm G} = 10^{-4} \, {\rm T},
\end{equation}
where 
$N_v$ is the total number of vortices Eq.~(\ref{eq:rot-vortex-num}), and
parameters like rotational period, baryon chemical potential or the core radius are set to typical values.

%% file: fermion-v9.tex
\section{Fermions in vortices}\label{sec:fermion}

We here discuss fermionic structures inside non-Abelian vortices.
Inside a vortex, fermions are important degrees of freedom at low
energies, since a part of superconducting gaps vanishes in the core and
fermions become massless.
We can investigate the fermion states by the Bogoliubov-de Gennes (BdG) equation.
In Sec.~\ref{sec:BdG}, we introduce the BdG equation for an {\it
Abelian} vortex, and find the solution of a zero-mode fermion which is
called a Majorana fermion.
In Sec.~\ref{sec:BdG2}, we discuss the BdG equation for a {\it
non-Abelian} vortex in the CFL phase, and find Majorana fermions which
belong to the triplet or singlet representation of the
$SU(2)_{\mathrm{C+F}} \times U(1)_{\mathrm{C+F}}$
symmetry.
In Sec.~\ref{sec:1+1_effective_theory}, we discuss the 1+1 dimensional
effective theory for Majorana fermions propagating along a vortex axis.
In Sec.~\ref{sec:supercurrent}, 
as a new result in this paper, 
we show the absence of supercurrent by fermion along a vortex, unlike the case of Witten's superconducting string.
In Sec.~\ref{sec:index}, we discuss the index theorem to count the
number of zero-mode fermions, and show that the result from the index
theorem is consistent with that from directly solving the BdG equation.

\subsection{The Bogoliubov-de Gennes equations and Majorana fermions for Abelian vortices}
\label{sec:BdG}

Here we analyze the internal fermionic structures of vortices in terms
of the BdG equations. 
So far we have discussed the structures of non-Abelian vortices based on
the GL equation, where the diquarks are dynamical degrees of freedom.
Such description is valid 
only at distances larger than the coherence length.
At short-distant scales, fermionic degrees of freedom become important.
The Bogoliubov-de Gennes (BdG) equation describes the dynamics of the fermions as well as the gap functions.
The BdG equation gives a self-consistent solution for the wave functions
of the fermion and the gap profile function. 
Note that the BdG equation is different from the Bogoliubov equation
or the Bardeen-Cooper-Schrieffer (BCS) equation.
The Bogoliubov or BCS equation describes only the plane wave for the
fermion, while the BdG equation allows more general wave functions which
is different from the plane wave.
This property is quite important for the analysis of vortices.
Inside vortices, the gap profile function $\Delta(r)$ is zero at the center
of the vortex ($\Delta(0)=0$), while it becomes the bulk gap (the gap in
the bulk space) at the position infinitely far from the vortex core
($\Delta(\infty)=\Delta_{\rm{bulk}}$).
Namely, the gap profile function for a vortex is position-dependent.
So we need to use the BdG equation in order to take the position
dependence of the gap of a vortex into account and to analyze the
fermion structure inside a vortex.

Let us comment on the previous studies on vortices in superconductors
using the BdG equations.
The BdG equation was first applied to a vortex inside a s-wave superconductor 
in Refs.~\cite{de1999superconductivity,Caroli1964307} in which fermions are
 non-relativistic.
In these first studies, the order 
parameter (the superconducting gap) was treated as a background field.
Successive studies were devoted to finding the self-consistent solutions
 of the BdG equation \cite{PhysRevLett.62.3089,PhysRevB.41.822}.
The complete self-consistent description was achieved by considering
both quasibound and scattering fermions in the vortex
\cite{PhysRevB.43.7609,PhysRevLett.90.077003}.
The description reviewed here is very similar to that of the
vortex-(relativistic)fermion system discussed in
Refs.~\cite{Jackiw:1981ee,Weinberg:1981eu,Witten:1984eb}.
The difference from these studies is that  we here consider a fermionic
matter at {\it finite densities}
(otherwise color superconductivity does not take place). 
The previous studies \cite{Jackiw:1981ee,Weinberg:1981eu,Witten:1984eb} were
formulated only in the vacuum.
The analysis of vortices in condensed matter systems via the BdG
equation
 predicted an enhanced local density of fermion states at
the Fermi level around the vortex core, which is experimentally observed
in various metals \cite{PhysRevLett.96.097006,PhysRevLett.101.166407}.
Recently, the analysis of vortices in terms of the BdG equation has been also
applied to the BEC-BCS crossover in fermionic cold atom systems
\cite{PhysRevLett.94.140401,PhysRevLett.96.090403}.
For the application of the BdG equation in condensed matter systems,
see also
Refs.~\cite{PhysRevLett.103.107002,PhysRevLett.103.020401,Volovik2009,PhysRevLett.102.187001,PhysRevB.79.184520,PhysRevLett.104.040502,2009arXiv0907.2681L,PhysRevB.81.125318,PhysRevLett.104.066404,PhysRevLett.104.067001,PhysRevB.81.205429,PhysRevB.85.165453}.

We discuss the BdG equation for a superconductor with a vortex.
Since a non-Abelian vortex has a complex structure because of the
internal degrees of freedom, 
we first review the results for an Abelian vortex, 
in which a single component massless fermion is trapped.
The fermionic structure of a non-Abelian vortex
is discussed in Sec.~\ref{sec:BdG2}.

We here consider a superconducting system made of a single species of
fermion. The fermions make pairs and form a Bose condensate.
The BdG equation in the Nambu-Gor'kov representation $\Psi=(\varphi,\eta)^{\mathrm{T}}$ (particle in the upper
component and hole in the lower component) is given by 
\begin{eqnarray}
{\mathcal H}\Psi = {\mathcal E}\Psi\, ,
\label{eq:B-dG}
\end{eqnarray}
where ${\mathcal E}$ is the energy measured from the chemical potential $\mu$
and  ${\mathcal H}$ is the Hamiltonian in the mean-field approximation, 
\begin{eqnarray}
{\mathcal H} \equiv \left(
\begin{array}{cc}
 -i\gamma_{0} \vec{\gamma} \cdot \vec{\nabla}-\mu  & \Delta(\vec{x}) \gamma_{0} \gamma_{5}     \\
 -\Delta^{\ast}(\vec{x}) \gamma_{0} \gamma_{5} & -i\gamma_{0} \vec{\gamma} \cdot \vec{\nabla}+\mu
\end{array}
\right)\, .
\label{eq:hamil_single}
\end{eqnarray}
The gap profile function $\Delta(\vec{x})$ (three dimensional coordinate
$\vec{x}=(x,y,z)$) is given as $\Delta(\vec{x}) \propto \langle
\Psi^{\rm T} \Psi \rangle$, where the expectation value is given by the
sum over all the fermion states in the ground state.
This Hamiltonian has the particle-hole symmetry, 
\begin{equation}
{\mathcal U}^{-1} {\mathcal H U}=-{\mathcal H}^{\ast}, \hspace{1em}
{\mathcal U}=
\left(
\begin{array}{cc}
 0 & \gamma_{2} \\
 \gamma_{2} & 0   
\end{array}
\right)\, .
\end{equation}
Thus, $\mathcal{H}\Psi = \mathcal E \Psi $ implies 
$
\mathcal{H} (\mathcal U\Psi) = -\mathcal E (\mathcal U\Psi) 
$ 
and the spectrum is symmetric above and below the Fermi sea.

If one considers a vortex solution, 
because of the translational invariance along the vortex
($z$) axis, the gap is a function of the distance
$r=\sqrt{x^2+y^2}$ from the center of the vortex and $\theta$ an angle
around the vortex; $\Delta(\vec{x})=|\Delta(r)|e^{i\theta}$.
The gap profile function also satisfies the following boundary 
conditions; $|\Delta(r=0)|=0$ at the center of the vortex and
$|\Delta(r=\infty)|=|\Delta_{\rm{bulk}}|$ at the position far from the
vortex with a bulk gap $\Delta_{\rm{bulk}}$.
Since the system is translationally invariant along the vortex axis,
we can always take the fermion states to be eigenstates of the momentum
in the $z$-direction $k_z$,
\begin{eqnarray}
\Psi_{\pm,m}^{k_{z}}(r,\theta,z)
=\Psi_{\pm,m}(r,\theta)\, {\rm e}^{i k_{z} z},
\end{eqnarray}
where $\Psi_{\pm,m}(r,\theta)$ is the wave function on the $x$-$y$ plane.
Here $\pm$ is for chirality, left and right, of the fermion and
an integer $m$ is related to the $z$ component of the total angular
momentum $J_{z}$.
In the Nambu-Gor'kov formalism, the wave function
$\Psi_{\pm,m}(r,\theta)$ is given as
\begin{eqnarray}
\Psi_{\pm,m}(r,\theta)=
\left(
\begin{array}{c}
 \varphi_{\pm,m}(r,\theta)  \\
 \eta_{\mp,m-1}(r,\theta)
\end{array}
\right),
\end{eqnarray}
with the particle component $\varphi_{\pm,m}(r,\theta)$ and the hole component $\eta_{\mp,m-1}(r,\theta)$.
The $z$ component of $J_{z}$ is $m+1/2$ for the particle and $(m-1)+1/2$ for the hole, respectively.
Note that the chirality $\pm$ of the particle is assigned in opposite to that of the hole.

The solution of the BdG equation (\ref{eq:B-dG}) gives all the fermion modes in the vortex.
They include the scattering states with energies $|{\mathcal E}|>|\Delta_{\rm{bulk}}|$ as well as the bound states with energies $|{\mathcal E}|<|\Delta_{\rm{bulk}}|$.
It is a non-trivial problem to obtain all the fermion solutions.
In the present discussion, we concentrate on the fermion states with the lowest energy inside
the vortex, which are the most important degrees of freedom at low energies.
Furthermore, we here regard the gap profile function $|\Delta(r)|$ as a background field and do not analyze the self-consistent solution for the gap profile function and the fermion wave functions.
Such study will be left for future works.

As a result of the particle-hole symmetry,  we find that the state with
${\mathcal E}=0$ is a ``Majorana fermion'', which has a special property
that a particle and a hole are equivalent. 
The explicit solution of the wave function of the Majorana fermion is given as, for the right mode 
($+$ for a particle, $-$ for a hole)  
\begin{eqnarray}
\varphi_{+,0}(r,\theta) &=& 
C \, {\rm e}^{-\int_{0}^{r}|\Delta(r')|\, dr'}
\left(
\begin{array}{c}
 J_{0}(\mu r) \\
 i J_{1}(\mu r)\, {\rm e}^{i \theta}
\end{array}
\right), \label{eq:one_flavor_left_particle}\\
\eta_{-,-1}(r,\theta) &=& 
C\, {\rm e}^{-\int_{0}^{r}|\Delta(r')|\, dr'}
\left(
\begin{array}{c}
 -J_{1}(\mu r) \, {\rm e}^{-i \theta}\\
 i J_{0}(\mu r) 
\end{array}
\right),\label{eq:one_flavor_left_hole}
\end{eqnarray}
and for the left mode ($-$ for a particle, $+$ for a hole)  
\begin{eqnarray}
\varphi_{-,0}(r,\theta) &=& 
C^{\prime}\, {\rm e}^{-\int_{0}^{r}|\Delta(r')|\, dr'}
\left(
\begin{array}{c}
 J_{0}(\mu r) \\
- i J_{1}(\mu r)\, {\rm e}^{i \theta}
\end{array}
\right), \label{eq:one_flavor_right_particle}\\
\eta_{+,-1}(r,\theta) &=& 
C^{\prime} \, {\rm e}^{-\int_{0}^{r}|\Delta(r')|\, dr'}
\left(
\begin{array}{c}
 J_{1}(\mu r) \, {\rm e}^{-i \theta}\\
 i J_{0}(\mu r) 
\end{array}
\right),\label{eq:one_flavor_right_hole}
\end{eqnarray}
where $C$ and $C^{\prime}$ are normalization constants, 
and $J_n(x)$ is the Bessel function. 
We have represented the solutions in the Weyl 
(2-component) spinors.
The derivation is described in detail in Appendix~\ref{sec:appendix-fermion}.

The solutions above satisfy a ``Majorana-like'' condition
($\kappa=\pm1$)\footnote{
$\kappa = 1$ is for the right mode and $\kappa = -1$ is for the left mode.
}
\begin{eqnarray}
\Psi = \kappa \, {\mathcal U}\, \Psi^{\ast}, 
\label{eq:Majorana}
\end{eqnarray}
which physically implies the equivalence between a particle 
and a hole.
We note the Majorana fermion is localized at around the center of the vortex.
This can bee seen by setting $|\Delta(r)|$ a constant
$|\Delta_{\mathrm{bulk}}|$, because the exponential functions in
Eqs.~(\ref{eq:one_flavor_left_particle}) and
(\ref{eq:one_flavor_left_hole}) or
Eqs.~(\ref{eq:one_flavor_right_particle}) and
(\ref{eq:one_flavor_right_hole}) exhibit the behavior like
$e^{-|\Delta_{\mathrm{bulk}}|r}$.
Fermion zero modes (Majorana fermions) in
relativistic theories
were found in the (Abelian) vortex in the vacuum \cite{Jackiw:1981ee}, where
the number of zero modes is determined by the index theorem to be $2n$
for vortices with winding number $n$ \cite{Weinberg:1981eu}.
The Majorana-fermion solution in a vortex in a p-wave superconductor was
found first by Fukui \cite{Fukui:2009mh} 
for non-relativistic
fermions, 
and later in Refs.~\cite{Nishida:2010wr,Yasui:2010yw} 
for relativistic fermions.
Although there exist Majorana-like solutions 
in vortices in p-wave superconductors both for  
non-relativistic and relativistic fermions,
they are absent for non-relativistic fermions in vortices in s-wave
superconductors \cite{Caroli1964307}.
We leave a comment that the fermions bound in the vortex are intuitively understood also from a view of the Andreev reflection.
When the fermions meet the interface between the normal phase (inside of the vortex) and the superconducting phase (outside of the vortex),
there appear the Cooper pairs created in the superconducting phase and the holes reflected in the normal phase.
This is called the Andreev reflection \cite{Andreev1965}.
The Andreev reflection was also considered in the CFL phase \cite{Sadzikowski:2002in}.
The multiple number of reflections of the fermions (particle and holes) at the interface make the bound state inside the vortex.

\subsection{Bogoliubov-de Gennes formalism and Majorana fermions for non-Abelian vortices}
\label{sec:BdG2}

Now let us investigate the fermion structure of a non-Abelian vortex in
the CFL phase.
Because of the internal symmetry, the BdG equation is a multicomponent
equation with $3(\rm{color}) \times 3(\rm{flavor})=9$ degrees of freedom
whose subspaces belong to multiplets of $ U(1)_{{\mathrm C} + {\mathrm
F}} \times SU(2)_{{\mathrm C} + {\mathrm F}}$ symmetry.
The gap structure of a non-Abelian vortex is given as \cite{Yasui:2010yw}
\begin{eqnarray}
\Phi(r,\theta) =
\left(
\begin{array}{ccc}
 \Delta_{0}(r) & 0  & 0  \\
 0 & \Delta_{0}(r)  & 0  \\
 0 & 0 & \Delta_{1}(r,\theta) 
\end{array}
\right)\, , \label{eq:gap_non-Abelian}
\end{eqnarray}
with the basis spanned by $\bar{\mathrm{u}}(\bar{\mathrm{r}})=\mathrm{ds}(\mathrm{gb})$, $\bar{\mathrm{d}}(\bar{\mathrm{g}})=\mathrm{su}(\mathrm{br})$ and $\bar{\mathrm{s}}(\bar{\mathrm{b}})=\mathrm{ud}(\mathrm{rg})$, where $\Delta_{1}(r,\theta) = |\Delta_{1}(r)|\, e^{i\theta}$ 
corresponds to the vortex configuration with winding number one, 
and $\Delta_{0}(r)$ does not have a winding number (though it is 
not necessarily spatially constant).
We remember that $\Delta_{1}(r)$ satisfies the boundary conditions, $\Delta_{1}(r=0)=0$ and $\Delta_{1}(r=\infty)=|\Delta_{\rm{CFL}}|$, with the gap $\Delta_{\rm{CFL}}$ in bulk space, while $\Delta_{0}(r)$ satisfies only $\Delta_{0}(r=\infty)=|\Delta_{\rm{CFL}}|$.
The value of $\Delta_{0}(r=0)$ depends the details of the equation of the gap profile function.
In the configuration in Eq.~(\ref{eq:gap_non-Abelian}), the $\mathrm{ud}$ diquark pair with green and blue has a non-trivial winding number, while the other pairs, i.e. $\mathrm{ds}$ diquark pair with blue and red and $\mathrm{su}$ diquark pair with red and green, have no winding. 
Correspondingly, the explicit form of the BdG equation is 
(for a similar representation in the homogeneous CFL phase, see Ref.~\cite{Alford:1999pa,Sadzikowski:2002in})
\begin{eqnarray}
\left(
\begin{array}{ccccccccc}
 \hat{\mathcal H}_0 & \hat{\Delta}_{1} & \hat{\Delta}_{0} & 0 & 0 & 0 & 0 & 0 & 0 \\
 \hat{\Delta}_{1} & \hat{\mathcal H}_0 & \hat{\Delta}_{0} & 0 & 0 & 0 & 0 & 0 & 0 \\
 \hat{\Delta}_{0} & \hat{\Delta}_{0} & \hat{\mathcal H}_0 & 0 & 0 & 0 & 0 & 0 & 0 \\
 0 & 0 & 0 & \hat{\mathcal H}_0 & -\hat{\Delta}_{1} & 0 & 0 & 0 & 0 \\
 0 & 0 & 0 & -\hat{\Delta}_{1} & \hat{\mathcal H}_0  & 0 & 0 & 0 & 0 \\
 0 & 0 & 0 & 0 & 0 & \hat{\mathcal H}_0 & -\hat{\Delta}_{0} & 0 & 0 \\
 0 & 0 & 0 & 0 & 0 & -\hat{\Delta}_{0} & \hat{\mathcal H}_0 & 0 & 0 \\
 0 & 0 & 0 & 0 & 0 & 0 & 0 & \hat{\mathcal H}_0 & -\hat{\Delta}_{0} \\
 0 & 0 & 0 & 0 & 0 & 0 & 0 & -\hat{\Delta}_{0} & \hat{\mathcal H}_0
\end{array}
\right)
\left(
\begin{array}{c}
 u_r \\
 d_g \\
 s_b \\
 d_r \\
 u_g \\
 s_r \\
 u_b \\
 s_g \\
 d_b
\end{array}
\right)
= {\mathcal E}
\left(
\begin{array}{c}
 u_r \\
 d_g \\
 s_b \\
 d_r \\
 u_g \\
 s_r \\
 u_b \\
 s_g \\
 d_b
\end{array}
\right),
\label{eq:CFL_eve} 
\end{eqnarray}
where we introduce the notation e.g., 
 $u_{r}$ for  the quark with flavor ``up" and color ``red" in
 the Nambu-Gor'kov representation.
The matrices $\hat{\mathcal H}_{0}$ and $\hat{\Delta}_{i}$ 
($i=0$ and 1) are given as follows: 
\begin{eqnarray}
\hat{\mathcal H}_{0} &=&
\left(
\begin{array}{cc}
 -i\gamma_{0} \vec{\gamma} \!\cdot\! \vec{\nabla} - \mu & 0 \\
 0 & -i\gamma_{0} \vec{\gamma} \!\cdot\! \vec{\nabla} + \mu
\end{array}
\right), \\
\hat{\Delta}_{i} &=&
\left(
\begin{array}{cc}
 0 & \Delta_{i} \gamma_{0} \gamma_{5} \\
 -\Delta_{i}^{\dag} \gamma_{0} \gamma_{5} & 0
\end{array}
\right).
\end{eqnarray} 

To find the solutions in the BdG equation (\ref{eq:CFL_eve}), it is useful to express the quark state as
 \begin{eqnarray}
\left(
\begin{array}{ccc}
 u_r & u_g  & u_b  \\
 d_r & d_g & d_b  \\
 s_r & s_g & s_b  
\end{array}
\right) =
\sum_{A=1}^{9}  \Psi^{(A)}\frac{\lambda_A}{\sqrt{2}}\, ,
\label{eq:CFL_base}
\end{eqnarray}
where $\lambda_A\, (A=1,\cdots,8)$ are the Gell-Mann 
matrices normalized as $\tr (\lambda_A \lambda_B) = 2 \delta_{AB}$
and $\lambda_9=\sqrt{2/3}\cdot {\bf 1}$.
We note that, because the non-Abelian vortex has $SU(2)_{\rm{C+R+L}}$ symmetry only,
the quark state should belong to the representation of $SU(2)_{\rm{C+R+L}}$ symmetry.
To clarify this, we define
\begin{eqnarray}
\Psi_{\rm t} &\equiv & \Psi^{(1)}\lambda_{1} + \Psi^{(2)}\lambda_{2} 
+ \Psi^{(3)}\lambda_{3}\, ,\label{eq:triplet_def}\\
\Psi_{\rm s} &\equiv & \Psi^{(8)}\lambda_{8} + \Psi^{(9)}\lambda_{9}\, .
\label{eq:singlet_def}
\end{eqnarray}
for triplet and singlet states, respectively.
Explicitly, $\Psi^{(i)}$ ($i=1,2,3,8,9$) are given as
\begin{eqnarray}
\Psi^{(1)} &=& (d_r+u_g)/\sqrt2, \\ 
\Psi^{(2)} &=& (d_r-u_g)/(\sqrt2 i), \\
\Psi^{(3)} &=& (u_r-d_g)/\sqrt2,
\end{eqnarray}
for triplet, and
\begin{eqnarray}
\Psi^{(8)} &=& (u_r+d_g-2s_b)/\sqrt6, \\
\Psi^{(9)} &=& (u_r+d_g+s_b)/\sqrt3,
\end{eqnarray}
for singlet.
The doublet states, $(u_b,d_b)$ and $(s_r, s_g)^{\rm T}$ do not couple to $\hat{\Delta}_{1}$, so they are irrelevant mode as the Majorana fermion.

Let us consider the transformation properties of the quark state $\Psi^{(i)}$ ($i=1,2,3,8,9$).
Under the $SU(2)_{\rm{C+R+L}}$ rotation, the quark state is transformed as
\begin{eqnarray}
\Psi \rightarrow \Psi'= U_{\mathrm{F}}\, \Psi \, U^{\rm T}_{\mathrm{C}}
\end{eqnarray}
where $U_{\mathrm{F}}={\rm e}^{i \vec{\theta} \cdot \vec{\lambda}/2}$ and 
$U_{\mathrm{C}}= {\rm e}^{i \vec{\phi} \cdot \vec{\lambda}/2}$ are the 
$SU(2)_{\rm L+R}$ and $SU(2)_{\mathrm{C}}$ rotations, respectively. 
Since we used the vector $\vec{\lambda}=(\lambda_1, \cdots, \lambda_8)$, 
the parameters $\vec{\theta}$ and $\vec{\phi}$ are defined only 
for the first three components 
$\vec{\theta}=(\theta_1, \theta_2, \theta_3, 0, 0, 0, 0, 0)$ and 
$\vec{\phi}=(\phi_1, \phi_2, \phi_3, 0, 0, 0, 0, 0)$. 
The locking for rotation in color and flavor space under $SU(2)_{\rm{C+R+L}}$ symmetry, 
the color and flavor rotations may be locked as $\phi_{1}=-\theta_{1}$, $\phi_{2}=\theta_{2}$, 
and $\phi_{3}=-\theta_{3}$ as the simplest choice.
We define
\begin{eqnarray}
\theta_{1} = \tilde{\theta}_{1}
\left(
\begin{array}{cc}
 1 & 0 \\
 0 & -1
\end{array}
\right), \hspace{1em}
\theta_{2} = \tilde{\theta}_{2}
\left(
\begin{array}{cc}
 1 & 0 \\
 0 & 1
\end{array}
\right),
\hspace{1em}
\theta_{3} = \tilde{\theta}_{3}
\left(
\begin{array}{cc}
 1 & 0 \\
 0 & -1
\end{array}
\right),
\end{eqnarray}
with $\tilde{\theta}_i$ $(i=1,2,3)$ being real numbers.
For infinitesimal $\tilde{\theta}_i$, we find that the quark state $\Psi_{\rm t}$ is transformed as triplet,
\begin{eqnarray}
\delta \Psi^{(1)} &=& \theta_{3} \Psi^{(2)} - \theta_{2} \Psi^{(3)}, \nonumber \\
\delta \Psi^{(2)} &=& \theta_{1} \Psi^{(3)} - \theta_{3} \Psi^{(1)},
\label{eq:symmetry_t}
 \\
\delta \Psi^{(3)} &=& \theta_{2} \Psi^{(1)} - \theta_{1} \Psi^{(2)},
\nonumber
\end{eqnarray}
and the singlet $\Psi_{\rm s}$ is invariant,
\begin{equation}
\delta \Psi^{(8)} = \delta \Psi^{(9)} = 0\, .
\label{eq:symmetry_s}
\end{equation}

With the setup for the BdG equation for a non-Abelian vortex with $SU(2)_{\rm{C+R+L}}$,
we can find fermion solution with ${\mathcal E}=0$.
Among several representations of the quark state in $SU(2)_{\rm{C+R+L}}$ symmetry,
the Majorana fermion is found only in the triplet sector.
The wave functions of the Majorana fermion (for right handed) is given in an analytic form as
\begin{eqnarray}
\Psi^{(1)}(r,\theta)= C_{1}
\left(
\begin{array}{c}
 \varphi(r,\theta) \\
 \eta(r,\theta)
\end{array}
\right), \ 
\Psi^{(2)}(r,\theta) = C_{2}
\left(
\begin{array}{c}
 \varphi(r,\theta) \\
 -\eta(r,\theta)
\end{array}
\right), \ 
\Psi^{(3)}(r,\theta) = C_{3}
\left(
\begin{array}{c}
 \varphi(r,\theta) \\
 \eta(r,\theta)
\end{array}
\right),
\end{eqnarray}
where $C_i$ are normalization constants and 
the particle ($\varphi$) and hole ($\eta$) components are 
\begin{eqnarray}
 \varphi(r,\theta) 
= {\rm e}^{-\int_{0}^{r} |\Delta_{1}(r')|\mbox{d}r'}
\left(
\begin{array}{c}
 J_{0}(\mu r) \\
 i J_{1}(\mu r)\, {\rm e}^{i\theta}
\end{array}
\right), \hspace{0.5em}
 \eta(r,\theta) = {\rm e}^{-\int_{0}^{r} |\Delta_{1}(r')|\mbox{d}r'}
\left(
\begin{array}{c}
 -J_{1}(\mu r)\, {\rm e}^{-i\theta} \\
 i J_{0}(\mu r) \\
\end{array}
\right), \label{tripletZMs}
\end{eqnarray}
for the vortex profile $|\Delta_{1}(r)|$.
The wave function is well localized around the center of the vortex,
because the wave function damps exponentially by
$e^{-|\Delta_{\rm{CFL}}|r}$ at large distances from the center of the vortex.
The wave functions for the triplet Majorana fermion are exactly the same to that for the single component Majorana fermion displayed in Eq.~(\ref{eq:one_flavor_left_particle}) and (\ref{eq:one_flavor_left_hole}), except for the minus sign in the hole in $\Psi^{(2)}$.

Concerning the singlet quark state, we obtain a solution for ${\mathcal E}=0$ also, as the asymptotic form for $r \rightarrow \infty$ was found in Ref.~\cite{Yasui:2010yw}.
We show the explicit form of the wave functions of the singlet solution for the right mode ($\gamma_{5}=+1$) in the Weyl representation
\begin{eqnarray}
\hat{u}_{r} \!=\!
\left(
\begin{array}{c}
 \varphi_{1}(r,\theta) \\
 \eta_{1}(r,\theta)
\end{array}
\right), \hspace{1em}
\hat{d}_{g} \!=\!
\left(
\begin{array}{c}
 \varphi_{2}(r,\theta) \\
 \eta_{2}(r,\theta)
\end{array}
\right), \hspace{1em}
\hat{s}_{b} \!=\!
\left(
\begin{array}{c}
 \varphi_{3}(r,\theta) \\
 \eta_{3}(r,\theta)
\end{array}
\right),
\end{eqnarray}
where
\begin{eqnarray}
\varphi_{i}(r,\theta) =
\left(
\begin{array}{c}
 f_{i}(r) \\
 i g_{i}(r) e^{i\theta} \\
 0 \\
 0
\end{array}
\right), \hspace{1em} %
\eta_{i}(r,\theta) =
\left(
\begin{array}{c}
 0 \\
 0 \\
 \bar{f}_{i}(r) e^{-i\theta} \\
 i \bar{g}_{i}(r)
\end{array}
\right),
\end{eqnarray}
for $\hat{u}_{r}$ ($i=1$) and $\hat{d}_{g}$ ($i=2$), and
\begin{eqnarray}
\varphi_{3}(r,\theta) =
\left(
\begin{array}{c}
 f_{3}(r) e^{-i\theta} \\
 i g_{3}(r) \\
 0 \\
 0
\end{array}
\right), \hspace{1em} %
\eta_{3}(r,\theta) =
\left(
\begin{array}{c}
 0 \\
 0 \\
 \bar{f}_{3}(r) \\
 i \bar{g}_{3}(r) e^{i\theta}
\end{array}
\right),
\end{eqnarray}
for $\hat{s}_{b}$. 
At large $r$, both $|\Delta_{0}|$ and $|\Delta_{1}|$ become a common constant $|\Delta_{\mathrm{CFL}}|$ given in the bulk state.
Then, with an approximation of small $|\Delta_{\mathrm{CFL}}|$ and large $\mu$, we find asymptotic forms of the wave functions with a condition of the convergence at large $r$.
The first solution is
\begin{eqnarray}
f_{i}(r) &=& \mathcal{N} e^{-|\Delta_{\mathrm{CFL}}|r/2} J_{0}(\mu r), \label{eq:singlet_f12} \\
g_{i}(r) &=& \mathcal{N} e^{-|\Delta_{\mathrm{CFL}}|r/2} J_{1}(\mu r), \label{eq:singlet_g12}
\end{eqnarray}
($i=1$, $2$) and
\begin{eqnarray}
f_{3}(r) &=& - \frac{\mathcal{N}}{2} e^{-|\Delta_{\mathrm{CFL}}|r/2} J_{0}(\mu r), \label{eq:singlet_f3} \\
g_{3}(r) &=& - \frac{\mathcal{N}}{2} e^{-|\Delta_{\mathrm{CFL}}|r/2} J_{1}(\mu r), \label{eq:singlet_g3}
\end{eqnarray}
with a normalization constant $\mathcal{N}$.
This is the solution which was given in
Ref.~\cite{Yasui:2010yw}.

As the second solution, we find a new asymptotic solution, 
which is very different from the solution in the triplet sector 
\begin{eqnarray}
f'_{i}(r) &=& \mathcal{N}' e^{-|\Delta_{\mathrm{CFL}}|r/2} \frac{\pi}{4} (\mu r)^{2} J_{1}(\mu r) 
 \big( J_{1}(\mu r) N_{0}(\mu r) - J_{0}(\mu r) N_{1}(\mu r) \big), \label{eq:singlet_f'12} \\
g'_{i}(r) &=& \mathcal{N}' e^{-|\Delta_{\mathrm{CFL}}|r/2} \frac{1}{4} \big( -\mu r J_{0}(\mu r) + J_{1}(\mu r) \big), \label{eq:singlet_g'12}
\end{eqnarray}
($i=1$, $2$) and
\begin{eqnarray}
f'_{3}(r) &=& \mathcal{N}' e^{-|\Delta_{\mathrm{CFL}}|r/2} \frac{\mu}{8|\Delta|} 
 \Big\{ 4 \left( \mu r J_{0}(\mu r) - J_{1}(\mu r) \right) \nonumber \\
&& - \pi (\mu r)^{2} \left( 2 J_{0}(\mu r) + \frac{|\Delta_{\mathrm{CFL}}|}{\mu} J_{1}(\mu r) \right)
 \big( J_{1}(\mu r) N_{0}(\mu r) - J_{0}(\mu r) N_{1}(\mu r) \big) \Big\}, \label{eq:singlet_f'3} \\
 g'_{3}(r) &=& \mathcal{N}' e^{-|\Delta_{\mathrm{CFL}}|r/2} \frac{\mu}{4|\Delta_{\mathrm{CFL}}|} \Big\{ \left( 2+|\Delta_{\mathrm{CFL}}|r \right) J_{0}(\mu r) 
 - \left( \frac{|\Delta_{\mathrm{CFL}}|}{\mu} + 2\mu r \right) J_{1}(\mu r) \nonumber \\
&&   + \pi (\mu r)^{2} J_{1}(\mu r)
 \big( J_{1}(\mu r) N_{0}(\mu r) - J_{0}(\mu r) N_{1}(\mu r) \big) \Big\}, \label{eq:singlet_g'3}
\end{eqnarray}
with a normalization constant $\mathcal{N}'$.
It should be emphasized that these asymptotic solutions are correct only at large $r$, at which $|\Delta_{0}|$ and $|\Delta_{1}|$ are constant.
However, these solutions may be divergent in small $r$ in general, 
because $|\Delta_{1}|$ becomes zero at $r=0$, and we find that it is the case.
The result that there is no normalizable zero mode fermion in the singlet is consistent with that of the index theorem as discussed in Sec.~\ref{sec:index}.
Although non-normalizable modes 
having a singular peak at the vortex core 
are usually considered to be unphysical, 
there is also a discussion in the context of 
cosmic strings that 
they may play some interesting roles 
such as baryogenesis \cite{Alford:1989ie}.

\subsection{Effective theory in $1+1$ dimension along vortex string}\label{sec:1+1_effective_theory}

\begin{figure}
\begin{center}
\includegraphics[height=5cm]{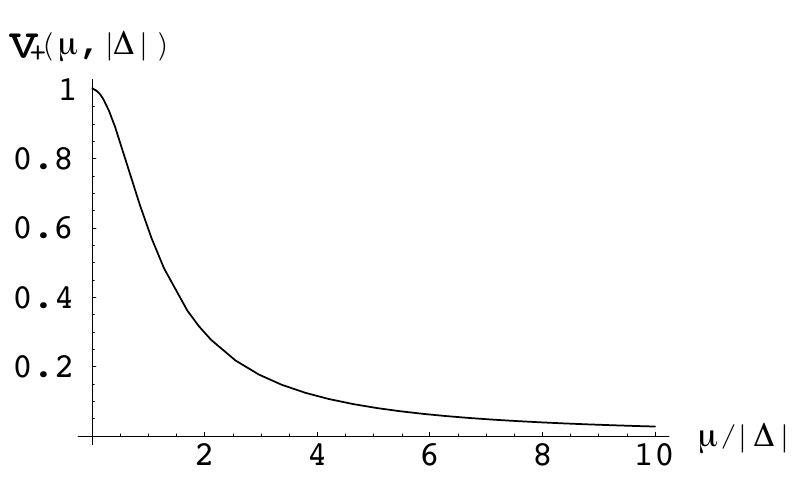}
\caption{Velocity of zero mode fermion propagating along the vortex axis. (Ref.~\cite{Yasui:2010yw}.)}
\label{fig:velocity}
\end{center}
\end{figure}

In the previous subsection, we have considered the transverse motion of the quark on the plane perpendicular to the vortex axis.
Now we discuss the longitudinal motion of the quark along the vortex.
Let us consider the case of a single flavor for an illustration.
The hamiltonian is written as a sum of the 
perpendicular part (with subscript $\perp$) and the longitudinal part (with subscript $z$),
\begin{eqnarray}
{\mathcal H} &=&
\left(
\begin{array}{cc}
 -i\vec{\alpha}_{\perp} \!\cdot\! \vec{\nabla}_{\perp} - \mu & 
|\Delta| {\rm e}^{i \theta} \gamma_{0} \gamma_{5} \\
 - |\Delta| {\rm e}^{-i \theta} \gamma_{0} \gamma_{5} & -i\vec{\alpha}_{\perp} \!\cdot\! \vec{\nabla}_{\perp} + \mu
\end{array}
\right)
+
\left(
\begin{array}{cc}
 -i \alpha_{z} \frac{\partial}{\partial z} & 0 \\
 0 &  -i \alpha_{z} \frac{\partial}{\partial z}
\end{array}
\right) \nonumber \\
&\equiv& {\mathcal H}_{\perp} + {\mathcal H}_{z}.
\label{eq:Hamiltonian_1+1}
\end{eqnarray}
Because the hamiltonian is a sum of the contributions from the transverse and longitudinal directions,
we separate the wave function as
\begin{eqnarray}
\Psi(t,z,r,\theta) = a(t,z) \Psi_{0}(r,\theta)\, ,
\label{eq:long_dep}
\end{eqnarray}
where $a(t,z)$ is a function with variables of the time $t$ and the
coordinate $z$ in the longitudinal direction,
 and $\Psi_{0}(r,\theta)$ is a function with variables of $(r, \theta)$
 on the transverse plane.
Concerning the transverse plane, we have found the solution of the
Majorana fermion as the lowest mode in the hamiltonian ${\mathcal
H}_{\perp} $, as shown in
Eqs.~(\ref{eq:one_flavor_left_particle})-(\ref{eq:one_flavor_right_hole}).
Since we are interested in the low-energy dynamics,
we can take 
$\Psi_{0}(r,\theta)$ to be the state of the
transverse zero mode. 
We can identify the field $a(t,z)$ with the low-energy degree of
freedom on the vortex.
The equation of motion for $a(t,z)$ can be derived in the following way.
We start with the original equation of motion, $i\partial/\partial t \Psi = {\mathcal H}\Psi$.
By integrating the transverse degree of freedom, 
namely by multiplying $\Psi_{0}^{\dag}$ from the left in the equation of
motion $i\partial/\partial t \Psi = {\mathcal H}\Psi$ 
and integrating over the transverse coordinate $r$ and $\theta$, 
we obtain the equation of motion for $a(t,z)$ as
\begin{eqnarray}
i \frac{\partial}{\partial t}a(t,z) = \int \Psi_{0}^{\dag}(r,\theta) 
{\mathcal H}_{z} \Psi_{0}(r,\theta) r dr d\theta a(t,z),
\end{eqnarray}
where we have used the normalization 
$\int \Psi_{0}^{\dag} \Psi_{0} rdr d\theta = 1$. 
On the right hand side, $\int \Psi_{0}^{\dag}(r,\theta) 
{\mathcal H}_{z} \Psi_{0}(r,\theta) rdr d\theta$  may be regarded as an effective hamiltonian for $a(t,z)$.
We can rewrite this equation as 
\begin{eqnarray}
i\left\{ \frac{\partial}{\partial t} + v_{+}(\mu,|\Delta|) \frac{\partial}{\partial z} \right\} a(t,z) = 0,
\label{eq:effective_theory}
\end{eqnarray}
with the ``velocity'' ${v}_{+}(\mu,|\Delta|)$ defined by 
\begin{eqnarray}
{v}_{+}(\mu,|\Delta|) \equiv \int  
\Psi_0^\dagger(r,\theta) 
\left(
\begin{array}{cc}
 \alpha_{z} & 0 \\
 0 &  \alpha_{z} 
\end{array}
\right) \Psi_0 (r,\theta)\, rdr d\theta \, .
\label{eq:effective_velocity}
\end{eqnarray}
The solution to Eq.~(\ref{eq:effective_theory}) is given by 
$a(t,z)\propto {\rm e}^{i{\mathcal E}t -ik_z z}$ with 
a linear (gapless) dispersion with respect to $k_z$:
\begin{eqnarray}
{\mathcal E}={v}_{+}(\mu,|\Delta|) \, k_{z}. \label{eq:dispersion}
\end{eqnarray}
where 
\begin{eqnarray}
{v}_{+}(\mu,|\Delta|) = \frac{\mu^2}{|\Delta|^2+\mu^2}
\frac{E\left(-\frac{\mu^2}{|\Delta|^2}\right)}{E\left(-\frac{\mu^2}{|\Delta|^2}\right)-K\left(-\frac{\mu^2}{|\Delta|^2}\right)} - 1,
\end{eqnarray}
where $K(x)$ and $E(x)$ 
 are the complete elliptic integrals of the first and second kinds, respectively;
\begin{eqnarray}
K(x) &=& \int_{0}^{\pi/2} \frac{1}{\sqrt{1-x^2 \sin^2 \theta}} d\theta, \\
E(x) &=& \int_{0}^{\pi/2} \sqrt{1-x^2 \sin^2 \theta} d\theta.
\end{eqnarray}
We plot ${v}_{+}(\mu,|\Delta|)$ as a function of $\mu/|\Delta|$ in Fig.~\ref{fig:velocity}.
(${v}_{+}=1$ is the speed of light.)
As a consequence, low-energy excitations inside the vortex 
are gapless (massless) fermions described by 
Eqs.~(\ref{eq:effective_theory}) and (\ref{eq:effective_velocity}). 
One can indeed express these fermions in terms of 
spinors in 1+1 dimensions, and write down an equation 
similar to the Dirac equation.

The effective theory is obtained also for the triplet Majorana fermion in the CFL phase.
The effective field with $t$ and $z$ has three components, each of which follows Eq.~(\ref{eq:effective_velocity}).
Only the change is to replace $|\Delta|$ to $|\Delta_{\mathrm{CFL}}|$.
The dispersion relation is not modified from Eq.~(\ref{eq:dispersion}).

\subsection{The absence of supercurrent along a vortex} \label{sec:supercurrent}

We here comment
on the absence of electromagnetic supercurrents 
induced by fermions 
along
vortices, even if we start from charged fermions in 3+1 dimensions.
This is anticipated from the ``Majorana'' nature of the
zero-mode fermions. 
We start with a single-component charged fermion described by the
following Hamiltonian
\begin{eqnarray}
{\mathcal H} &=&
\left(
\begin{array}{cc}
 -i\vec{\alpha}_{\perp} \!\cdot\! \vec{\nabla}_{\perp} - \mu & 
|\Delta| {\rm e}^{i \theta} \gamma_{0} \gamma_{5} \\
 - |\Delta| {\rm e}^{-i \theta} \gamma_{0} \gamma_{5} & -i\vec{\alpha}_{\perp} \!\cdot\! \vec{\nabla}_{\perp} + \mu
\end{array}
\right)
+
\left(
\begin{array}{cc}
 -i \alpha_{z} \left( \frac{\partial}{\partial z} -ieA_{z}\right)& 0 \\
 0 &  -i \left( \alpha_{z} \frac{\partial}{\partial z} +ieA_{z}\right)
\end{array}
\right) \nonumber \\
&\equiv& {\mathcal H}_{\perp} + {\mathcal H}_{z}^{A_{z}},
\end{eqnarray}
where the fermion is coupled to an electromagnetic gauge field 
$A^{\mu}(t,z)=(0,0,0,A_{z}(t,z))$. 
As we are interested in the low-energy dynamics, 
we can take the wave function as
\begin{eqnarray}
\Psi(t,\vec{x}) = a(t,z) \Psi_{0}(r,\theta),
\end{eqnarray}
where $\Psi_{0}(r,\theta)$ is the transverse zero-mode wave function, 
that satisfies
\begin{eqnarray}
 \mathcal{H}_{\perp} \Psi_{0}(r,\theta)=0.
\end{eqnarray}
Then, the dynamical equation for the effective dynamics along the vortex
axis is derived exactly in the same way as Eq.~(\ref{eq:effective_theory}),
\begin{eqnarray}
\frac{\partial}{\partial t} a(t,z)
&=& -i \int rdr d\theta \Psi_{0}^{\dag}(r,\theta) \mathcal{H}_{z}^{A_{z}}(t,z) \Psi_{0}(r,theta) \nonumber \\
&=& \mp \left[ v_{\pm} a(t,z) - ie' A_{z}(t,z) a(t,z) \right],
\end{eqnarray}
where we have defined 
``the effective velocity" as 
\begin{eqnarray}
v = \varphi_{\pm}^{\dag} \sigma_{3} \varphi_{\pm} - \eta_{\mp}^{\dag} \sigma_{3} \eta_{\mp},
\end{eqnarray}
and ``the effective charge" as 
\begin{eqnarray}
e' &=& e \left( \varphi_{\pm}^{\dag} \sigma_{3} \varphi_{\pm} + \eta_{\mp}^{\dag} \sigma_{3} \eta_{\mp} \right) \nonumber \\
&=& 0.
\end{eqnarray}
The final equation is obtained by substituting the wave function 
of the transverse zero mode. 
Because the effective charge is zero, 
 the Majorana fermion propagating along the vortex axis is 
neutral and does not couple to electromagnetic fields.
Therefore, there is no electromagnetic current  induced by the
Majorana fermions in vortices.
The same is true for 
the triplet Majorana fermions in
non-Abelian vortices, where the $SU(2)_{\mathrm{C+F}}$ non-Abelian gauge
field is switched on along the vortex axis as
$A^{i,\mu}(t,z)=(0,0,0,A_{z}^{i}(t,z))$ with $i=1,2,3$ for indices of
the adjoint representation (triplet) in $SU(2)_{\mathrm{C+F}}$ symmetry.

\subsection{Index theorem}\label{sec:index}

So far we have discussed zero mode fermions by explicitly solving the
BdG equation based on the mean-field Hamiltonian with a gap profile function.
In fact, the existence of zero modes is robust and does not depend on
the details of the system, since the massless modes on the edge of a
superconductor are closely related to the ``topology'' of the bulk matter.
A manifestation of this fact is the index theorem
\cite{Callias:1977kg,Weinberg:1981eu,Fukui:2009mh,Fujiwara:2011za}, which 
relates the number of zero modes and a topological invariant.
(The detailed description is given in Refs.~\cite{Fukui:2009mh,Fujiwara:2011za}.)
The index theorem has a long history and has elucidated topological
characteristics of anomalies in gauge fields \cite{Eguchi1980213}.
The index theorem has also been applied in the condensed matter
systems by many researchers
\cite{Fukui:2009mh,PhysRevB.81.075427,PhysRevB.81.184502,PhysRevB.82.165101,PhysRevB.82.094522,PhysRevB.81.214516,PhysRevB.82.184536,Roy:2013}.
We here review the application of the index theorem to the non-Abelian vortices in the CFL phase.

Let us first review the index theorem in the case of Euclidian Dirac
operators,  $H=i\Slash{D}$. 
We consider normalized eigenstates of this operator $\{ u_{i}(x)\}$ ,
which satisfy
\begin{eqnarray}
 \Slash{D} u_{i}(x) = \lambda_{i} u_{i}(x). 
\end{eqnarray}
Suppose we have an operator $\gamma_5$ which anticommute with $\Slash{D}$
\begin{eqnarray}
\gamma_{5} \Slash{D} + \Slash{D} \gamma_{5} = 0, \hspace{1em}
 (\gamma_{5})^2=1. 
\end{eqnarray}
The zero modes, which are the eigenfunctions for $\lambda_i = 0$, can
always be taken as the eigenstates of chirality, 
\begin{eqnarray}
\gamma_{5} u_{i}(x) = u_{i}(x) \hspace{0.5em}\mbox{or}\hspace{0.5em}
\gamma_{5} u_{i}(x) = -u_{i}(x).
\label{eq:chiral5_index_2}
\end{eqnarray}
Let us consider the analytical index of the operator $H$, which is
defined as the difference between the dimensions of the kernel and
co-kernel of the operator. 
The analytical index can be written as the difference
of the number of zero mode solutions for positive and negative
chiralities,  
\begin{equation}
 \mathrm{ind}\, H \equiv N_{+} - N_{-}. 
\end{equation}
where $N_{+}$ and $N_{-}$ are the numbers of zero modes 
for positive and negative chiralities. 
The index theorem relates the analytical index to the topological index.
To see this, we rewrite $\mathrm{ind}\, H $ as
\begin{eqnarray}
\mathrm{ind}\, H 
= \lim_{m \rightarrow 0} \mathrm{Tr}\, \gamma_{5} \frac{m^2}{H^{2}+m^2}.
\end{eqnarray}
As $m$ tends to zero, only the contribution of the zero eigenvalues
survives, and $+$ ($-$) states are accounted with plus (minus) sign due
to $\gamma_{5}$.
Furthermore, we introduce the current defined by
\begin{eqnarray}
J^{k} = \lim_{y \rightarrow x} \mbox{Tr}\, \gamma_{5} \gamma^{k} \left( \frac{1}{D\hspace{-0.7em}/+m} - \frac{1}{D\hspace{-0.7em}/+M} \right) \delta(x-y),
\end{eqnarray}
where $k$ indicates the spatial component and the second term is a regulator with large mass $M$ which is taken eventually to infinity.
By taking the divergence of the current, one can show the following relation \cite{Weinberg:1981eu}, 
\begin{eqnarray}
\mathrm{ind}\, H = -\frac{1}{2} \int dS_{k}J^{k} + c_{d},
\label{eq:index_theorem_0}
\end{eqnarray}
where $dS_{k}$ is the infinitesimal surface element on the boundary of
the space. 
It can be shown that the constant $c_d$ is the Chern number for even
$d$, 
\begin{eqnarray}
c_{d} \equiv \lim_{M \rightarrow \infty} \mathrm{Tr}\, \gamma_{5}
 \frac{M^{2}}{H^{2}+M^{2}}. 
\end{eqnarray}
Thus, the analytical index is related to the topological index defined in the right hand side in Eq.~(\ref{eq:index_theorem_0}).

We apply the index theorem to count the number of the zero mode Majorana fermions in non-Abelian vortices.
In the non-Abelian vortices,
 the analytical index of the operator $ \mathcal{H}_{\perp}$ is defined by
\begin{eqnarray}
 \mathrm{ind}\, \mathcal{H}_{\perp} = N_{+}(\mathcal{H}_{\perp}) - N_{-}(\mathcal{H}_{\perp}),
 \label{eq:index_1}
\end{eqnarray}
where $N_{\pm}(\mathcal{H}_{\perp})$ are the numbers of zero-energy states
of $\mathcal{H}_{\perp}$ with definite $\Gamma^3$ chiralities,
$\Gamma^3=\pm1$ (corresponding to $\gamma_{5}=\pm 1$ in Eq.~(\ref{eq:chiral5_index_2}).).
Here $\Gamma^3$ is defined as
\begin{eqnarray}
\Gamma^3 =
\left(
\begin{array}{cc}
 \alpha_{z} & 0 \\
 0 & \alpha_{z}
\end{array}
\right).
\end{eqnarray}

\begin{figure}
\begin{center}
\includegraphics[height=5cm]{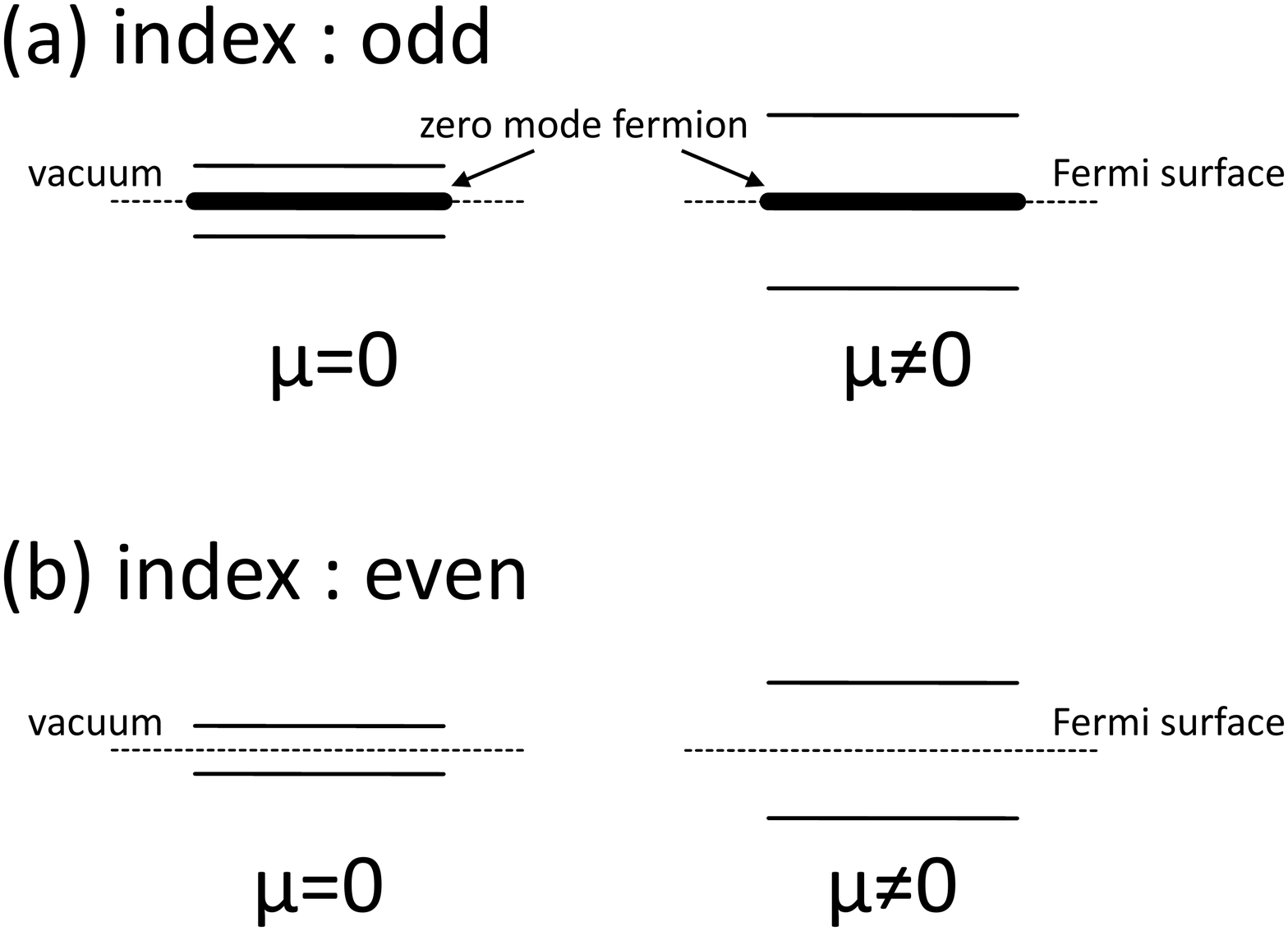}
\caption{A schematic picture for energy levels of fermions in vacuum ($\mu=0$) and in matter ($\mu \neq 0$). (a) when the index is odd, the zero mode fermion in vacuum still exist in the matter. (b) when the index is even, however, the zero mode fermions do not exist both in vacuum and in matter.}
\label{fig:index_fig}
\end{center}
\end{figure}

Let us review the derivation of the index theorem.
We note that the index is rewritten as
\begin{eqnarray}
\mathrm{ind} \, \mathcal{H}_{\perp} = \lim_{m \rightarrow 0} \mathrm{Tr} \, \Gamma^3 \frac{m^2}{\mathcal{H}_{\perp}^{\,2}+m^2},
 \label{eq:index_2}
\end{eqnarray}
where $\mathrm{Tr}$ stands for the trace over the two-dimensional space as well as the Dirac indices.
For $\mu=0$, $\mathrm{ind} \, \mathcal{H}_{\perp}$ is equal
to the number of Majorana fermions.
For $\mu\neq 0$ the relation does not hold, since the chiral
symmetry is broken ($\{\Gamma_{3},\mathcal{H}_{\perp}\} \neq 0$ ).
When $\mu$ is small, we can naturally assume that the states at finite $\mu$
 is smoothly connected to the states in vacuum (see Fig.~\ref{fig:index_fig}).
Note that the particle-hole symmetry is intact even at the finite
densities, $\mu \neq 0$.
Then we can expect that the zero modes can appear or disappear 
always in pairs, as one increases the chemical potential $\mu$. 
Thus, in the case of $\mu \neq 0$, $\mathrm{ind}\, \mathcal{H}_{\perp}$
gives the number of zero mode fermion modulo 2.
Here we introduce the axial current
\begin{eqnarray}
J^{j}(x,m,M) &=& \lim_{y \rightarrow x} \mathrm{Tr} \, \Gamma^{3} \Gamma^{j} \left( \frac{1}{-i\mathcal{H}_{\perp}+m} - \frac{1}{-i\mathcal{H}_{\perp}+M} \right) \delta^{(2)}(x-y) \nonumber \\
&=&
\lim_{y \rightarrow x} \mathrm{Tr} \, \Gamma^{3} \Gamma^{j} i \mathcal{H}_{\perp} \left( \frac{1}{\mathcal{H}_{\perp}^{2}+m^{2}} - \frac{1}{\mathcal{H}_{\perp}^{2}+M^{2}} \right) \delta^{(2)}(x-y),
\end{eqnarray}
with $j=1,2$ for the coordinate on the two dimensional plane, where $M$
is a large mass parameter for the Pauli-Villars regulator.
Then, the divergence of the current is written as 
\begin{eqnarray}
 \partial_{j} J^{j}(x,m,M) &=&
 2 \lim_{y \rightarrow x} \mathrm{Tr} \, \Gamma^{3} \Gamma^{j} i \mathcal{H}_{\perp} \left( \frac{m^2}{\mathcal{H}_{\perp}^{2}+m^{2}} - \frac{M^2}{\mathcal{H}_{\perp}^{2}+M^{2}} \right) \delta^{(2)}(x-y).
\end{eqnarray}
Therefore, we obtain
\begin{eqnarray}
\mathrm{ind}\, \mathcal{H}_{\perp} = c + \lim_{m \rightarrow 0, \, M \rightarrow \infty} \frac{1}{2} \oint_{|\vec{x}| \rightarrow \infty} \epsilon_{ij}
 J^{i} (x,m,M) dx^{j},
 \end{eqnarray}
where we have defined
\begin{eqnarray}
 c = \lim_{M \rightarrow \infty} \mathrm{Tr} \, \Gamma^{3} \frac{M^2}{\mathcal{H}_{\perp}^{2}+M^{2}}.
\end{eqnarray}
It is known that the $c$ is the Chern number associated with the gauge field.
In the present discussion, we do not consider the gauge field and hence we neglect $c$.
We note that the second term in the above equation is 
a topological invariant.
Thus, we have shown the index theorem; the index $\mathrm{ind}\, \mathcal{H}_{\perp}$ is related to the topologically invariant quantity.
We can know whether the number of zero mode fermion is even or odd when the index is even or odd, respectively.
In most cases, the number of zero mode fermion is 0 or 1, and the other
numbers can happen by chance.

The calculation of Eq.~(\ref{eq:index_2}) is performed by supplying the basis of the plane wave, as explicitly given in Ref.~\cite{Fujiwara:2011za}.
We use the basis of plane wave and give the axial current as
\begin{eqnarray}
J^{j}(x,m,M) = \int \frac{d^2 k}{(2\pi)^2} e^{-ik\cdot x} \mathrm{Tr} \, \Gamma^{3} \Gamma^{j} i {\mathcal H}_{\perp} \frac{1}{{\mathcal H}_{\perp}^2 + m^2} e^{ik\cdot x}, 
\end{eqnarray}
where the regularization of the current is neglected, because the above current is well-defined in the present discussion.
Indeed, the index theorem needs the current only at $|\vec{x}| \rightarrow \infty$, where $|\Delta| \rightarrow \Delta_{\mathrm{CFL}}$ and $\partial_{j} \Delta \sim {\mathcal O}(r^{-1})$.
The result is
\begin{eqnarray}
\mathrm{ind} \, \mathcal{H}_{\perp} =
\frac{1}{2\pi} \oint d\theta \partial_{\theta} \Theta_{k}(\theta) =
\left\{
\begin{array}{c}
 2Q \hspace{0.5em} \mathrm{(singlet)} \\
 Q \hspace{0.5em} \mathrm{(doublet)} \\
 q \hspace{0.5em} \mathrm{(triplet)}
\end{array}
\right.
\end{eqnarray}
for the generalized gap profile $\Delta(r,\theta) = \mathrm{diag}(\Delta_{Q},\Delta_{Q},\Delta_{q})$ with $\Delta_{k}(r,\theta) = |\Delta_{k}(r)| e^{i\Theta_{k}(\theta)}$ and the condition $\Theta_{k}(2\pi) = \Theta_{k}(0) + 2\pi k$.
The case of $Q=0$ and $q=1$ corresponds to the vortex configuration in Eq.~(\ref{eq:gap_non-Abelian}).
Therefore, the zero mode exist only in the triplet sector.

From the result of the index theorem, we find that the doublet zero mode
does not exist, and the singlet zero-mode does not exist either.
Although an asymptotic form of the wave function can be found at large $r \rightarrow \infty$, as shown in Eqs.~(\ref{eq:singlet_f12})-(\ref{eq:singlet_g'3}),
it turns out that the wave function is divergent at the origin ($r=0$).
Because the index theorem counts the number of the zero mode (precisely the difference between $N_{+}$ and $N_{-}$) which is normalizable over the space, the result does not contradict with that from the analysis of the BdG equation. 

The index of the Hamiltonian ${\mathcal H}_{\perp}$ is calculated also by analytical method \cite{PhysRevD.13.3398,Jackiw:1981ee} as presented in Ref.~\cite{Fukui:2009mh,Fujiwara:2011za}.
We consider the equation of motion
\begin{eqnarray}
 {\mathcal H}_{\perp}(r,\theta) \Psi(r,\theta) = 0
\end{eqnarray}
and analyze the eigen wave function $\Psi(r,\theta)$.
The advantage of this method is that it brings us with the number of zero-energy states for each $\Gamma^{3}$-chirality ($\Gamma^3=\pm1$), namely $N_{+}$ and $N_{-}$.

Let us see the doublet and triplet states.
We rewrite the Hamiltonian as
\begin{eqnarray}
\left(
\begin{array}{cc}
 0 & {\mathcal H}_{\perp,-} \\
 {\mathcal H}_{\perp,+} & 0 
\end{array}
\right)
\left(
\begin{array}{c}
 \Psi_{+} \\
 \Psi_{-}
\end{array}
\right) = 0,
\end{eqnarray}
where $\pm$ denotes the $\Gamma^{3}$-chirality, and
\begin{eqnarray}
{\mathcal H}_{\perp,\pm} =
\left(
\begin{array}{cc}
 -i\partial_{\pm} & -|\Delta_{q}(r)|e^{iq\theta} \\
 -|\Delta_{q}(r)|e^{iq\theta} & i\partial_{\mp}
\end{array}
\right),
\end{eqnarray}
with $\partial_{\pm}=e^{\pm i\theta}(\partial_{r}\pm i\partial_{\theta}/r)$.
This is the equation for the triplet.
If $q$ is replaced to $Q$, then we obtain the equation for the doublet.
We set the wave function with the variables $r$ and $\theta$ as
\begin{eqnarray}
\Psi_{m,\pm}(r,\theta) = 
\left(
\begin{array}{c}
 \alpha_{m}(r) e^{\pm i m \theta} \\
 i \beta_{m}(r) e^{i(\pm m - q \pm 1)\theta}
\end{array}
\right),
\end{eqnarray}
where $m$ is the quantum number associated with the angular momentum, and $\alpha_{m}(r)$ and $\beta_{m}(r)$ are radial components.
The equation of the radial components are given as
\begin{eqnarray}
\left( \frac{d}{dr} - \frac{M_{\pm}}{r} + \Omega \right) \psi_{m}(r) = 0,
\end{eqnarray}
for $\psi_{m}(r) = (\alpha_{m}(r),\beta_{m}(r))^{\mathrm T}$, $M_{\pm} = {\mathrm{diag}}(m,-m-1\pm q)$, and
\begin{eqnarray}
\Omega =
\left(
\begin{array}{cc}
 0 & |\Delta_{q}| \\
 |\Delta_{q}| & 0
\end{array}
\right).
\end{eqnarray}
The solutions of this equation were studied in Ref.~\cite{Jackiw:1981ee}.
Following their analysis, we naturally assume the asymptotic behavior of the gap profile function; $|\Delta_{q}| \sim r^{|q|}$ at $r \rightarrow 0$.
We also assume generically that $\Omega$ is expanded as
\begin{eqnarray}
\Omega = \sum_{n=0}^{\infty} \Omega_{n} r^{n},
\end{eqnarray}
where $\Omega_{0}=0$ if $q\neq 0$.
From mathematical point of view, we can assume that $|\Delta_{q}| \sim r$ at $r \rightarrow 0$.
However, it turns out that this behavior gives a logarithmic term ($\log r$) in the wave function.
Therefore, we neglect the case of $|\Delta_{q}| \sim r$ and consider the case of $|\Delta_{q}| \sim r^{|q|}$ only.

With this set up, we expand the wave function as
\begin{eqnarray}
\psi_{m}^{(i)}(r) = \psi_{m,n_{i}}^{(i)} r^{n_i} +  \psi_{m,n_{i}+1}^{(i)} r^{{n_i}+1} + \cdots,
\end{eqnarray}
for $i=1,2$.
The leading power $n_{i}$ is one of the diagonal elements of M; $n_{1}=m$ and $n_{2}=-m-1+q$ for $\Gamma^{3}=+1$, and $n_{1}=m$ and $n_{2}=-m-1-q$ for $\Gamma^{3}=-1$.
The normalizability of $\psi^{(i)}_{m}(r)$ at $r \rightarrow 0$ is guaranteed by $n_{i} \ge 0$.
It means that $0 \le m \le q-1$ for $\Gamma^{3}=+1$, and  $0 \le m \le -q-1$ for $\Gamma^{3}=-1$.
Then, we obtain (1) $N_{+}=q $ and $N_{-}=0$ for $1 \le q$, (2) $N_{+}=N_{-}=0$ for $q=0$, and (3) $N_{+}=0$ and $N_{-}=-q$ for $q\le -1$.
We finally conclude that ${\mathrm{ind}} {\mathcal H}_{\perp} = N_{+} - N_{-}=q$ in any case.
This result is consistent with one from the topological discussion.

As a final remark, we notice that another gap configuration $\Delta(r,\theta) = \mathrm{diag}(\Delta_{1},\Delta_{1},\Delta_{0})$ ($Q=1$ and $q=0$) corresponds to the M$_2$ vortex.
As discussed in Sec.~\ref{sec:LEEA}, this is unstable configuration decaying to two M$_1$ vortices.
Nevertheless, it will be interesting to consider the zero mode fermions (two zero modes in singlet and one zero mode in doublet), if the M$_2$ vortex could exist as quasi-stable state.
Especially, zero mode fermions in doublet form a Dirac fermion, which is still zero mode, inside the single vortex.
As the Majorana fermion in M$_1$ vortex contributes to the non-Abelian
statistics, the Dirac fermion in M$_2$ vortex provides us the
non-Abelian statistics, as discussed in Sec.~\ref{sec:na-statistics}.

\subsection{Topological superconductor}
\label{sec:topological}

Topological phases have been recently attracting much attention in
condensed-matter physics. 
Topological superconductors are characterized by a full pairing gap in
the bulk, and gapless surface states which consists of Majorana fermions. 
A state with a non-trivial topological number is called a topological
state. 
When there is a boundary between two states with different topological
numbers, the energy gap should be closed on the boundary so that the topological
number can change. This assures the existence of gapless surface states.

Then comes to the question: is a color superconductor topological?
This question is addressed and discussed by Nishida
\cite{Nishida:2010wr}. 
Whether a superconductor is topological or not is determined only by
discrete symmetries of the Hamiltonian and spatial dimensionality
according to the periodic table of topological superconductors
\cite{PhysRevB.78.195125}, at least for non-interacting Hamiltonians. 
As already discussed, the 
mean-field
Hamiltonian 
of a color superconductor
has a charge-conjugation symmetry. We can also define a
time-reversal symmetry for this Hamiltonian.
Then the Hamiltonian belongs to the symmetry class DIII, which is
topological in three spatial dimensions. 

We can define topological charges $N_{\rm R}$ and $N_{\rm L}$ for right-handed and left-handed
sectors independently in the chiral limit, in which the two sectors are
independent. For even(odd)-parity pairing\footnote{
For even-parity pairing, the order parameter is given by 
$
\Phi \sim \langle q^T C \gamma_5 q \rangle
$, while 
$
\Phi \sim \langle q^T C q \rangle
$ for odd-parity pairing.
}, $N_{\rm R} = \mp
N_{\rm L}$. 
Because of these non-zero topological charges, surface states exists. 
For example, a vortex in the superconductor supports right-handed and
left-handed massless fermions propagating in opposite
directions\footnote{
The vortex state discussed in Ref.~\cite{Nishida:2010wr} is
not the lowest-energy vortex solution, so it would be unstable against
decay.
}

At finite quark masses, right-handed and left-handed sectors mixes and
only the total topological charge $N = N_{\rm R}+ N_{\rm L}$ is well-defined.
What become of the surface states at finite quark masses? 
The results are qualitatively different depending on the parity of the
pairing. 
For even-parity pairing case, the total topological charge is zero, $N=0$.
In this case the presence of the fermion mass immediately opens up a gap
for localized fermions on a vortex. 
When the pairing gap is parity odd, 
the topological charge is unchanged for small fermion masses 
since the system remains gapped as long as $m^2 < \mu^2 + |\Delta|^2$.
If the fermion mass is increased
further, there will be a topological phase transition at $m^2 = \mu^2 +
|\Delta|^2$ and the fermions would acquire a mass gap.

%% file: na-statistics-v9.tex
\section{Non-Abelian exchange statistics of non-Abelian vortices}\label{sec:na-statistics}

The exchange statistics in the system of multiple numbers of quantum vortices with zero mode Majorana fermions provides a novel statistics, which is different from the conventional statistics, such as the Fermi-Dirac statistics for fermions, the Bose-Einstein statistics for bosons and the (Abelian) anyon statistics for anyons.
The exchange statistics of vortices with Majorana fermions obeys the non-Abelian statistics.
This is the subject in this section. 
In Sec.~\ref{sec:exchange}, we discuss general properties of exchanging vortices.
In Sec.~\ref{sec:na-statistics-1}, we review the non-Abelian statistics for {\it Abelian} vortices with Majorana fermions.
In Sec.~\ref{sec:na-statistics-2}, we discuss the non-Abelian statistics for {\it non-Abelian} vortices with Majorana fermions.
In Sec.~\ref{sec:na-statistics-3}, we comment on the case of (non-)Abelian vortices with Dirac fermions.

\subsection{Exchange of vortices}
\label{sec:exchange}

Let us suppose that there are multiple numbers of non-Abelian vortices.
The motion of the parallel vortices in $d=3+1$ dimensional space is regarded as 
the motion of the particles on 
a $d=2+1$ dimensional space, because the vortex has the translational invariance along the vortex axis.
Generally speaking, the exchange of two particles on plane obeys the braid group.
We introduce an operation $T_{k}$ which exchanges the $k$-th and $(k+1)$-th vortices, where the former vortex turns around the latter vortex in anticlockwise.
The braid group is given by the two rules;
\begin{eqnarray}
 &&\mathrm{(i)} \hspace{0.5em} T_{k}T_{k+1}T_{k} = T_{k+1}T_{k}T_{k+1}, \\
 &&\mathrm{(ii)} \hspace{0.5em} T_{k} T_{\ell} = T_{\ell} T_{k} \hspace{1em} (|k-\ell|>1).
\end{eqnarray}
It should be noted that the inverse of $T_{k}$ is not necessarily identical to $T_{k}$ ($(T_{k})^{-1} \neq T_{k}$) because the operation is directed.
This property from the braid group leads to the non-Abelian statistics \cite{PhysRevB.61.10267,PhysRevLett.86.268,PhysRevB.70.205338,PhysRevB.73.220502,PhysRevLett.99.037001,PhysRevB.75.212509}.

\begin{figure}
\begin{center}
\includegraphics[height=3.5cm]{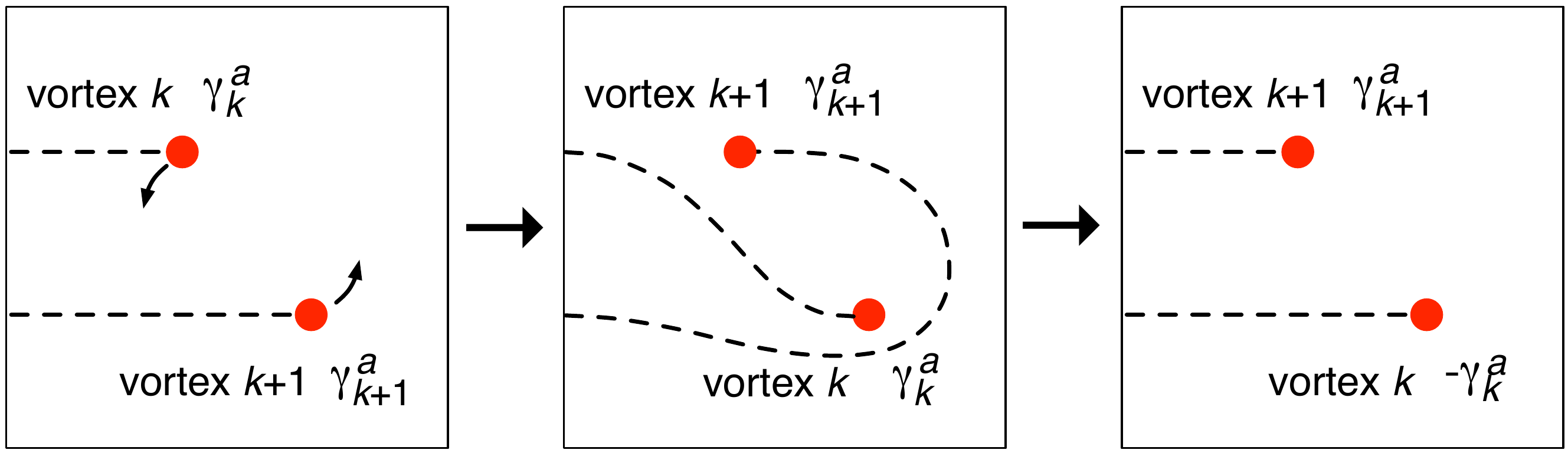}
\caption{Exchange of two vortices $k$ and $k+1$ with Majorana fermions described by $\gamma_{k}^{a}$ and $\gamma_{k+1}^{a}$ ($a=1,2,3$), respectively. The dashed lines are cut from the vortex to the infinitely distant point. (Ref.~\cite{Yasui:2010yh}.)}
\label{fig:figure_exchange_ver2}
\end{center}
\end{figure}

We consider that the vortices are exchanged adiabatically as shown in Fig.~\ref{fig:figure_exchange_ver2}.
Then, the Majorana fermion plays an important role because it is the most stable state protected topologically by perturbation from outside.
The wave function of the Majorana fermion is a 
double-valued function for the angle around the vortex axis 
This can be inferred from the form of the BdG equations 
(\ref{eq:CFL_eve}); 
when we shift the phase winding of a vortex as 
$\theta \to \theta + \alpha$, 
it can be canceled  
if the phases of particle and hole wave functions are 
shifted by $\alpha/2$ and $-\alpha/2$, respectively. 
It implies that when quasi-particles travel around a vortex 
at $\alpha = 2\pi$, 
both particle and hole wave functions receive minus sign. 
In order to regard the wave function as a single-valued function, we need to introduce a cut from the center of the vortex to infinitely 
far from the vortex. 
The directions of cuts are arbitrary and gauge dependent. 
It means that the wave function acquires a minus sign when the Majorana fermion goes across the cut.
Let us pickup the $k$-th and $(k+1)$-th vortices among the multiple number of vortices, and consider the exchange operation $T_{k}$ for those two vortices as in Fig.~\ref{fig:figure_exchange_ver2}.
By the operation $T_{k}$, the position of $k$-th and $(k+1)$-th vortices are exchanged, where 
the $k$-th vortex goes across the cut from 
the $(k+1)$-th vortex.
Then, the wave function of the Majorana fermion in the $(k+1)$-th vortex acquires a minus sign, while that in the $k$-th vortex dose not change.
The resulting exchange is expressed by the 
Majorana fermion operators
$\gamma_{k}^{a}$ and $\gamma_{k+1}^{a}$ for the  in the $k$-th and $(k+1)$-th vortices.
Here $a$ indicates the component of the Majorana fermion with internal symmetry.
The fermion operators $\gamma_{\ell}^{a}$ satisfy the anticommutation relation 
\beq
\{ \gamma_{k}^{a}, \gamma_{\ell}^{b} \}=2\delta_{k\ell}\delta^{ab}.
\eeq
We note the relation 
\beq 
(\gamma_{\ell}^{a})^{\dag} = \gamma_{\ell}^{a} 
\eeq
 because the particle state is equal to the hole state as the property of the Majorana fermion.

We find the operation $T_{k}$ gives the following exchange rule
\begin{eqnarray}
T_k : \left\{ 
\begin{array}{l}
 \gamma_{k}^{a}\quad \rightarrow\, \gamma_{k+1}^{a} \\
 \gamma_{k+1}^{a} \rightarrow -\gamma_{k}^{a}  
\end{array}
\right. .
\label{eq:exchange_Majorana}
\end{eqnarray}
Here $a=1$, 2, 3 denote the components in the triplet state, because the Majorana fermions belong to the triplet of $SU(2)_{\rm{C+F}} \simeq O(3)_{\rm{C+F}}$.
In the present discussion, we do not consider the singlet component. 

From Eq.~(\ref{eq:exchange_Majorana}), we find that the operation by two-time exchanges does not give the initial state; $T_{k}^{2} \neq 1$.
It means that the exchange of the vortices is different neither from the Fermi-Dirac statistics nor the Bose-Einstein statistics.

\subsection{Abelian vortices with Majorana fermions}
\label{sec:na-statistics-1}
The explicit expression of the operation $T_{k}$ in terms of Majorana fermions
was first found by Ivanov in case of the ``Abelian'' vortices with single component Majorana fermions ($a=1$ only)~\cite{PhysRevLett.86.268} 
in chiral $p$-wave superconductors.
See also the early discussions in other systems in Ref.~\cite{Sato:2003iz}.
The non-Abelian statistics by Majorana fermions in (Abelian) vortices, which explicit forms will be shown below, is attracting many researchers as devices for quantum computing \cite{Wilczek:2009,RevModPhys.80.1083,Kitaev:2006lla,Kitaev:1997wr}
They can be realized not only in the core of half-quantized vortices in chiral $p$-wave superconductors or $p$-wave superfluids \cite{volovik2009universe}, but also in the edge of topological superconductors and insulators \cite{PhysRevLett.100.096407} and even in the three dimensional systems in which Majorana fermions are trapped on monopole-like objects \cite{PhysRevLett.104.046401}.
Before going to the detailed discussion about the exchange in Eq.~(\ref{eq:exchange_Majorana}) 
for the multiple number of Majorana fermions,
in this subsection, 
let us study the case of the single fermion component 
found by Ivanov~\cite{PhysRevLett.86.268}. 
In this case, we have 
the Majorana fermion operators  
$\gamma_{k}$ and $\gamma_{k+1}$ at
the vortex $k$ and $k+1$, respectively.
Under the exchange of vortices, 
$\gamma_{k}$ and $\gamma_{k+1}$ are transformed as
\begin{eqnarray}
T_k : \left\{ 
\begin{array}{l}
 \gamma_{k}\quad \rightarrow\, \gamma_{k+1} \\
 \gamma_{k+1} \rightarrow -\gamma_{k}  
\end{array}
\right. .
\label{eq:exchange_Majorana_0}
\end{eqnarray}
If we define the operator
\begin{eqnarray}
\hat{\tau}_{k} = \frac{1}{\sqrt{2}} (1+\gamma_{k+1} \gamma_{k}),
\label{eq:exchange_Majorana_operator_0}
\end{eqnarray}
up to an overall phase,
it is easily to confirm that 
the transformation in Eq.~(\ref{eq:exchange_Majorana_0}) 
can be reproduced by 
\beq 
 \gamma_{\ell}   \to \hat{\tau}_{k} \gamma_{\ell} \hat{\tau}_{k}^{-1}.
\eeq 
Hence, the operator $\hat{\tau}_{k}$ represents the operation $T_{k}$.

Next, let us define the Hilbert (Fock) space.
It should be noted that the annihilation of the Majorana fermion is identical to the creation of the Majorana fermion ($\gamma_{k}=\gamma_{k}^{\dag}$).
Hence it is not possible to construct the Hilbert space on the single vortex.
With preparing a pair of vortices, 
we define the Dirac operator by 
\begin{eqnarray}
\hat{\Psi}_{k} \equiv \frac{1}{2} \left( \gamma_{2k-1} + i \gamma_{2k} \right).
\end{eqnarray}
The Dirac operator satisfies the usual anti-commutation relations for the creation ($\Psi_{k}^{\dag}$) and annihilation ($\Psi_{k}$) operators;
\begin{eqnarray}
 \left\{ \hat{\Psi}_{k}, \hat{\Psi}_{\ell}^{\dag} \right\} = \delta_{kl}, \hspace{0.5em}
 \left\{ \hat{\Psi}_{k}, \hat{\Psi}_{\ell} \right\} = \left\{ \hat{\Psi}_{k}^{\dag}, \hat{\Psi}_{\ell}^{\dag} \right\}=0.
\end{eqnarray}

We then define the Fock vacuum $| 0 \rangle$ as $\Psi_{k} | 0 \rangle=0$ for all $k$.
We can build the Hilbert space by acting the creation operators 
on the Fock vacuum. 

In the case of $n=2$ vortices, 
the Hilbert space can be constructed as 
$\{ | 0 \rangle, \Psi_{1}^{\dag} | 0 \rangle \}$.
With these basis, the operator $\hat{\tau}_{1}$ can 
be expressed by a matrix form as 
\begin{eqnarray}
\tau_{1}=
\left(
\begin{array}{cc}
 e^{-i\pi/4} & 0 \\
 0 & e^{i\pi/4}
\end{array}
\right). 
\end{eqnarray}
Because of a non-trivial phase factor, this gives rise to 
an Abelian anyon statistics. 

The non-Abelian statistics can be found in the case of 
$n \ge 4$ vortices.
For the case of $n=4$ vortices, we define the Hilbert space $\{ |0\rangle, i\Psi_{2}^{\dag}\Psi_{1}^{\dag} | 0 \rangle, \Psi_{1}^{\dag} | 0 \rangle, i \Psi_{2}^{\dag}| 0 \rangle \}$.
With these basis, 
$\hat{\tau}_{k}$ ($k=1,2,3$) can be expressed by matrix forms
\begin{eqnarray}
\tau_{1}=
\left(
\begin{array}{cccc}
 \omega^{\ast} & 0 & 0 & 0 \\
 0 & \omega & 0 & 0 \\
 0 & 0 & \omega & 0 \\
 0 & 0 & 0 & \omega^{\ast}
\end{array}
\right),
\tau_{2}=\frac{1}{\sqrt{2}}
\left(
\begin{array}{cccc}
 1 & -1 & 0 & 0 \\
 1 & 1 & 0 & 0 \\
 0 & 0 & 1 & -1 \\
 0 & 0 & 1 & 1
\end{array}
\right),
\tau_{3}=
\left(
\begin{array}{cccc}
 \omega^{\ast} & 0 & 0 & 0 \\
 0 & \omega & 0 & 0 \\
 0 & 0 & \omega^{\ast} & 0 \\
 0 & 0 & 0 & \omega
\end{array}
\right),
\end{eqnarray}
with $\omega=e^{i\pi/4}$ and $\omega^{\ast}=e^{-i\pi/4}$.
In these basis, $\tau_1, \tau_3$ are diagonal but $\tau_2$ is non-diagonal. One cannot diagonalize all of these at the same time.
We find the non-Abelian nature; $\tau_{1}\tau_{2} \neq \tau_{2}\tau_{1}$ and $\tau_{2}\tau_{3} \neq \tau_{3}\tau_{2}$.
This implies that the order of exchanging the $k$-th and $k+1$-th vortices  is not commutative.
It is straightforward to show the non-Abelian property
\begin{eqnarray}
\tau_{k}\tau_{k+1} \neq \tau_{k+1}\tau_{k},
\end{eqnarray}
for any even number $n$.
This is much different from the conventional statistics, such as the Fermi-Dirac statistics, the Bose-Einstein statistics and even the (Abelian) anyon statistics.
In those conventional statistics, the state after the exchange does not depend on the choice which particles are exchanged.

\subsection{Non-Abelian vortices with Majorana fermions}
\label{sec:na-statistics-2}

Now, let us discuss the property of the exchange of the non-Abelian vortices.
On this purpose, we introduce an operator $\hat{\tau}_{k}$ which representing the operation $T_{k}$ by 
\begin{eqnarray}
\hat{\tau}_{k} \equiv \prod_{a=1,2,3} \hat{\tau}_{k}^{a}, \hspace{0.5em}{\rm{with}} \hspace{0.5em} \hat{\tau}_{k}^{a} \equiv \frac{1}{\sqrt{2}} (1+\gamma_{k+1}^{a} \gamma_{k}^{a}). 
\label{eq:exchange_Majorana_operator}
\end{eqnarray}
This is invariant under the $SO(3)$ symmetry.
The operator $\hat{\tau}_{k}$ represents the operation $T_{k}$, because $\hat{\tau}_{k} \gamma_{\ell}^{a} \hat{\tau}_{k}^{-1}$ reproduces the exchange rule in Eq.~(\ref{eq:exchange_Majorana}).
We also introduce the non-local Dirac operators by
\begin{eqnarray}
\hat{\Psi}_{k}^{a} \equiv \frac{1}{2} \left( \gamma_{2k-1}^{a} + i \gamma_{2k}^{a} \right),
\end{eqnarray}
satisfying the anti-commutation relations 
\begin{eqnarray}
 \left\{ \hat{\Psi}_{k}^{a}, \hat{\Psi}_{\ell}^{b\,\dag} \right\} = \delta_{kl}\delta^{ab}, \hspace{0.5em}
 \left\{ \hat{\Psi}_{k}^{a}, \hat{\Psi}_{\ell}^{b} \right\} = \left\{ \hat{\Psi}_{k}^{a\,\dag}, \hat{\Psi}_{\ell}^{b\,\dag} \right\}=0.
\end{eqnarray}
Then, we can define the Hilbert (Fock) space.
The Fock vacuum state $|0\rangle$ is defined by $\hat{\Psi}_{k}^{a} |0\rangle=0$ for all $k$ and $a=1,2,3$, and multiparticle states are defined by operating successively $\hat{\Psi}_{k}^{a\,\dag}$ to the vacuum state.

As an example, we consider the {\it four} non-Abelian vortices ($n=4$).
Noting that the Dirac fermion is a triplet state, we obtain the Hilbert space with ${\mathcal M}$-plet (${\mathcal M}=\bf{1}$, $\bf{3}$, $\bf{5}$).
Furthermore, each ${\mathcal M}$-plet is classified by even or odd numbers of the Dirac fermions (${\mathcal P}={\mathcal E}$, ${\mathcal O}$).
Thus, the basis of the Hilbert space can be classified according to the representations $({\mathcal M}, {\mathcal P})$.
Their explicit forms are given as
\begin{eqnarray}
&&\hspace{-1em} |{\bf 1}_{00}\rangle = |0\rangle, \\
&&\hspace{-1em} |{\bf 1}_{33}\rangle 
   = i \frac{1}{3!} \epsilon^{abc} \frac{1}{3!} \epsilon^{def} 
     \hat{\Psi}_{1}^{a\dag} 
     \hat{\Psi}_{1}^{b\dag} 
     \hat{\Psi}_{1}^{c\dag} 
     \hat{\Psi}_{2}^{d\dag} 
     \hat{\Psi}_{2}^{e\dag} 
     \hat{\Psi}_{2}^{f\dag}  |0\rangle, \\
&&\hspace{-1em} |{\bf 1}_{11}\rangle 
   = i\frac{1}{\sqrt{3}} 
     \hat{\Psi}_{1}^{a\dag} 
     \hat{\Psi}_{2}^{a\dag} |0\rangle, \\
&&\hspace{-1em} |{\bf 1}_ {22}\rangle 
   = \frac{1}{\sqrt{3}} \frac{1}{2!} 
     \epsilon^{abc} \frac{1}{2!} \epsilon^{ade} 
     \hat{\Psi}_{1}^{b\dag} 
     \hat{\Psi}_{1}^{c\dag} 
     \hat{\Psi}_{2}^{d\dag} 
     \hat{\Psi}_{2}^{e\dag} |0\rangle,  
\end{eqnarray}
for the singlet-even (${\bf 1}$, $\mathcal{E}$) states and
\begin{eqnarray}
&&\hspace{-1em} |{\bf 1}_ {03}\rangle 
  = \frac{1}{3!} \epsilon^{abc} 
    \hat{\Psi}_{2}^{a\dag} 
    \hat{\Psi}_{2}^{b\dag} 
    \hat{\Psi}_{2}^{c\dag} |0\rangle, \\
&&\hspace{-1em} |{\bf 1}_{30}\rangle 
  = -i \frac{1}{3!} \epsilon^{abc} 
    \hat{\Psi}_{1}^{a\dag} 
    \hat{\Psi}_{1}^{b\dag} 
    \hat{\Psi}_{1}^{c\dag} |0\rangle, \\
&&\hspace{-1em} |{\bf 1}_{21}\rangle 
  = -\frac{1}{\sqrt{3}} \frac{1}{2!} \epsilon^{abc} 
    \hat{\Psi}_{1}^{a\dag} 
    \hat{\Psi}_{1}^{b\dag} 
    \hat{\Psi}_{2}^{c\dag} |0\rangle, \\
&&\hspace{-1em} |{\bf 1}_{12}\rangle 
  = i \frac{1}{\sqrt{3}} \frac{1}{2!} \epsilon^{abc} 
    \hat{\Psi}_{1}^{a\dag} 
    \hat{\Psi}_{2}^{b\dag} 
    \hat{\Psi}_{2}^{c\dag} |0\rangle, 
\end{eqnarray}
for the singlet-odd (${\bf 1}$, $\mathcal{O}$) states.
There are six bases,
\begin{eqnarray}
&&\hspace{-1.5em} |{\bf 3}_{02}\rangle 
  = \frac{1}{2!} \epsilon^{abc} 
    \hat{\Psi}_{2}^{b\dag} 
    \hat{\Psi}_{2}^{c\dag} |0\rangle, \\
&&\hspace{-1.5em} |{\bf 3}_ {31}\rangle 
  = -i \frac{1}{3!} \epsilon^{bcd} 
    \hat{\Psi}_{1}^{b\dag} 
    \hat{\Psi}_{1}^{c\dag} 
    \hat{\Psi}_{1}^{d\dag} 
    \hat{\Psi}_{2}^{a\dag} |0\rangle, \\
&&\hspace{-1.5em} |{\bf 3}_ {22}\rangle 
  = \frac{1}{\sqrt{2}} \epsilon^{abc} \frac{1}{2!} \epsilon^{bde} 
    \frac{1}{2!} \epsilon^{cfg} 
    \hat{\Psi}_{1}^{d\dag} 
    \hat{\Psi}_{1}^{e\dag}  
    \hat{\Psi}_{2}^{f\dag} 
    \hat{\Psi}_{2}^{g\dag} |0\rangle, \\ 
&&\hspace{-1.5em} |{\bf 3}_{11}\rangle 
  = i \frac{1}{\sqrt{2}} \epsilon^{abc} 
    \hat{\Psi}_{1}^{b\dag} 
    \hat{\Psi}_{2}^{c\dag}  |0\rangle, \\
&&\hspace{-1.5em} |{\bf 3}_{20}\rangle 
  = -\frac{1}{2!} \epsilon^{abc} 
    \hat{\Psi}_{1}^{b\dag} 
    \hat{\Psi}_{1}^{c\dag} |0\rangle, \\
&&\hspace{-1.5em} |{\bf 3}_ {13}\rangle 
  = i \frac{1}{3!} \epsilon^{bcd} 
    \hat{\Psi}_{1}^{a\dag} 
    \hat{\Psi}_{2}^{b\dag} 
    \hat{\Psi}_{2}^{c\dag} 
    \hat{\Psi}_{2}^{d\dag} |0\rangle, 
\end{eqnarray}
for the triplet-even (${\bf 3}$, $\mathcal{E}$) states and
\begin{eqnarray}
&&\hspace{-1em} |{\bf 3}_{01}\rangle 
  = \hat{\Psi}_{2}^{a\dag} |0\rangle, \\
&&\hspace{-1em} |{\bf 3}_ {32}\rangle 
  = i \frac{1}{3!} \epsilon^{bcd} 
    \hat{\Psi}_{1}^{b\dag} 
    \hat{\Psi}_{1}^{c\dag} 
    \hat{\Psi}_{1}^{d\dag} 
    \frac{1}{2!} \epsilon^{aef} 
    \hat{\Psi}_{2}^{e\dag} 
    \hat{\Psi}_{2}^{f\dag} |0\rangle, \\
&&\hspace{-1em} |{\bf 3}_ {21}\rangle 
  = \frac{1}{\sqrt{2}} \epsilon^{abc} \frac{1}{2!} \epsilon^{bde} 
    \hat{\Psi}_{1}^{d\dag} 
    \hat{\Psi}_{1}^{e\dag}  
    \hat{\Psi}_{2}^{c\dag} |0\rangle, \\
&&\hspace{-1em} |{\bf 3}_{12}\rangle 
  = -i \frac{1}{\sqrt{3}} \epsilon^{abc} \frac{1}{2!} \epsilon^{cde} 
    \hat{\Psi}_{1}^{b\dag} 
    \hat{\Psi}_{2}^{d\dag} 
    \hat{\Psi}_{2}^{e\dag} |0\rangle, \\
&&\hspace{-1em} |{\bf 3}_{23}\rangle 
  = \frac{1}{2!} \epsilon^{abc} 
    \hat{\Psi}_{1}^{b\dag} 
    \hat{\Psi}_{1}^{c\dag} 
    \frac{1}{3!} \epsilon^{def} 
    \hat{\Psi}_{2}^{d\dag} 
    \hat{\Psi}_{2}^{e\dag} 
    \hat{\Psi}_{2}^{f\dag} |0\rangle, \\
&&\hspace{-1em} |{\bf 3}_{10}\rangle 
  = i \hat{\Psi}_{1}^{a \dag} |0\rangle, 
\end{eqnarray}
for the triplet-odd (${\bf 3}$, $\mathcal{O}$) states.
There are two bases,
\begin{eqnarray}
&&\hspace{-1em} |{\bf 5}_{22}\rangle
  = i {\mathcal N} 
    \left\{\frac{1}{2} \left( \frac{1}{2!} \epsilon^{acd} 
                               \hat{\Psi}_{1}^{c\dag} 
                               \hat{\Psi}_{1}^{d\dag} 
                              \frac{1}{2!} \epsilon^{bef} 
                               \hat{\Psi}_{2}^{e\dag} 
                               \hat{\Psi}_{2}^{f\dag}      \right. \right. \nonumber \\
&&\hspace{7em}  + \left. \frac{1}{2!} \epsilon^{bcd} 
                               \hat{\Psi}_{1}^{c\dag} 
                               \hat{\Psi}_{1}^{d\dag} 
                              \frac{1}{2!} \epsilon^{aef} 
                               \hat{\Psi}_{2}^{e\dag} 
                               \hat{\Psi}_{2}^{f\dag} 
                      \right)
     \nonumber \\
&&\hspace{4em} 
     \left. - \frac{\delta^{ab}}{3} \frac{1}{2!} \epsilon^{cde} 
               \hat{\Psi}_{1}^{d\dag} 
               \hat{\Psi}_{1}^{e\dag} 
              \frac{1}{2!} \epsilon^{cfg} 
               \hat{\Psi}_{2}^{f\dag} 
               \hat{\Psi}_{2}^{g\dag} 
     \right\} |0\rangle, \\
&&\hspace{-1em} |{\bf 5}_{11}\rangle 
  = -{\mathcal N} 
     \left\{ \frac{1}{2} \left( \hat{\Psi}_{1}^{a\dag} 
                                \hat{\Psi}_{2}^{b\dag}    
                              + \hat{\Psi}_{1}^{b\dag} 
                                \hat{\Psi}_{2}^{a\dag} 
                        \right)
 - \frac{\delta^{ab}}{3} \hat{\Psi}_{1}^{c\dag} 
                                  \hat{\Psi}_{2}^{c\dag} 
     \right\} |0\rangle, \nonumber \\
\end{eqnarray}
for the quintet-even (${\bf 5}$, $\mathcal{E}$) states and
\begin{eqnarray}
&&\hspace{-1em} |{\bf 5}_{21}\rangle 
  = -i {\mathcal N} 
    \left\{ \frac{1}{2} \left( \frac{1}{2!} \epsilon^{acd} 
                                 \hat{\Psi}_{1}^{c\dag} 
                                 \hat{\Psi}_{1}^{d\dag} 
                                 \hat{\Psi}_{2}^{b\dag} 
+ \frac{1}{2!} \epsilon^{bcd} 
                                 \hat{\Psi}_{1}^{c\dag} 
                                 \hat{\Psi}_{1}^{d\dag} 
                                 \hat{\Psi}_{2}^{a\dag} 
                       \right) \right. \nonumber \\
&&\hspace{5em} \left. - \frac{\delta^{ab}}{3} \frac{1}{2!} \epsilon^{cde} 
               \hat{\Psi}_{1}^{c\dag} 
               \hat{\Psi}_{1}^{d\dag} 
               \hat{\Psi}_{2}^{e\dag} 
     \right\} |0\rangle, \\
&&\hspace{-1em} |{\bf 5}_{12}\rangle 
  = -{\mathcal N} 
    \left\{ \frac{1}{2} \left(   \hat{\Psi}_{1}^{a\dag} 
                                \frac{1}{2!}\epsilon^{bcd} 
                                 \hat{\Psi}_{2}^{c\dag} 
                                 \hat{\Psi}_{2}^{d\dag}
+ \hat{\Psi}_{1}^{b\dag} 
                                \frac{1}{2!} \epsilon^{acd} 
                                 \hat{\Psi}_{2}^{c\dag} 
                                 \hat{\Psi}_{2}^{d\dag} 
                       \right) \right. \nonumber \\
&&\hspace{5em} \left.  - \frac{\delta^{ab}}{3} \frac{1}{2!} \epsilon^{cde}
                    \hat{\Psi}_{1}^{c\dag} 
                    \hat{\Psi}_{2}^{d\dag} 
                    \hat{\Psi}_{2}^{e\dag}
    \right\} |0\rangle, 
\end{eqnarray}
for the quintet-odd (${\bf 5}$, $\mathcal{O}$) states with 
${\mathcal N}=\sqrt{3/2}$ for $a=b$ and ${\mathcal N}=\sqrt{2}$ for $a\neq b$.

According to the classification by $({\mathcal M}, {\mathcal P})$, 
the operator $\hat{\tau}_{k}$ is given as matrices with the bases of the Hilbert space.
We find that the matrix representations 
of the operators $\hat{\tau}_{k}$ as
\begin{eqnarray}
\tau_{k}^{{\mathcal M}, {\mathcal P}} = \sigma_{k}^{{\mathcal M}} \otimes h_{k}^{{\mathcal P}},
\label{eq:decomposition_matrix}
\end{eqnarray}
i.e. a tensor product of the ${\mathcal M}$-dependent term $\sigma_{k}^{{\mathcal M}} $ and the ${\mathcal P}$-dependent term $h_{k}^{{\mathcal P}}$.
Their matrix forms are 
\begin{eqnarray}
\sigma_{1}^{{\bf 1}} =
\left(
\begin{array}{cc}
 -1 & 0 \\
 0 & 1
\end{array}
\right),\quad 
\sigma_{2}^{{\bf 1}} = \frac{1}{2}
\left(
\begin{array}{cc}
 1 & \sqrt{3} \\
 \sqrt{3} & -1
\end{array}
\right),\quad 
\sigma_{3}^{{\bf 1}} = \sigma_{1}^{{\bf 1}},
\label{eq:sigma1}
\end{eqnarray}
for ${\mathcal M}={\bf 1}$, 
\begin{eqnarray}
&&\sigma_{1}^{{\bf 3}} =
\left(
\begin{array}{ccc}
 -1 & 0 & 0 \\
 0 & 1 & 0 \\
 0 & 0 & 1
\end{array}
\right)\! , 
\sigma_{2}^{{\bf 3}} = \frac{1}{2}\!
\left(
\begin{array}{ccc}
 1 & \sqrt{2} & 1 \\
 \sqrt{2} & 0 & -\sqrt{2} \\
 1 & -\sqrt{2} & 1
\end{array}
\right)\! , 
\sigma_{3}^{{\bf 3}} =
\left(
\begin{array}{ccc}
 1 & 0 & 0 \\
 0 & 1 & 0 \\
 0 & 0 & -1
\end{array}
\right)\! ,
\label{eq:sigma3}
\end{eqnarray}
for ${\mathcal M}={\bf 3}$, and 
\begin{eqnarray}
\sigma_{1}^{{\bf 5}} = \sigma_{2}^{{\bf 5}} = \sigma_{3}^{{\bf 5}} = 1,
\label{eq:sigma5}
\end{eqnarray}
for ${\mathcal M}={\bf 5}$. 
On the other hand, $h_{k}^{\mathcal P}$ are common to all the ${\mathcal M}$-plets, and are given as
\begin{eqnarray}
&&
 h_{1}^{{\mathcal E}}\! =\! h_{1}^{{\mathcal O}}\! =\!
\left(
\begin{array}{cc}
 {\rm e}^{i\frac{\pi}{4}} & 0 \\
 0 & {\rm e}^{-i\frac{\pi}{4}}
\end{array}
\right)\! , \quad
h_{2}^{{\mathcal E}}\! =\! h_{2}^{{\mathcal O}}\! =\! \frac{1}{\sqrt{2}}\!
\left(
\begin{array}{cc}
 1 & -1 \\
 1 & 1
\end{array}
\right)\!, \quad 
 h_{3}^{{\mathcal E}}\! =\! h_{3}^{{\mathcal O}\dag} 
\!=\! h_{1}^{{\mathcal E}}.
\label{eq:h}
\end{eqnarray}
Those matrices give the mixing of the wave functions of the Dirac fermions in exchanging the vortices.
We note that they are non-Abelian matrices; 
\begin{eqnarray}
 \hat{\tau}_{\ell}^{{\mathcal M}, {\mathcal P}} \hat{\tau}_{\ell+1}^{{\mathcal M}, {\mathcal P}} \neq \hat{\tau}_{\ell+1}^{{\mathcal M}, {\mathcal P}} \hat{\tau}_{\ell}^{{\mathcal M}, {\mathcal P}}.
\end{eqnarray}
This is confirmed by $\hat{\sigma}_{\ell}^{{\mathcal M}} \hat{\sigma}_{\ell+1}^{{\mathcal M}} \neq \hat{\sigma}_{\ell+1}^{{\mathcal M}} \hat{\sigma}_{\ell}^{{\mathcal M}}$ and $\hat{h}_{\ell}^{{\mathcal P}} \hat{h}_{\ell+1}^{{\mathcal P}} \neq \hat{h}_{\ell+1}^{{\mathcal P}} \hat{h}_{\ell}^{{\mathcal P}}$.
It means that the simultaneous diagonalization is not possible.
In other words, the different exchange of the neighboring vortices can lead to the different state.
Therefore it obeys the non-Abelian statistics.

\begin{figure}
\begin{center}
\includegraphics[height=5cm]{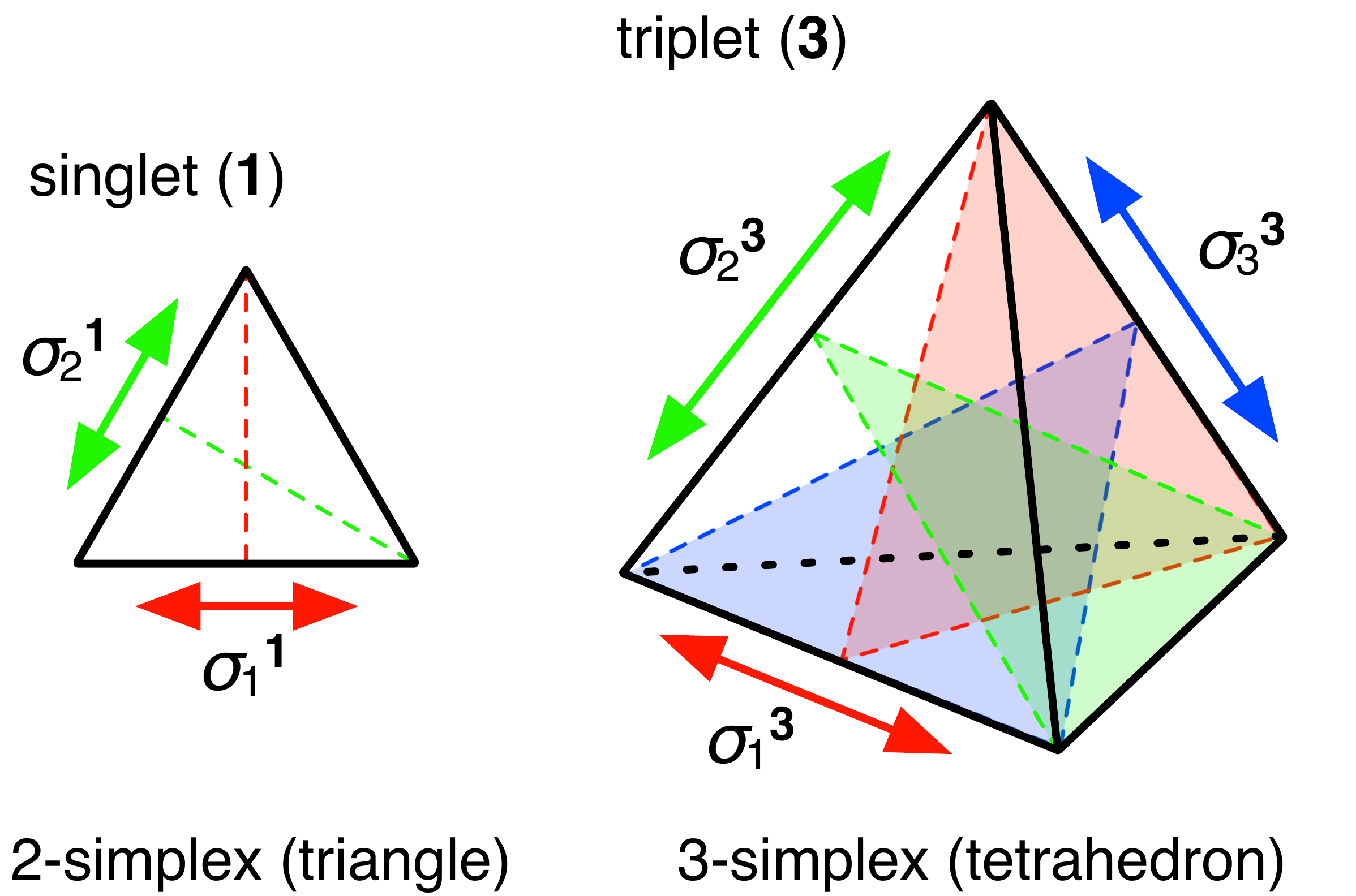}
\caption{The triangle for ${\mathcal M}={\bf 1}$ and the tetrahedron for ${\mathcal M}={\bf 3}$ are shown. (Ref.~\cite{Yasui:2010yh}.)}
\label{fig:triangle_tetrahedron_ver7}
\end{center}
\end{figure}

An important remark is in order.
We note that $h_{k}^{\mathcal P}$ is nothing but the matrices in the non-Abelian statistics for ``Abelian'' vortices which contains a single component Majorana fermions found by Ivanov.
What is new for ``non-Abelian'' vortices is the existence of the multiple component Majorana fermions with $SO(3)$ symmetry.
Then, the non-Abelian statistics of the non-Abelian vortices have been obtained  as tensor product of 
$\sigma_{k}^{\mathcal M}$ and $h_{k}^{\mathcal P}$, 
where $\sigma_{k}^{\mathcal M}$ is a new ingredient from 
the $SO(3)$ symmetry.
Interestingly, it is found that $\sigma_{k}^{\mathcal M}$ obeys the Coxeter group~\cite{1934,humphreys1992reflection}.
The Coxeter group gives a symmetry of the polytopes in high-dimensional space.
The Coxeter group was studied by Harold Scott MacDonald (``Donald") Coxeter, one of the great mathematician of the 20th century.
The generators $s_{i}$ ($i=1,2,...$) of the Coxeter group are given by two conditions;
\begin{eqnarray}
 &&\mathrm{(a)} \hspace{0.5em} s_{i}^2=1, \\
 &&\mathrm{(b)} \hspace{0.5em} (s_{i}s_{j})^{m_{ij}}=1,
\end{eqnarray}
with positive integers $m_{ij} \ge 2$.
The conditions (a) and (b) represent the mirror reflection and the rotation of the polytopes, respectively.
In our present case, it is confirmed that $\sigma_{k}^{\mathcal M}$ satisfies these two relations; (a) $\sigma_{i}^2=1$ and (b) $(\sigma_{i}\sigma_{j})^{m_{ij}}=1$ with $m_{ij}=3$ for $|i-j|=1$ and $m_{ij}=2$ for $|i-j|>1$.
The matrices $\sigma_{k}^{\bf 1}$ for ${\mathcal M}={\bf 1}$ in Eq.~(\ref{eq:sigma1}) represent the symmetry for a triangle, and the matrices $\sigma_{k}^{\bf 3}$ for ${\mathcal M}={\bf 3}$  in Eq.~(\ref{eq:sigma3}) represent the symmetry for a tetrahedron in Fig.~\ref{fig:triangle_tetrahedron_ver7}.
As a result, it is found that the structure of high dimensional polytopes exists in the Hilbert space for the non-Abelian vortices.

Finally, we comment that the decomposition in Eq.~(\ref{eq:decomposition_matrix}) can be obtained for the operator $\hat{\tau}_{k}$ itself.
The operator $\hat{\tau}_{k}$ is expressed as a product of two $SO(3)$ invariant unitary 
operators, 
\begin{eqnarray}
\hat{\tau}_{k} = \hat{\sigma}_{k} \hat{h}_{k}\, ,
\end{eqnarray}
where
\begin{eqnarray}
\hat{\sigma}_{k} 
  = \frac{1}{2} 
    \Big( 1 - \gamma_{k+1}^{1}\gamma_{k+1}^{2}\gamma_{k}^{1}\gamma_{k}^{2} 
            - \gamma_{k+1}^{2}\gamma_{k+1}^{3}\gamma_{k}^{2}\gamma_{k}^{3} 
- \gamma_{k+1}^{3}\gamma_{k+1}^{1}\gamma_{k}^{3}\gamma_{k}^{1}
     \Big)
\end{eqnarray}
and
\begin{eqnarray}
\hat{h}_{k} 
  = \frac{1}{\sqrt{2}} 
    \Big( 1- \gamma_{k+1}^{1}\gamma_{k+1}^{2}\gamma_{k+1}^{3}
             \gamma_{k}^{1} \gamma_{k}^{2} \gamma_{k}^{3}
    \Big).
\end{eqnarray}
It is easily verified that operators $\hat{\sigma}_{k}$ 
satisfy relations (a) and (b) of the Coxeter group, such that
\begin{eqnarray}
\hat{\sigma}_{k}^{2} &=& 1, \nonumber \\
(\hat{\sigma}_{k} \hat{\sigma}_{l})^{3} &=& 1 \quad {\rm for}\quad |k-l|=1,
  \nonumber \\
(\hat{\sigma}_{k} \hat{\sigma}_{l})^{2} &=& 1 \quad {\rm for}\quad |k-l|>1.
\end{eqnarray}
We confirm that $\hat{\sigma}_{k}$ obeys the Coxeter group 
for an arbitrary number of non-Abelian vortices.
Therefore, for arbitrary large number of non-Abelian vortices, the non-Abelian statistics contains the polytope structures in high dimensional space.
This is a useful method, because it can be applied to the vortices with $SO(N)$ symmetry ($N$ : odd number) as demonstrated in Ref.~\cite{Hirono:2012ad}.

\subsection{Abelian/Non-Abelian vortices with Dirac fermions}
\label{sec:na-statistics-3}

So far we have discussed the exchange statistics of vortices, where either a single Majorana fermion or three Majorana fermions is contained.
They are called Majorana vortices.
There, the Dirac fermion is defined by picking up two Majorana fermions which are located in different vortices.
In this sense, the Dirac fermions are defined {\it non-locally}.

However, the situation is much different, when there are even number of Majorana fermions~\cite{Yasui:2011gk,Yasui:2012zb}.
In this case, we can construct the Dirac operators 
by picking up two Majorana fermions at the same vortex.
Namely, the Dirac fermions are defined {\it locally} 
at the single vortex.
Vortices containing the Dirac fermions can be 
called Dirac vortices. 

In the CFL phase, an M$_1$ vortex contains 
three Majorana fermions and can be regarded 
as a Majorana vortex as discussed above, 
while an M$_2$ vortex contains 
two Dirac fermions \cite{Fujiwara:2011za} 
and can be regarded as a Dirac vortex. 
Although ${\mathrm M}_{2}$ vortices are energetically  
unstable against the decay into the two ${\mathrm M}_{1}$ 
as illustrated in Fig.~\ref{fig:split}, 
there is a possibility that 
the ${\mathrm M}_{2}$ vortices might be 
metastable when gluons are much heavier 
than other scalar fields.  
As presented in the Majorana vortex, we can consider the exchange of the Dirac vortices. 
The Dirac fermions in the M$_2$ vortices can have the internal degrees of freedom from $U(2)$ symmetry \cite{Yasui:2012zb}.
It was found that 
while exchanges of vortices with a single Dirac fermion 
do {\it not} give non-Abelian statistics \cite{Yasui:2011gk}, 
exchanges of vortices with multiple Dirac fermions 
such as  the M$_2$ vortices 
do give non-Abelian statistics
\cite{Yasui:2012zb}. 
We also note that non-Abelian vortices 
in supersymmetric $U(N)$ QCD 
contain $N-1$ Dirac fermions in their cores as summarized in 
Appendix \ref{sec:susy}.  

%% file: global-v9.tex
\section{Topological objects associated with 
chiral symmetry breaking}\label{sec:global}

In the previous sections, we have studied 
the (non-)topological solitons 
in the order parameter manifold 
$U(3)_{\rm C-(L+R)+B} \simeq [SU(3)_{\rm C} \times SU(3)_{\rm L+R} \times U(1)_{\rm B}] / [SU(3)_{\rm C+L+R} \times {\mathbb Z}_{\rm 3}]$. 
In this section, we study 
topological objects in the chiral symmetry breaking 
sector $U(3)_{\rm L-R+A} \simeq [SU(3)_{\rm L} \times SU(3)_{\rm R}\times U(1)_{\rm A}]/[SU(3)_{\rm L+R} \times {\mathbb Z}_3]$.
The $U(1)_{\rm A}$ symmetry is explicitly broken 
by the instanton effect (the chiral anomaly) and 
the corresponding  $U(1)_{\rm A}$ Nambu-Goldstone mode, the $\eta'$ meson, acquires a potential term.
In Sect.~\ref{sec:dw_chiral}, 
we first discuss $U(1)_{\rm A}$ axial domain walls 
interpolating ground states of the instanton-induced potential, 
with massless and massive quarks.  
After giving a linear sigma model in Sect.~\ref{sec:gv_non_inst},
we discuss $U(1)_{\rm A}$ Abelian axial vortices and 
non-Abelian axial vortices in the absence of 
the instanton-induced potentials 
in Sect.~\ref{sec:global_vortex}.  
In Sect.~\ref{sec:composite-wall-vortex}, 
we discuss composite states of axial domain walls and 
axial vortices in the presence of the instanton-induced potentials.
In Sect.~\ref{sec:wall-decay}, 
we discuss the quantum decay of axial domain walls. 
Skyrmions as qualitons are briefly discussed in Sec. \ref{sec:skyrmion}.

\subsection{Axial domain walls}\label{sec:dw_chiral}

\subsubsection{Fractional axial   
domain walls in the chiral limit} \label{sec:fractional-SG}

Instantons flip the chiralities of quarks and thus break the
$U(1)_{\rm A}$ symmetry (the chiral anomaly).
Instanton effects in the CFL phase are parametrically small at 
asymptotically high densities and controlled calculations are possible.
In the chiral limit with massless quarks, 
the leading contribution to the mass of the $\eta'$ meson arises from
two-instanton diagrams
\cite{Schafer:2002ty}.
As shown in Eq.~(\ref{eq:V_mass_instanton}), the instanton-induced potential takes the form
\beq
V_{2\text{-inst}} 
= - 2 C \cos 3\varphi_{\rm A}.
\label{eq:v_inst_chiral}
\eeq
Here the $U(1)_{\rm A}$ phase mode 
$\ph_{\rm A}$ is arg($\det \Sigma$) 
in terms of the gauge invariant field $\Sigma$ in Eq.~(\ref{eq:sigma0}). 
Thus, the effective Lagrangian for the $\eta'$ meson 
is given by
\beq
\Lag = \frac{3f_{\eta'}^2}{4}\left((\p_0\varphi_{\rm A})^2 - v_{\eta'}^2 (\p_i \varphi_{\rm A})^2\right) + 2 C \cos 3\varphi_{\rm A},
\label{eq:axion_3}
\eeq
where the elementary meson field is defined as $\varphi_{\rm A} = 2\eta'/\sqrt{6}f_{\eta'}$.
The decay constant $f_{\eta'}$ 
and the velocity $v_{\eta'}$ were found as 
\cite{Son:1999cm}
\beq
f_{\eta'} = \frac{3\mu^2}{2\pi^2},\quad v_{\eta'} = \frac{1}{\sqrt3}.
\eeq

The above Lagrangian is the sine-Gordon model with a period  $\varphi_{\rm A} \sim \varphi_{\rm A} + 2\pi/3$, 
which allows sine-Gordon domain wall solutions 
\cite{Skyrme:1961vr}. 
Since the potential is periodic in $\varphi_{\rm A}$ with the period $2\pi/3$, there are three
different minima $\varphi_{\rm A} = 0,2\pi/3,4\pi/3$ in the period $\varphi_{\rm A} \in [0,2\pi)$.
One of the minimal configurations is a domain wall  
which interpolates between $\varphi_{\rm A} = 0$ at $x = -\infty$ 
and $\varphi_{\rm A} = 2\pi/3$ at $x=\infty$. 
Assuming that the field depends only on 
one space direction, say $x$, 
an exact solution of a single static domain wall 
can be available: 
\beq
\varphi_{\rm A} (x) = \frac{4}{3} \arctan e^{\frac{m_{\eta'}}{v_{\eta'}} (x-x_0)}, \label{eq:fractional-SG}
\eeq
where 
\beq 
 m_{\eta'} = {2\sqrt{3C} \over f_{\eta'}} \label{eq:eta-mass}
\eeq 
is the mass of the $\eta'$ meson, and $x_0$ denotes the position 
of the domain wall. 
The tension of the domain wall is given by
\beq
T = \frac{8\sqrt{C}f_{\eta'}v_{\eta'}}{\sqrt3}.
\label{eq:tension_dw}
\eeq
The other two domain walls interpolating between 
$\varphi_{\rm A} = 2\pi/3$ and $\varphi_{\rm A} = 4\pi/3$, 
and between $\varphi_{\rm A} = 4\pi/3$ and $2\pi$ 
are simply obtained by phase shifts. 
All three domain walls wind 
the $U(1)_{\rm A}$ phase $1/3$ times,   
unlike the unit winding for the usual sine-Gordon domain walls. 
Therefore, these domain walls can be called 
{\it fractional} axial (sine-Gordon) domain walls.

Two fractional sine-Gordon domain walls 
repel each other 
(the repulsion $\sim e^{-2R}$ with distance $2R$) 
\cite{Perring:1962vs}. 
If one considers an integer sine-Gordon domain wall 
winding $U(1)_{\rm A}$ once, it splits into 
three fractional sine-Gordon domain walls. 

Note that 
if we include the amplitude mode $|\det \Sigma|$ in addition to 
the effective theory for $\ph_{\rm A}$ in Eq.~(\ref{eq:axion_3}), 
it is similar to the so-called $N=3$ axion model  \cite{Peccei:1977hh,Peccei:1977ur,Dine:1981rt}
which is an elegant extension of the standard model for solving the strong CP problem \cite{Peccei:1977hh,Peccei:1977ur,Weinberg:1977ma,Wilczek:1977pj}. The domain walls in the axion models were first studied in Ref.~\cite{Sikivie:1982qv}.
Domain walls in the $N>1$ axion models are stable
because they connect two vacua with different phases $\ph_{\rm A}$.
Consequently, 
the domain wall energy dominates 
the energy density of the universe,  
causing the cosmological domain wall problem.
Only the $N=1$ case is known to be cosmologically viable, 
in which the domain wall can decay, as is explained 
in Sect.~\ref{sec:integer-SG-decay}, 
and the domain wall energy does not dominate 
the energy density of the universe.  
In our case of dense QCD, 
one might think that  
the axial domain walls would be stable because 
they interpolate between two disconnected points 
among $\varphi_{\rm A} = 0,2\pi/3,4\pi/3$.  
This is, however, not the case. 
The axial domain walls in dense QCD are metastable 
as we discuss in Sect.~\ref{sec:fractional-SG-decay}.

\subsubsection{Integer axial domain walls with massive quarks}\label{sec:domain-wall-massive}

Let us next consider an axial domain wall in the case that 
the quarks masses are not all zero \cite{Son:2000fh}.
Here, we assume $m_{\rm u} = m_{\rm d} = m_{\rm s} = m$ for simplicity, 
so that the mixing in the neutral mesons $(\pi^0,\eta,\eta')$ simply vanishes 
\cite{Manuel:2000wm,Bedaque:2001je,Schafer:2001za,Schafer:2002ty}. As shown in Eq.~(\ref{eq:V_mass_instanton}), 
the potential for the $\eta'$ meson receives 
two additional contributions
to the one in Eq.~(\ref{eq:v_inst_chiral}).
One is from the quark mass term \cite{Son:1999cm} 
and the other is the one-instanton contribution  \cite{Son:2000fh,Schafer:2002ty}, 
\beq
V = - 6 m A \cos \varphi_{\rm A} - 12 B m^2 \cos \varphi_{\rm A} - 2 C \cos 3\varphi_{\rm A},
\label{eq:pot_sG_gene}
\eeq
where the first term stands for the one-instanton contribution, 
which is a contribution from the chiral condensates.
The coefficients $A,B$ were obtained in Ref.~\cite{Schafer:2002ty}. 

In the limit where the chemical potential is infinite 
while keeping the quark masses fixed, we have $m^2 B \gg m A \gg C$.
So we can omit the terms whose coefficients are $A$ and $C$, then the effective theory becomes
\beq
\Lag = \frac{3f_{\eta'}^2}{4}\left((\p_0\varphi_{\rm A})^2 - v_{\eta'}^2 (\p_i \varphi_{\rm A})^2\right) + 12 B m^2 \cos \varphi_{\rm A} .
\label{eq:vinst_mass}
\eeq
This is the conventional sine-Gordon model.
A single static domain wall solution is known to be \cite{Sikivie:1982qv,Son:2000fh}
\beq
\varphi_{\rm A}(x) = 4\arctan e^{\frac{{\tilde m}_{\eta'}}{v_{\eta'}}(x-x_0)},  \label{eq:wall1}
\eeq
with ${\tilde m}_{\eta'} = 2m\sqrt{2B}/f_{\eta'}$. 
The tension of the domain wall is given by
\beq
T = 24\sqrt{2B}f_{\eta'}v_{\eta'}m.
\eeq 
The difference between axial domain walls in 
Eqs.~(\ref{eq:fractional-SG}) 
and (\ref{eq:wall1}) is only the coefficients; 
In contrast to the fractional axial domain wall 
in Eq.~(\ref{eq:fractional-SG}), 
the domain wall in Eq.~(\ref{eq:wall1})
winds the $U(1)_{\rm A}$ once, 
therefore it can be referred to as 
an {\it integer} axial (sine-Gordon) domain wall 
characterized by the first homotopy group\footnote{
\label{footnote:texture}
Sine-Gordon kinks should be regarded as textures 
more like defects. So they are classified by 
an approximate order parameter manifold 
in the absence of the potential, 
while the exact order parameter is lifted by the potential. 
When the spatial infinities are identified as a point, 
topological textures in ${\mathbb R}^{d}$ are classified by 
a map from $S^{d}$ to the (approximate) order parameter 
manifold $M$, and the corresponding homotopy group is 
$\pi_d(M)$.
Fractional sine-Gordon domain walls discussed 
in the last subsection cannot be characterized by 
the first homotopy group. 
}  
\beq
 \pi_1[U(1)_{\rm A}] \simeq {\mathbb Z} .
  \label{eq:SG-homotopy2}
\eeq
The classical stability of the integer axial domain walls was 
studied in Ref.~\cite{Buckley:2001bm}.

\begin{figure}[ht]
\begin{center}
\includegraphics[width=16cm]{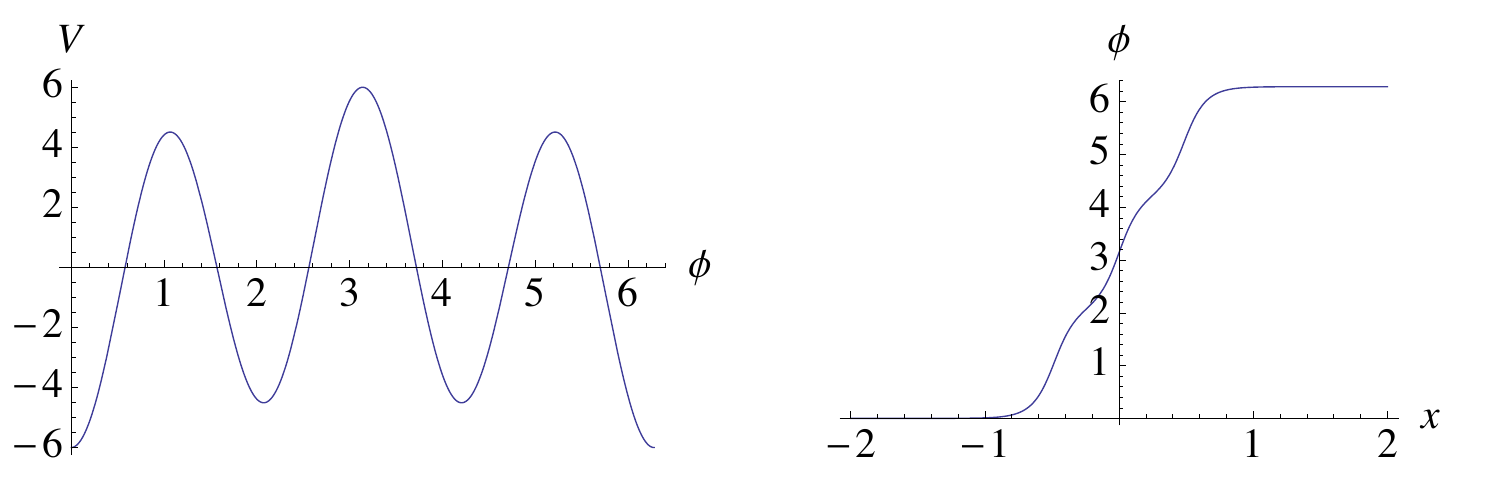}
\caption{
A composite of fractional axial domain walls.
An example of a numerical solution for a generic sine-Gordon potential $V = - \alpha \cos \varphi_{\rm A} - \beta \cos 3\varphi_{\rm A}$
shown in the left panel.
We set $\alpha=1$, $\beta=5$ and $f_\eta' =1$ to show a typical configuration.
The domain wall solution is shown in the right panel. 
The integer axial domain wall 
consists of three fractional axial domain walls. 
}
\label{fig:multi_sG}
\end{center}
\end{figure}

Finally, let us consider the most generic potential in Eq.~(\ref{eq:pot_sG_gene}).
If we include the two-instanton contributions, 
the potential includes $\cos\varphi_{\rm A}$ and $\cos3\varphi_{\rm A}$ as in Eq.~(\ref{eq:pot_sG_gene}). 
This potential is known as that of the triple sine-Gordon model.
For generic parameter choices, 
the minimum of the potential is unique,  
$\varphi_{\rm A} = 0$, 
and 
only an integer axial domain wall 
winding the $U(1)_{\rm A}$ phase once is allowed.
A numerical solution is shown in Fig.~\ref{fig:multi_sG}. 
The figure shows substructures of  three peaks, 
and this integer axial domain wall  
can be interpreted as a composite of three 
fractional axial domain walls 
with 1/3 $U(1)_{\rm A}$ windings in Eq.~(\ref{eq:wall1}).
Without two-instanton contribution $\cos\varphi_{\rm A}$, 
these three fractional axial domain walls repel each other 
and the integer axial domain wall 
decays completely into  three fractional axial domain walls, 
as mentioned above.  
On the other hand, 
the two-instanton-induced potential $\cos\varphi_{\rm A}$ 
introduces the energy between domain walls, 
resulting in a constant attractive force between them. 
Therefore, the linear confinement of domain walls occurs.

\subsection{Linear sigma model}\label{sec:gv_non_inst}

As discussed above, 
instanton effects are parametrically small 
in the asymptotically high density limit. 
Then the $\U(1)_{\rm A}$ symmetry 
becomes an exact symmetry 
of the system. In the CFL phase, the diquarks are condensed and 
the $\U(1)_{\rm A}$ is spontaneously broken. 
As explained in Sect.~\ref{sec:U(1)A}, therefore, there exist 
topologically stable global vortices, $\U(1)_{\rm A}$ global vortices. 
However, this is not the whole story. One should be careful that not only the $\U(1)_{\rm A}$ symmetry, but 
the full axial symmetry $[\SU(3)_{\rm L-R}\times U(1)_{\rm A}]/\mathbb{Z}_3$ of QCD
is spontaneously broken completely 
in the CFL phase.
As we have seen, this kind of spontaneously broken non-Abelian symmetry gives rise
to non-Abelian vortices. 
One crucial difference from the previous sections is that no local symmetries take part in here. 
Thus, the $U(1)_{\rm A}$ vortices that we study 
in this section are {\it global} non-Abelian vortices.

We start with a generic linear sigma model in the chiral limit: 
\beq
\Lag &=& \Tr\left[\p_0 \Sigma^\dagger \p^0 \Sigma + u^2\p_i\Sigma^\dagger \p^i\Sigma - \lambda_2(\Sigma^\dagger \Sigma)^2 + \eta^2 \Sigma^\dagger \Sigma\right]
- \lambda_1 
 \left(\Tr\left[\Sigma^\dagger\Sigma\right]\right)^2 \non
&& 
- \kappa (\det \Sigma + {\rm c.c.})
- \frac{3\eta^4}{4\left(3\lambda_1 + \lambda_2\right)},
\label{eq:lag_lsm}
\eeq
where $\Sigma$ is a 
$3\times 3$ complex matrix scalar field without any constraints
as is defined in Eq.~(\ref{eq:sigma0}). 
The speed of the modes is taken to be 
$u \sim v_{\pi} \sim v_{\eta'}$.
The coefficient $\kappa$ of the anomaly term 
is related to the one $C$ of the chiral Lagrangian 
as $\kappa \left<\Sig\right>^3 = C$.
The field $\Sigma$ is invariant under $\SU(3)_{\rm C}$ and $\U(1)_{\rm B}$, 
while it transforms under $\SU(3)_{\rm L}\times\SU(3)_{\rm R} \times \U(1)_{\rm A}$ as
\beq
\Sigma \to e^{i\theta_{\rm A}} g_{\rm L}^\dagger \Sigma g_{\rm R},\quad \left(e^{i\theta_{\rm A}},g_{\rm L},g_{\rm R}\right) \in U(1)_{\rm A} \times SU(3)_{\rm L} \times SU(3)_{\rm R}.
\eeq 
Here, we have used different notation for $\theta_{\rm A}$ from Eq.~(\ref{eq:sigma-trans}).
Since we are interested in topological vortices in this model, we do not specify values of the coupling constants
$\lambda_{1,2}, \eta$ but leave them as the parameters of the model. 
By taking the heavy mass limit, 
one can relate
this linear sigma model with 
the chiral Lagrangian of the light fields 
in Eq.~(\ref{eq:chiral_lag}) with the anomaly term 
in Eq.~(\ref{eq:V_mass_instanton}). 
We also ignore the electromagnetic interaction here.

Taking into account discrete symmetries, 
the full chiral symmetry of the Lagrangian in Eq.~(\ref{eq:lag_lsm}) is 
$G_{\rm F}$ given in Eq.~(\ref{eq:GF}). 
Let us consider the case in which the anomaly effect is absent,  $\kappa=0$. 
The ground states are stable
when the coupling constants in Eq.~(\ref{eq:lag_lsm}) satisfy 
\beq
\eta^2 > 0,\quad 3 \lambda_1 + \lambda_2 > 0.
\eeq
 Up to the flavor rotation, one can choose the ground state value as
\beq
\left<\Sigma\right> = v {\bf 1}_3,\quad v = \sqrt{\frac{\eta^2}{2(3\lambda_1 + \lambda_2)}}.
\eeq
The vacuum expectation value $v$ should be identified with 
$v \sim f_\pi \sim f_{\eta'}$.
In the ground state, the chiral symmetry $G_{\rm F}$ breaks down to
its diagonal subgroup $H_{\rm F} = \SU(3)_{\rm V}/\mathbb Z_{3{\rm V}}$ given in Eq.~(\ref{eq:HF}). 
Then, the order parameter manifold is 
$G_{\rm F}/H_{\rm F} \simeq  U(3)_{\rm L-R+A}$ given in Eq.~(\ref{eq:GFHF}). 

The mass spectra are as follows: there are $8+1$ NG bosons associated with the spontaneously 
broken $\U(3)_{\rm L-R+A}$ and the same number of massive bosons, whose masses are
\beq
m_1^2 = 2\eta^2,\quad
m_8^2 = 4\lambda_2v^2.
\eeq

Note that the linear sigma model (\ref{eq:lag_lsm}) can be regarded as that for the chiral symmetry breaking 
in the low density hadronic phase.
There are a number of works studying various topological solitons in the hadronic phase.
For example, $\eta'$ strings, which are 
the $\U(1)_{\rm A}$  global vortices, were studied
in Refs.~\cite{Zhang:1997is,Balachandran:2001qn}. 
Non-topological vortices called pion strings with a 
trivial topology $\pi_1 [SU(2)]=0$ 
were studied in Refs.~\cite{Brandenberger:1998ew,Zhang:1997is}. 

Non-Abelian global vortices in the hadronic phase were 
first found by Balachandran and Digal in Ref.~\cite{Balachandran:2002je}.
Various aspects of the non-Abelian global vortices were studied 
in the subsequent papers \cite{Nitta:2007dp,Nakano:2007dq,Eto:2009wu}.
Since the linear sigma model for the chiral symmetry breaking in the CFL phase 
shares many similar properties with that for the hadronic phase,
there should exist the same kind of non-Abelian global vortices
in dense QCD.

\subsection{Abelian and non-Abelian axial vortices}
\label{sec:global_vortex}

\subsubsection{Abelian axial vortices}\label{sec:gv_A}

Since the order parameter manifold 
$G_{\rm F}/H_{\rm F} \simeq \U(3)_{\rm L-R+A}$ is not simply connected, the first homotopy group is nontrivial: 
\beq 
 \pi_1 [U(1)_{\rm A}] \simeq \mathbb{Z}. \label{eq:A-gl-vor-homotopy}
\eeq
Therefore, there exist topologically stable vortices.
In order to generate a nontrivial loop in 
the order parameter manifold, 
one may use only a $T_0$ generator.
Such a loop corresponds to the $\eta'$ string \cite{Zhang:1997is,Balachandran:2001qn}
for which the order parameter $\Sigma$ behaves as
\beq
\Sigma(r,\theta) = v f(r) e^{i\theta}{\bf 1}_3.
\eeq
The equation of motion for the amplitude function $f(r)$ is of the form
\beq
u^2\left(f'' + \frac{f'}{r} - \frac{f}{r^2}\right) - \frac{m_1^2}{2}f(f^2-1)=0.
\eeq
This is exactly the same equation as the equation of motion for the familiar $\U(1)$ global vortices.
Solutions can be numerically obtained with the suitable boundary condition
$f(0)=0$ and $f(\infty) = 1$; see {\rm e.g.} Ref.~\cite{Vilenkin:1994}.
The tension of the Abelian global string is \cite{Nitta:2007dp}
\beq
T_{\U(1)_{\rm A}} = 3 \times 2\pi v^2u^2 \log\frac{L}{\xi} + \text{const.}\,.
\eeq
with the size of the system $L$ and the size of the axial vortex 
$\xi = m_1^{-1}$.

\subsubsection{Non-Abelian axial vortices}\label{sec:gv_NA}
One can construct a smaller loop inside the order parameter manifold by combining the $U(1)_{\rm A}$ generator 
$T_0 \sim {\bf 1}_3$ and non-Abelian generators 
$T_a$ ($a=1,2,\cdots,8$) of $SU(3)$ \cite{Balachandran:2002je}. 
They correspond to non-Abelian axial vortices 
characterized by the homotopy group
\beq 
 \pi_1 [\U(3)_{\rm L-R+A}] \simeq \mathbb{Z}. \label{eq:NA-gl-vor-homotopy}
\eeq
The three typical configurations are given by 
\beq
&& \Sigma = v\, {\rm diag}\left(e^{i\theta}f(r),g(r),g(r)\right)
\xrightarrow[]{r\to\infty} 
v\,{\rm diag}\left(e^{i\theta},1,1\right)
=v\,e^{i\frac{\theta}{3}} {\rm diag}\left(e^{i\frac{2\theta}{3}},e^{-i\frac{\theta}{3}},e^{-i\frac{\theta}{3}}\right), 
 \non
&& \Sigma = v\, {\rm diag}\left(g(r),e^{i\theta}f(r),g(r)\right)
\xrightarrow[]{r\to\infty} 
v\,{\rm diag}\left(1,e^{i\theta},1\right)
=v\,e^{i\frac{\theta}{3}} {\rm diag}\left(e^{-i\frac{\theta}{3}},e^{i\frac{2\theta}{3}},e^{-i\frac{\theta}{3}}\right),
\non
&& \Sigma = v\, {\rm diag}\left(g(r),g(r),e^{i\theta}f(r)\right)
\xrightarrow[]{r\to\infty} 
v\,{\rm diag}\left(1,1,e^{i\theta}\right)
=v\,e^{i\frac{\theta}{3}} {\rm diag}\left(e^{-i\frac{\theta}{3}},e^{-i\frac{\theta}{3}},e^{i\frac{2\theta}{3}}\right).\non
\label{eq:ansz_ngv}
\eeq
with the profile functions $f(r)$ and $g(r)$ satisfying 
the boundary conditions
\beq
 f(r\to \infty) = g(r\to \infty) = 1 ,
\quad 
f(0) = g'(0)=0.
\eeq
The boundary condition at spatial infinities 
in Eqs.~(\ref{eq:ansz_ngv}) 
clearly shows that the corresponding loops 
wind 1/3 of the $\U(1)_{\rm A}$ phase, 
and are generated by 
non-Abelian generators of $SU(3)_{\rm L-R}$ 
at the same time. 
They are called fractional vortices 
because of the fractional winding of the $\U(1)_{\rm A}$ phase
and non-Abelian vortices  
because of the contribution of the non-Abelian generators. 
Because of the fractionality of the $\U(1)_{\rm A}$ winding, 
the tension of a single non-Abelian axial vortex is one-third of that 
for an Abelian axial vortex \cite{Nitta:2007dp}:
\beq
T_{\U(3)_{\rm L-R+A}} = 2\pi v^2 u^2 \log\frac{L}{\xi} + \text{const.}\,,
\eeq
with $\xi\sim {\rm max}(m_8^{-1},m_1^{-1})$.

\begin{figure}[ht]
\begin{center}
\begin{tabular}{cc}
\includegraphics[width=7cm]{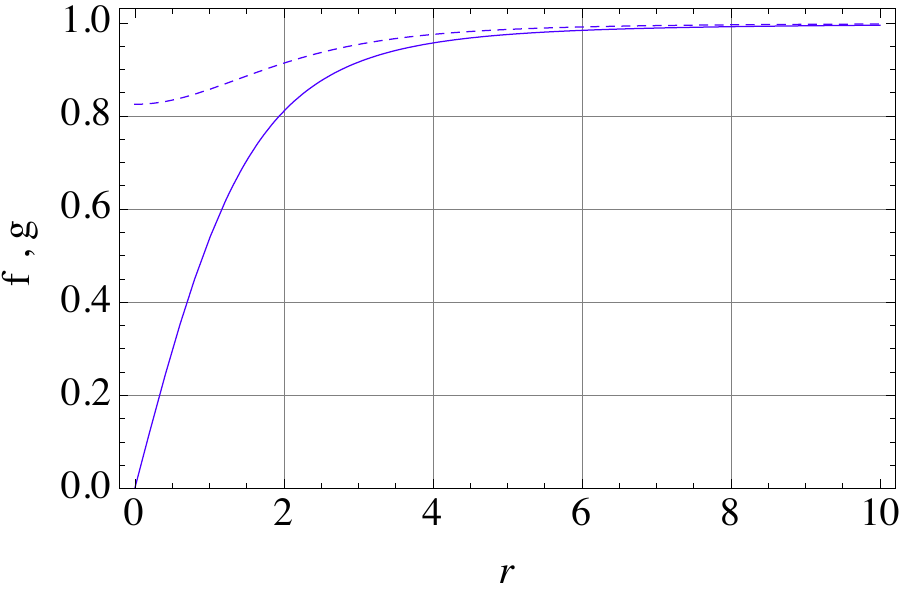}
&
\includegraphics[width=7cm]{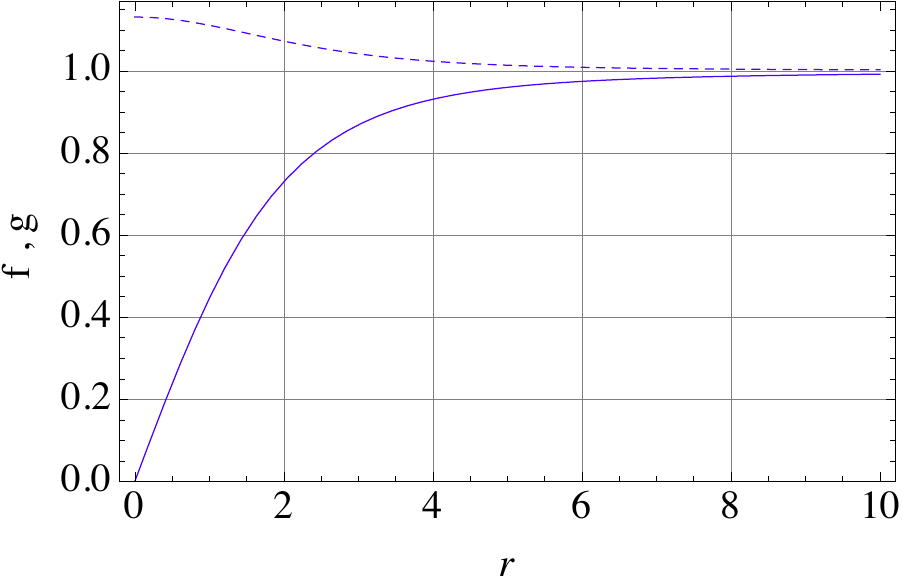}
\end{tabular}
\caption{Typical profile functions of $f(r)$ (solid line) and $g(r)$ (broken line).
The parameters are set to be $m_1=1$ and $m_8 = 2$ for the left panel and
$m_1=2$ and $m_8=1$ for the right panel.}
\label{fig:nag_su3}
\end{center}
\end{figure}

The equation of motion for the amplitudes $f(r)$ and $g(r)$ are
\beq
&&u^2\left(f'' + \frac{f'}{r} - \frac{f}{r^2}\right) - \frac{m_1}{6}f(f^2+2g^2-3) - \frac{m_8^2}{3}f(f^2 - g^2) = 0,\\
&&u^2\left(g'' + \frac{g'}{r}\right) - \frac{m_1^2}{6}g(f^2+2g^2-3) + \frac{m_8^2}{6}g(f^2-g^2) = 0,
\eeq
where $\tau \equiv m_8/m_1$. These should be solved with the boundary conditions $f(0) = g'(0) = 0$.
The asymptotic behavior of a profile function of a $\U(1)$ global vortex
is established; see {\rm e.g.} Ref.~\cite{Vilenkin:1994}. 
On the other hand, the profile functions for 
a non-Abelian axial vortex were studied in Refs.~\cite{Nitta:2007dp,Eto:2009wu} in detail. 
Typical profile functions for the cases with $m_1 < m_8$ and $m_1 > m_8$ are shown in Fig.~\ref{fig:nag_su3}.
The asymptotic behaviors depend on the two mass scales $m_1$ and $m_8$.
For instance, the profile functions near the center of the vortex and far
from the vortex were found \cite{Nitta:2007dp,Eto:2009wu} to be 
\beq
f(r) = c_1 r + c_3  r^3 + \cdots, \quad
g(r) = d_0 + d_2 r^2 + \cdots,\qquad \left(r \ll \min\{m_1^{-1},m_8^{-1}\}\right),
\eeq
and
\beq
f(r) = 1 + \frac{a_2}{r^2} + \frac{a_4}{r^4} + \cdots,\quad
g(r) = 1 + \frac{b_2}{r^2} + \frac{b_4}{r^4} + \cdots,\qquad
\left(r \gg \min\{m_1^{-1},m_8^{-1}\}\right).
\eeq
While numerical study is needed to determine the coefficients $c_i$ and $d_i$,
the coefficients $a_i$ and $b_i$ are analytically determined \cite{Eto:2009wu} as
\beq
a_2 = - \frac{1}{3}\left(\frac{1}{m_1^2}+\frac{2}{m_8^2}\right),\quad
b_2 = - \frac{1}{3}\left(\frac{1}{m_1^2} - \frac{1}{m_8^2}\right).
\eeq
Full numerical profile functions were also obtained in 
Refs.~\cite{Nitta:2007dp,Eto:2009wu}.

The configurations in Eq.~(\ref{eq:ansz_ngv}) 
are connected by the flavor rotation $\SU(3)_{\rm L+R}$ 
and all of them form infinitely degenerate family 
\cite{Nitta:2007dp}. 
Since the configurations in Eq.~(\ref{eq:ansz_ngv}) 
are invariant under the little group $\U(2)_{\rm L+R}$,
the physically independent solutions are labeled by
\beq
\frac{\SU(3)_{\rm L+R}}{\U(1)_{\rm L+R} \times \SU(2)_{\rm L+R}} \simeq \mathbb CP^2.
\eeq
Therefore, $\mathbb CP^2$ Nambu-Goldstone modes appear 
as semi-superfluid non-Abelian vortices.
However, these $\mathbb CP^2$ Nambu-Goldstone modes 
are {\it not} localized around a vortex. 
Instead, the wave functions of the $\mathbb CP^2$ modes 
extend to spatial infinity, because 
$\SU(3)_{\rm L+R}$ transformations change the boundary conditions 
at spatial infinities, 
as can be seen in Eq.~(\ref{eq:ansz_ngv}). 
They are non-normalizable and should be regarded as 
more like bulk modes.
\footnote{
If $g_{\rm L}$ were the gauge transformation as 
the case in Sect.~\ref{sec:vortices} $(g_{\rm L} \to g_{\rm C} \in \SU(3)_{\rm C})$, 
the boundary conditions 
are physically identical, 
because they 
can be transformed to each other
by a suitable gauge transformation in Eq.~(\ref{twist2}). 
Consequently, 
the $\mathbb CP^2$ modes would become normalizable.
}

The intervortex force between two well separated 
non-Abelian axial vortices was obtained in Ref.~\cite{Nakano:2007dq}.
The details of the calculations are quite similar to those in Sect~\ref{sec:intervortex-force}.
As explained in Sect.~\ref{sec:intervortex-force}, as far as two string interaction is concerned, only
an $\SU(2)_{\rm L+R}$ rotation is enough to set up two strings whose relative orientation is generic.
In the following, we fix the amplitudes $f(r)$ and $g(r)$ at one, 
which is valid for a vortex distance $2R$ much larger than 
the mass scales $m_1^{-1}, m_8^{-1}$. 
Thus, we take two well separated strings parallel to the $z$-axis separated with the distance $2R$ into the $x$-axis as in Fig.~\ref{fig:interaction}: 
\beq
\Sigma_1 = v {\rm diag}(e^{i\theta_1},1,1),
\eeq
and
\beq
\Sigma_2 =v \left(
\begin{array}{cc}
g\left(
\begin{array}{cc}
e^{i\theta_2} & \\
& 1\\
\end{array}
\right)g^{-1} & 0\\
0 & 1
\end{array}
\right),
\eeq
where $g$ is an element of $\SU(2)_{\rm L+R}$:
\beq
g = \cos \frac{\alpha}{2} + i \vec n \cdot \vec \sigma \sin \frac{\alpha}{2}.
\eeq
The interaction energy density of the two strings is obtained by subtracting two individual
string energies from the total configuration energy for the Abrikosov ansatz $\Sigma_{\rm tot} = \Sigma_1\Sigma_2$ as
\beq
F(r,\theta,R,\alpha) &=& \Tr\left(|\p \Sigma_{\rm tot}|^2 - |\p \Sigma_1|^2 - |\p \Sigma_2|^2\right) \non
&=& \pm (1+\cos\alpha)\left(\frac{-R^2 + r^2}{R^4+r^4 - 2R^2r^2\cos 2\theta}\right),
\eeq
where the sign $\pm$ is for the vortex-vortex and 
vortex-anti-vortex pairs. 
Integrating this over the $x$-$y$ plane, one gets the 
interaction energy 
\beq
E(a,\alpha,L) = \pm \pi (1+\cos\alpha)\left[-\ln4 -2\ln R + \ln(R^2+L^2)\right].
\eeq
Finally, the intervortex force can be obtained by differentiating $E$ by the interval $2R$ as
\beq
f(R,\alpha,L) = \pm \pi (1+\cos\alpha) \left(\frac{1}{R}-\frac{R}{R^2+L^2}\right) \simeq \pm (1+\cos\alpha) \frac{\pi}{R}. 
\label{eq:inter-vortex-force-global}
\eeq

Although the calculations here are quite similar to those in 
Sect.~\ref{sec:intervortex-force}, there is a sharp contrast
in comparison with the intervortex forces. 
While the intervortex force of the semi-superfluid vortices is independent of the orientation at the leading order,
one of the non-Abelian axial vortices depends on the relative orientation $\alpha$ at the leading order.
This difference reflects the fact that the orientational modes $\mathbb CP^2$ are normalizable or non-normalizable
for semi-superfluid vortices or non-Abelian axial vortices, respectively. 

The intervortex force 
at the leading order in Eq.~(\ref{eq:inter-vortex-force-global}) 
vanishes at $\alpha = \pi/2$.
For instance, the three configurations in Eq.~(\ref{eq:ansz_ngv}) 
do not interact at this order. 
A $U(1)_{\rm A}$ vortex can be marginally separated to
three non-Abelian axial vortices as
\beq 
{\rm diag.}\,(e^{i\theta},e^{i\theta},e^{i\theta})
\to {\rm diag.}\,(e^{i\theta_1},1,1) \times 
 {\rm diag.}\,(1,e^{i\theta_2},1) \times
 {\rm diag.}\,(1,1,e^{i\theta_3}) \label{eq:U(1)A-decay}
\eeq  
at this order, 
where $\theta_{1,2,3}$ denotes an angle coordinate  
at each vortex center. 

However, when the leading order term vanishes in general,
one needs to consider the next-leading order. 
The next-leading order of interaction between 
two vortices winding around different components 
was obtained in the context 
of two-component Bose-Einstein condensates in Ref.~\cite{Eto:2011wp}. 
In terms of the parameters in Eq.~(\ref{eq:lag_lsm}), 
the interaction energy and 
force  between vortices at a distance $2R$ much larger than 
$m_1^{-1},m_8^{-1}$ are
\beq
&& E_{\rm next} \sim  \lambda_2 {\log (2R/\xi) \over (2R)^2}, \non
&& f_{\rm next} \sim  \lambda_2 
 \1{(2R)^3} \left(\log {2R \over \xi} -\1{2}\right),
\eeq
respectively, 
which is repulsive. 

\subsection{
Composites of axial domain walls and 
axial vortices
}\label{sec:composite-wall-vortex}

Let us next take  into account  the instanton effects
in the presence of vortices. 
We first consider  the chiral limit with massless quarks 
in Sect.~\ref{sec:composite-chiral}, 
followed by 
the case with massive quarks 
in Sect.~\ref{sec:composite-nonchiral}.

\subsubsection{Abelian and non-Abelian axial vortices attached by  fractional axial domain walls in the chiral limit}\label{sec:composite-chiral}

\begin{figure}[ht]
\begin{center}
\begin{tabular}{cc}
\includegraphics[height=6cm]{string_vortex_3d}
&
\includegraphics[height=6cm]{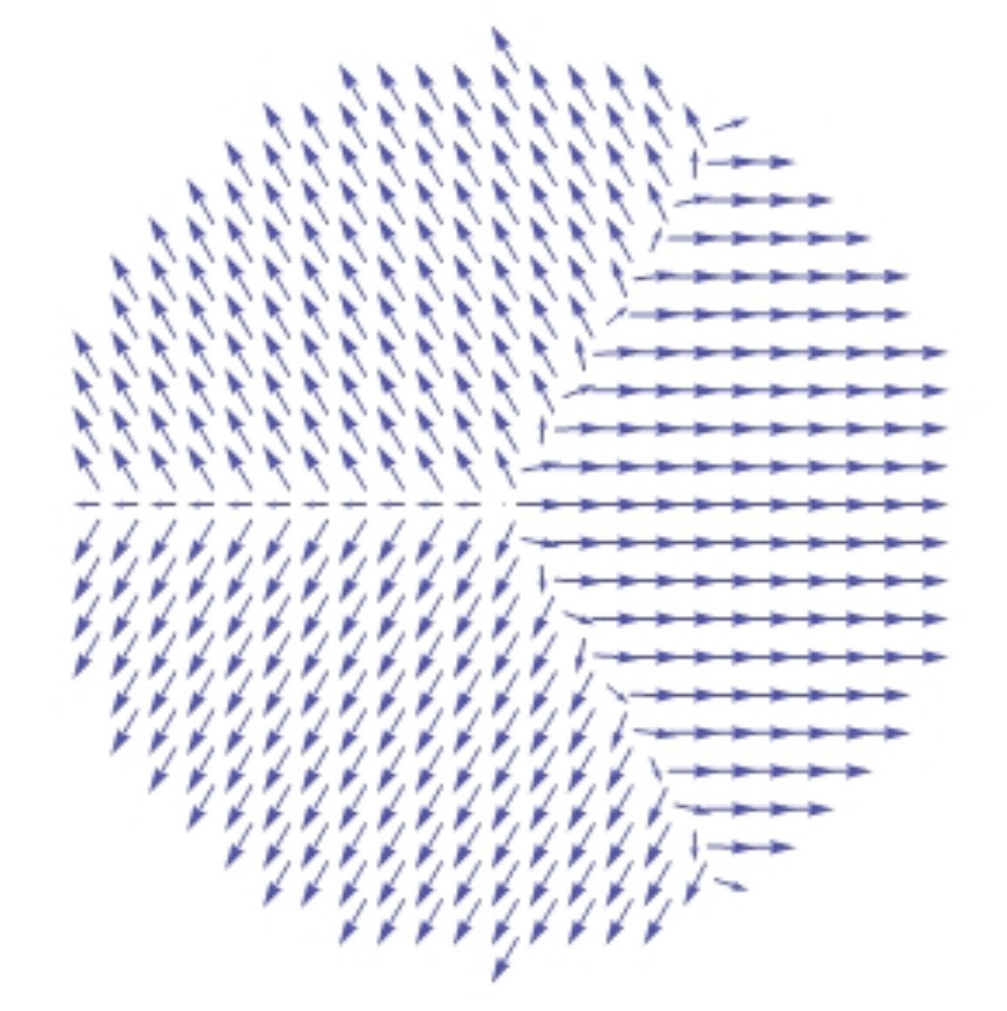}
\end{tabular}
\caption{
A three-pronged fractional axial domain wall junction is formed in the presence of an Abelian axial vortex in the chiral limit. 
The effective potential is $V_{\eta'} = -C\cos3\varphi_A$ which is generated by two instanton effect. 
An energy density is shown in the right figure and the $U(1)_{\rm A}$ phase  
$(\cos \varphi_{\rm A},\sin \varphi_{\rm A})$ is shown in the right figure.
}
\label{fig:three_dw_junction}
\end{center}
\end{figure}

In the chiral limit, the instanton-induced potential for the $\eta'$-boson is given in Eq.~(\ref{eq:v_inst_chiral}). 
While we have studied domain walls in the chiral 
Lagrangian in Sect.~\ref{sec:dw_chiral}, 
they remain  almost the same with slight modifications 
in the linear sigma model introduced Sect.~\ref{sec:gv_non_inst}. 
Similarly to axion strings, 
Abelian axial vortices
are attached by axial domain walls in 
the presence of the instanton-induced potential $\sim \cos 3\varphi_{\rm A}$. 
Since the phase changes from $\varphi_{\rm A}=0$ to
$\varphi_{\rm A} = 2\pi$ around a single Abelian axial vortex, three different domain walls forming a three-pronged  junction attach to it,  
as illustrated in Fig.~\ref{fig:three_dw_junction}.
Since the domain walls repel each other, 
the configuration becomes a 
${\mathbb Z}_3$ symmetric domain wall junction.
A numerical solution for this configuration 
was first obtained  in the chiral phase transition in
the low baryon density region in Ref.~\cite{Balachandran:2001qn}.

\begin{figure}[ht]
\begin{center}
\includegraphics[height=4cm]{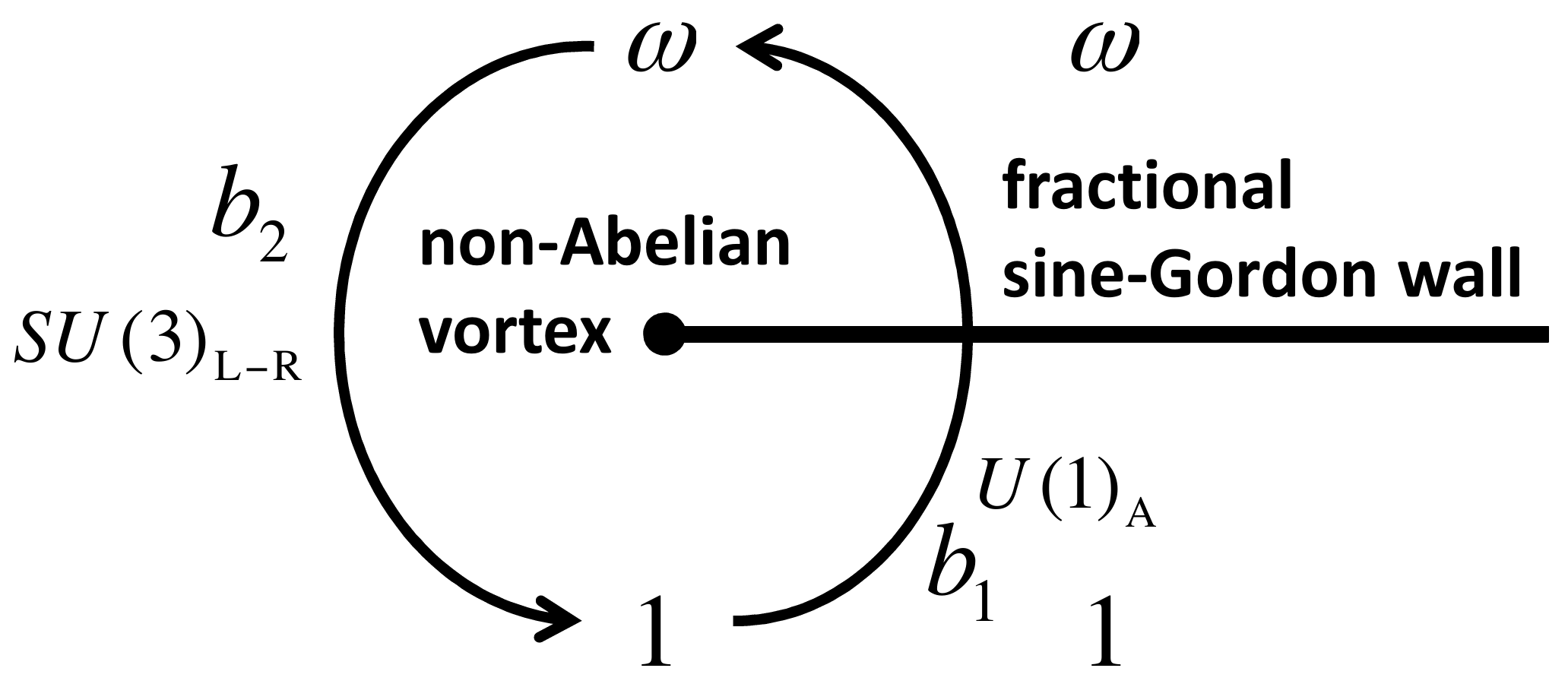}
\caption{
A non-Abelian axial vortex attached by a fractional axial 
domain wall with
the instanton-induced potential $V_{\eta'} = -C\cos3\varphi_A$. 
Along the path $b_1$, only the $U(1)_{\rm A}$ phase is rotated 
by $2\pi/3$. Then, the $SU(3)_{\rm L-R}$ transformation 
$\exp [( i /3) (\theta-\pi/2) \diag (2,-1,-1)]$ is performed 
along the path $b_2$, where $\theta$ ($\pi/2 \leq \theta \leq 3\pi/2$) is the angle of the polar coordinates  
at the black point.\label{fig:NAvortex-wall}
}
\includegraphics[width=0.7\linewidth]{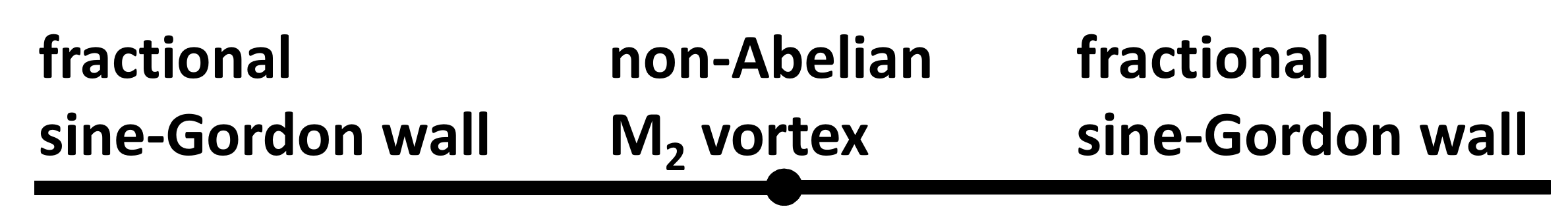}
\caption{
An M$_2$ non-Abelian axial vortex attached by two fractional axial 
domain walls with
the instanton-induced potential $V_{\eta'} = -C\cos3\varphi_A$. 
\label{fig:NAvortex-wall2}
}
\end{center}
\end{figure}

Next, let us consider non-Abelian axial vortices 
in the presence of  the instanton-induced potential $\sim \cos 3\varphi_{\rm A}$ \cite{Eto:2013bxa}. 
The $\U(1)_{\rm A}$ phase changes by $2\pi/3$ 
around a non-Abelian axial vortex.
Therefore, one fractional axial wall attaches 
to one non-Abelian axial vortex, 
as illustrated in Fig.~\ref{fig:NAvortex-wall}.
Let us consider it in more detail, focusing on 
the configuration of the type $\diag (e^{i \theta},1,1)$. 
In the vicinity of the vortex, 
let us divide a closed loop encircling the vortex 
to  the paths $b_1$ and $b_2$, as in  Fig.~\ref{fig:NAvortex-wall}.
Then, along the path $b_1$ and $b_2$, 
the order parameter receives the transformation by 
the group elements  
\beq
 &&  g(\theta) 
=\Bigg\{
\begin{array}{c}
\exp \left[{2i \over 3} \left(\theta + {\pi \over 2}\right) \diag (1,1,1)\right] 
\in U(1)_{\rm A} ,\quad \quad
b_1:
-{\pi \over 2} \leq \theta \leq {\pi \over 2} \cr
 \exp \left[{ 2i \over 3} \left(\theta-{\pi\over 2}\right) \diag (2,-1,-1)\right] \omega \in SU(3)_{\rm L-R}, \quad 
b_2 :
{\pi \over 2} \leq \theta \leq {3 \over 2}\pi .
\end{array}
\eeq 
Only the $U(1)_{\rm A}$ phase is rotated in the path $b_1$  
while only the $SU(3)_{\rm L-R}$ transformation 
acts along the path $b_2$. 
This configuration was discussed in the chiral phase transition in
the low baryon density region in Ref.~\cite{Balachandran:2002je}.
This is of course unstable because it is pulled 
by the domain wall tension to spatial infinity \cite{Eto:2013bxa}.

As for non-Abelian semi-superfluid vortices, 
there is an M$_2$ non-Abelian vortex, 
which is
\beq
&& \Sigma = v\, {\rm diag}\left(g(r),e^{i\theta}f(r),e^{i\theta}f(r)\right)\non
&& \xrightarrow[]{r\to\infty} 
v\,{\rm diag}\left(1,e^{i\theta},e^{i\theta}\right)
=v\,e^{i\frac{2\theta}{3}} {\rm diag}\left(e^{-i\frac{2\theta}{3}},e^{i\frac{\theta}{3}},e^{i\frac{\theta}{3}}\right), 
\eeq
in the absence of the instanton-induced potential. 
In the presence of the instanton-induced potential, 
the $U(1)_{\rm A}$ phase rotates by $-2\pi/3$. 
Therefore, two axial domain walls are attached,  
as illustrated in Fig.~\ref{fig:NAvortex-wall2}.
A numerical solution was constructed in Ref.~\cite{Balachandran:2002je}
with an approximation.

\begin{figure}[ht]
\begin{center}
\includegraphics[height=6cm]{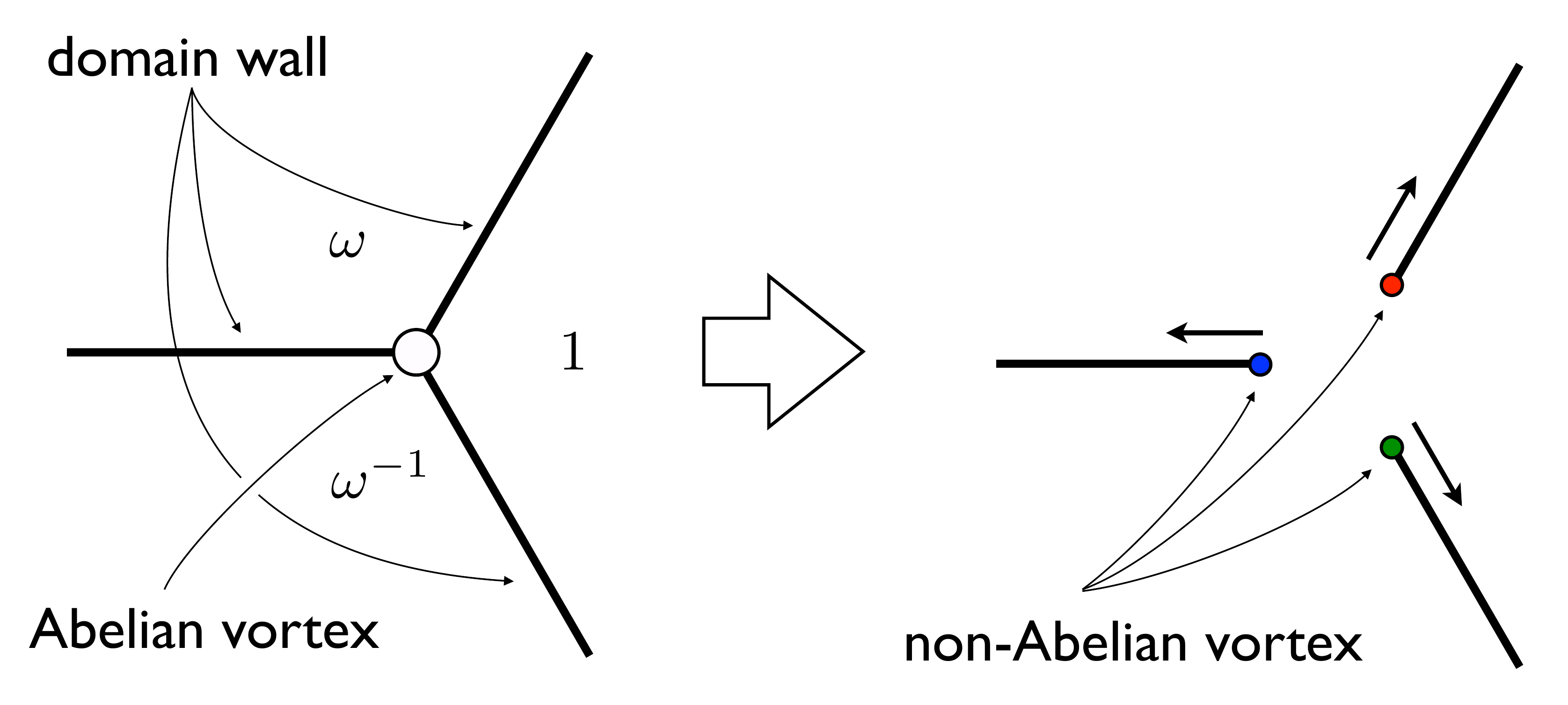}\\\ \\
\begin{tabular}{cc}
\includegraphics[height=6cm]{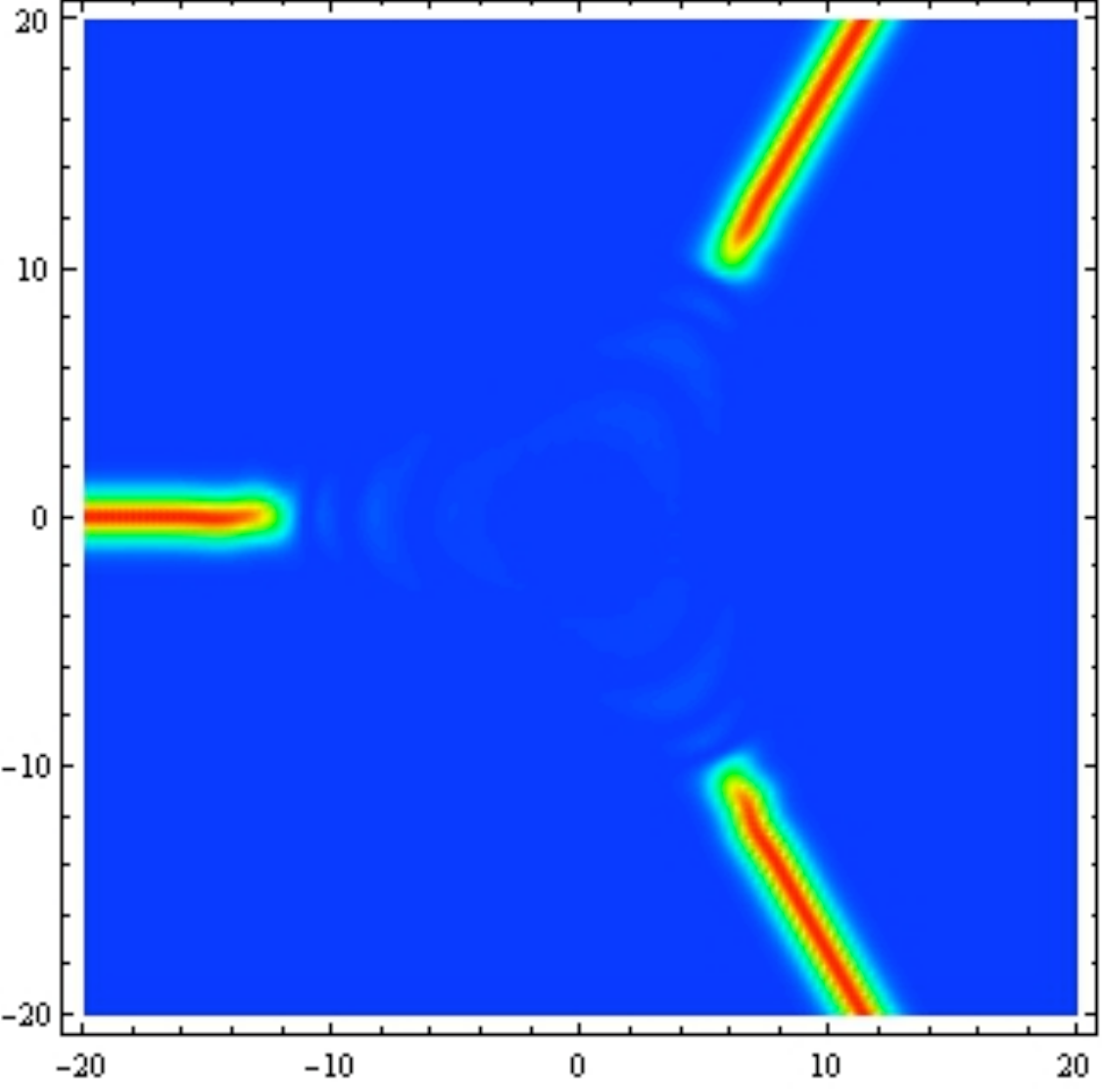}
&
\includegraphics[height=6cm]{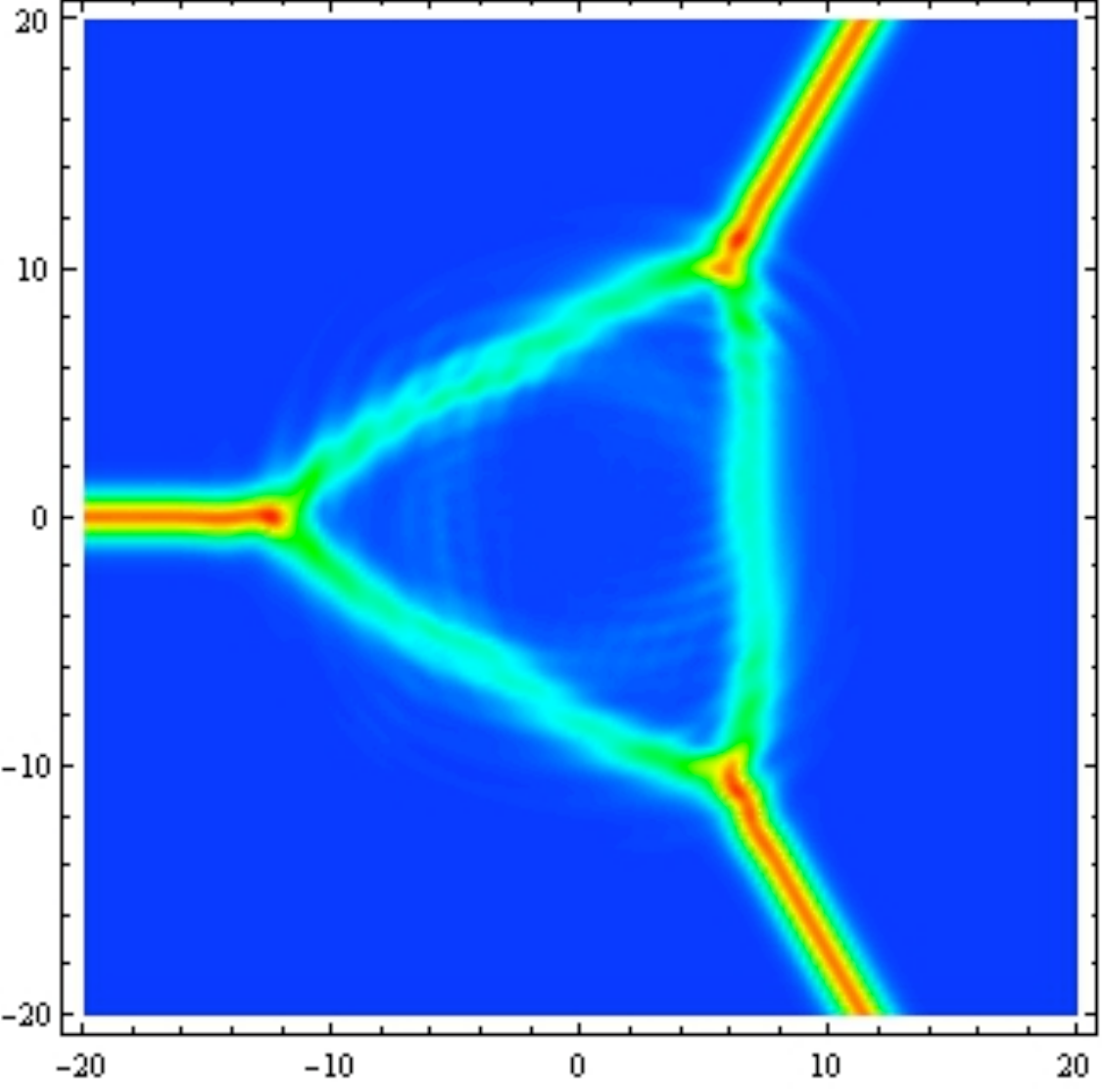}
\end{tabular}
\caption{
Decay of a three-prolonged fractional axial domain wall junction 
in the chiral limit. 
The junction is unstable to decay three non-Abelian axial vortices which are pulled by three semi-infinitely long 
fractional axial domain walls. 
The figures (the potential energy density in the left panel and the total energy density in the right panel) 
in the second line are snap shots at an instance. These figures clearly show three domain walls teared off from the
junction point. An initial configuration is the junction solution given in Fig.~\ref{fig:three_dw_junction}.}
\label{fig:three_dw_junction2}
\end{center}
\end{figure}
\begin{figure}[ht]
\begin{center}
\includegraphics[height=6cm]{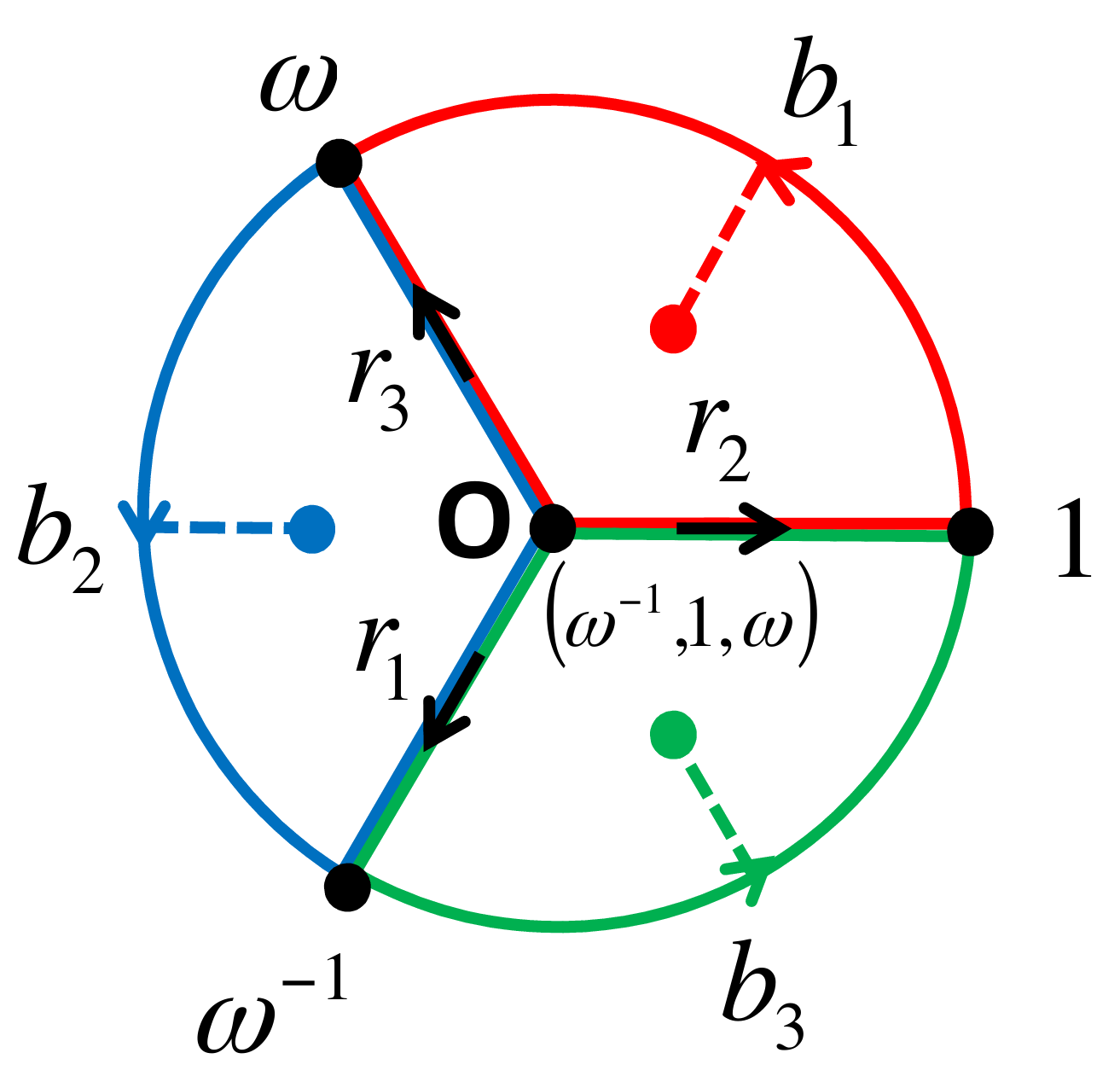}
\caption{
Classical decay of an axial domain wall junction.
The three ground states are $\Sigma ={\bf 1}_3$, $\omega {\bf 1}_3$, and $\omega^{-1} {\bf 1}_3 = \omega^2 {\bf 1}_3$, 
and the origin O is $\Sigma = \diag (\omega^{-1},1,\omega)$. 
The Abelian axial vortex initially located at the origin O decays 
int to three non-Abelian axial vortices, denoted by 
the red, green, blue dots. 
The three fractional axial domain walls 
 denoted by the red, blue and green dotted lines 
initially separate 
${\bf 1}_3$ and $\omega {\bf 1}_3$, $\omega {\bf 1}_3$ and 
$\omega^{-1} {\bf 1}_3$, and $\omega^{-1} {\bf 1}_3$ and 
${\bf 1}_3$,  respectively. 
The $b_1$, $b_2$ and $b_3$ are the paths with angles $2\pi/3$ 
at the boundary at the spatial infinity, 
and $r_1$, $r_2$, $r_3$ denote the paths from the origin O to spatial 
infinities. 
For $b_1$, $b_2$ and $b_3$, 
the $U(1)_{\rm A}$ phase is rotated by 
$\exp [i \theta \diag (1,1,1)]$ with the angle $\theta$ 
of the polar coordinates. 
Along the paths $r_1$, $r_2$ and $r_3$,
$SU(3)_{\rm L-R}$ group transformations occur 
$\exp [i u(r) \diag(0,-1,1)]$, 
$\exp [i u(r) \diag(1,0,-1)]$ and 
$\exp [i u(r) \diag(-1,1,0)]$, respectively, 
with a monotonically increasing function $u(r)$ 
with the boundary conditions 
$u(r=0)=0$ and $u(r=\infty)=2\pi/3$. 
}
\label{fig:wall-junction-decay}
\end{center}
\end{figure}

Let us point out that 
an axial vortex attached by three fractional axial 
domain walls 
decays into a set of three non-Abelian axial vortices, 
each of which is attached by a fractional axial domain wall \cite{Eto:2013bxa}. 
Let us remember that, in the absence of the instanton induced potential,
an axial vortex can be separated into three non-Abelian axial vortices without binding force at the leading order as 
in Eq.~(\ref{eq:U(1)A-decay}), 
since the force in Eq.~(\ref{eq:inter-vortex-force-global}) 
vanishes at $\alpha=\pi$. 
Therefore, the configuration of 
the single Abelian axial vortex 
attached by three domain walls is unstable to decay, 
as illustrated in the right of Fig.~\ref{fig:three_dw_junction2},
because each non-Abelian axial vortex is pulled by the tension of 
a semi-infinitely long domain wall.   
The $\U(1)_{\rm A}$ phase changes by $2\pi/3$ 
around each of the non-Abelian axial vortices
attached by fractional axial domain walls. 
We show a detailed configuration of this decaying configuration in Fig.~\ref{fig:wall-junction-decay}.
The Abelian axial vortex initially located at the origin O decays 
into three non-Abelian axial vortices, denoted by 
the red, green and blue dots. 
The three fractional axial domain walls 
 denoted by the red, blue and green dotted lines 
initially separate 
$\Sig \sim {\bf 1}_3$ and $\omega {\bf 1}_3$, $\omega {\bf 1}_3$ and 
$\omega^{-1} {\bf 1}_3$, and $\omega^{-1} {\bf 1}_3$ and 
${\bf 1}_3$,  respectively. 
The red, blue and green non-Abelian axial vortices 
are encircled by the paths
\beq
 b_1 - r_3 + r_2, \quad 
 b_2 - r_1 + r_3, \quad 
 b_3 - r_2 + r_1 \label{eq:paths2}
\eeq
respectively, 
as in Eq.~(\ref{eq:paths}) 
for non-Abelian semi-superfluid vortices. 
At the boundary of the spatial infinity, 
the $U(1)_{\rm A}$ phase is rotated by 
$\exp [i \theta \diag (1,1,1)]$ with the angle $\theta$ 
of the polar coordinates from the origin O.  
Therefore,  
the $U(1)_{\rm A}$ phase is rotated 
by $2\pi/3$ along each of the paths 
$b_1$, $b_2$, and $b_3$. 
Let us suppose that the three paths in Eq.~(\ref{eq:paths2}) 
enclose the three configurations in Eq.~(\ref{eq:ansz_ngv}), respectively. Then, we find that 
the transformations $g(r) \in SU(3)_{\rm L-R}$ occur 
along the paths $r_1$, $r_2$, and $r_3$ as 
\beq 
r_1: && g(r) 
= \exp [i u(r) \diag(0,-1,1)]
=\bigg\{\begin{array}{c}
  \diag (1,1,1), \quad r=0 \cr 
  \diag (1,\omega^{-1},\omega) , \quad r=\infty
\end{array}, \non
r_2: && g(r) 
= \exp [i u(r) \diag(1,0,-1)]
=\bigg\{\begin{array}{c}
  \diag (1,1,1), \quad r=0 \cr 
  \diag (\omega,1,\omega^{-1}) , \quad r=\infty
\end{array}, \label{eq:decay-path2}\\
r_3: && 
g(r) = \exp [i u(r) \diag(-1,1,0)]
=\bigg\{\begin{array}{c}
  \diag (1,1,1), \quad r=0 \cr 
  \diag (\omega^{-1},\omega,1) , \quad r=\infty
\end{array}, \nonumber
\eeq
respectively, 
as in Eq.~(\ref{eq:path-decay}) for non-Abelian 
semi-superfluid vortices, 
where  $u(r)$ is a monotonically increasing function  
with the boundary conditions 
$u(r=0)=0$ and $u(r=\infty)=2\pi/3$. 
We find that the origin O is consistently given by 
\beq
 \Sigma = v \diag (\omega^{-1},1,\omega).
\eeq 
From a symmetry, permutations of each component 
are equally possible as the case of decay 
of a $U(1)_{\rm B}$ vortex.

The M$_2$ non-Abelian axial vortex in Fig.~\ref{fig:NAvortex-wall2}
also decays into two non-Abelian axial vortices 
for the same reason.  

Note that there is a sharp contrast to the axion string. 
Though an axion string in the $N=3$ axion model 
also gets attached by three domain walls, 
the domain walls cannot tear off the axion string into three fractional strings \cite{Vilenkin:1994}.

\subsubsection{Abelian axial vortices attached by 
a composite axial domain wall with massive quarks}
\label{sec:composite-nonchiral}

\begin{figure}[ht]
\begin{center}
\includegraphics[width=12cm]{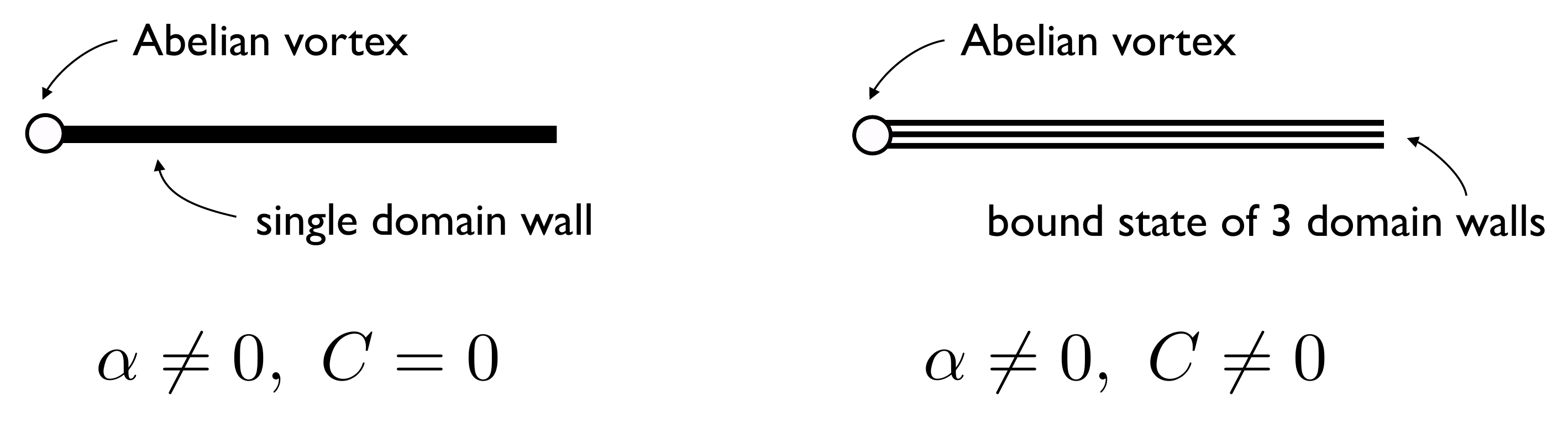}
\caption{
An Abelian axial vortex attached by axial domain walls for the effective potential $V_{\eta'} = -\alpha \cos\varphi_A - C \cos3\varphi_A$
with $\alpha \equiv 6Am + 12 B m^2$.
When $C = 0$, a fat single axial domain wall attaches to an axial vortex. When both $\alpha$ and $C$ are not zeros,
three thin domain walls are bounded 
as in Fig.~\ref{fig:multi_sG}
and attach to a vortex. 
The case with $\alpha = 0$ and $C \neq 0$ is 
shown in Fig.~\ref{fig:three_dw_junction}.
}
\label{fig:vw}
\end{center}
\end{figure}

What happens when we include an effect caused by the quark masses? The dominant contribution to the potential for the
$\eta'$ meson in the high density limit is $\sim  \cos \varphi_{\rm A}$, as in Eq.~(\ref{eq:vinst_mass}). 
One interesting consequence is that this potential does not allow the existence of a non-Abelian axial vortex alone. 
For the single-valuedness of $\ph_{\rm A}$,
a set of three non-Abelian axial vortices must appear at the same time 
as an integer Abelian axial vortex, 
and 
a single integer axial domain wall attaches to it, 
as illustrated in the left panel of Fig.~\ref{fig:vw}. 

When both $\cos\varphi_{\rm A}$ and $\cos3\varphi_{\rm A}$ exist, 
there appears an integer axial domain wall 
as a composite of three fractional axial domain walls, 
as discussed in Sect.~\ref{sec:domain-wall-massive}. 
Again, only an Abelian axial vortex can exist,  
which is attached by an integer axial domain wall 
as a composite of three fractional axial domain walls,
as illustrated in the right panel of Fig.~\ref{fig:vw}.

\subsection{Quantum decays of axial domain walls}\label{sec:wall-decay}

\subsubsection{Decay of integer axial domain walls} 
\label{sec:integer-SG-decay}

\begin{figure}[ht]
\begin{center}
\includegraphics[height=7cm]{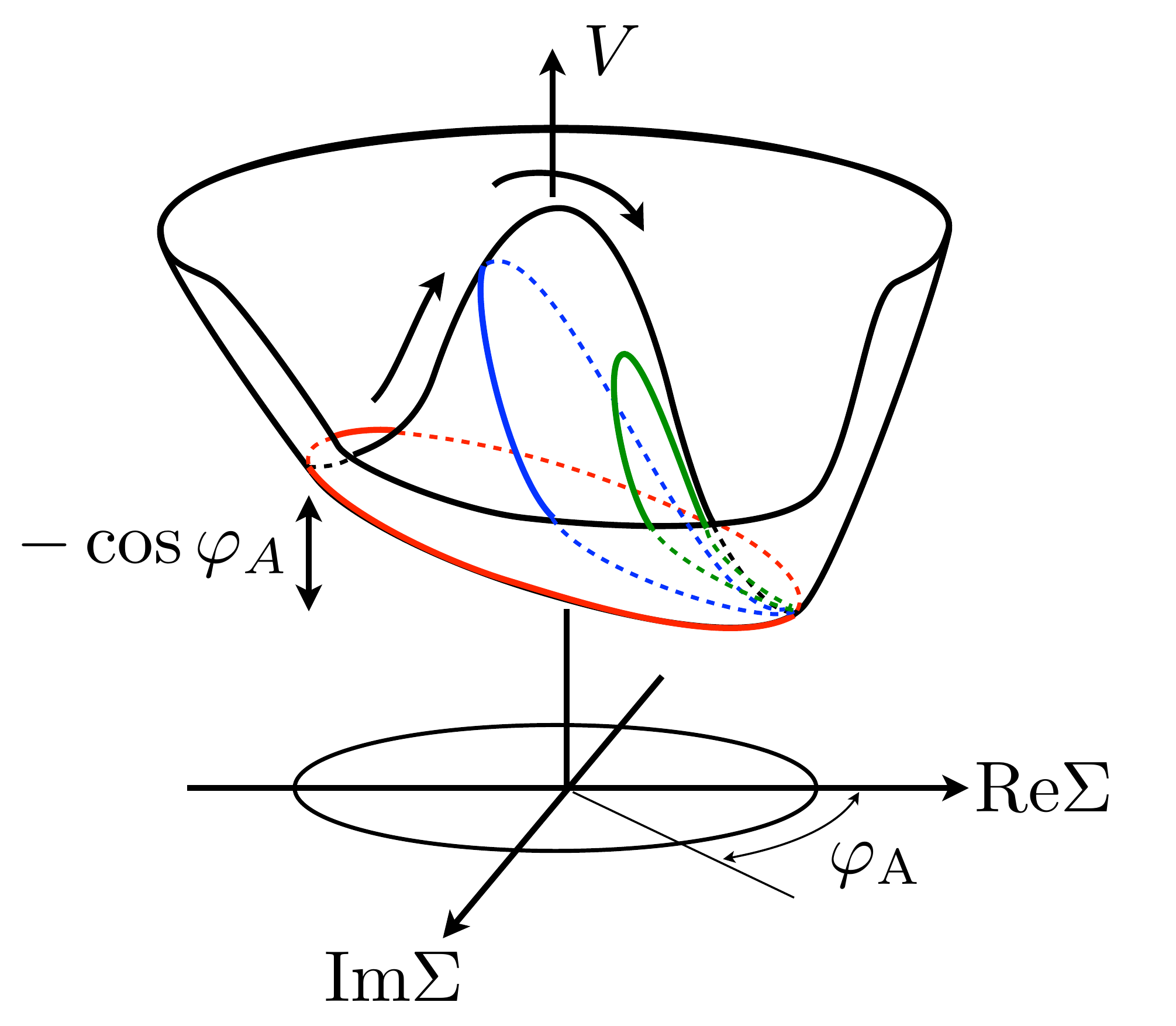}
\caption{\label{fig:decay_potential}
Unwinding an integer axial domain wall. 
The potential is composed of a Mexican hat potential 
plus a linear potential. 
A metastable axial domain wall is represented by a red loop. 
This can be unwound through blue and green loops by the quantum tunneling. 
}
\begin{tabular}{cc}
\includegraphics[width=6cm]{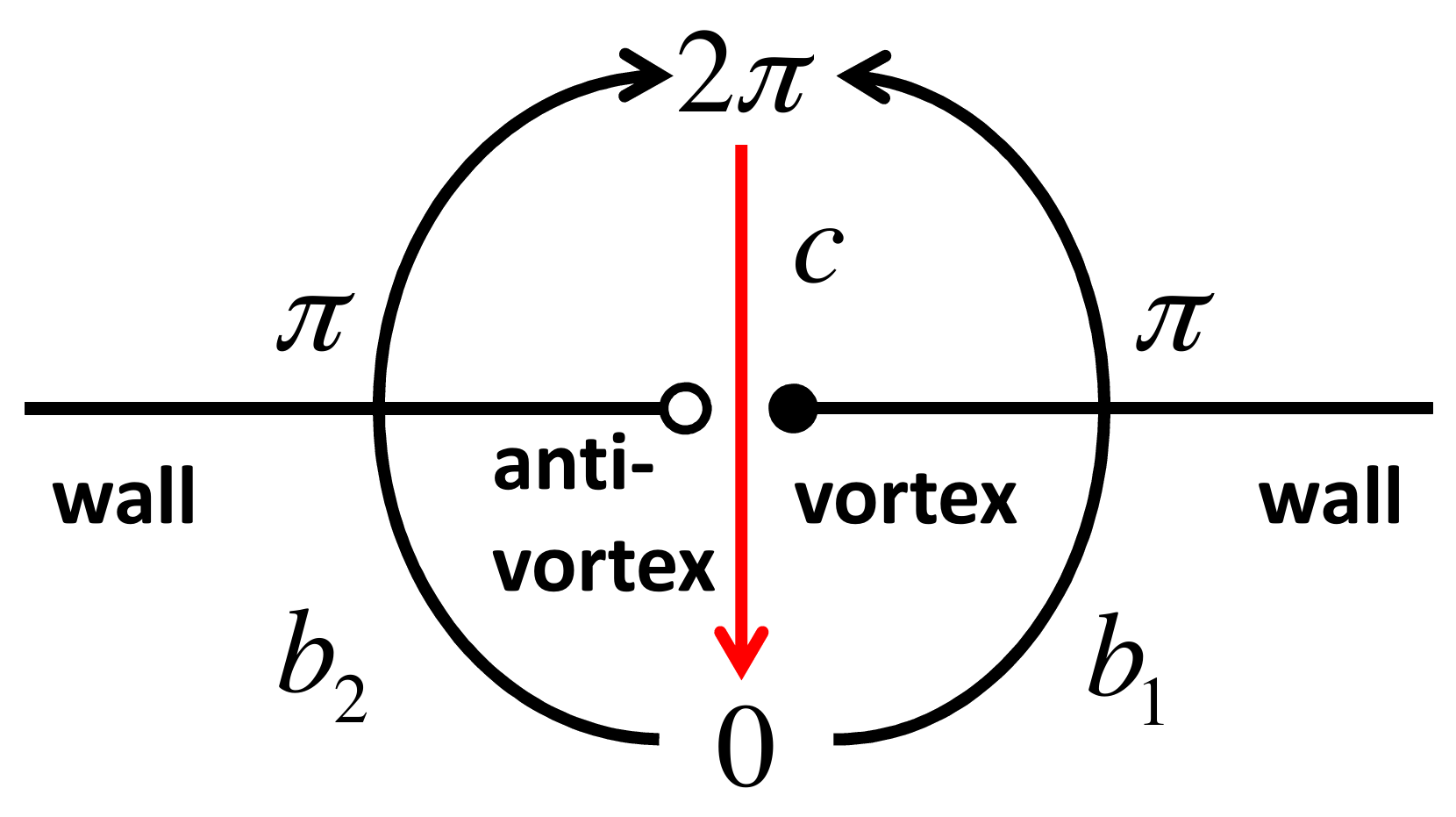}&
\includegraphics[width=5cm]{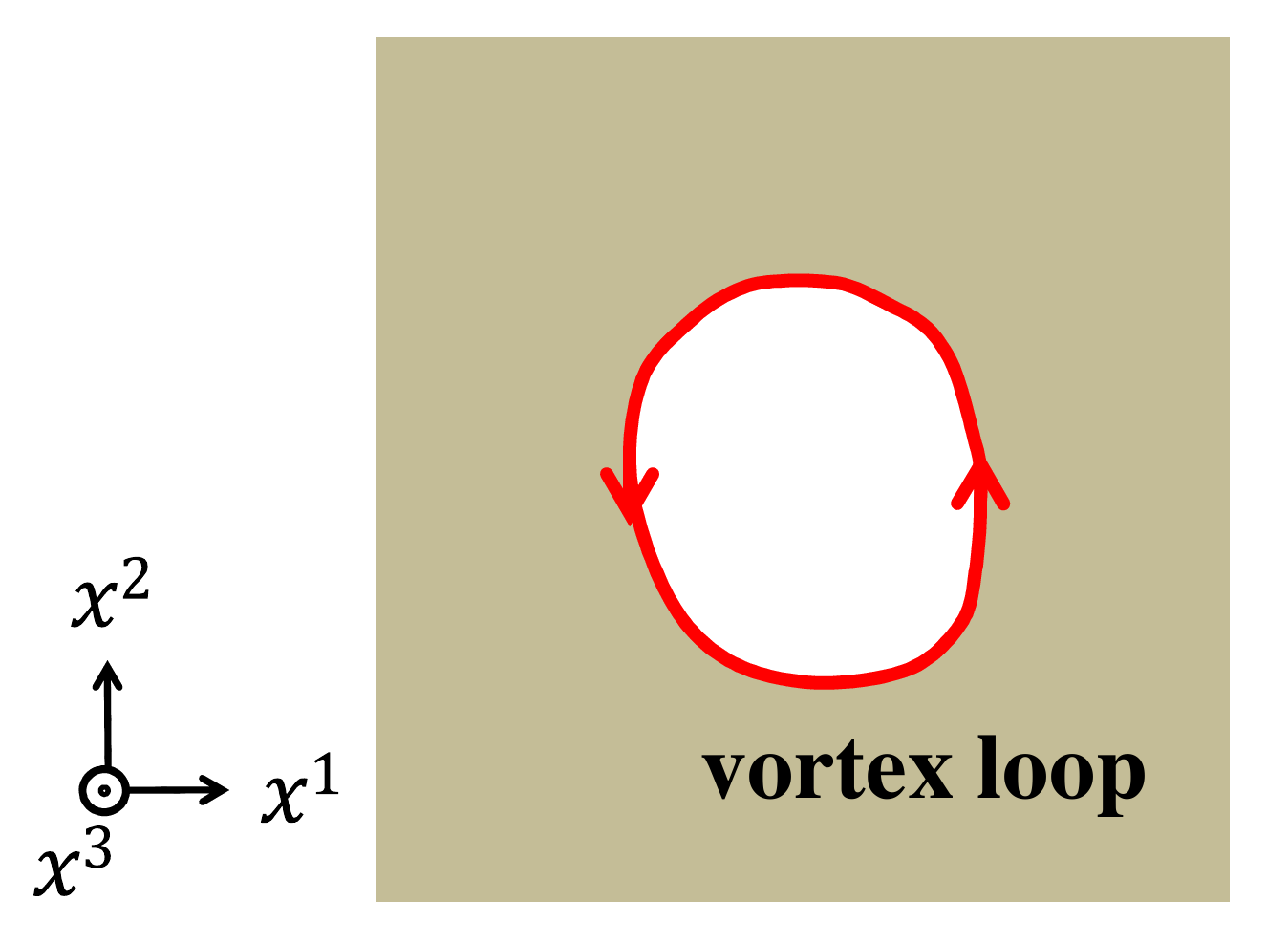}\\
(a) &(b)
\end{tabular}
\caption{Quantum decay of an integer axial domain wall. 
(a) A hole bounded by a pair of an Abelian axial vortex and an Abelian 
axial anti-vortex is created in $d=2+1$. 
The numbers denote the phase of $\ph_{\rm A}$. 
(b) A two-dimensional hole bounded by a closed Abelian axial vortex 
is created in $d=3+1$. \label{fig:integer-wall-decay}
}
\end{center}
\end{figure}

In the $U(1)_{\rm A}$ model with the 
potential $V \sim \cos \ph_{\rm A}$, 
the minimum domain wall is an integer axial 
domain wall interpolating between 
$\ph_{\rm A} = 0$ and $\ph_{\rm A} = 2\pi$, 
given in Eq.~(\ref{eq:wall1}). 
While this domain wall is classically stable \cite{Buckley:2001bm},
it is metastable if one takes into account the quantum tunneling effect, 
and it can decay into the ground state 
quantum mechanically (or thermally), 
as illustrated in Fig.~\ref{fig:decay_potential}. 
This is also discussed in Ref.~\cite{Bachas:1994ds},
where it is called as a ribbon soliton.
Since the axial domain wall separates $\ph_{\rm A}=0$ and 
 $\ph_{\rm A}=2\pi$, which are equivalent, 
a hole can be created as in Fig.~\ref{fig:integer-wall-decay}(a), 
where  
the order parameter remains the same 
along the path $c$ through the hole.
The $U(1)_{\rm A}$ phase is rotated counterclockwise 
along the closed counterclockwise loop $b_1+c$;
it encloses an Abelian axial vortex, denoted by the black point 
in Fig.~\ref{fig:integer-wall-decay}(a). 
On the other hand, 
the $U(1)_{\rm A}$ phase is rotated 
clockwise along
the counter-clockwise loop $-c-b_2$,    
which encloses a $U(1)_{\rm}$ anti-vortex, 
denoted by the white point 
in Fig.~\ref{fig:integer-wall-decay}(a).  
Therefore, the hole is  bounded by a pair of a vortex and an anti-vortex. 
In $d=3+1$ dimensions,  a two dimensional hole bounded by 
a closed Abelian axial vortex loop is created, 
as illustrated in Fig.~\ref{fig:integer-wall-decay}(b).  
The energy of the axial domain wall turns to the radiation of 
the $\eta'$ mesons. 

Let us calculate the decay probability of axial domain walls. 
Similar calculations were done for 
the $U(1)_{\rm A}$ axial domain wall in the 2SC phase in Ref.~\cite{Son:2000fh} and an application of their
calculations to the CFL phase was also pointed out there.
Once the hole is created on the integer axial wall, it will expand if the size of this hole 
is larger than a critical value, and the axial domain wall decays. 
The quantum tunneling probability of this
process can be calculated following Ref.~\cite{Preskill:1992ck}. 
Let $R$ be the initial radius of the hole created on the axial domain wall.
Then, the bounce action of this tunneling process is 
\beq
B = 4\pi R^2 T_{\rm v} - \frac{4}{3}\pi R^3 T_{\rm w},
\label{eq:decay-prob1}
\eeq
where $T_{\rm v}$ and $T_{\rm w}$ stand for the tensions of the vortex and the axial domain wall, respectively.
The critical radius $R_{\rm c}$ is the one that minimizes this bounce action, given by 
\beq
R_{\rm c} = \frac{2T_{\rm v}}{T_{\rm w}} .
\eeq
Thus, the decay probability is
\beq
P \sim e^{-B}\big|_{R=R_{\rm c}} = \exp\left(- \frac{16\pi}{3}\frac{T_{\rm v}^3}{v_{\eta'}T_{\rm w}^2}\right).\label{eq:decay-prob2}
\eeq
The factor $1/v_{\eta'}$ in the exponent reflects the fact that $v_{\eta'}$ is the speed of the modes in this system \cite{Son:2000fh}. 
Since we have $T_{\rm v} \sim f_{\eta'}^2\log L/\Delta_{\rm CFL} \sim \mu^2 \log L/\Delta_{\rm CFL}$ and
$T_{\rm w} \sim \sqrt{B} f_{\eta'}m \sim \Delta_{\rm CFL} \mu m$, the decay probability is roughly estimated as
$P \sim \exp[-(\mu\log L/\Delta_{\rm CFL})^4 / (\Delta_{\rm CFL}\mu)^2]$. 
Here, $L$ is an IR cutoff scale and we choose $\Delta_{\rm CFL}$ as a UV cutoff. 
Thus, at the high baryon density limit, the decay probability becomes parametrically small and
the integer axial domain walls have  a long lifetime.

\subsubsection{Decay of fractional axial domain walls}\label{sec:fractional-SG-decay}
The $U(1)_{\rm A}$ model with the 
potential $V \sim \cos 3 \ph_{\rm A}$ 
allows 
the ground states $\ph_{\rm A} = 0$, 
$\ph_{\rm A} = 2\pi/3$ and $\ph_{\rm A} = 4\pi/3$, 
and  
the minimum domain wall is a fractional axial 
domain wall interpolating between two of them, say 
$\ph_{\rm A} = 0$ and $\ph_{\rm A} = 2\pi/3$,  
as given in Eq.~(\ref{eq:fractional-SG}). 
As described in Sect.~\ref{sec:fractional-SG}
minimum axion domain walls in the $N=3$ axion models are stable
because they connect the vacua with different phases 
$\ph_{\rm A}$.

\begin{figure}[ht]
\begin{center}
\includegraphics[width=12cm]{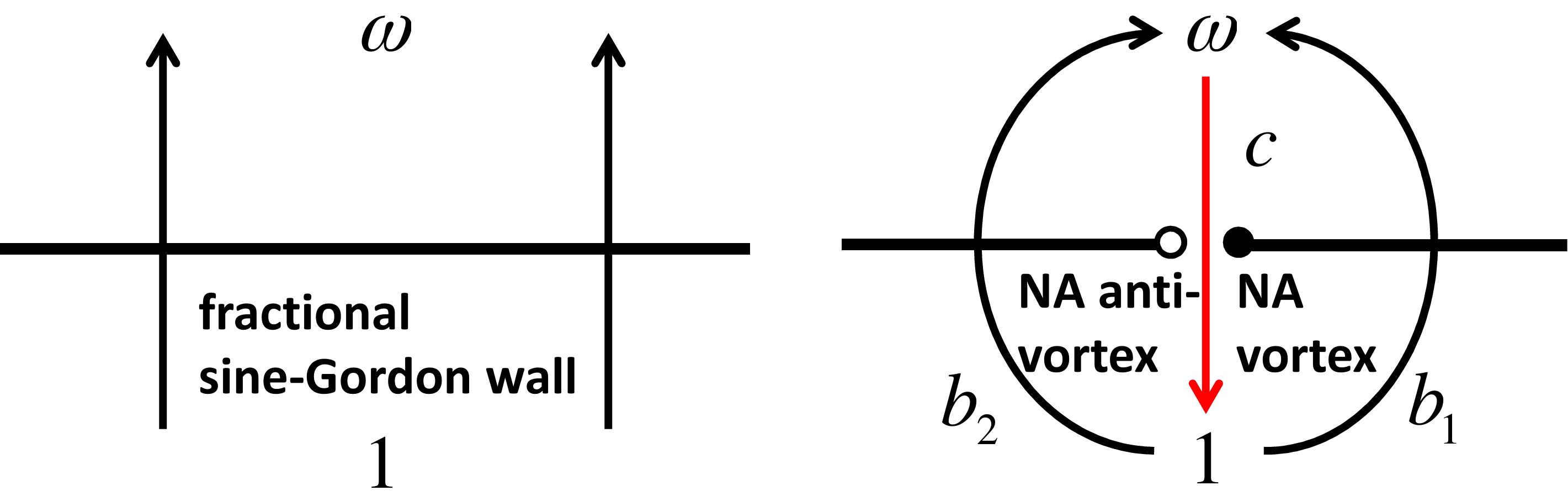}
\caption{Quantum decay of a fractional axial domain wall. 
A pair of a non-Abelian axial vortex and a non-Abelian 
axial anti-vortex is created. 
}
\label{fig:wall-decay}
\end{center}
\end{figure}

On the other hand, 
fractional axial domain walls in the CFL phase also 
interpolate between two different ground states, say 
$\ph_{\rm A} = 0$ and $\ph_{\rm A} = 2\pi/3$, 
separated by the potential term 
$V \sim \cos 3 \ph_{\rm A}$ for the phase $\ph_{\rm A}$. 
However, they are not stable, unlike the case of 
the $N=3$ axion model, but they are  
metastable and can decay quantum mechanically 
or thermally \cite{Eto:2013bxa}. 
The point is that these ground states can be connected 
by a path in the $SU(3)_{\rm L-R}$ group 
without a potential
inside the whole order parameter space 
$U(3)_{\rm L-R+A} 
= [SU(3)_{\rm L-R} \times U(1)_{\rm A}]/{\mathbb Z}_3$. 
To see this, we note that 
the ground states $\ph_{\rm A} = 0$, 
$2\pi/3$, and $4\pi/3$ 
are $\Sigma \sim {\bf 1}_3$,  $\omega {\bf 1}_3$ 
and  $\omega^2 {\bf 1}_3 = \omega^{-1} {\bf 1}_3$, respectively,  
with $\omega = e^{ 2 \pi i/3}$. 
Let us first consider $d=2+1$ dimensions for simplicity. 
For example, let us consider an axial domain wall interpolating 
between $\Sigma \sim {\bf 1}_3$ and 
$\Sigma \sim \omega {\bf 1}_3$ 
as in the left panel of Fig.~\ref{fig:wall-decay}. 
This wall can decay by creating 
the path $c$ in the right panel of 
 Fig.~\ref{fig:wall-decay}, 
along which the two ground states $1$ and $\omega$ 
are connected by 
\beq 
 \exp \left[ {i\over 3} \left(\theta-{\pi\over 2}\right) \diag (2,-1,-1)\right] \omega  
 =
\bigg\{\begin{array}{c}
  \omega, \quad \theta = {\pi \over 2}\cr 
  1         , \quad  \theta ={3\over 2} \pi
\end{array}
\quad
({\pi \over 2} \leq \theta \leq {3 \over 2}\pi) 
\eeq
in the $SU(3)_{\rm L-R}$ group.  
Here $\theta$ represents the angle from the black point. 
Then, one finds that the counter-clockwise loop $b_1+c$ 
encloses a non-Abelian axial vortex 
of the type diag. $(e^{i\theta},1,1)$
(represented by the black point).
This is nothing but the configuration in Fig.~\ref{fig:NAvortex-wall}.  
The clockwise closed loop $-b_2+c$ also encloses a non-Abelian axial vortex (denoted by the white point), 
which implies that it is a non-Abelian axial anti-vortex. 
Therefore, a hole bounded by a pair of  
a non-Abelian axial vortex and a non-Abelian axial anti-vortex
is created.

In $d=3+1$ dimensions, a two dimensional hole bounded by 
a closed non-Abelian axial vortex loop is created as the Abelian case. 
When one deforms the path $b_1$ to $-c$ in Fig.~\ref{fig:wall-decay}, 
one must create a non-Abelian vortex, 
implying an energy barrier between these two paths.
Therefore, 
the calculation of the decaying probability is the same 
as Eqs.~(\ref{eq:decay-prob1})--(\ref{eq:decay-prob2})
in the case of an integer axial domain wall,  
with replacing the tension of vortices and domain walls 
with non-Abelian ones.
Through this decaying process, the domain wall energy turns to 
radiation of the $U(3)_{\rm L-R+A}$ Nambu-Goldstone modes 
(the $\eta'$ meson and the CFL pions).

We conclude this subsection by noting that, 
while 
integer axial domain walls can decay 
without $SU(3)_{\rm L-R}$ degrees of freedom, 
fractional axial domain walls 
can decay once $SU(3)_{\rm L-R}$ degrees of freedom 
are taken into account, 
unlike the case of the $N=3$ axion model, 
in which domain walls are stable. 

\subsection{Quantum anomalies and transport effects on topological defects} \label{sec:anomaly}

\begin{figure}[ht]
\begin{center}
\includegraphics[width=12cm]{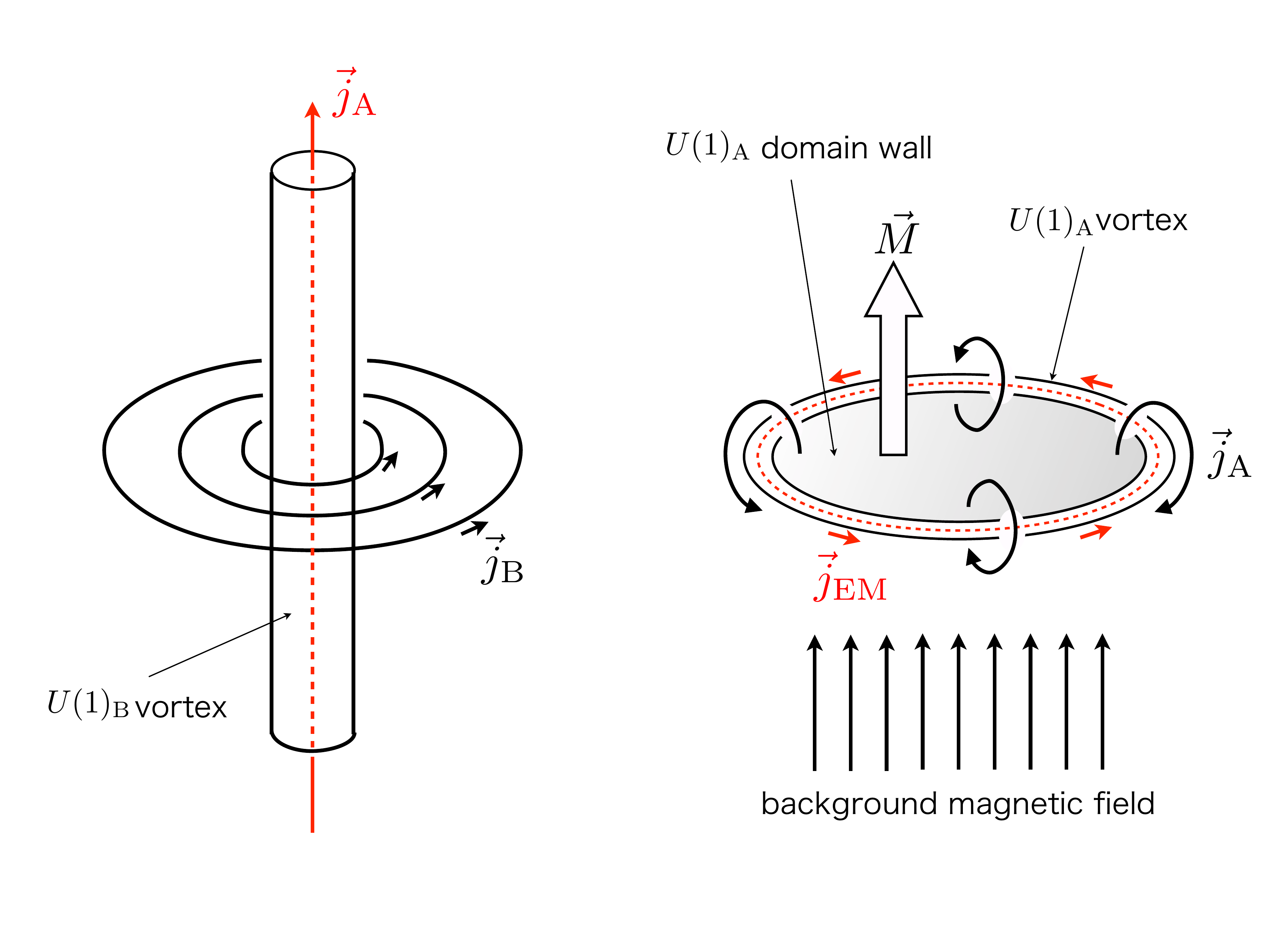}
\caption{Schematic pictures for the anomaly effects to a  $U(1)_{\rm B}$ superfluid vortex,
an axial domain wall and an axial vortex (drum vorton)\cite{Son:2004tq}.}
\label{fig:anomalous}
\end{center}
\end{figure}

It is well known that the axial $U(1)_{\rm A}$ symmetry is explicitly broken by
the quantum anomaly in the QCD vacuum. Even at finite density, it was
found that the anomaly  
plays important roles, especially in the presence of topological solitons \cite{Son:2004tq,Son:2007ny}. 
Because the effective theory has to reproduce the anomaly relation of QCD, the effective theory given
in Sect.~\ref{sec:CFL_meson} should be corrected. The additional terms were found in Ref.~\cite{Son:2004tq} as
\beq
\mathcal L_{\rm anom} &=& \frac{1}{16\pi^2}\p_\mu \varphi_{\rm A} \bigg[e^2 C_{{\rm A}\gamma\gamma}A_\nu \tilde F^{\mu\nu}
-2 e C_{{\rm AB}\gamma}\left(\mu n_\nu-\frac{1}{2}\p_\nu\varphi_{\rm B}\right)\tilde F^{\mu\nu} \non
&-& \frac{1}{2}C_{\rm ABB}\epsilon^{\mu\nu\alpha\beta}\left(\mu n_\nu - \frac{1}{2}\p_\nu \varphi_{\rm B}
\right)\p_\alpha\p_\beta\varphi_{\rm B}\bigg],
\eeq
with $n_\nu = (\mu,0,0,0)$. The first term describes two-photon decay like $\eta' \to 2\gamma$.
For the CFL phase, the coefficients were found as
\beq
C_{\rm A\gamma\gamma} = \frac{4}{3},\quad
C_{\rm AB\gamma } = \frac{2}{3},\quad
C_{\rm ABB} = \frac{1}{3}.
\eeq
Although the anomalous terms do not contribute to the equations of motion since they are total derivatives,
they give extraordinary effects if a topological soliton exist. According to Ref.~\cite{Son:2004tq}, here
we review two phenomena below. The first anomalous phenomenon is axial current on a $U(1)_{\rm B}$ superfluid vortex.
Let us consider a straight $U(1)_{\rm B}$ superfluid vortex extending along the $z$-axis. Then, rotation of $U(1)_{\rm B}$ phase is
$j_{\rm B}^3 = (\p_x\p_y - \p_y\p_x) \varphi_{\rm B} = 2\pi \delta(x)\delta(y)$; see Eqs.~(\ref{eq:U(1)Bsuperflow})
and (\ref{eq:circulation-U1}). 
Therefore, in the presence of $U(1)_{\rm B}$ superfluid vortex,
the anomaly term reduces to
\beq
\mathcal L_{\rm anom} = \frac{\mu}{12\pi}\int dtdz\, \p_z\varphi_{\rm A}.
\eeq
From Noether's theorem, this anomalous term gives the non-zero axial current
\beq
j_\mu^{\rm A} = \frac{\mu}{12\pi}(0,0,0,1).
\eeq
Thus, there is an axial current running on the superfluid vortex; see Fig.~\ref{fig:anomalous}.
The second anomalous phenomenon is magnetization of axial domain walls under a background magnetic field $\vec B$.
Let us consider an axial domain wall perpendicular to the $z$-axis. Since the $U(1)_{\rm A}$ phase $\varphi_{\rm A}$ changes
from $0$ to $2\pi$ along the $z$-axis, we have non-zero $\p_z \varphi_{\rm A}$. Then the anomaly term reduces to
\beq
\mathcal L_{\rm anom} = \frac{e\mu}{12\pi^2} \vec B \cdot \vec\nabla \varphi_{\rm A}.
\eeq
This equation implies that the axial domain wall is magnetized with a finite magnetic moment per unit area 
equal to 
\beq
M = \frac{e\mu}{6\pi}. 
\label{eq:magnetic_moment}
\eeq
The magnetic moment is oriented perpendicular to the axial domain wall.

This phenomenon can be understood in a different way. 
Since an axial domain wall is meta-stable, 
it is bounded by an axial vortex, see Fig.~\ref{fig:anomalous}. 
It is called a drum vorton \cite{Carter:2002te} which was firstly found in a nonlinear sigma model
at finite temperature.
Since $\epsilon_{ijk}\p_j\p_k \varphi_{\rm A} = 2\pi \delta^2(x_\perp)$ on the axial vortex,
the anomalous action can be written as a line integral along the vortex as
\beq
S_{\rm anom} = \frac{e\mu}{3\pi}\int d\vec\ell\cdot \vec A.
\eeq
This means that an electric current runs along the core of the axial vortex equal to
\beq
j^{\rm EM} = \frac{e\mu}{3\pi}.
\eeq
When the axial vortex surrounds an axial domain wall, it gives rise to a magnetic moment per unit area equal to
$j^{\rm EM}/2 = e\mu/(6\pi)$, which is exactly the same as Eq.~(\ref{eq:magnetic_moment}).

These anomalous effects were further investigated in
Ref.~\cite{Son:2007ny}, in which it was found that 
a strong magnetic field of order $10^{17}$-$10^{18}$ [G] transforms the CFL phase into
a new phase containing multiple $\eta$ (or $\eta'$) domain walls, which are magnetized due to the anomalous effects. 
It was also proposed in Ref.~\cite{Son:2007ny} 
that  enormously strong magnetic fields of magnetars ($10^{14 \sim 15}$ [G]) are due to 
ferromagnetism (spontaneous magnetization) by the non-zero gradient of the chiral field in the meson current 
(Goldstone boson current) phase \cite{Kryjevski:2005qq,Schafer:2005ym}. 
There are also a number of works studying anomalous phenomena.
The so-called chiral magnetic effects \cite{Kharzeev:2007jp,Fukushima:2008xe} are similar phenomena that occur
when a system with a non-zero chiral chemical potential 
is under a strong magnetic field.
Electric charge separation of Q-balls by the chiral magnetic effect was studied in Ref.~\cite{Eto:2010vi}, and
seemingly anomaly induced electric charges on a baryon as a Skyrmion under a strong background magnetic field
were discussed in Refs.~\cite{Eto:2011id,Kharzeev:2011sq,Wakamatsu:2011xd}.
It was found in Ref.~\cite{Gorsky:2010dr}
that fermions escape during the decay process of metastable defects through the anomaly term. 
It was found 
for the low-density nuclear matter 
that a stack of magnetized pionic domain walls is energetically more favorable
than nuclear matter \cite{Son:2007ny}. The relation between the ferromagnetism of neutron stars and a stack of 
magnetized pionic walls was also discussed in Ref.~\cite{Eto:2012qd}.
It has also been proposed that an anomalous global vortex may generate primordial magnetic fields in galaxies \cite{Brandenberger:1998ew}.
\subsection{Skyrmions as qualitons} \label{sec:skyrmion}

As discussed in Sect.~\ref{sec:CFL_meson}, 
there are Nambu-Goldstone boson modes (CFL mesons),
namely $SU(3)_{\mathrm{C+L+R}}$ octet and singlet modes, in the ground state of the CFL phase.
The nonlinear realization of the Nambu-Goldstone bosons gives the effective Lagrangian whose forms are restricted by $SU(3)_{\mathrm{C+L+R}}$ symmetry.
The simplest form of the effective Lagrangian for the pion field is
constituted by the kinetic term, and the four-derivative term (the
Skyrme term), and the Wess-Zumino-Witten term from the chiral anomaly:
\begin{eqnarray}
{\mathcal L}_{\mathrm{eff}} &=& 
\frac{F^2}{4} \mathrm{Tr} \left( \D_{0}\Phi_{\mathrm L} \D_{0}\Phi_{\mathrm L}^{\dag} \right) 
- \frac{F'{}^2}{4} \mathrm{Tr} \left( \D_{i}\Phi_{\mathrm L} \D_{i}\Phi_{\mathrm L}^{\dag} \right)
+ {\mathcal L}_{\mathrm{Skyrme}}+ n_{\mathrm L} {\mathcal L}_{\mathrm{WZW}} \non
&&+ (\mathrm{L} \leftrightarrow \mathrm{R}) + \cdots, 
\end{eqnarray}
where $\Phi_{\rm L,R}$ is restricted by $\left|\det \Phi_{\rm L,R}\right| = 1$,
$n_{\mathrm L}=1$, and the ellipsis denotes the terms with the higher order derivative.\footnote{The Skyrme term which stabilizes the hedgehog solution was given in Ref.~\cite{Jackson:2003dk}.}
$F$ is the ``pion decay constant" in the CFL phase.
Because of the nontrivial third homotopy group 
of the order parameter manifold, 
\beq 
  \pi_{3}[U(3)_{\rm L-R+A}] \simeq {\mathbb Z}, \label{eq:skyrme}
\eeq 
one expects the existence of Skrmions 
\cite{Skyrme:1962vh,Skyrme:1961vq}
as three-dimensional textures, 
as explained in footnote \ref{footnote:texture}.   
As the solution, there is a ``hedgehog" configuration as the most stable (soliton) state from the energy balance between the kinetic term and the Skyrme term \cite{Hong:1999dk}.
The Wess-Zumino-Witten term guarantees the correct quantization of the soliton as a spin 1/2 object. 
The winding number in Eq.~(\ref{eq:skyrme}) carries the baryon charge $(1 \,{\mathrm{mod}\,2})/3$. 
We note that $U(1)_{\mathrm{B}}$ symmetry for baryon number conservation in the QCD Lagrangian is broken to ${\mathbb Z}_2$ symmetry in the CFL phase, because the transformation $q \rightarrow -q$ for the quark field $q$ leaves the condensate invariant.
This stable object is called ``qualiton" (or ``superqualiton") \cite{Hong:1999dk}.
Since the low-energy constants as the coefficients in the effective Lagrangian in the CFL phase are calculable by matching with QCD in high density limit,
we obtain the hedgehog configuration and hence the mass of the qualiton.
They were given as $\alpha\, 4\pi F$ \cite{Hong:1999dk,Hong:2000ff}, with $\alpha = {\mathcal O}(0.1-1)$ and the pion decay constant $F$ in the CFL phase.
We note that the mass $\alpha\, 4\pi F$ is proportional to the chemical potential $\mu$ because $F \sim \mu$ (see Eq.~(\ref{eq:decay_const_speed_of_meson})).
However, this is different from the quark mass (\ref{eq:spectra}) in the CFL phase.
This discrepancy will be an interesting problem.
The mass of the qualiton is important, because the qualiton can be produced easily in the ground state and can cause the rearrangement of the ground state, if the mass is smaller than the gap \cite{Hong:1999dk}.

We leave a comment that the word ``qualiton'' was originally used in the quark soliton model to analyze the properties of hadrons in vacuum at zero chemical potential.
We have to be careful to strictly distinguish the ``qualiton'' in the CFL phase from that in vacuum.
The qualiton in the CFL phase corresponds rather to the Skyrmion in vacuum.
Thus we observe that the state of ``qualiton'' in the CFL phase corresponds to the baryon state in vacuum.
In fact, there is a discussion that the qualiton is identical to the quark \cite{Hong:1999dk}.
This may support the idea of the ``quark-hadron duality'', which indicates the correspondence of the low-energy modes in the CFL phase and those in vacuum \cite{Jackson:2003dk,Lee:2009dpa}.

%% file: other-v9.tex
\section{Topological objects in other phases}
\label{sec:other}

We review topological solitons in phases other
than the CFL phase in this section.
The CFL phase discussed so far is the ground state  at high densities
where masses of the three flavors are small compared to the baryon
chemical potential.
If one lowers the density gradually, the effect of finite strange quark
mass comes in. 
Then it is expected that kaons form a condensate in addition to the
CFL condensates. 
This is called the CFL+K phase \cite{Bedaque:2001je}.
If the density is further decreased, only the light flavors (u and
d) contribute to the condensate, which is called the 2SC phase \cite{Bailin:1983bm}.
In Sect.~\ref{sec:2SC}, we discuss the domain walls and color magnetic
fluxes in the 2SC phase.
In Sect.~\ref{sec:CFLK}, we review the strings, vortons, domain
walls, and drum vortons in the CFL+K phase.

\subsection{2SC phase}\label{sec:2SC} 

\subsubsection{$U(1)_{\rm A}$ domain walls }

The CFL phase changes into the so-called 2SC phase at a lower density
\cite{Bailin:1979nh,Bailin:1983bm,Alford:1997zt,Rapp:1997zu}. Due to
asymmetry among the strange quark mass
and the masses of the up and down quarks, 
only up and down quarks form Cooper pairs in the 2SC phase. 
The 2SC pairing pattern is thus given by
\begin{equation}
(\Phi_{\rm L,R})_a^A \sim \delta^A_3\delta_a^3, 
\end{equation}
where the symmetry breaking pattern is
\beq
SU(3)_{\rm C} \times SU(2)_{\rm L} \times SU(2)_{\rm R} \times U(1)_{\rm B} \times U(1)_{\rm S} \times U(1)_{\rm A} \non
\to
SU(2)_{\rm C} \times SU(2)_{\rm L} \times SU(2)_{\rm R} \times \tilde U(1)_{\rm B} \times U(1)_{\rm S}.
\eeq
Here we assume that the up and down quarks are massless.
The axial $U(1)_{\rm A}$ symmetry is not an exact symmetry but it is
explicitly broken by instanton effects.
$\tilde U(1)_{\rm B}$ is a linear combination of the original $U(1)_{\rm B}$ and the broken $U(1)_8 \in SU(3)_{\rm C}$
generated by $T_8$.
The $U(1)_{\rm S}$ is the phase rotation of the strange quark.
The order parameter space is of the form
\beq
\mathcal M_{\rm 2SC} =
\frac{
SU(3)_{\rm C}  \times U(1)_{\rm B}
}{
SU(2)_{\rm C}  \times \tilde U(1)_{\rm B}
}  \times U(1)_{\rm A}.
\eeq

It was observed that metastable $U(1)_{\rm A}$ domain walls \cite{Son:2000fh} also exist in the 2SC phase.
Although domain walls are very familiar in field theory, it is widely believed the standard model has no domain walls.
It was pointed out that metastable domain walls could exist in QCD at zero temperature and density \cite{Forbes:2000et},
but no definite statement can be made because the theory is not under theoretical control.
In contrast, QCD at high density is weakly coupled theory due to asymptotic freedom, so that relevant physics
are under theoretical control. In Ref.~\cite{Son:2000fh}, it was found that QCD at high density, especially in the 2SC phase, 
must have domain walls.

The domain walls in the 2SC phase are similar to those explained in Sect.~\ref{sec:domain-wall-massive}.
In the 2SC phase, the chiral symmetry is not broken but the $U(1)_{\rm A}$ is spontaneously broken by the diquark
condensate. Then the relevant order parameter manifold is $U(1)_{\rm A}$.
The corresponding mode $\varphi_{\rm A}$ is a pseudo Nambu-Goldstone field, which gets a finite mass
by instanton effects. The effective Lagrangian is 
\beq
\mathcal L = f^2\left[(\p_0 \varphi_{\rm A})^2 - v^2 (\p_i\varphi_{\rm A})^2 \right] - V_{\rm inst},
\label{eq:lag_u1a_2sc}
\eeq
where the constants $f$ and $v$ were determined \cite{Beane:2000ms} as
$f^2 = \mu^2/8\pi^2$ and $v^2 = 1/3$.
$V_{\rm inst}$ stands for an instanton-induced effective potential that was calculated \cite{Son:2000fh} as
\beq
V_{\rm inst} = - a \mu^2 \Delta_{\rm CFL}^2 \cos \varphi_{\rm A}.
\eeq
The dimensionless coefficient $a$ that vanishes at $\mu \to \infty$ is given by
\beq
a = 5 \times 10^4\left(\log\frac{\mu}{\Lambda_{\rm QCD}}\right)^7 \left(\frac{\Lambda_{\rm QCD}}{\mu}\right)^{\frac{29}{3}}.
\eeq
The Lagrangian (\ref{eq:lag_u1a_2sc}) is the same form as Eq.~(\ref{eq:vinst_mass}), so that integer sine-Gordon
domain walls exist. The domain wall tension was found  as
\beq
T_{\rm w}^{\rm 2SC} = 8\sqrt{2a}vf\mu\Delta_{\rm CFL}.
\eeq
Its decay probability was also calculated \cite{Son:2000fh} as
\beq
\Gamma \sim \exp\left\{-\frac{\pi^4}{3}\frac{v^3}{a}\frac{\mu^2}{\Delta_{\rm CFL}^2} \left(\log \frac{1}{\sqrt a}\right)^3\right\}.
\eeq
Since $a$ decreases with increasing $\mu$, the decay rate is suppressed
and the domain walls are long lived \cite{Son:2000fh}.

\subsubsection{Color magnetic flux tubes}

The interesting problem of the color Aharonov-Bohm scattering of fermions off
the color magnetic flux tubes \footnote{Since they are non-topological
solitons and their stability is not ensured by the symmetry, they would be unstable against decaying. }
in the 2SC phase and forces exerted on vortices are
studied in Ref.~\cite{Alford:2010qf}.
The authors calculated the cross section for Aharonov-Bohm scattering of gapless
fermions off the flux tubes, and the associated collision time and
frictional force on a moving flux tube.
Since the CFL phase has no gapless quarks, 
a color magnetic flux tube in the 2SC phase is considered where several
massless fermions are present and the effect would be more significant
than in the CFL phase.

The Aharonov-Bohm effect turns out to lead to a strong interaction
between a charged particle and a magnetic flux tube. 
In the case of a single U(1) gauge group, the cross section per unit
length is given by \cite{Alford:1988sj}
\begin{equation}
 \frac{d \sigma}{d \theta } = 
\frac{\sin^2(\pi \tilde \beta)}
{2 \pi k \sin^2 (\theta/2)}, \quad \tilde \beta = \frac{q_p}{q_c},
\end{equation}
where $q_p$ and $q_c$ are the charges of the scattering particle and
condensed field respectively, 
$k$ is the momentum in the plane perpendicular to the vortex, 
and $\theta$ is the scattering angle.
The Aharonov-Bohm interaction has several important characteristics:
\begin{itemize}
 \item The cross section vanishes if $\tilde\beta$ is an integer.
 \item The cross section is independent of the thickness of the flux
       tube. The scattering occurs even at energies much smaller
       than the symmetry breaking scale.
 \item The cross section diverges for small $k$ and forward scattering
       $\theta \sim 0$.
\end{itemize}
Thus, it is important to determine the values of $\tilde \beta$ for
gapless fermions in the 2SC phase, which is done in Ref.~\cite{Alford:2010qf}.

The authors also calculated the characteristic timescale for a perturbation from
equilibrium to relax, due to scattering of the fermions oﬀ the color
magnetic flux tubes.
The timescale is roughly the mean free time of collisions of fermions
off fluxes, which is evaluated for a fermion species $i$ as 
\begin{equation}
 \tau_{if}^{-1} = \frac{n_v}{p_{Fi}} \sin^2 (\pi \tilde \beta_i),
\end{equation}
where $n_v$ is the vortex area density and $p_{Fi}$ is the Fermi
momentum of the fermion denoted by $i$.
The timescale are compared with the relaxation time due to Coulomb
interactions, 
\begin{equation}
\tau^{-1}_{qq} = \frac{6 \zeta (3)}{\pi^2} \tilde \alpha T
\end{equation}
where $\zeta(3) = 1.202$, $T$ is the temperature and $\tilde \alpha$ is
the fine structure constant for the rotated electromagnetism.
The temperature $T_f$ below which flux
tubes dominate the relaxation of deviations from thermal equilibrium is
determined as 
\begin{equation}
 T_f = \frac{\pi^2}{6 \zeta(3)} \frac{\sin^2 (\pi \beta_{bu})}{\tilde
  \alpha }\frac{n_v}{\mu_q},
\label{eq:tf}
\end{equation}
where $\beta_{bu}$ is the $\beta$ for up blue quark (which is most
abundant) and  $\mu_q$ is the quark chemical potential.
Below the temperature (\ref{eq:tf}) the Aharonov-Bohm scattering plays
the main role in thermal relaxations.

\subsection{CFL+K} \label{sec:CFLK}

In this section, we will review solitons in the CFL+K phase.
We will ignore instanton effects in Sect.~\ref{sec:cflK_no_instanton}
and 
take them into account in Sect.~\ref{sec:cflK_instanton}.

\subsubsection{Superconducting strings and vortons}
\label{sec:cflK_no_instanton}

The CFL phase is the ground state in the massless three-flavor limit in
high-density QCD.
If one considers a finite strange quark mass, 
it puts a stress on the ground state to reduce the strange quark
density compared to up and down quarks.
As long as the strange quark mass is small, 
it cannot overcome the diquark pairing energy and the ground state is in the
CFL phase.
At sufficiently large strange quark masses, a kaon condensate would be
formed \cite{Schafer:2000ew,Bedaque:2001je,Bedaque:2001at}, 
since kaons $K^0$ (d$\bar{\rm s}$) and $K^+$ (u$\bar{\rm s}$) are the lightest mesons \cite{Son:1999cm} that can reduce
the strange quark content of the ground state.
Kaon condensation is characterized by the spontaneous breaking of
$U(1)_{\rm Y}$ symmetry or $U(1)_{\rm EM}$ symmetry, or both.
Let us denote the kaon field as $K = (K^+,K^0)^T$ .
The field $K$ is transformed under $U(1)_{\rm Y}$ and $U(1)_{\rm EM }$ symmetry as 
\begin{equation}
\begin{pmatrix}
K^+  \\
K^0
\end{pmatrix}
\rightarrow 
e^{i \alpha}
\begin{pmatrix}
K^+  \\
K^0
\end{pmatrix}, 
\quad 
\begin{pmatrix}
K^+  \\
K^0
\end{pmatrix}
\rightarrow 
\begin{pmatrix}
e^{i \beta(x)} K^+  \\
K^0
\end{pmatrix},
\end{equation}
respectively, where $\alpha$ and $\beta(x)$ are parameters characterizing
$U(1)_{\rm Y}$ and $U(1)_{\rm EM}$ transformations.
We here consider the phase in which $U(1)_{\rm Y}$ symmetry is broken in
the bulk.
The diquark condensate take the form
\beq
\Sigma_{K^0} = \left(
\begin{array}{ccc}
1 & 0 & 0 \\
0 & \cos \theta_0 & e^{-i\psi} \sin \theta_0\\
0 & - e^{i\psi}\sin\theta_0 & \cos\theta_0
\end{array}
\right),
\eeq
where $\theta_0$ represents the strength of the $K^0$ condensation and the phase
$\psi$ corresponds to the Nambu-Goldstone mode associated with the
spontaneous breaking of the hypercharge symmetry $U(1)_{\rm Y}$.
The strength of condensation was found \cite{Bedaque:2001je} as
\beq
\cos\theta_0 = \frac{m_0^2}{\mu_{\rm eff}^2},\quad
m_0^2 = \frac{3\Delta_{\rm CFL}^2}{\pi^2 f_\pi^2}\,m_{\rm u}(m_{\rm d}+m_{\rm s}),\quad
\mu_{\rm eff} = \frac{m_{\rm s}^2}{2p_F}.
\eeq

We can expect the appearance of topological vortices which
are characterized by $U(1)_{\rm Y}$ winding numbers \cite{Forbes:2001gj,
Son:2001xd}. 
Furthermore, a particularly interesting possibility is pointed out: the
cores of $U(1)_{\rm Y}$ vortices can be electromagnetically charged and 
superconducting \cite{Kaplan:2001hh}. 
Namely, $U(1)_{\rm EM}$ symmetry is broken only inside vortices.
We here briefly review these topological vortices following the argument
given in  Ref.~\cite{Buckley:2002mx}:
Firstly, we expand the chiral Lagrangian (\ref{eq:chiral_lag}) 
to the fourth order in the lightest fields $K = (K^+,K^0)$ as
\beq
\Lag_K = |\p_0K|^2 - v^2|\p_iK|^2 - \lambda\left(|K|^2 - \frac{\eta^2}{2}\right)^2 - \delta m^2 K^\dagger \sigma_3K,
\label{eq:lag_kaon}
\eeq
with
\beq
\lambda \simeq \frac{4\mu_{\rm eff}^2 - m_0^2}{6f_\pi^2},\quad
\lambda \eta^2 = \mu_{\rm eff} - \frac{m_0^2 + m_+^2}{2},\quad
\delta m^2 = \frac{m_+^2 - m_0^2}{2},\\
m_0^2 = \frac{3\Delta_{\rm CFL}^2}{\pi^2 f_\pi^2}\,m_{\rm u}(m_{\rm d}+m_{\rm s}),\quad
m_+^2 = \frac{3\Delta_{\rm CFL}^2}{\pi^2 f_\pi^2}\,m_{\rm d}(m_{\rm s}+m_{\rm u}).
\eeq
The reduced theory (\ref{eq:lag_kaon}) has a stable topological vortex solution whose profile is
described by the standard ansatz \cite{Buckley:2002mx}
\beq
K^0(r,\theta) = \sqrt{\frac{\mu_{\rm eff}^2 - m_0^2}{2\lambda}}\, f(r) e^{i\theta},\quad
K^+(r,\theta) = \sigma g(r),
\eeq
where the profile functions $f(r)$ and $g(r)$ satisfy the boundary conditions
$f(\infty) = 1$, $f(0)=0$, $g(\infty) = 0$ and $g'(0)=0$. At the center of the vortex core,
there is a non-zero $K^+$ condensation. Since $K^+$ is charged, the vortex is superconducting string: 
A persistent electromagnetic current can flow on a $U(1)_{\rm Y}$ vortex.
In contrast, $U(1)_{\rm Y}$  symmetry is restored inside vortices  
since $K^0$ condensate vanishes.
Such vortices are dense-QCD realizations of the superconducting cosmic
strings, that have been for a long time \cite{Witten:1986qx}. 

Vortex loops that carry persistent currents are called ``vortons'' in
the context of cosmic strings \cite{Davis:1988ij}.
The vortons may exist in the CFL+K$^0$ phase \cite{Kaplan:2001hh, Buckley:2002ur,Buckley:2002mx,Bedaque:2011fg,Bedaque:2011ry}.
For the vortons, we should add a time and $z$ dependence in the phase of $K^+$ as
\beq
K^0 = K^0_{\rm string}(r,\theta),\quad
K^+ = K^+_{\rm cond}(r,\theta) e^{-i\omega t + ikz},
\eeq
where $z$ is the coordinate along the vortex loop. Substituting this into the original Lagrangian (\ref{eq:chiral_lag}) and
picking up all the terms to fourth order in the fields, one finds \cite{Buckley:2002mx}
\beq
\tilde \Lag_K &=& -v_\pi^2|\p_iK^0|^2 - v_\pi^2|\p_iK^+|^2 \non
&+& M_0^2|K^0|^2 + M_+^2|K^+|^2
- \lambda |K^0|^4 - \lambda_+ |K^+|^4 - \zeta |K^0|^2 |K^+|^2,
\eeq
with the parameters defined by
\beq
\omega_{\rm eff} = \omega + \mu_{\rm eff},\quad 
M_0^2 = \mu_{\rm eff}^2 - m_0^2,\quad 
M_+^2 = \omega_{\rm eff}^2 - v_\pi^2k^2 - m_+^2,\qquad\\
\lambda_+ = \frac{4(\omega_{\rm eff}^2-v_\pi^2k^2)-m_+^2}{6f_\pi^2},\quad
\zeta = \frac{(\omega_{\rm eff}+\mu_{\rm eff})^2+4\omega_{\rm eff}\mu_{\rm eff}-v_\pi^2k^2-m_+^2-m_0^2}{6f_\pi^2}.
\eeq
The vorton is characterized by two additional conserved changes; a topological charge and a Noether charge
\beq
N &=& \oint_C \frac{dz}{2\pi} \arg \log K^+ = k R,\\
Q &=& \int d^3x\, j_+^0 \sim R \omega_{\rm eff} S,
\eeq
with $R$ being the radius of the vorton and $S \equiv \int d^2x\, |K^+|^2$.
The stability of the vorton can be seen by finding the $R$ dependence of energy of the vorton
\beq
E_{\rm vorton}(R) \simeq \left\{\frac{2\pi^2(\mu_{\rm eff}^2 - m_0^2)}{\lambda}v^2 \log \frac{L}{\xi}\right\}R + 
\left(\frac{1}{2\pi S} + 2\pi v_\pi^2N^2 S\right)\frac{1}{R}.
\eeq
The existence of a global minimum implies that the vorton is stable and its  size is given by
\beq
2\pi R = \sqrt{
\frac{Q^2 + (2\pi)^2 v_\pi^2 N^2 S^2}{\pi S v^2 (\log L/\xi)(\mu_{\rm eff}^2 - m_0^2)/\lambda}}.
\eeq
Note that there are no constraints on the size of vortons.
The vortons can be any size as the cosmic string vortons.

\subsubsection{Domain walls and drum vortons}
\label{sec:cflK_instanton}

It was pointed out in Ref.~\cite{Son:2001xd} that the CFL+K phase may have metastable domain walls.
When weak interactions are taken into account, $U(1)_{\rm Y}$ symmetry is explicitly broken. So the $U(1)_{\rm Y}$ 
Nambu-Goldstone mode $\psi$ gets a small mass. This is seen in the effective theory in Eq.~(\ref{eq:lag_kaon})
by adding an effective potential \cite{Son:2001xd}
\beq
V(\psi) &=& f_\pi^2 m^2 \cos \psi,
\label{eq:eff_pot_K}\\
m^2 &=& \frac{162 \sqrt{2}\,\pi}{21-8\log 2} \frac{G_{\rm F}}{g_{\rm s}} \cos\theta_{\rm C}\sin\theta_{\rm C} m_{\rm u}m_{\rm s}\Delta_{\rm CFL}^2,
\eeq
where $G_{\rm F}$ is the Fermi constant and $\theta_{\rm C}$ is the Cabbibo angle.
Thus the effective theory for the quasi Nambu-Goldstone particle $\psi$ is the sine-Gordon model, and
there exists a meta-stable domain wall solution that we call the kaon domain wall
\beq
\psi(z) = 4\arctan e^{mz/v_\pi},
\eeq
where $z$ is a coordinate perpendicular to the domain wall.
The tension of the domain wall was calculated \cite{Son:2001xd} as
\beq
T_{\rm dw}^K = 8 v_\pi f_\pi^2 m
\eeq
It was also found that the decay rate of the kaon domain wall with respect to hole nucleation is parametrically small
at high density and zero temperature
\beq
\Gamma \sim \exp\left(-\frac{\pi^4v_\pi^3}{12}\frac{f_\pi^2}{m^2}\log \frac{m_K}{m}\right),
\eeq
with $f_\pi^2/m^2 \sim \mu^2/m^2$.

Let us next consider the superconducting vortices explained in
Sect.~\ref{sec:cflK_instanton} in 
the presence of the instanton induced potential.  As we have seen in Sect.~\ref{sec:global}, 
a kaon domain wall must be attaches to a kaon superconducting string due to the potential in
Eq.~(\ref{eq:eff_pot_K}). When a kaon domain wall decays with a hole nucleation, a kaon superconducting
string attached to the edge of the hole.

Let us next turn to kaon vortons 
in the presence of the potential (\ref{eq:eff_pot_K}).
Since any kaon superconducting strings are attached by the kaon domain walls, the kaon vortons also
have domain walls stretched across their surfaces as drums \cite{Buckley:2002mx}.
They are called drum vortons \cite{Carter:2002te}, and were first found in a nonlinear sigma model
at finite temperature.
It has been argued that the domain wall gives an upper bound on the size of vortons in contrast to the vortons
without  domain walls \cite{Buckley:2002mx}.

%% file: conclusion-v9.tex
\section{Summary and discussions}\label{sec:conclusion}

Quark matter at extremely high density 
becomes a color superconductor due to 
the formation of diquark pairings.
It exhibits superfluidity as well as 
color superconductivity because of 
the spontaneously broken baryon symmetry 
as well as color symmetry.  
When a color superconductor is rotating, 
as is the case if it is realized in the core of neutron stars, 
non-Abelian vortices are created along the rotation axis.
Non-Abelian vortices are superfluid vortices
carrying 1/3 quantized circulation and a color magnetic flux. 
Corresponding to the color of the magnetic flux, 
a non-Abelian vortex carries orientational zero modes 
${\mathbb C}P^2$.  
The properties of non-Abelian vortices have been 
studied in 
the time-dependent GL effective theory 
for high density QCD 
and the BdG equation. 
A superfluid $U(1)_{\rm B}$ vortex 
decays into a set of three non-Abelian vortices, 
because the interaction between two non-Abelian vortices 
at large distance is repulsive, 
independent of orientational modes.
Two kinds of bosonic gapless modes propagate 
along a non-Abelian vortex string. 
One is a translational or Kelvin mode with 
a quadratic dissipation 
(of the type II Nambu-Goldstone mode), 
and the other is orientational ${\mathbb C}P^2$ zero modes
with a linear dissipation 
(of the type I Nambu-Goldstone mode). 
The dynamics of these modes can be described by 
the low-energy effective theories  
on the $1+1$ dimensional vortex world-sheet, 
a free complex scalar field with 
the first derivative with respect to time 
and a ${\mathbb C}P^2$ model, respectively 
in $d=1+1$ dimensions.
The $SU(3)$ isometry of 
the ${\mathbb C}P^2$ space is exact, 
neglecting quark masses and 
electromagnetic interactions. 
The effect of strange quark mass can be taken into account as an effective potential in the ${\mathbb C}P^2$ vortex effective theory, 
which shows 
that all vortices decay into one kind immediately. 
On the other hand,
the electromagnetic interaction contributes 
a finite correction to
the tension of the non-Abelian vortex, 
which also induces an effective potential 
in the ${\mathbb C}P^2$ vortex effective theory.
As stationary solutions 
of the effective potential, there exist 
the BDM vortex as 
the ground state in the absence of strange quark masses, 
metastable ${\mathbb C}P^1$ vortices, 
and unstable pure color vortices. 
Metastable ${\mathbb C}P^1$ vortices decay 
into the BDM vortex through quantum tunneling. 
Another effect of the electromagnetic interactions is 
that the vortex effective theory becomes 
a $U(1)$ gauged ${\mathbb C}P^2$ model.

Neutron vortices and proton vortices exist in the $npe$ phase 
because of neutron superfluidity and 
proton superconductivity 
under rotation and magnetic field, respectively.  
The existence of colorful boojums has been 
predicted in the interface between the CFL phase 
and the $npe$ phase, between which 
there may be other phases such 
as the CFL+K , 2SC phases, and so on.
At a colorful boojum, 
three non-Abelian vortices with the total color flux canceled out 
join in the CFL phase, 
and three neutron vortices and three proton vortices
meet in the $npe$ phase.
There appear a Dirac monopole of the massless gauge field  
and a surface current of the massive gauge field.
Two kinds of colored monopoles appear 
at non-Abelian vortices when strange quark mass 
is taken into account.

At extremely high density limit 
in which the strange quark mass can be neglected, 
there appears a quantum mechanically induced potential 
in the low-energy  ${\mathbb C}P^2$ vortex effective theory through non-perturbative effects.
Consequently, there appear 
quantum monopoles 
confined by non-Abelian vortices 
as kinks on the vortex,  
which are relevant to show 
a duality between 
the confining phase, 
where quarks are confined and monopoles are condensed, 
and the CFL phase, where  
monopoles are confined and quarks are condensed.   
Yang-Mills instantons are 
trapped inside a non-Abelian vortex 
and stably exist 
as lumps or sigma model instantons
in the $d=1+1$ dimensional ${\mathbb C}P^2$ model 
in the vortex world-sheet.

The interactions between a non-Abelian vortex 
and phonons and gluons have been obtained by
 a dual transformation in which 
the phonon field and the gluon field  are 
dualized into an Abelian two-form field 
and a non-Abelian massive 
two-form field, respectively. 
On the other hand,  
the interaction between the mesons and 
a non-Abelian vortex can be described 
in the chiral Lagrangian. 
The interaction between a non-Abelian vortex 
and photons can be described by a $U(1)$ gauged 
${\mathbb C}P^2$ model. 
One of interesting consequences of 
the electromagnetic interactions 
is that a lattice of non-Abelian vortices behaves 
as a polarizer. 

It has been shown in the BdG equations that 
there exist 
localized and normalizable 
triplet Majorana fermion zero modes 
and a localized but non-normalizable 
singlet Majorana fermion zero mode  
in the core of a non-Abelian vortex.
The low-energy effective theory 
of the localized gapless fermions propagating along the vortex string has been constructed 
and  the chemical potential dependence of the 
velocity of gapless modes has been obtained.
The index theorem for the fermion zero modes 
in the background of a non-Abelian vortex 
ensures the existence of such fermion zero modes.
As a result of the Majorana property, 
localized fermions do not carry a current along 
a vortex string.  
A characterization of color superconductors as topological superconductors has been also discussed. 
As a novel application of Majorana fermion zero modes 
trapped inside a non-Abelian vortex, 
the exchange statistics of non-Abelian vortices in 
$d=2+1$ dimensions has been studied 
and a new non-Abelian statistics has been found. 

In the CFL phase, 
the chiral symmetry is also spontaneously broken, 
and there appear various kinds of topological 
objects, such as 
axial domain walls, Abelian and non-Abelian axial strings,  
and Skyrmions.  
In the chiral limit with massless quarks, 
the instanton-induced potential shows that 
a non-Abelian axial string is attached by 
an axial domain wall. 
An Abelian axial string is attached by 
three axial domain walls 
and decays into three non-Abelian axial vortices, 
each of which is attached by an axial domain wall.
In the presence of quark masses, 
three axial domain walls attract each other to be 
combined as a composite wall. 
Consequently, three axial domain walls attached to 
an Abelian axial string are also combined. 
While integer and fractional axial domain walls 
are metastable classically, they decay 
by quantum tunnelings. 
The quantum anomaly induces an axial current along 
$U(1)_{\rm B}$ superfluid vortices, and  
an electric current along $U(1)_{\rm A}$ axial strings 
and magnetic field perpendicular to axial domain walls 
in the presence of a background magnetic field.

Finally, we have reviewed 
topological solitons in the phases other
than the CFL phase.
The CFL phase is the ground state  at high densities
where the masses of the three flavors are small compared to the baryon
chemical potential.
If one lowers the density gradually, the effect of finite strange quark
mass comes in. 
Then it is expected that kaons form a condensation in addition to the
CFL condensates (CFL+K phase).
If the density is further decreased, only the light flavors ($u$ and
$d$) contribute to the condensate, which is called the 2SC phase.
We have discussed the domain walls and color magnetic
fluxes in the 2SC phase, 
and the K-strings, vortons, domain
walls and drum vortons in the CFL+K phase.

Here we make comments on the other phases 
that we did not discussed the last section. 
The magnetic CFL (MCFL) phase was proposed as 
the CFL phase under a strong magnetic field 
\cite{Ferrer:2005vd,Ferrer:2011ig,Ferrer:2012wa}; 
it may be relevant in the core of neutron stars.
At moderate densities, 
charged gluons are condensed 
due to the 
chromomagnetic instability in the ground state, 
known as 
the Nielsen-Olesen instability \cite{Nielsen:1978rm}. 
Gluon vortices can be formed inducing a magnetic field of 
a rotated magnetic field inside a superconductor \cite{Ferrer:2006ie}. 
The gluon vortices 
are different from non-Abelian 
vortices discussed in this paper. 
When gluon vortices are generated in dense quark matter under a strong magnetic field, anti-screening occurs, boosting the magnetic field to values higher than the applied one 
\cite{Ferrer:2006ie}, 
a phenomenon similar to that in magnetized electroweak theory 
\cite{Ambjorn:1988fx,Ambjorn:1988tm}.
See also Refs.~\cite{Ferrer:2006vw,Ferrer:2006wc,
Ferrer:2007iw,Ferrer:2007uw}.

The different type of vortices for the gluonic phase in the dense two-flavor QCD was discussed in Ref.~\cite{Gorbar:2005pi}.

The gap function of metallic superconductors  
is proposed to have a spatial modulation 
under strong magnetic field 
\cite{Machida:1984zz}, i.e., 
in the Fulde-Ferrell-Larkin-Ovchinnikov 
(FFLO) states
\cite{Fulde:1964zz,larkin:1964zz}. 
Recently, exact self-consistent solutions 
of the FFLO states have been found 
\cite{Basar:2008im,Basar:2008ki,Basar:2009fg} 
(see also \cite{Yoshii:2011yt}).  
A similar modulation was discussed in 
color superconductors of dense QCD 
\cite{Nickel:2008ng}. 
The crystalline superconducting phase has been 
proposed in dense QCD matter; 
see Refs.~ \cite{Casalbuoni:2003wh,Anglani:2013gfu} 
for a review.
In this case, the modulation is not  only along 
one spatial direction but also along three dimensions, 
forming a crystal.  
If this is realized, for instance in dense stars, 
it should significantly affect on various dynamics.
For instance, vortices created by a rotation
should be trapped in nodes of the modulations. 
It is an interesting problem whether the coexistence of a vortex lattice and a crystalline structure is possible.

\newpage
Before closing this review paper, 
let us summarize future problems. 
We first make a list of problems 
and explain each of them subsequently. 
\begin{enumerate}

\item
Dynamics of vortices.

\begin{enumerate}[(a)]

\item \label{dis:anomaly}
What does the $U(1)_{\rm A}$ anomaly 
do for non-Abelian semi-superfluid vortices?

\item \label{dis:ring}
The stability of vortex rings. 

\item \label{dis:nonlinear}
Nonlinear/higher order effects 
on the low-energy effective theory. 

\item \label{dis:interaction}
The vortex-vortex interaction at short distances.

\item \label{dis:metastability}
The metastability of $U(1)_{\rm B}$ (and M$_2$) vortices.

\item \label{dis:reconnection}
The reconnection of non-Abelian semi-superfluid vortices. 

\end{enumerate}

\item 
Generation of vortices. 
\begin{enumerate}[(a)]
\item \label{dis:rotation}
Rotation of the CFL matter. 

\item \label{dis:Kibble-Zurek}
The Kibble-Zurek mechanism at phase transitions.

\end{enumerate}

\item
Vortex lattices, vortex phase diagram, and vortex states. 

\begin{enumerate}[(a)]
\item  \label{dis:lattice-transition}
Transitions from a colorful vortex lattice 
to a colorless $U(1)_{\rm B}$ vortex lattice. 

\item    \label{dis:disordered-lattice}
Disordered colorful vortex lattices.

\item  \label{dis:vortex-phase}
Vortex matter/vortex phase diagram.

\item  \label{dis:turbulence}
The quantum turbulence. The Kolmogorov's law.

\end{enumerate}

\item 
Dynamics of orientational modes.

\begin{enumerate}[(a)]
\item  \label{dis:duality}
Quantum monopoles, instantons, and 
the quark-hadron duality.

\end{enumerate}

\item 
Interaction between vortices and quasi-particles.

\begin{enumerate}[(a)]

\item  \label{dis:transportation}
Effects to transportation properties of various quasi-particles.

\end{enumerate}

\item 
Fermions.

\begin{enumerate}[(a)]
\item \label{dis:fermion-boson}
Coupling between 
bosonic 
and 
fermionic zero modes. 

\item  \label{dis:Callan-Rubakov}
Fermions scattering off vortices. 
The Callan-Rubakov effect. 

\item  \label{dis:singlet}
The use of non-normalizable singlet Majorana fermion.

\item  \label{dis:self-consistent}
Self-consistent solutions of vortices. 

\item \label{dis:crossover}
Vortex core structures in BEC/BCS crossover.

\item  \label{dis:statistics}
Do the non-Abelian statistics affect the state of matter?

\item \label{dis:quantum-comp}
Topological quantum computation. 

\end{enumerate}

\item
Chiral symmetry breaking.

\begin{enumerate}[(a)]

\item  \label{dis:inta-axialvort-pion}
The interaction between 
non-Abelian axial vortices and CFL pions.

\item  \label{dis:int-Skrm-vor-wall}
The interaction between Skyrmions and 
non-Abelian axial vortices and axial domain walls.

\item  \label{dis:int-semisuper-axial}
The interaction between 
non-Abelian axial vortices and non-Abelian semi-superfluid vortices. 

\end{enumerate}

\item 
How do we detect the CFL phase? 
Hadron colliders and neutron stars. 
\begin{enumerate}[(a)]

\item \label{dis:colliders}
Heavy-ion collisions.

\item \label{dis:pulsar}
Physics of neutron stars 
such as 
pulsar glitch phenomena,
strong magnetic fields (magnetars),
cooling of neutron stars, 
and gravitational waves from neutron stars.

\end{enumerate}

\end{enumerate}
 
\newpage

(\ref{dis:anomaly})
A production of magnetic field 
by axial domain walls was studied in Sec.~\ref{sec:anomaly}.
However, the anomaly effects on non-Abelian semi-superfluid vortices 
have not been studied. 
The $U(1)_{\rm A}$ axial current should exist along 
non-Abelian semi-superfluid vortices as for 
$U(1)_{\rm B}$ superfluid vortices. 
There should be a significant role of such 
a current in 
a colorful vortex lattice in rotating CFL matter 
in Sec.~\ref{sec:lattice}.

(\ref{dis:ring})
In the absence of dissipation 
at zero temperature, 
a superfluid vortex ring moving at constant velocity 
is stable because of the inertial force. 
A non-Abelian semi-superfluid vortex ring is also 
stable. 
In the presence of dissipation, 
the size of ring decreases in time because of 
the Magnus force. 
However,  
due to (\ref{dis:anomaly}),
the $U(1)_{\rm A}$ current 
is present along the vortex ring 
and would make it stable.
The stability of vortex rings should be 
important for states of superfluids 
such as quantum turbulence.
In the case of a superfluid vortex ring,
the dynamics can be described by 
the sine-Gordon model \cite{Lund:1976ze}.
This may be extended to the case of 
non-Abelian vortex rings.

(\ref{dis:nonlinear}) 
For translational zero (Kelvin) modes  $X$ and $Y$ 
of a vortex,  
a higher order term for  $X$ and $Y$
in the low-energy effective Lagrangian 
was studied in the context of fluid mechanics 
\cite{hasimoto1972soliton}, 
and a nonlinear Schr\"odinger equation was found.
Nonlinear soliton waves, called Hasimoto solitons, 
were studied extensively. 
While higher derivative correction terms 
should be in the form of the Nambu-Goto action 
in the case of relativistic field theories, 
it is not clear whether such action is relevant 
for non-relativistic cases. 
On the other hand, higher derivative corrections 
were found for the ${\mathbb C}P^2$ orientational zero modes in the context of supersymmetric theories 
\cite{Eto:2012qda}.
A more general form of the possible higher derivative correction 
was given in Ref.~\cite{Liu:2009rz} by using nonlinear realizations.
A similar term should be present for non-Abelian 
semi-superfluid vortices in the CFL phase.

(\ref{dis:interaction})
The interaction between two non-Abelian semi-superfluid
 vortices is essentially the same 
as that between Abelian superfluid vortices, 
at large distances 
much larger than the core size.   
The interaction is mediated by the massless $U(1)_{\rm B}$ 
Nambu-Goldstone boson (phonon), 
giving a long-range force. 
At short distances, two vortices begin to exchange massive modes such as gluons and massive scalar (Higgs) fields. 
In Sec.~\ref{sec:ori-vor-vor-int}, 
the contribution from gluons was found to 
depend on relative $\mathbb CP^2$ orientations, or color fluxes. 
It is attractive when the two vortices have different colors. 
There remains an open question of what is 
a contribution from the Higgs fields.
Full numerical calculations or an analytic estimation from asymptotic forms are needed, as was done for 
a related model \cite{Auzzi:2007wj}. 

(\ref{dis:metastability})
A $U(1)_{\rm B}$ vortex has been suggested 
to decay into three non-Abelian vortices because 
of energetic at large distances and repulsion between 
non-Abelian vortices. 
However, the interaction at short distances  
is not known thus far as explained in (\ref{dis:interaction}). 
In fact, $U(1)_{\rm B}$ vortices do not carry fluxes while 
non-Abelian vortices do. 
Therefore, the gauge field gives an attraction among 
non-Abelian vortices with different colors and 
a binding energy among them so that 
there is a possibility of metastability of 
$U(1)_{\rm B}$ vortices.  
Metastability was found for vortices in a three component BEC, 
which is the same as the diagonal configuration of dense QCD 
except for the absence of gauge fields \cite{Cipriani:2013wia}.

\begin{figure}[ht]
\begin{center}
\includegraphics[width=6cm]{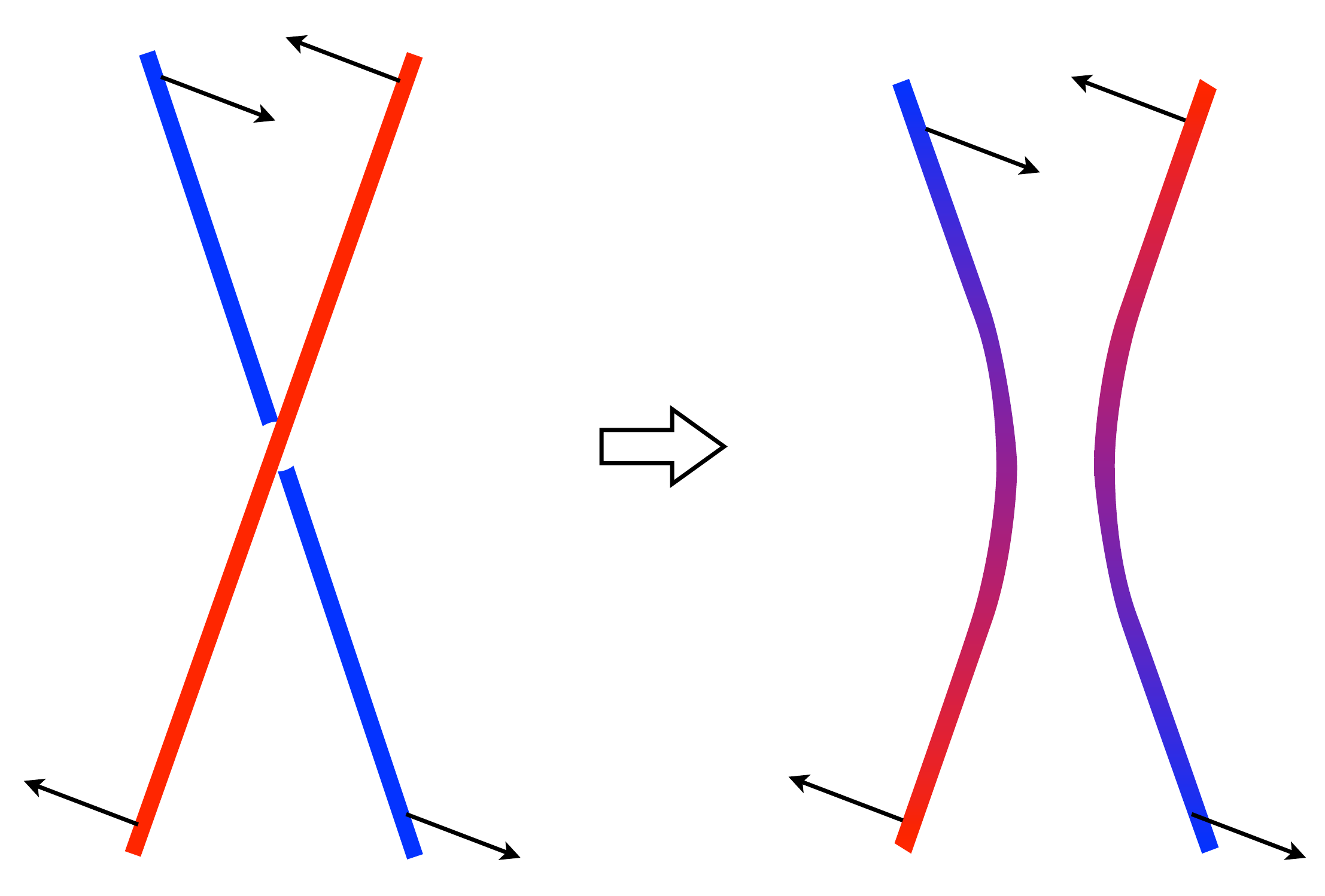}
\caption{Reconnection of two non-Abelian vortices.
}
\label{fig:reconnection}
\end{center}
\end{figure}

(\ref{dis:reconnection})
When two vortices collide in superfluids such as 
helium superfluids or atomic BECs, 
they reconnect with each other
\cite{PhysRevLett.71.1375,Nazarenko}, 
as was confirmed numerically by  
the Gross-Pitaevskii equation \cite{PhysRevA.62.011602}
and experimentally 
by direct observations in helium superfluids \cite{Bewley03092008}. 
The reconnection process is very important in 
quantum turbulence 
\cite{Vinen:2007,0953-8984-21-16-164207,Tsubota2013191,
doi:10.1146/annurev-conmatphys-062910-140533}
where vortices reduce their length 
through the reconnection.
In the case of CFL matter, 
it is quite nontrivial whether two non-Abelian semi-superfluid vortices
 reconnect when they collide 
because they have internal orientations in the $\mathbb CP^2$ space;
when two vortices have different $\mathbb CP^2$ modes, 
they may pass through without reconnection because 
the two different orientations could not be connected 
if they reconnected.  
This question was addressed in the context of cosmic strings; 
whether two cosmic strings produced at a phase transition 
in the early universe can reconnect or not 
is important for how many strings remain in our universe.
This problem was solved for relativistic non-Abelian strings 
\cite{Eto:2006db}; even when two non-Abelian vortices 
have two different orientations (color fluxes) 
in the internal space, 
they always reconnect at least when the colliding speed 
is not large, in which two different orientations 
are smoothly connected at the collision point,  
as in Fig.~\ref{fig:reconnection}.  
Although this result was shown for relativistic strings, 
we expect that the same holds for semi-superfluid vortices in 
 the non-relativistic case, because the $\mathbb CP^2$ effective theory is relativistic (of the type I NG modes).
After a reconnection, Kelvin modes are induced 
and propagate along the two strings  
as in the case of usual superfluid vortices.
In addition, the $\mathbb CP^2$ modes are also induced 
and propagate along the two strings.  
Strings would emit phonons and gluons 
through the interactions studied in 
Sec.~\ref{sec:int-phonon-gluon}.  

(\ref{dis:rotation})
As for creation vortices, one mechanism is rotation.
Since dense stars rotate rapidly, this is the more plausible scenario 
if the CFL matter is realized in their core. 
However, the minimum vortices in the CFL phase are non-Abelian 
semi-superfluid vortices carrying color magnetic fluxes. 
Since it is unlike the case that color fluxes are created from nothing, 
one plausible scenario is that $U(1)_{\rm B}$ superfluid vortices are created at first as usual and they decay into non-Abelian vortices 
with total fluxes canceled out (however, see (\ref{dis:metastability})). 
A simulation for the formation of non-Abelian vortices in rotation 
is one of important future problem.
As a similar simulation, creation of  superfluid vortices 
with formation of Abrikosov lattices 
and split into fractional vortices with formation of colorful 
vortex lattices was simulated  in a three-component BEC 
\cite{Cipriani:2013wia}. 

(\ref{dis:Kibble-Zurek})
The other scenario of a creation of vortices is the Kibble-Zurek mechanism at the phase transition \cite{Kibble:1976sj,Hindmarsh:1994re,Zurek:1985qw,Zurek:1996sj}.
In cosmology, this is an almost unique mechanism of a creation 
of cosmic strings and other defects \cite{Hindmarsh:1994re}. 
The Kibble-Zurek mechanism has been experimentally confirmed 
in various condensed matter systems. 
An estimation of the number of non-Abelia vortices produced 
at phase transitions should be made.

(\ref{dis:lattice-transition})
We have introduced colorful vortex lattices in Sec.~\ref{sec:lattice}.
As the rotation speed is larger, the distance between 
the vortices become shorter. 
At some point, there may be a transition to 
a $U(1)_{\rm B}$ vortex lattice because 
there should be attraction between vortices 
with different colors, as discussed in (\ref{dis:interaction}), 
and $U(1)_{\rm B}$ vortices may be metastable 
(\ref{dis:metastability}). 
 
(\ref{dis:disordered-lattice})
Another interesting topic is what happens 
when a colorful vortex lattice is not ordered 
in color as in Fig.~\ref{fig:lattice}(b). 
If the colors of a vortex lattice are ordered as in Fig.~\ref{fig:lattice}(a),  
it is easily transformed into a $U(1)_{\rm B}$ vortex lattice 
and connected to superfluid vortices in the confining phase, 
as supposed in Sec.~\ref{sec:boojum}. 
However, if the colors of a vortex lattice are disordered as in Fig.~\ref{fig:lattice}(b), there should appear an unpaired   
triplet when the lattice is transformed to 
 a $U(1)_{\rm B}$ vortex lattice  or 
connected to superfluid vortices in the confining phase.  
There may be characterization of disordered lattice by 
frustration or entropy.

(\ref{dis:vortex-phase})
Vortex matter, like 
vortex-solid or vortex-liquid, 
is expected to exist in high $T_{\rm c}$ superconductors
 \cite{RevModPhys.66.1125}.
Such kinds of vortex matter could also exist in the CFL phase and 
it is important to determine what kind of vortex state is realized 
at finite rotation speeds and temperatures.
In particular, at finite temperatures, 
vortices fluctuate and Kelvin modes are induced. 
When the temperature is further increased, 
a vortex lattice would eventually melt into a vortex liquid. 
At even higher temperatures, 
vortices start to reconnect with each other frequently
as in (\ref{dis:reconnection}).
In this state the CFL condensates vanish and 
the matter is no longer in the CFL phase.
This is a view of the quark-gluon-plasma phase from the vortex point of view.

(\ref{dis:turbulence}) 
The quantum turbulence is a state of superfluids 
governed by quantized vortices.
The energy flows from a larger scale to a smaller scale 
and the Kolmogorov's power law $E \sim k^{-5/3}$ holds. 
This energy transfer is carried by reconnection of 
vortices and emission of phonons from Kelvin waves 
\cite{Vinen:2007,0953-8984-21-16-164207,Tsubota2013191,
doi:10.1146/annurev-conmatphys-062910-140533}.
It is interesting whether such a law holds for 
CFL matter 
and whether there is a state which can be called as 
``non-Abelian'' quantum turbulence.

(\ref{dis:duality})
In Sec.~\ref{sec:monopoles}, we constructed 
quantum monopoles on an infinitely long 
non-Abelian vortex string. 
However, we do not want to have a 
monopole-anti-monopole meson 
with an infinite string. 
Therefore, we may consider 
monopole-anti-monopole mesons in a vortex ring, 
which may be stable 
as discussed in [\ref{dis:ring}]. 
Such monopole mesons with finite mass 
may be more relevant for quark-hadron duality. 
It is still unclear 
whether the quantum potential can be 
explained by an effect of Yang-Mills instantons 
trapped in the vortex world-sheet 
discussed in Sec.~\ref{sec:instanton-in-vortex}.

(\ref{dis:transportation})
The transportation of phonons was studied in Ref.~\cite{Mannarelli:2008jq}. 
Bulk viscosity was studied in \cite{Mannarelli:2009ia}. 
We may take into account the presence of 
 colorful vortices. 
Phonons are scattered by vortices 
through the vortex-phonon interaction 
studied in Sec.~\ref{sec:int-phonon-gluon}.  
In particular, a vortex reconnection 
(\ref{dis:reconnection})
induces Kelvin modes and enhances 
the dissipation 
through the phonon-vortex interaction. 
Interactions of vortices with other quasi-particles 
such as the CFL mesons and gluons studied 
in Sec.~\ref{sec:int} 
would also affect transportation of these quasi-particles.

(\ref{dis:fermion-boson})
Thus far, bosonic zero modes and 
fermionic zero modes have been studied independently 
in the GL theory and the BdG equations, respectively.
However, we have not discussed a relation between them 
or if they are coupled to each other. 
In this regard,
fermions cannot be dealt with in the GL theory 
because fermions are integrated out.
Therefore, we should use the BdG equation 
to study both fermions and boson in a unified way.
It is, however, not easy task to deal with bosons 
in the BdG equation not only technically but also practically.
As a compromise, we can use nonlinear realizations 
\cite{Coleman:1969sm,Callan:1969sn}
to construct coupling between 
the triplet Majorana fermion zero modes 
and the $\mathbb CP^2$ zero modes, 
since the CFL symmetry $SU(3)_{\rm C+F}$ 
is spontaneously broken into $SU(2)\times U(1)$ 
and fermions belong to the triplet of the unbroken symmetry.
The $\mathbb CP^2$ zero modes are frozen
in the presence of strange quark mass, and therefore
this coupling is not relevant for the study of fermions 
in the energy scale below strange quark mass.

(\ref{dis:Callan-Rubakov})
The Callan-Rubakov effect 
was first proposed for 
the scattering of fermions off 
magnetic monopoles in 
Refs.~\cite{Callan:1982ah,Callan:1982au,Callan:1982ac,Rubakov:1982fp},
where the $s$-wave component of fermion wave function is greatly enhanced, 
and it was later extended to the fermion scattering off 
cosmic strings and domain walls 
\cite{Brandenberger:1988rp}. 
The idea is that when fermions in the bulk interact with 
topological solitons only fermion zero modes can reach the core of solitons. The solitons give the boundary condition for scattering fermions and, for instance, enforces the condensates of fermions 
close to the core. 
This mechanism was applied for instance 
to proton decays in grand unified theories and 
baryogenesis by cosmic strings. 
The fermions zero modes found in this paper 
may give a similar effect. 

(\ref{dis:singlet})
Singlet fermion zero mode was found to be 
non-normalizable and cannot be counted 
by the index theorem as shown in Sec.~\ref{sec:BdG2}. 
Although non-normalizable modes 
having a singular peak at the vortex core 
are usually considered to be unphysical, 
there is also a discussion in the context of 
cosmic strings that 
they may play some interesting roles 
such as baryogenesis \cite{Alford:1989ie}.

(\ref{dis:self-consistent})
We have constructed only fermion zero modes 
that are expressed in terms of the gap function. 
To know the precise form, we need the profiles of 
the gap function. 
For this, one needs to construct the gap function 
and fermion solutions self-consistently. 
Self-consistent vortex solutions were 
constructed numerically for s-wave superconductors 
\cite{PhysRevB.41.822,PhysRevB.43.7609}. 
One also obtains all massive modes trapped inside the 
core of a vortex. 
Self-consistent vortex solutions in the CFL phase 
remain as a future problem. 

(\ref{dis:crossover})
Superfluidity of ultracold Fermi gasses 
was established in all the regions of BEC/BCS crossover  
by experimentally observing a vortex lattice 
\cite{Zwierlein:2005}.  
Theoretically, for instance, 
the presence (absence) of fermion modes in the vortex cores in 
the BCS (BEC) region \cite{PhysRevLett.94.140401}
is one of interesting phenomena in BEC/BEC crossover. 
Crossover of BCS (weak coupling) 
at high density and BEC (strong coupling) at low density 
in QCD was also studied  
in Refs.~\cite{Matsuzaki:1999ww,Abuki:2001be,Nishida:2005ds}.
As in ultracold atomic gasses, the vortex core structure 
of the BEC/BCS crossover in dense QCD 
is an interesting problem.

(\ref{dis:statistics})
Non-Abelian statics of non-Abelian vortices have been 
studied \cite{Yasui:2010yh,Hirono:2012ad},  
as reviewed in Sec.~\ref{sec:na-statistics}. 
This makes vortices non-Abelian anyons when 
restricted to $d=2+1$ dimensions. 
When promoted to $d=3+1$ dimensions, 
vortices are strings.
The statistics still remains when one considers adiabatic exchanges of parallel vortices. 
Another possibility is vortex rings. 
As discussed in (\ref{dis:ring}), non-Abelian vortex rings 
are stable in the absence of dissipations 
at zero temperature, or
they may be stable even at finite temperature 
once axial anomaly is taken into account.
In this case, one may consider exchange 
of two vortex rings where a smaller one passes through 
inside a larger one. 
It is an open question whether such a statistics affects dynamics of 
CFL matter. 

(\ref{dis:quantum-comp})
Apart from QCD matter, 
non-Abelian exchange statistics 
can be used for topological quantum computations 
\cite{RevModPhys.80.1083}, 
which were proposed as a quantum computation 
robust against perturbations. 
It is an open question whether triplet Majorana fermions 
can be useful for it.  

(\ref{dis:inta-axialvort-pion})
The interaction between a non-Abelian semi-superfluid vortex 
and gluons was achieved \cite{Hirono:2010gq} 
by dualizing gluons to massive 
two-form fields \cite{Seo:1979id} 
in Sec.~\ref{sec:int-phonon-gluon}. 
In this case, the order parameter space, 
$U(3)_{\rm C-F+B}$, around which semi-superfluid vortices wind,
 is gauged by the $SU(3)_{\rm C}$ color gauge symmetry, 
and so we have massive two-form fields.
On the other hand, 
the order parameter space $U(3)_{\rm L-R+A}$ is not gauged. 
Accordingly, the interaction between 
a non-Abelian axial vortex and the $U(3)$ fields, 
i.~e. the CFL pions and the $\eta'$ meson, 
should be described by dualizing the $U(3)$ fields to 
massless two-form fields, known as 
the Freedman-Townsend model \cite{Freedman:1980us} 
(see also Ref.~\cite{Furuta:2001kx}).

(\ref{dis:int-Skrm-vor-wall})
Skyrmions can interact with axial domain walls and axial vortices. 
Presumably the interaction is attractive. 
After a Skyrmion is absorbed into an axial domain wall, 
it may be stable as a lump ($2+1$-dimensional Skyrmion)  
or unstable depending on the models \cite{Nitta:2012wi,Nitta:2012rq}. 
On the other hand, if a Skyrmion is absorbed into 
an axial vortex, it may become a kink on a vortex.

(\ref{dis:int-semisuper-axial})
In the CFL phase, two kinds of non-Abelian vortices are present 
as discussed in this review, non-Abelian semi-superfluid vortices 
and non-Abelian axial vortices. 
While the order parameter spaces around which these vortices 
wind are different, it is an open question whether these two kinds of vortices interact non-trivially.

(\ref{dis:colliders})
In heavy-ion collision experiments, where two relativistically
accelerated nuclei are collided, it might be possible that some kinds of
superfluid phases are realized, like the neutron superfluid or the CFL
matter, especially at low-energy collisions. If a superfluid is produced
in heavy-ion collisions, quantum vortices would appear since the created
matter would have finite angular momentum at off-central collisions. It
is an interesting problem to find the experimental signatures of such
phenomena.

(\ref{dis:pulsar}) 
A more plausible candidate for  
dense QCD matter in nature may be the core of 
dense stars such as neutron stars (or quark stars) 
 \cite{shapiro2008black,glendenning2000compact}. 
It has been argued that  
the recent discovery of a massive neutron star 
\cite{Demorest:2010bx} 
tends to deny the existence of exotic matter 
such as hyperon or quark matter in their cores 
because of the equation of state for neutron stars, 
but there is also an argument that 
exotic matters are still possible 
(see e.g. Refs.~\cite{Weissenborn:2011ut,Masuda:2012kf,Taranto:2013gya,Sasaki:2013mha}
and Refs.~\cite{Blaschke:2007ri,Yasutake:2009kj} for recent and 
earlier works) 
and the situation is not conclusive. 
Although the evidence of the presence of such quark matter 
is elusive, we may give observation limits to the existence 
from rotating QCD matter in particular vortices. 
For reviews of dense matter in compact stars,   
see Refs.~\cite{Page:2006ud,Schmitt:2010pn,Yasutake:2012dw}. 
Here we make brief comments on this issue.
 
Pulsar glitch phenomena may be explained by 
unpinning of pinned vortices,  
as suggested by Anderson and Itoh \cite{Anderson:1975zze}.
While some unpinning mechanism were suggested,  
there has been no agreement thus far. 
Non-Abelian vortices in the CFL phase 
may play some role to explain this mechanism 
in particular colorful vortex lattices and boojums.

Strong magnetic fields of neutron stars 
are still a very important unsolved  problem 
in particular magnetars. 
As discussed in Sec.~\ref{sec:anomaly}, 
magnetic fields are produced 
from axial domain walls and axial vortices 
by the axial anomaly. 
This mechanism may explain strong
magnetic fields of neutron stars.


The emission of neutrino is the main process of the cooling of 
compact stars \cite{shapiro2008black,glendenning2000compact}.  
Neutrino emission in the CFL phase 
was studied in Ref.~\cite{Reddy:2002xc,Jaikumar:2002vg}.
It turns out 
that the decays of phonons into neutrinos are the dominant process of
cooling (phonons couple to $Z^0$ weak bosons).
The astrophysical implication of the 
cooling of compact stars in the presence of 
color superconducting phases was studied; 
for instance, see Ref.~\cite{Noda:2011ag}.
One should take into account the presence 
of a colorful vortex lattice, 
since the phonons interact vortices. 
Reheating in the presence of vortices 
was also studied in Ref.~\cite{Niebergal:2009yb}, 
without specifying the kinds of vortices.

Gravitational wave detectors are now one of 
the hottest topic in astrophysics. 
Gravitational waves in the presence of 
vortex lattices in the CFL phase 
(and the 2SC phase)
were studied, assuming vortex lattices 
of unstable color magnetic fluxes 
and compared with neutron star  observations 
in Ref.~\cite{Glampedakis:2012qp}.
The calculation should be revised 
by considering colorful vortex lattices.  
Apart from vortices, the r-mode instability of neutron stars 
and gravitational waves related to this mode was studied  
assuming the CFL phase \cite{Mannarelli:2008je,Andersson:2010sh}. 
This calculation should be modified in the presence of  a colorful vortex lattice.

Sole observations of gravitational waves will tell us only the stiffness of 
the equation of state of neutron stars, while these observations  
combined with observations of neutrinos 
gamma ray and X ray will tell us more detailed information of 
cooling process of neutron stars, 
which should contain information of the presence of absence of 
exotic matters and vortices in such matter. 

\vspace{0.5cm}
We believe that the problems listed above are important and should be
investigated in the future.
We hope that this list is useful for those who 
are interested in this field. 

\section*{Acknowledgments}\label{sec:ack}

We would like to thank our collaborators, 
Mattia Cipriani, 
Takanori Fujiwara, 
Takahiro Fukui,  
Noriko Shiiki, 
Kazunori Itakura, 
Takuya Kanazawa, 
Michikazu Kobayashi, 
Taeko Matsuura, 
Eiji Nakano, 
Walter Vinci, 
and 
Naoki Yamamoto 
for extensive discussions and collaborations. 
Many parts of this review result from these collaborations.
M.N. are especially grateful to Taeko Matsuura and Eiji Nakano, 
who invited him to this subject. 
Without them, this review paper would have been impossible.
We further thank Eiji Nakano, Motoi Tachibana and Naoki Yamamoto for careful reading of the manuscript and crucial comments, and Kei Iida and Nobutoshi Yasutake for useful comments.
 
This work is supported in part by 
Grant-in-Aid for Scientific Research (Grants
No. 23740198 (M.~E.), No. 23740198 and No. 25400268 (M.~N.)).
The work of 
Y.~H. is partially supported by the Japan Society for the Promotion of
Science for Young Scientists and partially by JSPS Strategic Young
Researcher Overseas Visits Program for Accelerating Brain Circulation (No.R2411). 
The work of M.~N. is also supported in part by 
the ``Topological Quantum Phenomena'' 
Grant-in-Aid for Scientific Research 
on Innovative Areas (Grants No. 23103515 and No. 25103720)  
from the Ministry of Education, Culture, Sports, Science and Technology (MEXT) of Japan. 
The work of S.~Y. is supported by a Grant-in-Aid for
Scientific Research on Priority Areas ``Elucidation of New
Hadrons with a Variety of Flavors (E01: 21105006).''

%% file: susy-v9.tex
\section{Non-Abelian vortices in supersymmetric QCD}\label{sec:susy}

In the context of supersymmetric (SUSY) QCD, 
non-Abelian vortices were discovered 
independently in Refs.~\cite{Hanany:2003hp,Auzzi:2003fs} 
as a generalization of Abrikosov-Nielsen-Olesen (ANO) 
vortices \cite{Abrikosov:1956sx,Nielsen:1973cs}
in the Abelian-Higgs model.
Since then,  
much progress has been made in recent years; 
see Refs.~\cite{Tong:2005un,Eto:2005sw,Eto:2006pg,Shifman:2007ce,
2009supersymmetric,Tong:2008qd,Konishi:2008vj,Konishi:2010zh} 
for a review. 
They are topologically the same objects as non-Abelian vortices in
dense QCD discussed in this review paper. 
It is interesting that such similar objects 
were found in different contexts, dense QCD and SUSY QCD, 
almost at the same time. 
Here, we summarize differences and similarities 
of non-Abelian vortices in dense QCD and SUSY QCD, 
and summarize properties of non-Abelian vortices in 
SUSY QCD, which may be useful 
in the study of those in dense QCD. 
One of the most important common feature is 
the existence of bosonic orientational gapless modes 
${\mathbb C}P^2$ and ${\mathbb C}P^{N-1}$ 
in the core of a non-Abelian vortex
in dense QCD and SUSY $U(N)$ QCD, respectively.

Three crucial differences are present 
between non-Abelian vortices in 
dense QCD and SUSY QCD, as summarized in the following:
\begin{enumerate}
\item
The overall $U(1)$ symmetry is global 
$U(1)_{\rm B}$ symmetry in dense QCD, 
while it is gauged in SUSY QCD. 
Consequently, 
\begin{enumerate}[(a)]
\item
vortices are global vortices in 
 in dense QCD, 
while they are local vortices 
in SUSY QCD, 

\item
the energy of vortices in dense QCD is logarithmically 
divergent with the infinite system size, while 
that of local vortices in SUSY QCD 
is finite,

\item
the interaction between two vortices at a distance $R$
is $1/R$ for dense QCD, 
while it is zero for two local BPS vortices in SUSY QCD, 
and $\pm e^{- c R}$ for two local vortices in non-SUSY QCD.
\end{enumerate}

\item
Dense QCD matter is non-relativistic
while SUSY QCD is relativistic. 
Consequently, 
\begin{enumerate}[(a)]
\item 
Two translational zero modes are not independent from 
each other, one is the momentum conjugate to the other, 
and there exists only one Kelvin mode (Kelvon) 
for vortices in dense QCD, as shown in Sec.~\ref{sec:dynamics}. 
On the other hand, two translational zero modes are independent
in vortices in SUSY QCD. 
The former are so-called type II Nambu-Goldstone modes
and the latter are so-called type I Nambu-Goldstone modes  
\cite{Watanabe:2012hr,Hidaka:2012ym,Nitta:2013mj}, 
as summarized 
in the footnote \ref{footnote:NG-modes}. 
\end{enumerate}

\item
The bosons are not independent degrees of freedom, 
and fermions couple to 
bosons (the order parameter $\Delta$)  
in the BdG equation in dense QCD, 
while bosons and fermions are degrees of freedom  independent from each other in SUSY QCD,
and they are related by SUSY transformations. 
Consequently, 
\begin{enumerate}[(a)]
\item
  {\it Majorana} fermions belonging to {\it triplet (the adjoint representation)} of the unbroken  $SU(2)$ symmetry in the vortex core are localized in a non-Abelian vortex in dense QCD, 
while {\it Dirac} fermions in the {\it fundamental}  representation of the unbroken $SU(N-1)$  symmetry in the vortex core are are localized in a non-Abelian vortex 
in SUSY QCD.

\item
Fermion zero modes and bosonic zero modes 
are not related to each other in dense QCD, 
while fermion zero modes are ``tangent bundle" to bosonic zero modes.  On the other hand, in SUSY QCD, the linearized equations of motion of the bosonic fields coincide with the equation of motion of the fermions, 
and therefore the index theorems for fermions and bosons 
coincide \cite{Hanany:2003hp}. 

\end{enumerate}

\end{enumerate}
Although these three are major differences, 
both vortices have many common features.

Unlike ANO vortices in the Abelian-Higgs model, 
non-Abelian vortices have 
non-Abelian internal orientations 
and associated conserved charges.
In the case of $U(N)$ gauge 
theory with $N$ flavors of Higgs fields in the 
fundamental representation, 
the internal orientation is the complex projective space 
${\mathbb C}P^{N-1}$,  
which corresponds to Nambu-Goldstone modes associated with 
the $SU(N)_{\rm C+F}$ color-flavor locked global symmetry 
spontaneously broken in the presence of vortices.
Because of non-Abelian internal orientations, 
we can expect that the dynamics of the non-Abelian vortices is much richer and more interesting compared to the ANO vortices. 

Since there exists no net force between 
multiple BPS vortices, 
a space of the whole solutions 
is characterized by multiple collective coordinates 
called moduli.  The number of the moduli parameters 
was determined to be $2kN$ for 
$k$ vortices in $U(N)$ gauge theory \cite{Hanany:2003hp}.
The moduli space of multiple vortices with full moduli parameters was completely determined without metric 
by partially solving BPS vortex equations 
\cite{Isozumi:2004vg,Eto:2005yh,Eto:2006pg,Eto:2006cx};  
The moduli space for $k$ separated vortices is a $k$-symmetric product 
\beq
 {\mathcal M}_k^{\rm sep} 
\simeq ({\mathbb C} \times {\mathbb C}P^{N-1})^k/{\mathcal S}_k 
  \;\;\subset {\mathcal M}_k
\eeq
of the single vortex moduli space \cite{Eto:2005yh} 
while the whole space ${\mathcal M}_k$ is regular.

First, let us consider vortex particles in 
$d=2+1$ dimensions or parallel vortices in $d=3+1$ dimensions. 
Although there is no force between BPS vortices at rest, 
vortices scatter nontrivially when they are moving. 
The low-energy dynamics of BPS solitons 
can be described as geodesics of 
a proper metric on the moduli space \cite{Manton:1981mp}. 
By using a general formula for the moduli space metric 
given in \cite{Eto:2006uw}, 
the explicit moduli space metric was obtained for 
the moduli subspace ${\mathcal M}_k^{\rm sep}$ of $k$ well-separated 
vortices which is valid when the 
separation of vortices are much larger than 
the length scale of the vortices, i.~e., 
the Compton wave length of massive vector bosons \cite{Fujimori:2010fk}.
This metric describes low-energy scattering of 
two slowly moving vortices \cite{Eto:2011pj}. 
The moduli space metric of the moduli subspace 
for two coincident vortices 
\cite{Eto:2006db,Eto:2010aj,Eto:2006dx}, 
which is supplement to ${\mathcal M}_{k=2}^{\rm sep}$ inside 
the whole space ${\mathcal M}_{k=2}$,  
was also found, which 
shows that 
two non-Abelian vortices scatter with 90 degree 
in head-on collision even though they have different internal orientations 
${\mathbb C}P^{N-1}$ as the initial conditions \cite{Eto:2006db}. 

Going back to $d=3+1$ dimensions, 
one of nontrivial and important dynamics is 
the reconnection 
of two vortex strings colliding with an angle.
It was shown in  \cite{Eto:2006db} that 
when two vortex strings collide with an angle, 
they always reconnect each other.

There exists a static force between 
non-BPS vortices. 
Static force between two non-BPS non-Abelian local 
vortices was shown to depend on internal orientations \cite{Auzzi:2007wj}. 
The dynamics of 
non-BPS vortices can be described 
by the moduli space approximation 
plus the potential term corresponding to the static force.
Therefore, the dynamics of non-Abelian vortices 
in dense QCD is also expected to be described by 
the moduli approximation and the potential term 
of the static interaction $1/r$ studied in Sec.~\ref{sec:intervortex-force}.  
In particular, 
the reconnection dynamics should hold for 
non-Abelian vortices in dense QCD,
because its possibility depends only on topology. 
Since the reconnection dynamics is an essential process 
for instance in quantum turbulence in superfluids 
and ultracold atomic gasses, 
it should be important if the CFL phase is 
realized in the core of dense stars.

Non-BPS non-Abelian vortices 
were studied for instance in Refs.~\cite{Gorsky:2004ad,Auzzi:2007wj,Gorsky:2007kv}.
It was also claimed that vortices in Ref.~\cite{Gorsky:2007kv}  describe those in dense QCD, but it is not the case 
because the $U(1)_{\rm B}$ symmetry is gauged there. 

%% file: toric-v9.tex
\section{Toric diagram}\label{sec:toric}

Here we would like to explain a geometrical aspect of the complex projective space $\mathbb C P^{N-1}$,
which is a complex manifold of the complex dimension $N-1$.
The $\mathbb CP^{N-1}$ manifold is known as a toric geometry, on which there is a $\U(1)^{N-1}$ action
allowing several fixed points. One can find a good explanation of the toric geometry in Ref.~\cite{Leung:1997tw}.

\begin{figure}[ht]
\begin{center}
\includegraphics[height=5cm]{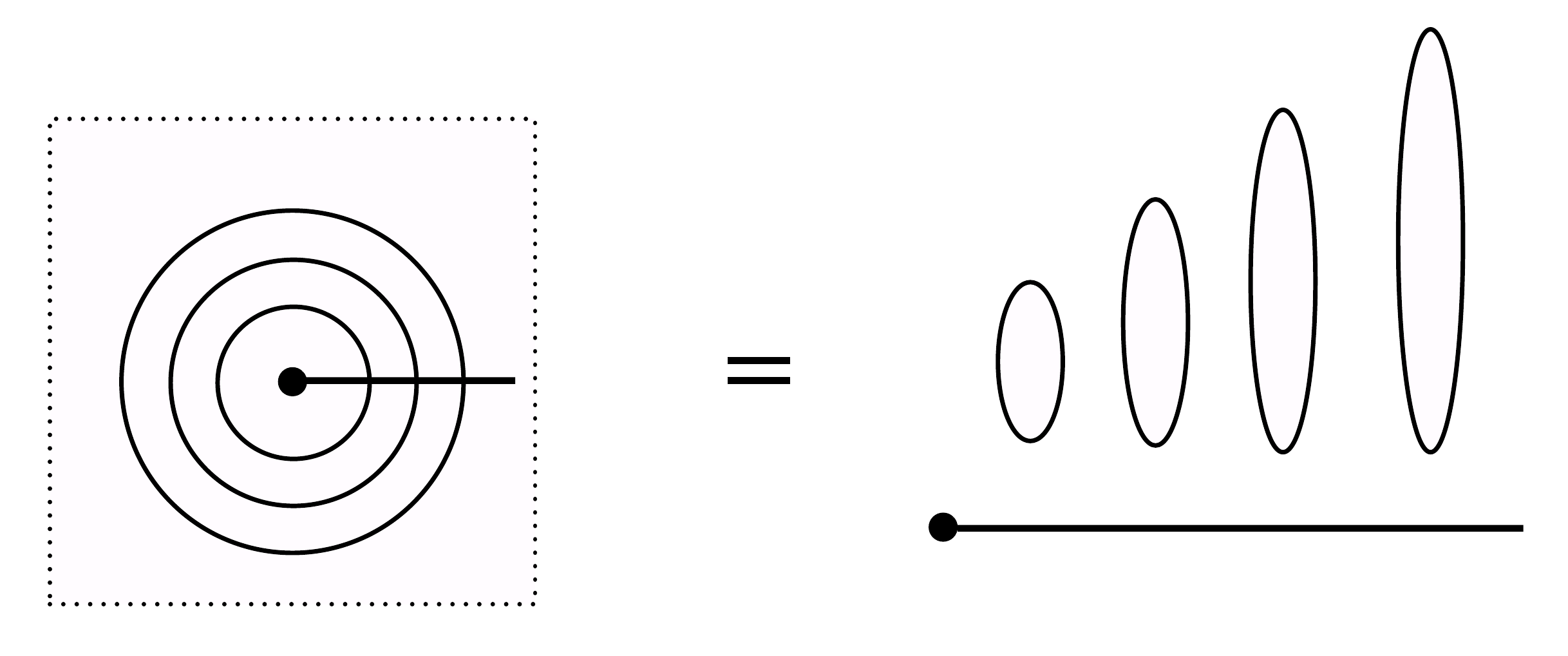}
\caption{A toric diagram for the complex plane $\mathbb C$.}
\label{fig:toric_C}
\end{center}
\end{figure}

The simplest example of the toric geometry is the complex plane $\mathbb C$. Let $z$ be a coordinate
on $\mathbb C$. There exists a $\U(1)$ action
\beq
z \to e^{i\theta} z.
\eeq 
A fixed point of the $\U(1)$ is the origin $z=0$. The complex plane can be seen as a half-line with a circle on top.
The circle shrinks at the fixed point; see Fig.~\ref{fig:toric_C}.

\begin{figure}[ht]
\begin{center}
\includegraphics[height=5cm]{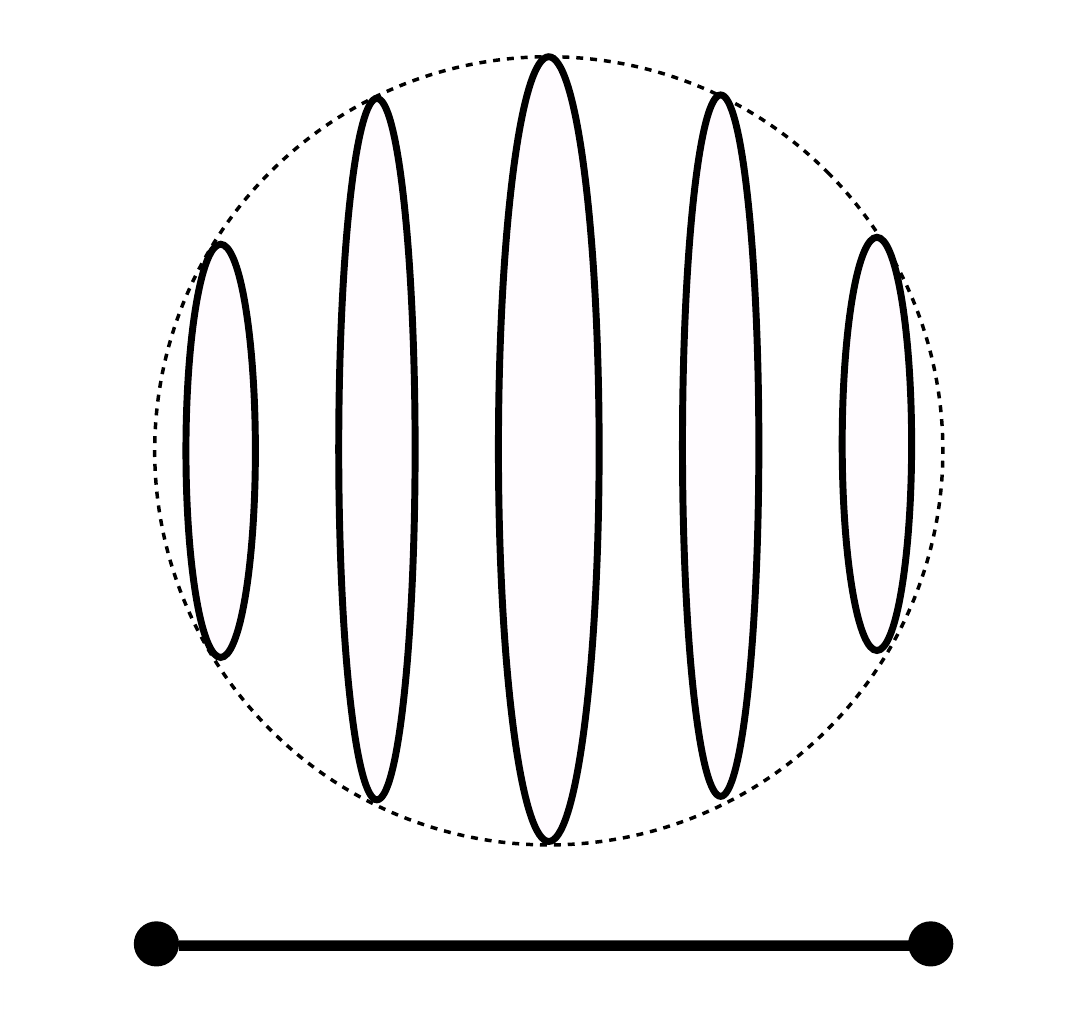}
\caption{A toric diagram for the complex plane $\mathbb CP^1$.}
\label{fig:toric_cp1}
\end{center}
\end{figure}

The next example is $\mathbb CP^1$.
As is well known $\mathbb CP^1$ is the space of two complex numbers $(\phi_1,\phi_2)$ which
are not all zero. The two complex numbers are themselves identified  up to multiplying by a non-zero
complex number
\beq
(\phi_1,\phi_2) \sim \lambda (\phi_1,\phi_2),\quad \lambda \in \mathbb C^*.
\eeq
One can fix the equivalence relation by setting
\beq
(\phi_1,\phi_2) \to (1,z)\qquad \text{for}\quad \phi_1\neq0,\\
(\phi_1,\phi_2) \to (z',1)\qquad \text{for}\quad \phi_2\neq0.
\eeq
$z$ and $z'$ are complex numbers, which are related by $zz'=1$, except for $z = 0$ and $z'=0$ cases.
There is again a $\U(1)$ action
\beq
(1,z) \to (1,e^{i\theta}z),\quad (z',1) \to (e^{-i\theta}z',1).
\eeq
The fixed points are $z=0$ and $z' = 0$.
Let us take a real basis which is invariant under the $\U(1)$ action
\beq
X = \frac{|z|^2}{1 + |z|^2} = \frac{1}{1 + |z'|^2}.
\eeq
This $X$ takes values in the segment $[0,1]$: $X=0$ and $X=1$ corresponds to 
$z=0$ and $z'=0$, respectively. Thus, $\mathbb CP^1$ can be viewed as the interval with
a circle on top, as shown in Fig.~\ref{fig:toric_cp1}.

\begin{figure}[ht]
\begin{center}
\includegraphics[height=7cm]{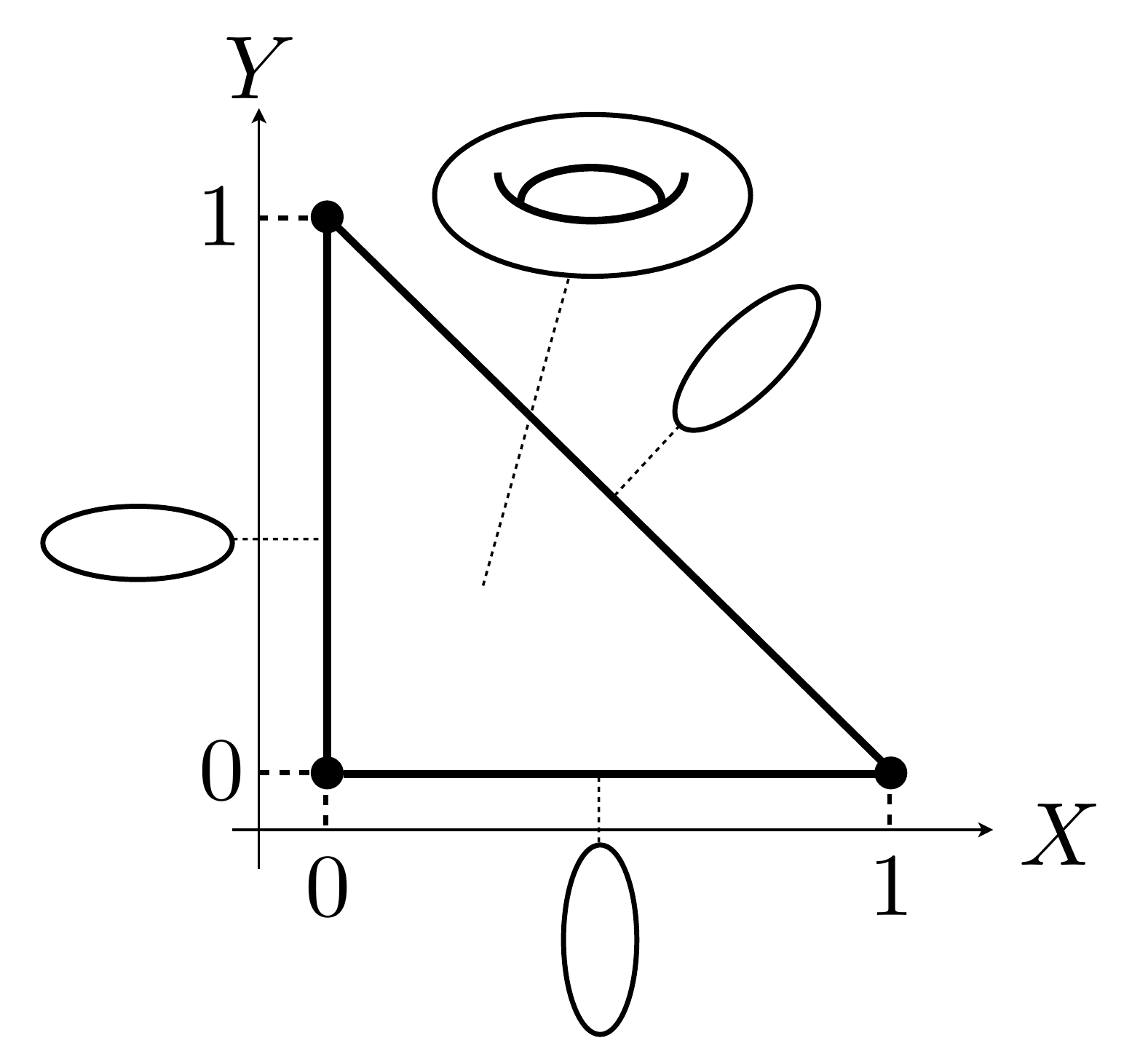}
\caption{A toric diagram for the complex plane $\mathbb CP^2$.}
\label{fig:toric_cp2}
\end{center}
\end{figure}

$\mathbb CP^2$ is a straightforward extension of $\mathbb CP^1$.
It is defined as the space of three complex numbers, $(\phi_1,\phi_2,\phi_3)$, not all zeros
with the following identification
\beq
(\phi_1,\phi_2,\phi_3) \sim \lambda (\phi_1,\phi_2,\phi_3),\quad \lambda \in \mathbb C^*.
\label{eq:def_cp2}
\eeq
One can fix this equivalence relation by setting the first component $\phi_1 = 1$ when $\phi_1 \neq 0$ as
\beq
(\phi_1,\phi_2,\phi_3) \to (1,z,w).
\eeq
There is $\U(1)^2$ action except for the trivial $\U(1)$. Without loss of generality, one can
take the basis of $\U(1)^2$ as
\beq
(1,z,w) \to (1,e^{i\theta} z, e^{i\phi}w).
\eeq
As in the case of $\mathbb CP^1$, let us take a real basis that is invariant under $\U(1)^2$.
It is of the form
\beq
X = \frac{|z|^2}{1+|z|^2+|w|^2},\quad
Y = \frac{|w|^2}{1+|z|^2+|w|^2}.
\eeq
Both $X$ and $Y$ take their value in the segment $[0,1]$ and $X+Y \le 1$ holds.
Therefore, $\mathbb CP^2$ is expressed as a rectangular equilateral triangle, as
shown in Fig.~\ref{fig:toric_cp2}.
At a generic point, we have a $T^2$ fiber $(\theta,\phi)$ since the action of $\U(1)^2$ is not free. 
The $T^2$ fiber shrinks to a $T^2$ at the edges of the triangle. For instance, $Y=0$ $(w=0)$ is the fixed point
of $\phi$. Therefore, the $T^1$ fiber of $\phi$ shrinks there. Note that the submanifold at $Y=0$ is $\mathbb CP^1$,
which can be easily understood by setting $\phi_3 = 0$ in Eq.~(\ref{eq:def_cp2}).
The other edges correspond to the other $\mathbb CP^1$s.
At the vertices of the triangle, the $\U(1)^2$ action becomes completely free. Therefore, the $T^2$ fiber shrinks.

%% file: eff_act_appendix-v9.tex
\section{Derivation of the low-energy effective theory of orientational zero modes}\label{sec:deriv_LET}

Here we explain how to derive the low-energy effective  field theory (\ref{eq:eff_lag}) 
of the unit winding non-Abelian vortex. 
The effective action consists of a linear combination
of $\Tr[F_{\alpha i}F^{\alpha i}]$ and $\Tr[\D_\alpha \Phi \D^\alpha \Phi^\dagger]$ ($\alpha = 0,3$).
Therefore, we only need to calculate these terms one by one in terms of the 
orientational moduli field $\phi(t,x^3)$, which appears via the gauge field $A_\alpha$
given in Eq.~(\ref{eq:ansatz_SY}). To this end, it will turn out that it is convenient to introduce
a $3\times 3$ matrix valued function $\mathcal F_\alpha(a,b)$ of two complex variables $(a,b)$ by
\beq
\mathcal F_\alpha (a,b) \equiv
a \phi \p_\alpha \phi^\dagger + b \p_\alpha \phi \phi^\dagger + (a-b)\phi^\dagger\phi\p_\alpha \phi \phi^\dagger.
\eeq
One finds that this function $\mathcal F_\alpha(a,b)$ satisfies the following relations:
\beq
\mathcal F_\alpha(a,b)^\dagger &=& \mathcal F(b^*,a^*),\\
\Tr \mathcal F_\alpha(a,b) &=& 0,\\
\alpha \mathcal F_\alpha(a,b) &=& \mathcal F_\alpha(\alpha a, \alpha b),\\
\mathcal F_\alpha(a,b) + \mathcal F_\alpha(a',b') &= &
\mathcal F_\alpha(a+a',b+b').
\eeq
Then the gauge field given in Eq.~(\ref{eq:ansatz_SY}) can be expressed as
\beq
A_\alpha = \frac{i\rho_\alpha}{g_{\rm s}}\mathcal F_\alpha(1,-1).
\eeq
One can easily check the following equation by making use of the constraint $\phi^\dagger \phi = 1$
\beq
\Tr[ \mathcal F_\alpha(a,b)^\dagger \mathcal F^\alpha(a,b)] = (|a|^2+|b|^2) \Lag_{\mathbb CP^2}^{(\alpha)},
\eeq
where no summation is taken over $\alpha$ and $ \Lag_{\mathbb CP^2}^{(\alpha)}$ is given in Eq.~(\ref{eq:lag_cp2}).
In terms of the function $\mathcal F_\alpha$, the covariant derivative and the field strength are expressed by
\beq
\D_\alpha \Phi &=& e^{\frac{i\theta}{3}} \Delta_{\rm CFL} 
\mathcal F_\alpha(f-g+\rho_\alpha g, f-g-\rho_\alpha f),\\
F_{\alpha i} &=& \frac{1}{g_{\rm s}}\epsilon_{ij}\frac{x_j}{r^2}(1-\rho_\alpha) \mathcal F_\alpha(1,1)
- \frac{i}{g_{\rm s}}\frac{x_j}{r}\rho'_\alpha\mathcal F_\alpha(1,-1).
\eeq
With these expression at hand, we can now calculate the effective Lagrangian.
Let us first calculate the $F^2$ term
\beq
\Tr[F_{i\alpha}F^{i\alpha}] &=& \frac{1}{g_{\rm s}^2}\frac{x_ix^i}{r^4}h^2g(1-\rho_\alpha)^2\mathcal G_\alpha(1,1|1,1)
- \frac{1}{g_{\rm s}^2}\frac{x_ix^i}{r^2}(\rho'_\alpha)^2\mathcal G_\alpha(1,-1|1,-1) \non
&=& - \frac{2}{g_{\rm s}^2}\left[\rho'_\alpha{}^2 + \frac{h^2(1-\rho_\alpha)^2}{r^2}\right] \Lag_{\mathbb CP^2}^{(\alpha)},
\label{eq:FF_ap}
\eeq
where we have defined
\beq
\mathcal G_\alpha(k,l|m,n) \equiv \Tr[\mathcal F_\alpha(k,l)^\dagger \mathcal F^\alpha(m,n)],
\eeq
with no summation for $\alpha$. Note that we have eliminated the term proportional to
$\mathcal G_\alpha(1,1|1,-1)$ in Eq.~(\ref{eq:FF_ap}) because $\mathcal G_\alpha(1,1|1,-1)$
is zero. Let us next calculate the $|\D_\alpha \Phi|^2$ term. To this end, let us rewrite $\D_\alpha \Phi$ as
\beq
\D_\alpha \Phi &=& - \mathcal F_\alpha\left(\frac{(f-g)\rho_\alpha}{2},\frac{(f-g)\rho_\alpha}{2}\right)\non
&&+ \mathcal F_\alpha\left(f-g + \rho_\alpha g + \frac{(f-g)\rho_\alpha}{2}, f-g -\rho_\alpha f + \frac{(f-g)\rho_\alpha}{2}\right)\non
&=&  - \mathcal F_\alpha\left(\frac{(f-g)\rho_\alpha}{2},\frac{(f-g)\rho_\alpha}{2}\right)
+ \mathcal F_\alpha\left(f-g+\frac{(f+g)\rho_\alpha}{2},f-g-\frac{(f+g)\rho_\alpha}{2}\right)\non
&=& - \frac{(f-g)\rho_\alpha}{2}\mathcal F_\alpha(1,1) + \mathcal F_\alpha(f-g,f-g) + \mathcal F_\alpha\left(\frac{(f+g)\rho_\alpha}{2},
-\frac{(f+g)\rho_\alpha}{2}\right)\non
&=& \left(1-\frac{\rho_\alpha}{2}\right)(f-g)\mathcal F_\alpha(1,1) + \frac{(f+g)\rho_\alpha}{2} \mathcal F_\alpha(1,-1).
\eeq
Then we have
\beq
\Tr[\D_\alpha \Phi^\dagger \D^\alpha \Phi] &=& 
 \left(1-\frac{\rho_\alpha}{2}\right)^2(f-g)^2 \mathcal G_\alpha(1,1|1,1) + \frac{(f+g)^2\rho_\alpha^2}{4}\mathcal G_\alpha(1,1|1,-1)\non
 &=& 2\Delta_{\rm CFL}^2\left[(1-\rho_\alpha)(f-g)^2+\frac{\rho_\alpha^2}{2}(f^2+g^2)\right] \Lag_{\mathbb CP^2}^{(\alpha)}.
 \label{eq:DPDP_ap}
\eeq
Plugging Eqs.~(\ref{eq:FF_ap}) and (\ref{eq:DPDP_ap}) into Eqs.~(\ref{eq:eff0}) and (\ref{eq:eff3}), 
it is straightforward to find the K\"ahler classes $C^{0,3}$ given in Eqs.~(\ref{eq:C0}) and (\ref{eq:C3}).

%% file: dual-v9.tex
\section{Derivation of the dual Lagrangians for phonons and gluons}\label{sec:appendix_dual}

In this appendix, we derive the dual Lagrangians for phonons and
gluons. 
After a dual transformation, massive gluons are described by massive
non-Abelian antisymmetric tensor fields \cite{Seo:1979id} and $\U(1)_\B$
phonons 
are described by massless antisymmetric tensor fields.
In the dual description, vortices appear as sources that
can absorb or emit these particles.

\subsection{Low-energy effective theory of the CFL phase}\label{sec:LEET}

We start with a time-dependent Ginzburg-Landau(GL) effective Lagrangian
for the CFL phase, which is given in Eq.~(\ref{eq:tdgl}), 
\begin{eqnarray}
  \Lcal(x) 
  &=&
  \frac{\varepsilon_3}{2} \left(\textbf{E}^a \right)^2 
  - \frac{1}{2\lambda_3}\left(\textbf{B}^a\right)^2
  + K_0 \Tr \left[\left(D_0 \Phi\right)^\+ D^0 \Phi \right]
  + K_3 \Tr \left[\left(D_i \Phi\right)^\+ D^i \Phi \right]
  \nonumber
  \\ 
  &&   -4 i \gamma \Tr \left[ \Phi^\+ D_0 \Phi\right]
   - V(\Phi), 
  \label{eq:cfl-lagrangian-0} 
\end{eqnarray}
where
$E^a_i = F^a_{0i}$, 
$B^a_i = \frac{1}{2}\epsilon_{ijk}F^a_{jk}$. 
The effect of $\U(1)_{\rm EM}$ electromagnetism is neglected here. 
The parameters $\varepsilon_3$ and $\lambda_3$ are the color dielectric constant and
the color magnetic permeability.
Lorentz symmetry does not have to be maintained in general since 
superconducting matter exists.
However, the kinetic term of gluons has a modified Lorentz symmetry in
which the speed of
light is replaced by $1/\sqrt{\varepsilon_3 \lambda_3}$.
It is always possible to restore the Lorentz invariance of the
kinetic term of the gauge fields by rescaling $x^0$, $A^a_0$, $K_0$, 
$\gamma$ and $K_3$.
Therefore, we can start with the Lagrangian in which $\epsilon$ and $\lambda$
are taken to be unity.
For notational convenience, we introduce a vector $K_\mu \equiv
(K_0, K_3,K_3,K_3)$.
Thus our starting point is the following GL Lagrangian: 
\begin{eqnarray}
  \Lcal(x) &=& -\frac{1}{4} \left(F^a_{\mu\nu} \right)^2 
  + K_\mu \Tr \left[\left(D_\mu \Phi\right)^\+ D^\mu \Phi \right]
  -4 i \gamma \Tr \left[ \Phi^\+ D_0 \Phi\right]
  - V(\Phi)
  \label{eq:cfl-lagrangian}.
\end{eqnarray}

\subsection{The dual transformation}\label{sec:dual}

Here we perform dual transformations within the path integral formalism
to derive a dual Lagrangian for the CFL phase.
After the transformation, massive gluons are described by massive
non-Abelian antisymmetric tensor fields \cite{Seo:1979id} and $\U(1)_\B$
phonons 
are described by massless antisymmetric tensor fields.
We show that in the dual description vortices appear as sources that
can absorb or emit these particles.

\subsubsection{The dual transformation of massive gluons}

The partition function of the CFL phase can be written as
\begin{equation}
  Z = \int \Dcal A^a_\mu(x) \Dcal \Phi(x) \exp \left\{ i\int
  d^4x \Lcal(x)  \right\},
  \label{partition-function}
\end{equation}
with the Lagrangian defined in Eq.~(\ref{eq:cfl-lagrangian}).
We shall impose the gauge fixing condition on the field $\Phi$ rather than
on the gauge fields since they are integrated out in the end.
The gauge fixing condition is taken care of 
when we consider a concrete vortex solution.

We introduce non-Abelian antisymmetric tensor fields $B^a_{\mu\nu}$ by 
a Hubbard-Stratonovich transformation, 
\begin{equation}
  \exp \left[ i\int \diff^4 x \left\{ -\frac{1}{4}(F^a_{\mu\nu})^2
				  \right\} \right]
  \propto \int \Dcal B^a_{\mu\nu}\exp \left[ i\int \diff^4x
  \left\{ -\frac{1}{4} \left[ m^2 (B^a_{\mu\nu})^2 - 2m \tilde{B}^a_{\mu\nu}
		      F^{a,\mu\nu}) \right]  \right\}
  \right],
  \label{dual-transformation}
\end{equation}
where 
$\tilde{B}^a_{\mu\nu} \equiv \frac{1}{2} \epsilon_{\mu\nu\rho\sigma} B^{a, \rho \sigma}$. 
The parameter $m$ introduced above is a free parameter at this stage.
We will choose $m$ later so that the kinetic term of $B^a_{\mu\nu}$ is
canonically normalized. 

Substituting (\ref{dual-transformation}) into (\ref{partition-function}), 
we can now perform the integration over the gauge fields $A^a_\mu$.
The degrees of freedom of gluons are expressed by $B^a_{\mu\nu}$
after this transformation.
Each term in the Lagrangian is transformed as follows:
\begin{eqnarray}
  && K_\mu \Tr\{ (D_\mu \Phi)^\+(D^\mu \Phi)\} 
  -4 i \gamma \Tr \left[\Phi^\+ D_0 \Phi\right] 
  \nonumber\\
  &&= K_\mu \Tr \left\{ \Phi^\+ (\overleftarrow{\p}_\mu + ig_{\rm s} A^a_\mu T^a )
  (\overrightarrow{\p}^\mu - ig_{\rm s} A^{b,\mu} T^b) \Phi \right\} 
    -4 i \gamma \Tr \left[\Phi^\+ (\p_0 -ig_{\rm s} A^a_\mu T^a) \Phi\right] 
  \nonumber\\
  &&=K_\mu \Tr \{(\p_\mu\Phi)^\+(\p^\mu\Phi)\}  
    -4 i \gamma \Tr \left[\Phi^\+ \p_0 \Phi\right]  
  + g_{\rm s} A^a_\mu J^{a,\mu} \nonumber\\
  && \quad + g_{\rm s}^2 g_{\mu\nu} \sqrt{K_\mu K_\nu} A^{a,\mu} A^{b,\nu}  
  \Tr \left[ \Phi^ \+ T^aT^b \Phi \right],
\end{eqnarray}
with $ 
J^a_\mu \equiv -iK_\mu \Tr \left[ \Phi^\+ (\overleftarrow{\p}_\mu -
		   \overrightarrow{\p}_\mu)T^a \Phi  \right]
+  4 \gamma \Tr \left[\Phi^\+  T^a \Phi\right]
$, and
\begin{equation}
 \begin{split}
   -\frac{1}{2}m \tilde{B}^a_{\mu\nu} F^{a,\mu\nu} &= -\frac{1}{2}m
   \tilde{B}^a_{\mu\nu} ( 2\p_\nu A^a_\mu+g_{\rm s} f^{abc}A^b_\mu A^c_\nu )\\
   &= m A^a_\mu \p_\nu \tilde{B}^a_{\mu\nu} + \frac{1}{2}m g_{\rm s} f^{abc}
   A^a_\mu A^b_\nu \tilde{B}^c_{\mu\nu} .
 \end{split}
\end{equation}
Performing the integration over $A^a_\mu$, the following part
of the partition function is rewritten as
\begin{equation}
 \begin{split}
  & \int \Dcal A^a_\mu \exp \left\{ i \int \diff^4 x \left[ \frac{1}{2}g_{\rm s}^2 A^{a,\mu}
  K^{ab}_{\mu\nu} A^{b,\nu} - m\left( \p^\nu \tilde{B}^a_{\mu\nu}
  - \frac{g_{\rm s}}{m}J^a_\mu \right) A^{a,\mu} \right] \right\} 
  \\
  & \propto (\det K^{ab}_{\mu\nu})^{-1/2} \exp \left\{
  i\int \diff^4 x \left[ -\frac{1}{2} \left(\frac{m}{g_{\rm s}} \right)^2
  \left(\p_\rho \tilde{B}^{a,\mu\rho} - \frac{g_{\rm s}}{m}J^{a,\mu} \right)
  \left(K^{-1} \right)^{ab}_{\mu\nu}
  \left(\p_\sigma \tilde{B}^{b,\nu\sigma} - \frac{g_{\rm s}}{m}J^{b,\nu} \right)
  \right] \right\},
 \end{split}
\end{equation}
where $K^{ab}_{\mu\nu}$ is defined by
\begin{equation}
  \begin{split}
    K^{ab}_{\mu\nu} &= \frac{1}{2} g_{\mu\nu} \sqrt{K_\mu K_\nu} \Tr \left[\Phi^\+ T^aT^b \Phi \right] -
    \frac{m}{g_{\rm s}}f^{abc} \tilde{B}^c_{\mu\nu} 
    \\
    & \equiv \bm{\Phi}^{ab}_{\mu\nu} - \frac{m}{g_{\rm s}} \hat{B}^{ab}_{\mu\nu},
  \end{split}
\end{equation}
with
$
\bm{\Phi}^{ab}_{\mu\nu} \equiv \frac{1}{2} g_{\mu\nu} \sqrt{K_\mu K_\nu} \Tr
\left[\Phi^\+ T^aT^b \Phi \right]
$ and 
$
\hat{B}^{ab}_{\mu\nu} \equiv f^{abc} \tilde{B}^c_{\mu\nu}
$. We define the inverse of $K^{ab}_{\mu\nu}$ by the power-series expansion in $1/g_{\rm s}$
\begin{equation}
 \begin{split}
  K^{-1} 
  = \left(\bm{\Phi} - \frac{m}{g_{\rm s}}\hat{B} \right)^{-1} 
  = \bm{\Phi}^{-1} \sum_{n=0}^\infty \left( \frac{m}{g_{\rm s}} \hat{B}
  \bm{\Phi}^{-1}\right)^n.
 \end{split}
\end{equation}
As a result, we obtain the following partition function:
\begin{equation}
  Z \propto
  \int \Dcal B^a_{\mu\nu}(\det K^{ab}_{\mu\nu})^{-1/2} \exp \left\{
  i\int \diff^4 x \Lcal^\ast_{\rm G}(x) \right\}
  ,
\end{equation}
where $\Lcal^\ast_{\rm G}$ denotes the gluonic part of the dual Lagrangian 
\begin{equation}
  \Lcal^\ast_{\rm G} =  -\frac{1}{2} \left(\frac{m}{g_{\rm s}} \right)^2
  \left(\p_\rho \tilde{B}^{a,\mu\rho} - \frac{g_{\rm s}}{m}J^{a,\mu} \right)
  \left(K^{-1} \right)^{ab}_{\mu\nu}
  \left(\p_\sigma \tilde{B}^{b,\nu\sigma} - \frac{g_{\rm s}}{m}J^{b,\nu} \right)
  -\frac{1}{4}m^2(B^a_{\mu\nu})^2
  .
  \label{eq:dual-lagrangian}
\end{equation}

Now we define the non-Abelian vorticity tensor $\omega^a_{\mu\nu}$ 
as the coefficient of the term linearly
proportional to $B^a_{\mu\nu}$. Collecting the relevant terms in the above
Lagrangian, the coupling between massive gluons and 
the vorticity is given by
\begin{eqnarray}
  \Lcal^{\ast}_{\rm G} 
  &\supset&
  \frac{1}{2} \frac{m}{g_{\rm s}} \left[ 
  \p_\rho \tilde{B}^{a,\mu\rho} (\bm{\Phi}^{-1})^{ab}_{\mu\nu}J^{b,\nu} +
  J^{a,\mu} (\bm{\Phi}^{-1})^{ab}_{\mu \nu} \p_\rho \tilde{B}^{b,\nu\rho}
  \right]
  -\frac{1}{2}\left(\frac{m}{g_{\rm s}}\right) J^{a,\mu}[\bm{\Phi}^{-1}\hat{B}\bm{\Phi}^{-1}]^{ab}_{\mu\nu}J^{b,\nu} 
  \nonumber \\
  & \equiv& -\frac{1}{2} \left( \frac{m}{g_{\rm s}} \right)B^a_{\lambda \sigma}
  \omega^{a,\lambda \sigma},
\end{eqnarray}
where we have defined the vorticity tensor $\omega^a_{\mu\nu}$ as 
\begin{equation}
  \omega^{a,\lambda \sigma}
  \equiv \epsilon^{\lambda\sigma\mu\nu}\left[ \p_\nu \left\{
  (\bm{\Phi}^{-1})(^{ab}_{\mu\rho})
  J^{b,\rho} \right\} +
  J^{e,\alpha}(\bm{\Phi}^{-1})^{ec}_{\alpha\mu}f^{cda}(\bm{\Phi}^{-1})^{db}_{\nu\beta}J^{b,\beta}
				     \right].
  \label{eq:vorticity}
\end{equation}
Here $A{(^{ab}_{\mu\nu})}$ is a symmetrized summation defined by
$A{(^{ab}_{\mu\nu})}\equiv A^{ab}_{\mu\nu}+A^{ba}_{\nu\mu}$.
This expression for the non-Abelian vorticity is valid for general
vortex configurations.
The information of vortex configuration is included in $\bm{\Phi}$ and $J^a_\mu$.

\subsubsection{The dual transformation of $\U(1)_\B$ phonons}

In the following, we perform a dual transformation of the NG boson 
associated with the breaking of $\U(1)_\B$ symmetry.
This mode corresponds to the fluctuation of the overall phase of $\Phi$ 
which can be parametrized as $\Phi(x) = \e^{i\pi(x)}\psi(x)$, 
where $\pi(x)$ is a real scalar field.
Substituting this into the following part in the Lagrangian
(\ref{eq:cfl-lagrangian}) leads to \footnote{
The term $\Tr \left[\p_\mu \psi^\+ \psi -  \psi^\+ \p_\mu \psi \right]$
automatically vanishes since $\psi$ can be decomposed as $\psi = (\Delta+\rho)
\bm{1}_N + (\chi^a + i \zeta^a  ) T^a
$
and the modes $\zeta^a$ are absorbed by gluons.
}
\begin{equation}
 \begin{split}
  K_\mu  \Tr\{ (\p_\mu \Phi)^\+ (\p^\mu \Phi) \} 
  -&4 i \gamma \Tr \{ \Phi^\+ \p_0 \Phi \} 
   = \\
& K_\mu \left(\p_\mu \pi \right)^2 M^2
  - \p^\mu \pi J^0_\mu + K_\mu \Tr (\p_\mu \psi)^2
  -4 i \gamma \Tr \{ \psi^\+ \p_0 \psi \}
  ,
 \end{split}
\end{equation}
with 
$
J^0_\mu \equiv -4 \delta_{\mu 0} \gamma M^2
$
and $M^2 \equiv \Tr\left[\psi^\+ \psi\right]$.
We will transform the $\U(1)_\B$ phonon field $\pi(x)$ into a massless two-form
field $B^0_{\mu\nu}$. 
Note that the field $\pi(x)$ has a multivalued part in general; 
since $\pi(x)$ is the phase degree of freedom, 
$\pi(x)$ can be multivalued without violating the
single-valuedness of $\Phi(x)$. 
In fact the multivalued part of $\pi(x)$
corresponds to a vortex. 
Let us denote the multivalued part of $\pi(x)$ as $\pi_{\rm MV}(x)$. 

The dual transformation of this $\U(1)_{\B}$ phonon field is essentially 
the same as the case of a superfluid.
We basically follow the argument of Ref.~\cite{Lee:1993ty}.
Let us introduce an auxiliary field $C_\mu$ 
by linearizing the kinetic term of $\pi(x)$ 
in the partition function as follows
\begin{equation}
 \begin{split}
  Z & \propto 
  \int \Dcal \pi \Dcal \pi_{\rm MV} \exp i
  \left[
  \int \diff^4x \left(
   M^2 K_\mu \left\{\p_\mu (\pi + \pi_{\rm MV}) \right\}^2
  - \p^\mu (\pi + \pi_{\rm MV}) J^0_\mu
  \right)
  \right]  
  \\
  & \propto 
  \int \Dcal \pi \Dcal \pi_{\rm MV} \Dcal C_\mu \exp i
  \left[
  \int \diff^4x \left(
    -\frac{C^2_\mu}{M^2} - 2 C_\mu \sqrt{K_\mu} \p^\mu (\pi+\pi_{\rm MV})
    - \p^\mu(\pi + \pi_{\rm MV}) J^0_\mu
  \right)
  \right] .
 \end{split}
\end{equation}
Integration over $\pi(x)$ gives a delta function
\begin{equation}
  \int \Dcal \pi 
  \exp i
  \left[
  \int \diff^4x \left(
   -2  C_\mu \sqrt{K_\mu} \p^\mu \pi
   +\pi \p^\mu J^0_\mu
  \right)
  \right] 
  =  
  \delta
  \left\{
  \p^\mu \left(
   2  C_\mu \sqrt{K_\mu}
   +  J^0_\mu
  \right)
  \right\}.
\end{equation}
Then let us introduce the dual antisymmetric tensor field $B_{\mu\nu}^0$ by
\begin{equation}
  \int \Dcal C_\mu   
  \delta
    \left\{
     \p^\mu \left(
      2  C_\mu\sqrt{K_\mu} +  J^0_\mu
    \right)
  \right\} \cdots 
  = 
  \int \Dcal C_\mu \Dcal B_{\mu\nu}^0
  \delta
    \left(
      2  C_\mu \sqrt{K_\mu} +  J^0_\mu - m^0 \p^\nu \tilde{B}^0_{\mu\nu}
  \right) \cdots
\end{equation}
where the dots denote the rest of the integrand and 
$m^0$ is a parameter.
By this change of variables we have introduced an infinite gauge volume, 
corresponding to the transformation $\delta B^0_{\mu\nu}  = \p_\mu
\Lambda_\nu - \p_\nu \Lambda_\mu$ with a massless vector field $\Lambda_\mu$.
This can be taken care of by fixing the gauge later.
There is no nontrivial Jacobian factor as the change of
variables is linear.
Integrating over $C_\mu$, and transforming a resultant term in the Lagrangian as
\begin{equation}
  \begin{split} 
   m^0 \p^\nu \tilde{B}^0_{\mu\nu} \p^\mu \pi_{\rm MV} &= 
   -m^0 {B}^{0,\rho\sigma} \epsilon_{\mu\nu\rho\sigma} \p^\nu \p^\mu \pi_{\rm MV} 
   \\
   & \equiv 
   - 2\pi m^0 B^{0,\rho\sigma}\omega^0_{\rho\sigma} ,
  \end{split}
\end{equation}
where the first equality holds up to a total derivative and 
we have defined
\beq
  \omega^0_{\rho\sigma} \equiv \frac{1}{2\pi}\epsilon_{\mu\nu\rho\sigma} \p^\nu \p^\mu
  \pi_{\rm MV}\,.
\eeq
We thus obtain the dual Lagrangian for the $\U(1)_\B$ phonon part
\begin{equation}
  \begin{split}
   \Lcal^\ast_{\rm Ph}&= 
    -\left(\frac{1}{2M}\right)^2 K_\mu
     ( m^0 \p_\nu \tilde{B}^0_{\mu\nu} - J^0_\mu )^2
   -2\pi m^0 B^{0,\mu\nu} \omega^0_{\mu\nu} 
   \label{eq:Ph}.
  \end{split}
\end{equation}
Note that the term linear in $B^0_{\mu\nu}$ coming from the first term of
(\ref{eq:Ph}) is a total derivative and does not contribute to the
equation of motion. 
The partition function is proportional to 
\begin{equation}
  Z \propto \int \Dcal \pi_{\rm MV} \Dcal B^0_{\mu\nu} \exp i\left[\int \diff^4
  x \Lcal^\ast_{\rm Ph}\right].
\end{equation}
The $\U(1)_\B$ phonons are now described by a massless two-form field 
$B^0_{\mu\nu}$ and vortices appear as sources for $B^0_{\mu\nu}$.

\subsection{The dual Lagrangian}

We have shown that the partition function $Z$ of the CFL phase 
is proportional to $Z^\ast$ with the dual Lagrangian $\Lcal^\ast$:
\begin{equation}
  Z \propto Z^\ast = 
  \int \Dcal B^a_{\mu\nu} \Dcal \pi_{\rm MV} \Dcal B^0_{\mu\nu} \Dcal \psi~
  (\det K^{ab}_{\mu\nu})^{-1/2} \exp \left\{
  i\int \diff^4 x \Lcal^\ast (x) \right\},
\end{equation}
where
\begin{equation}
  \Lcal^\ast = \Lcal^\ast_{\rm G} + \Lcal^\ast_{\rm Ph} + K_\mu \Tr
  (\p_\mu \psi)^2 
  -4 i \gamma \Tr \{\psi^\+ \p_0 \psi\} 
  - V(\psi) .
\end{equation}
Here $\Lcal^\ast_{\rm G}$ and $\Lcal^\ast_{\rm Ph}$ are given 
in (\ref{eq:dual-lagrangian}) and (\ref{eq:Ph}), respectively.
The result above is valid for general vortex configurations.
We can discuss the interaction between vortices and quasiparticles
in terms of the dual Lagrangian.
Vortices are expected to appear as a source term for gluons and $\U(1)_\B$ phonons.

%% file: appendix-fermion-v9.tex
\section{Derivation of fermion zero modes}\label{sec:appendix-fermion}

We consider the BdG equation for the single component fermion.
The Hamiltonian is given by the particle-hole basis as
\begin{eqnarray}
{\mathcal H} =
\left( 
\begin{array}{cc}
 H_0-\mu & |\Delta(r)|e^{i\theta} \gamma_0 \gamma_5 \\
 -|\Delta(r)|e^{-i\theta} \gamma_0 \gamma_5 & H_0+\mu
\end{array}
\right),
\label{eq:BdGeq_1}
\end{eqnarray}
where we define
\begin{eqnarray}
H_0 =
\left( 
\begin{array}{cc}
 -\mu & -i\vec{\sigma} \cdot \vec{\nabla} \\
 -i\vec{\sigma} \cdot \vec{\nabla} \gamma_5 & -\mu
\end{array}
\right).
\end{eqnarray}
The BdG equation is then given by
\begin{eqnarray}
{\mathcal H} \Psi_{n} = E \Psi_{n},
\end{eqnarray}
with the wave function
\begin{eqnarray}
\Psi_{n} =
\left(
\begin{array}{c}
 \psi_{n+1} \\
 \eta_{n}
\end{array}
\right),
\end{eqnarray}
with the particle component ($\psi_{n+1}$) and the hole component ($\eta_{n}$).
Then, the BdG equation is given explicitly as
\begin{eqnarray}
\left( 
\begin{array}{cc}
 H_0-\mu & |\Delta(r)|e^{i\theta} \gamma_0 \gamma_5 \\
 -|\Delta(r)|e^{-i\theta} \gamma_0 \gamma_5 & H_0+\mu
\end{array}
\right)
\left(
\begin{array}{c}
 \psi_{n+1} \\
 \eta_{n}
\end{array}
\right)
=
E
\left(
\begin{array}{c}
 \psi_{n+1} \\
 \eta_{n}
\end{array}
\right).
\label{eq:BdGeq_2}
\end{eqnarray}
The particle and hole components are written as
\begin{eqnarray}
\psi_{n+1} =
\left(
\begin{array}{c}
 f(r) \Phi_{n} \\
 i\,g(r) \Phi_{n+1} \\
 \pm f(r) \Phi_{n} \\
 \pm i\,g(r) \Phi_{n+1}
\end{array}
\right) e^{-ik_{z}z}, \hspace{0.5em}
\eta_{n} =
\left(
\begin{array}{c}
 \bar{f}(r) \Phi_{n} \\
 i\,\bar{g}(r) \Phi_{n+1} \\
 \mp \bar{f}(r) \Phi_{n} \\
 \mp i\,\bar{g}(r) \Phi_{n+1}
\end{array}
\right)e^{ik_{z}z},
\end{eqnarray}
where we introduce four functions $f(r)$, $g(r)$, $\bar{f}(r)$ and $\bar{g}(r)$, and define $\Phi_{n}=e^{in\theta}$.
Here $\pm$ ($\mp$) refers to the eigenvalue of $\gamma_{5}$.
From Eq.~(\ref{eq:BdGeq_2}), we obtain
\begin{eqnarray}
(-\mu \pm k_{z})f \pm \left( \frac{\partial}{\partial r} + \frac{n+2}{r} \right) g \mp |\Delta| \bar{f} &=& E f, \\
\mp \left( \frac{\partial}{\partial r} - \frac{n+1}{r} \right) f + (-\mu \mp k_{z}) g \mp |\Delta| \bar{g} &=& E g, \\
(\mu \mp k_{z}) \bar{f} \mp \left( \frac{\partial}{\partial r} + \frac{n+1}{r} \right) \bar{g} \mp |\Delta| f &=& E \bar{f}, \\
\pm \left( \frac{\partial}{\partial r} - \frac{n}{r} \right) \bar{f} + (\mu \pm k_{z}) \bar{g} \mp |\Delta| g &=& E \bar{g}.
\end{eqnarray}

Now, we consider the solution of zero-mode fermion with $E=0$.
Here, we impose the ``Majorana condition",
\begin{eqnarray}
\Psi = U \Psi^{\ast},
\end{eqnarray}
with the unitary matrix
\begin{eqnarray}
U=
\left(
\begin{array}{cc}
 0 & \gamma_{2} \\
 \gamma_{2} & 0
\end{array}
\right),
\end{eqnarray}
which satisfies
\begin{eqnarray}
U^{-1} {\mathcal H}U = -{\mathcal H}^{\ast}.
\end{eqnarray}
We find then that the Majorana condition leads to the solution with $E=0$.
From the BdG equation, ${\mathcal H}\Psi = E \Psi$, we obtain ${\mathcal H}^{\ast} \Psi^{\ast} = E \Psi^{\ast}$ which in turn gives ${\mathcal H}\Psi = - E\Psi$ from $\Psi = U \Psi^{\ast}$ and $U^{-1} {\mathcal H}U = -{\mathcal H}^{\ast}$.
Therefore, we obtain $E=0$.
The Majorana condition restricts on the wave functions, $\psi_{n+1}$ and $\eta_{n}$, as
\begin{eqnarray}
n=-1,
\end{eqnarray}
and
\begin{eqnarray}
\bar{f} = \mp g, \\
\bar{g} = f.
\end{eqnarray}
Thus, we find the BdG equation simplified to
\begin{eqnarray}
-\mu f \pm \left( \frac{\partial}{\partial r} + \frac{1}{r} \right) g + |\Delta| g &=& 0, \\
\mp \frac{\partial}{\partial r} f  - \mu g - |\Delta| f &=& 0, \\
\mp \mu g - \frac{\partial}{\partial r} f \mp |\Delta| f &=& 0, \\
- \left( \frac{\partial}{\partial r} + \frac{1}{r} \right) g \mp \mu f \mp |\Delta| g &=& 0,
\end{eqnarray}
with $k_{z}=0$, which turns out to be the two independent equations,
\begin{eqnarray}
\mp \frac{\partial}{\partial r} f - |\Delta| f - \mu g &=& 0, \\
\pm \left(  \frac{\partial}{\partial r} + \frac{1}{r} \right) g + |\Delta|g - \mu f &=& 0.
\end{eqnarray}
When we eliminate $g(r)$, we obtain the equation only for $f(r)$,
\begin{eqnarray}
\frac{\partial^2}{\partial r^2} f + \frac{1}{r} \frac{\partial}{\partial r} \pm 2 |\Delta| \frac{\partial}{\partial r} f \pm \frac{|\Delta|}{r} f + (\mu^2 + |\Delta|^2)f = 0.
\end{eqnarray}
The solution of this equation is given as
\begin{eqnarray}
f(r) = C e^{-\int_{0}^{r}|\Delta(r')|{\mathrm d}r'} J_{0}(\mu r),
\end{eqnarray}
with the condition that it is regular at $r=0$ and infinitely large $r$.
As a consequence, we find the other components as
\begin{eqnarray}
g(r) &=& \pm C e^{-\int_{0}^{r}|\Delta(r')|{\mathrm d}r'} J_{1}(\mu r), \\
\bar{f}(r) &=& \mp C e^{-\int_{0}^{r}|\Delta(r')|{\mathrm d}r'} J_{1}(\mu r), \\
\bar{g}(r) &=& C e^{-\int_{0}^{r}|\Delta(r')|{\mathrm d}r'} J_{0}(\mu r).
\end{eqnarray}
This is the wave function of the Majorana fermion.